\RequirePackage{amsmath}

\documentclass[10pt]{iopart}

\usepackage{iopams}  
\usepackage[utf8]{inputenc}
\usepackage[OT4]{fontenc}
\usepackage[colorlinks=true,linkcolor=blue,urlcolor=red,citecolor=red]{hyperref} 
\usepackage{graphicx}
\usepackage{mathtools}
\usepackage{cite}

\makeatletter
\newcommand{\vast}{\bBigg@{3}} 
\newcommand{\Vast}{\bBigg@{4}}
\makeatother 
\begin{document}

\article{}{Quantum dark solitons in  ultracold one-dimensional  Bose and Fermi gases}

\author{Andrzej Syrwid}
\address{
Department of Physics, The Royal Institute of Technology, Stockholm SE-10691, Sweden
}
\address{
Instytut Fizyki Teoretycznej, 
Uniwersytet Jagiello\'nski, ulica Profesora Stanis\l{}awa \L{}ojasiewicza 11, PL-30-348 Krak\'ow, Poland
}
\vspace{10pt}
\begin{indented}
\item[]July 2020
\end{indented}

\begin{abstract}
Solitons are ubiquitous phenomena that appear, among others, in the description of tsunami waves, fiber-optic communication and ultracold atomic gases. The latter systems turned out to be an excellent playground for investigations of matter-wave solitons in a quantum world. This Tutorial provides a general overview of the ultracold contact interacting Bose and Fermi systems in a one-dimensional space that can be described by the renowned Lieb-Liniger and Yang-Gaudin models. Both the quantum many-body systems are exactly solvable by means of the Bethe ansatz technique, granting us a possibility for investigations of quantum nature of solitonic excitations. We discuss in details a specific class of quantum many-body excited eigenstates called yrast states and show that they are strictly related to quantum dark solitons in the both considered Bose and Fermi systems. 
\end{abstract}

\vspace{2cm}
\fbox{
\begin{minipage}{45em}
\large
This is the version of the article before peer review or editing, as submitted by an author to Journal of Physics B: Atomic, Molecular and Optical Physics. IOP Publishing Ltd is not responsible for any errors or omissions in this version of the manuscript or any version derived from it. The Version of Record is available online at \url{https://doi.org/10.1088/1361-6455/abd37f}
\end{minipage}
}

%
%
%
%
\ioptwocol
\eqnobysec
\section{Introduction}

Everything that surrounds us consists of an enormous number of particles. The properties of quantum many-body systems they constitute are determined, among others, by the temperature, density, type of atoms and character of their mutual interactions. Due to extreme complexity of such systems it is often very difficult to extract even basic information about their features. A rapid development of experimental techniques in cooling and trapping atomic gases has led to a realization of Bose-Einstein condensation in a gas of bosons \cite{CornellBEC, BEC2,pethick}. In such an ultracold Bose system a macroscopic fraction of particles occupies the lowest energy level. Consequently, the gas behaves collectively and the description can often be reduced to the problem of a single atom living in an effective averaged potential generated by the milieu of other atoms \cite{pethick}. Despite this simplification, investigations of properties of such systems, and in particular their quantum nature, are still very challenging -- both from the experimental and the theoretical point of view.

A great progress in the ultracold laboratory methods granted us an opportunity to study systems in lower dimensions. Indeed, the shape of the atomic trap can be manipulated so that one or two directions are tightly confined. At very low temperatures, when the lowest excitation in the confined direction requires much more energy than the thermal energy $k_B T$, the corresponding degrees of freedom are frozen and the particles live effectively in a lower dimensional space \cite{pethick}. This experimental achievement entailed a sudden growth of interest in theoretical models of ultracold atomic gases in one-dimensional space introduced a few decades earlier. Here, we focus on the Lieb-Liniger and Yang-Gaudin models describing ultracold one-dimensional (1D) Bose and Fermi gases, respectively. In both cases it is assumed that the interparticle interactions can be modeled by a zero-range interaction potential. It turns out that such 1D many-body systems are exactly solvable by means of the so-called Bethe ansatz, which opens a great opportunity for theoretical studies of quantum many-body phenomena.

In 1834 John Scott Russell noticed a stable solitary wave packet propagating with a constant velocity along the Union Canal he inspected \cite{Russell,Allen}. Such an extraordinary phenomenon became a subject of theoretical investigations, which led to the formulation of the Korteweg and de Vries nonlinear equation describing the object observed by Russell \cite{KdV,Allen}. Later research shown that the phenomenon, called a soliton, is very ubiquitous and appears in many different physical systems \cite{Allen,Kivshar2003Book}. Today, the term soliton refers to specific solutions of nonlinear equations, for which a balance between the dispersive and nonlinear effects allows for existence of stable solitary wave structure. Ultracold Bose gases turned out to be excellent systems for the studies of matter-wave solitons. While the experimental techniques for realization of solitons in Bose-Einstein condensate (BEC) are well developed, the investigations of their quantum nature constitutes a great challenge. In addition, the relation between the so-called dark soliton state, being a result of some collective excitation, and quantum many-body eigenstates of the system is very puzzling. Here, we show that there is a specific class of quantum many-body eigenstates in the Lieb-Liniger model that are unequivocally connected with dark solitons. Thanks to the full quantum many-body description employed by us, we also explore a quantum nature of these extraordinary objects \cite{Syrwid15, Syrwid16, Syrwid2017HW}.

In analogy to the Bose case, there was an attempt to realize dark solitons in a superfluid Fermi system. Nevertheless, the method that works well for bosons, fails when it is applied to a spin-$\frac{1}{2}$ Fermi gas in a superfluid state. In 1D such a system is described by the exactly solvable Yang-Gaudin model. Employing the methods developed during the investigations of ultracold Bose system, we demonstrate that dark soliton signatures are encoded in a specific class of the system eigenstates and their emergence can be observed by analyzing the wave function describing a $\downarrow$-$\uparrow$ pair of fermions \cite{SyrwidFermi2018}.


\section{Lieb-Liniger model: One-dimensional gas of ultracold bosons}
\label{Lieb-Liniger_model}

Nonrelativistic gas of identical bosons in 1D can be in general described by the following Hamiltonian
\begin{eqnarray}
\begin{array}{ll}
\displaystyle{
\mathrm{\hat{H}}=}
&
\displaystyle{\int \mathrm{d}x \, \hat{\Psi}^{\dagger}(x)\left( - \frac{\hbar^2}{2m}\frac{\partial^2}{\partial x^2} \right) \hat{\Psi}(x)}
\\
&
\displaystyle{+ \frac{1}{2}\int \mathrm{d}x \mathrm{d}y  \, \hat{\Psi}^{\dagger}(x) \hat{\Psi}^{\dagger}(y) \mathcal{V}(x,y) \hat{\Psi}(y) \hat{\Psi}(x)
,}
\end{array}
\label{HamGeneralBosons1D}
\end{eqnarray}
where $m$ is the single particle mass, $\mathcal{V}(x,y)$ describes the time-independent boson-boson interaction and $\hat{\Psi}(x)$ denotes the canonical quantum Bose field, which
in the Heisenberg $(H)$ picture, satisfies the following canonical equal-time commutation relations
\begin{eqnarray}
&\left[ \hat{\Psi}_H(x,t), \hat{\Psi}_H^\dagger (y,t)  \right] =  \delta(x-y), 
\label{BoseFieldsCummutators01}
\\
& \left[ \hat{\Psi}_H(x,t), \hat{\Psi}_H (y,t)  \right]= \left[ \hat{\Psi}_H^\dagger(x,t), \hat{\Psi}_H^\dagger (y,t)  \right] =0.
\label{BoseFieldsCummutators}
\end{eqnarray}
State $\left| 0 \right> $ that we destroy by acting $\hat{\Psi}(x)$
\begin{eqnarray}
\displaystyle{
\begin{array}{llllllll} \hat{\Psi}(x)\left|0 \right>=0, &&& \left< 0 \right|\hat{\Psi}^{\dagger}(x)=0,  &&&  x\in \mathbb{R},\\  \left|0 \right>^{\dagger}=\left< 0 \right|, &&& \left< 0 |0 \right>=1, && \end{array}
}
\label{FockVacuum}
\end{eqnarray}
is often dubbed the \emph{Fock vacuum}.
 
When the Bose gas  is dilute and ultracold,
it is convenient to substitute $\mathcal{V}(x,y)$ potential (generally very complicated) with the simplified $V(x,y)$. If $\mathcal{V}(x,y)=\mathcal{V}(x-y)$ is short range and the interparticle interaction can be restricted to the case of $s$--wave scattering only, then the only requirement we need to impose on $V(x,y)$ is the reproduction of the value of the scattering length. Therefore, we can choose $V(x-y)=2c\, \delta(x-y)$, where $c$ plays the role of the coupling strength (see Refs.~\cite{pethick,CastinArxiv, SachaBook}).  By this simplification the Hamiltonian in Eq.~(\ref{HamGeneralBosons1D}) reduces to the renowned Lieb-Liniger Hamiltonian, for which bosons interact via contact $\delta$-potential \cite{LiebLiniger1,BogoliubovKorepinInverseScattering,Gaudin_BetheWF}
\begin{eqnarray}
\begin{array}{ll}
\displaystyle{
\mathrm{\hat{H}}_{L-L}}
&
\displaystyle{=\int \mathrm{d}x  \,\hat{\Psi}^{\dagger}(x)\left( - \frac{\hbar^2}{2m}\frac{\partial^2}{\partial x^2} \right) \hat{\Psi}(x)} 
\\
&
\displaystyle{+ \int \mathrm{d}x  \, c \, \hat{\Psi}^{\dagger}(x) \hat{\Psi}^{\dagger}(x) \hat{\Psi}(x) \hat{\Psi}(x).
}
\end{array}
\label{LiebLinigerHamiltonian}
\end{eqnarray}
The Hamiltonian $\mathrm{\hat{H}}_{L-L}$ commutes with the total momentum and the total number of particles operators
\begin{eqnarray}
\mathrm{\hat{P}}&=-\frac{i\hbar}{2}\int \!\!\mathrm{d}x \,\Big[\hat{\Psi}^{\dagger}(x)\partial_{x}\hat{\Psi}(x)-  \left(\partial_{x}\hat{\Psi}^{\dagger}(x)\right) \hat{\Psi}(x)  \Big]
 \nonumber
\\ \label{LLSymmetries}
&=i\hbar\int \!\!\mathrm{d}x\left(\partial_{x}\hat{\Psi}^{\dagger}(x)\right) \hat{\Psi}(x),
\\  \nonumber
\mathrm{\hat{N}}&=\int \!\!\mathrm{d}x\, \hat{\Psi}^{\dagger}(x)\hat{\Psi}(x), \,\,\,\,\,\, \left[\mathrm{\hat{H}}_{L-L},\mathrm{\hat{P}}\right] = \left[\mathrm{\hat{H}}_{L-L},\mathrm{\hat{N}}\right] =0,
\end{eqnarray}
so the number of particles and the total momentum are conserved quantities. It also means that $\mathrm{\hat{H}}_{L-L}$ can be partitioned into blocks corresponding to different total momenta.

The Lieb-Liniger model has an analytical solution, which can be found with the help of the so-called Bethe ansatz method \cite{Bethe1}. Therefore, in principle when no boundary conditions are imposed, one can find analytical formulae for all the eigenfunctions and the corresponding eigenvalues. Let us restrict to a sector with a fixed number of particles $N$. 
An arbitrary eigenstate of the considered system can be cast into the following form
\begin{eqnarray}
\begin{array}{ll}
\displaystyle{
\left| \psi_N(\{k\}_{N})\right >=\frac{1}{\sqrt{N!}}\int\mathrm{d}^N x\,   \Phi_N(\{x\}_N,\{k\}_N) }
\\
\qquad\qquad\qquad\qquad
\displaystyle{\times\hat{\Psi}^\dagger(x_1)\ldots \hat{\Psi}^\dagger(x_N)\left| 0 \right>, 
}
\end{array}
\label{LLEigenstate2ndQuantizedForm}
\end{eqnarray}
where $\{\xi\}_N=\{\xi_1,\xi_2,\ldots,\xi_N\}, \, \xi=x\lor k$. By the assumption, $\left| \psi_N(\{k\}_{N})\right >$ is simultaneously an eigenstate of  $\mathrm{\hat{H}}_{L-L}$, $\mathrm{\hat{P}}$ and $\mathrm{\hat{N}}$ with eigenvalues $E_N$, $P_N$ and $N$, respectively.
$\Phi_N(\{x\}_N,\{k\}_N)$ is a symmetric function with respect to all $x_{j=1,2,\ldots,N}$ and is an eigenstate of the following quantum mechanical energy and total momentum operators 
\begin{eqnarray}
\displaystyle{\hat{\mathcal{H}}_N=- \frac{\hbar^2}{2m}\sum_{j=1}^N \frac{\partial^2}{\partial x_j^2}  +2c \sum_{N\geq j>s\geq 1}\delta(x_j-x_s)},
\label{LLHamiltonian1stQuantized}
\\
\displaystyle{\hat{\mathcal{P}}_N=-i\hbar\sum_{j=1}^N \frac{\partial}{\partial x_j}  },
\label{LLMomentum1stQuantized}
\\
\mathcal{\hat{H}}_N\Phi_N(\{x\}_N,\{k\}_N)=E_N\Phi_N(\{x\}_N,\{k\}_N),\\
  \mathcal{\hat{P}}_N\Phi_N(\{x\}_N,\{k\}_N)=P_N\Phi_N(\{x\}_N,\{k\}_N).
\label{LL1stEigenEqs}
\end{eqnarray}
The mysterious $k_j$ numbers parameterize the state unequivocally and will be called \textbf{\emph{quasimomenta}}. For convenience, we introduce the following parameter of an inverse length dimension
\begin{align}
\bar{c}=\frac{2m}{\hbar^2} c.
\label{cbar}
\end{align}
 The dimensionless intensive parameter that measures the interactions strength reads \cite{LiebLiniger1,Lieb2,pethick}
\begin{align}
\displaystyle{
\gamma = \frac{\bar{c}}{\rho},
}
\label{GammaLL}
\end{align}
where $\rho =\frac{N}{L}$ denotes the average density of particles in the system of size $L$. The weak (strong) interaction limit refers to $|\gamma|\ll 1$ ($|\gamma| \gg 1$).

Eigenstate $\Phi_N(\{x\}_N,\{k\}_N)$ describes the system of $N$ identical bosons and is symmetric with respect to all permutations of the particles' positions. For convenience, at this point we are going to operate in the domain $\mathcal{T}: x_1<x_2<\ldots<x_{N-1}<x_N$,
in which particles will never touch and thus we can neglect the point-like interactions, i.e.
\begin{align}
\displaystyle{
\mathcal{\hat{H}}_N\stackrel{\mathcal{T}}{\longrightarrow}\mathcal{\hat{H}}_N^0  =-\frac{\hbar^2}{2m}\sum_{j=1}^N \frac{\partial^2}{\partial x_j^2}.
}  \vspace{-0.1cm}
\label{LLHamiltonianDomainT}
\end{align}
 Such an apparent simplification entails the necessity to introduce additional boundary conditions 
 \begin{align}
\displaystyle{
\left(\frac{\partial}{\partial x_{j+1}}- \frac{\partial}{\partial x_{j}}-\bar{c} \right)\Phi_N\Bigg|_{x_{j+1}-x_j\rightarrow 0}=0, 
}
\label{LLBoundConditions}
\end{align} 
which are induced by the $\delta$-potential at $x_s=x_j, \, s\neq j$.

An eigenstate of both $\hat{\mathcal{H}}_N^0$ and $\hat{\mathcal{P}}_N$ is simply given by $\Phi_N^0\propto \mathrm{det}\left[ \mathrm{e}^{i k_j x_s}\right]_{j,s=1,2,\ldots,N}$,
for which
\begin{align}
\displaystyle{
\begin{array}{lll}
\displaystyle{\hat{\mathcal{H}}_N^0\Phi_N^0=E_N^0 \Phi_N^0}, & & \hat{\mathcal{P}}_N\Phi_N^0=P_N^0\Phi_N^0,  \vspace{0.2cm} \\ 
\displaystyle{E_N^0 =\frac{\hbar^2}{2m}\sum_{j=1}^N k_j^2}, & & \displaystyle{P_N^0 =\hbar \sum_{j=1}^N k_j},
\end{array}
} 
\label{LLDet0SolutionEIGENEQSDomainT}
\end{align}
where  $k_j$ are the above-mentioned quasimomenta. 
It turns out that the wave function fulfilling conditions in Eq.~(\ref{LLBoundConditions}) can be cast into the following form ($\mathrm{sgn}(\pi)=\pm 1$ denotes the sign of a permutation $\pi$) 
\begin{eqnarray}
\Phi_N^\mathcal{T}&(\{x\}_N,\{k\}_N) 
\nonumber \\ \label{LLBetheWaveFunctionInTDomain}
&= \mathcal{C} \sum_{\pi\in \mathcal{S}_N} \Bigg[ \mathrm{sgn}(\pi)\,&\mathrm{exp} \left( i\sum_{n=1}^N k_{\pi(n)}x_n \right)
\\ \nonumber
&&\times\prod_{j>s}\left( k_{\pi(j)}-k_{\pi(s)}-i\bar{c}\right) \Bigg],
\end{eqnarray}
with  $\mathcal{C}=\left( N! \prod_{j>s}\left[ (k_j-k_s)^2+\bar{c}^2 \right] \right)^{-1/2}$ \cite{BogoliubovKorepinInverseScattering}.
The derivation of the boundary conditions in Eq.~(\ref{LLBoundConditions}) as well as the construction of the general solution, Eq.~(\ref{LLBetheWaveFunctionInTDomain}), can be found in Ref.~\cite{BogoliubovKorepinInverseScattering}.

The above solution, Eq.~(\ref{LLBetheWaveFunctionInTDomain}), is valid only in the domain $\mathcal{T}$ and can be extended to the whole $\mathbb{R}^N$ thanks to the method suggested by M. Gaudin \cite{BogoliubovKorepinInverseScattering,Gaudin_BetheWF}. That is, $i \frac{2mc}{\hbar^2}$ present in the product in Eq.~(\ref{LLBetheWaveFunctionInTDomain}) has to be replaced by $i \frac{2mc}{\hbar^2} \, \mathrm{sign}(x_j-x_s)$, where $\mathrm{sign}(x)=x/|x|=\pm 1, \text{for } x\neq 0$  and $\mathrm{sign}(0)=0$. Ultimately, the analytical solution of the Lieb-Liniger model reads
\begin{eqnarray}
\Phi_N(&\{x\}_N,\{k\}_N) \nonumber
\\ 
&=\mathcal{C}  \sum_{\pi\in \mathcal{S}_N}  \Bigg[ \mathrm{sgn}(\pi)\,\mathrm{exp}  \left(  i\sum_{n=1}^N k_{\pi(n)}x_n  \right) \label{LLBetheWaveFunction}
\\ \nonumber
&\qquad\quad\times \prod_{j>s} \left(  k_{\pi(j)}  - k_{\pi(s)}  - i\bar{c}\,  \mathrm{sign}(x_j - x_s)  \right)  \Bigg],
\end{eqnarray}
with corresponding eigenvalues
 \begin{align}
\displaystyle{
\begin{array}{lll}
\displaystyle{\mathcal{\hat{H}}_N\Phi_N=E_N \Phi_N}, & & \mathcal{\hat{P}}_N\Phi_N=P_N \Phi_N,
\vspace{0.2cm} \\ 
\displaystyle{E_N =\frac{\hbar^2}{2m}\sum_{j=1}^N k_j^2}, & & \displaystyle{P_N =\hbar \sum_{j=1}^N k_j}.
\end{array}
}
\label{LLBetheWaveFunctionEigs}
\end{align}
It is easy to show that $\Phi_N$ has two symmetries 
 \begin{equation}
 \displaystyle{
\Phi_N(\sigma\{x\}_N,\{k\}_N) = \Phi_N(\{x\}_N,\{k\}_N)
},
\label{LLBetheWaveFunctionSymX}
\end{equation}
\begin{equation}
 \displaystyle{
\Phi_N(\{x\}_N,\rho\{k\}_N) = \mathrm{sgn}(\rho) \Phi_N(\{x\}_N,\{k\}_N)
},
\label{LLBetheWaveFunctionSymK}
\end{equation}
where $\sigma,\rho \in \mathcal{S}_N$ are arbitrary permutations of the particles' positions and quasimomenta respectively. While the first one represents the Bose exchange symmetry, the second symmetry at first sight resembles the Pauli exclusion principle.
 However, such an observation is entirely unjustifiable in the considered case. Although the quasimomenta are closely related to the momenta of individual particles, one should distinguish one from the other.

\subsection{Periodic boundary conditions}
\label{periodic_boundary_conditions}

 Let us now assume that we deal with the Lieb-Liniger system confined in a ring of length $L$.
Consequently, the solution in Eq.~(\ref{LLBetheWaveFunction}) has to satisfy the following cyclicity conditions (for all $j=1,2,\ldots,N$)
\begin{eqnarray} 
\Phi_N(x_1,\ldots, x_j+L,\ldots,x_N,\{k\}_N)  \label{LLPeriodicBounds} 
\\ \nonumber
\qquad\qquad\qquad=\Phi_N(x_1,\ldots, x_{j},\ldots,x_N,\{k\}_N).  
\end{eqnarray} 
This leads to the well-known system of the so-called \textbf{\emph{Bethe equations}} (see~\ref{appendixLL_periodic_theorems}) \cite{LiebLiniger1,BogoliubovKorepinInverseScattering,Gaudin_BetheWF,Jiang15},  whose solutions represent the permitted values of quasimomenta in the presence of periodic boundary conditions
\begin{eqnarray}
\displaystyle{
  \mathrm{e}^{ik_j L} = -\prod_{s=1}^N\frac{k_{j}-k_{s}+i\bar{c}}{k_{j}-k_{s}-i\bar{c}},  \qquad j=1,2,\ldots, N.
} 
\label{LLBetheEqsPeriodic}
\end{eqnarray}
 Note that the structure of the Bethe Eqs.~(\ref{LLBetheEqsPeriodic}) indicates that if the set $\{k_j\}$ is a solution, then the set $\{k'_j\}$, where $ k'_j=k_j +\frac{2\pi}{L} r$ ($r=\pm1,\pm2,\ldots$),
represents the another solution but corresponding to the total momentum shifted by $\Delta P_N =\frac{2\pi}{L} N r$. 

Sometimes it is convenient to rewrite the Bethe Eqs.~(\ref{LLBetheEqsPeriodic})
 to the logarithmic form 
\begin{align}
 k_j L+\sum_{s=1}^N \theta (k_j-k_s)= 2\pi I_j^\text{p},
\label{LLBetheEqsPeriodicLog}
\end{align} 
where
 \begin{align}
\theta\left(\xi \right)=i\ln \left[ \frac{i\bar{c} +\xi }{i\bar{c} -\xi } \right] 
= i\ln \left[ \frac{ \xi+ i\bar{c} }{ \xi-i\bar{c} } \right] + \pi= -\theta\left(-\xi \right),
\label{ThetaFunction}
\end{align}
is the so-called \textbf{\emph{scattering phase}}. The numbers $I_j^\text{p}$, which are arbitrary integers (half-integers) when $N$ is odd (even), parameterize the solutions of the above Bethe equations. 
Nevertheless, we still do not know if the solutions exist. It is also not yet clear whether the parameterization given by $ I_j^\text{p}$ numbers is unique or not. While in general we do not have answers for these questions, there are 3 theorems concerning the solutions of the Bethe equations when $\bar{c}>0$ \cite{BogoliubovKorepinInverseScattering, Gaudin_BetheWF, YangYang1969}. 

\vspace{0.1cm}

\textbf{\emph{Theorem 1.}} All the solutions $k_j$ of the Bethe Eqs.~(\ref{LLBetheEqsPeriodic}) are real numbers, when $\bar{c}>0$.

\vspace{0.1cm}
\textbf{\emph{Theorem 2.}} For $\bar{c}>0$ the solutions of the Bethe Eqs.~(\ref{LLBetheEqsPeriodicLog}) exist and can be uniquely parameterized by a set of integer (half-integer) numbers $\{ I_j^\text{p}\}$.

\vspace{0.1cm}
\textbf{\emph{Theorem 3.}} For $\bar{c}>0$, the solutions $k_j$ and $k_s$ of the Bethe Eqs.~(\ref{LLBetheEqsPeriodicLog}) satisfy  the relation $k_j>k_s$ ($k_j=k_s$) if the corresponding parameterizing numbers $ I_j^\text{p}> I_s^\text{p}$ ($ I_j^\text{p}= I_s^\text{p}$).

\vspace{0.1cm}

 Due to the antisymmetry with respect to exchange of any pair of quasimomenta, Eq.~(\ref{LLBetheWaveFunctionSymK}), the wave function $\Phi_N=0$ when at least two quasimomenta $k_j$ and $k_s$  are equal for $j\neq s$. Therefore, in agreement with \emph{Theorems  2} \& \emph{3}, for $\bar{c}>0$ physically relevant solutions of the Bethe Eqs.~(\ref{LLBetheEqsPeriodicLog}) are uniquely parameterized by distinct numbers $ I_j^\text{p}$ only. This also implies that in the case of the repulsively interacting  system $E_N$ is always positive.
It should also be mentioned that for $\bar{c}>0$ (when $k_j\in \mathbb{R}$) 
\begin{align}
\displaystyle{
 \theta (\xi) = 2\,\mathrm{arctan} \left(\frac{\xi}{\bar{c}}\right), \qquad \xi\in \mathbb{R}.
}
\label{thetaToarctan}
\end{align}
If so, the Bethe equations read
\begin{align}
\displaystyle{
 k_j L+2\sum_{s=1}^N \mathrm{arctan} \left(\frac{k_j-k_s}{\bar{c}}\right) = 2\pi I_j^\text{p}.
} 
\label{LLBetheEqsPeriodicArcTan} 
\end{align}
Note that due to the antisymmetry of $\theta(\xi)$ (or equivalently $\mathrm{arctan}(\xi)$ for $\bar{c}>0$), one finds 
\begin{align}
\displaystyle{ 
 P_N=\frac{2\pi \hbar}{L}\sum_{j=1}^N  I_j^\text{p}.
}  
\label{LLBetheEqsPeriodicTotMomentum}
\end{align}

\begin{figure}[h!]
\includegraphics[scale=0.173]{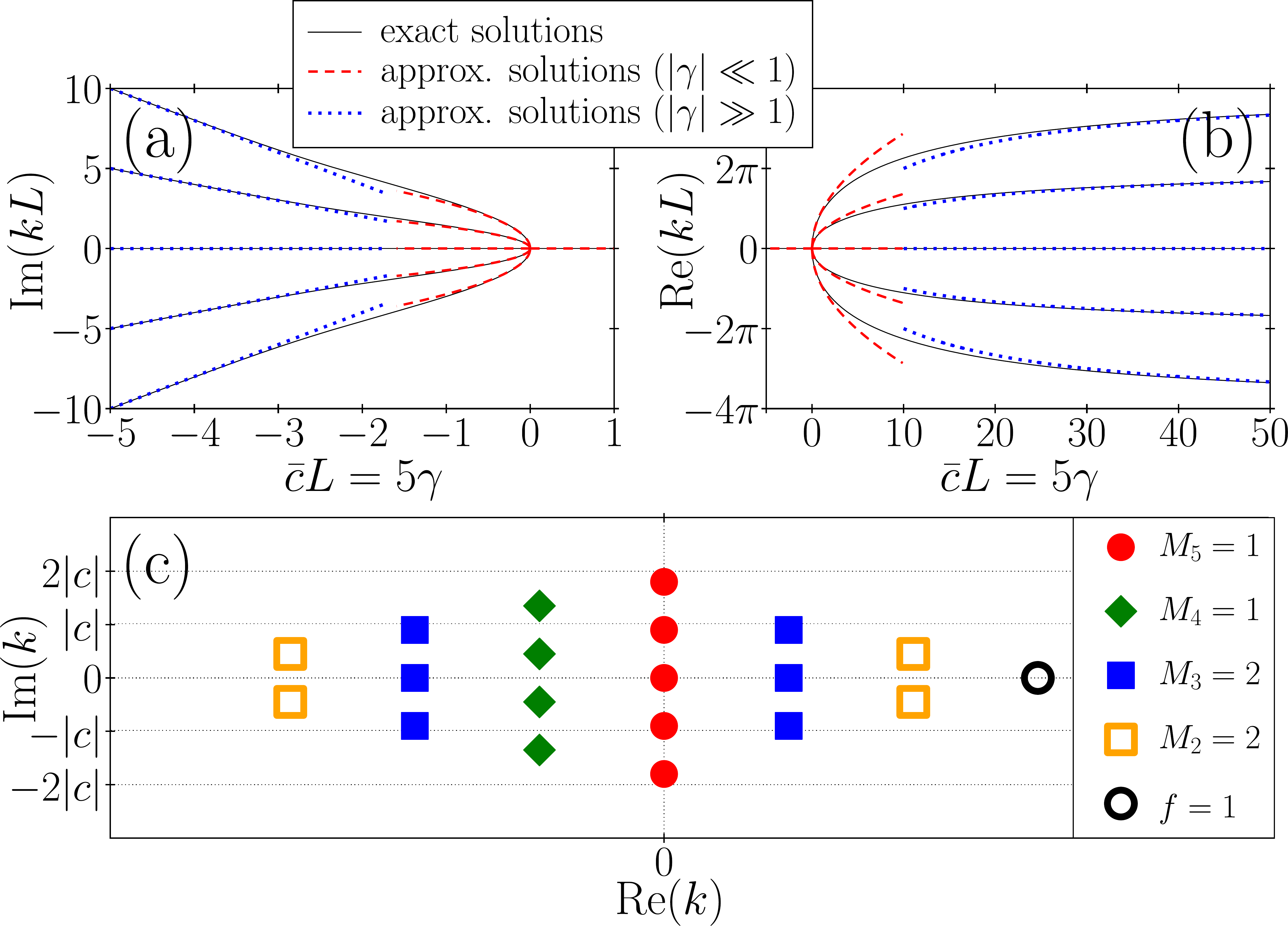} 
\vspace{-0.5cm}
\caption{  
Upper panels  (a) and (b)  present imaginary and real parts of $N=5$ particle ground state solutions of Eqs.~(\ref{LLBetheEqsPeriodicLog}) parameterized by  $\{ I_j^\text{p}\}=\{-2,-1,0,1,2\}$ for different values of the interaction strength $\gamma$. The exact solutions (solid black lines) of Eqs.~(\ref{LLBetheEqsPeriodicLog}) are compared with the approximate solutions in the weakly (dashed red lines) and the strongly interacting regimes (dotted blue lines). The solutions of Eqs. (\ref{BetheSolsWeaklyInteracting}) and (\ref{BetheEqsStrongIndividualSolution}) were parameterized by the numbers $d_{j=1,2,\ldots,5 }=0$ and $\{ I_j^\text{p}\}=\{-2,-1,0,1,2\}$. The corresponding string solutions are obtained for $k_0=0$ and $j=5$, see Eq.~(\ref{LLPeriodicString}). Panel (c) shows a scheme of a multi-string state of the 20-particle system in the presence of very strong  attraction $\gamma \rightarrow -\infty$ ($L<\infty$).    }
\label{f2}
\end{figure}

To get some intuition we analyze Eqs.~(\ref{LLBetheEqsPeriodicLog}) in the limiting cases of interparticle interaction strengths (see also Refs.~\cite{Batchelor2004, Batchelor2005a, Sakmann2005, Batchelor2006, Calabrese2007}). For $N=2$ it is convenient to investigate the difference $\delta k=k_2-k_1$, for which $\delta k L+2\theta(\delta k)=2\pi \Delta I^\text{p}$,
where $\Delta  I^\text{p}= I^\text{p}_2- I^\text{p}_1$. Assuming a finite particle density ($\frac{N}{L}<\infty$), one expands the scattering phase $\theta$ around $\bar{c}=0$ (in fact $\gamma = 0$)
getting $\theta(\xi)\approx -\pi -2 \bar{c}/\xi + 2 \pi r$ ($r\in \mathbb{Z}$). 
Basing on \emph{Theorem 3} we find that for small $|\bar{c}|L$ (small $|\gamma|$) $|\delta k|$ is minimized when  $k_2 =- k_1=\pm\sqrt{\bar{c}/L}$. This simple result shows us not only that the minimal difference between distinct Bethe solutions vanishes slower than $\bar{c}$ implying $\bar{c}/\delta k\stackrel{\bar{c}\rightarrow 0}{\longrightarrow} 0$, but also indicates that in the limit $\bar{c}\rightarrow 0_-$ we can expect paired complex solutions, i.e. $k_{\pm}=k_0\pm i \sqrt{|\bar{c}|/L}$, where $k_0 \in \mathbb{R}$. In general, by using the same expansion of the scattering phase $\theta$ one can show that in the limit of weak interparticle interactions ($\gamma \rightarrow 0$) the quasimomenta $k_{j=1,2,\ldots,N}$  should satisfy the following approximate equations
\begin{align}
k_j \stackrel{\gamma \rightarrow 0_\pm}{=}
\frac{2\pi}{L}d_j+\frac{2\bar{c}}{L} \sum_{\substack{s=1 \\ s\neq j}}^N\frac{1}{k_j-k_s},
\label{BetheSolsWeaklyInteracting}
\end{align}
where the numbers $d_j=0,\pm 1, \pm 2, \ldots$ do not have to be distinct and denote excitations. For example,  while the numbers $d_j=0$ for all $j=1,2,\ldots, N$ correspond to the ground state, each $d_j \neq 0$ indicates a quasiparticle excitation. Such approximate equations works fine also for the weakly attractive system ($\bar{c}<0$) and their complex solutions can be successfully used as a starting point for  numerical investigations of the solutions of the Bethe Eqs.~(\ref{LLBetheEqsPeriodic}). The comparison between the solutions of the approximate Eqs.~(\ref{BetheSolsWeaklyInteracting}) and the exact ones, Eqs. ~(\ref{LLBetheEqsPeriodic}), is shown in Fig.~\ref{f2}(a)\&(b). 

We already know that in the weakly interacting regime ($\gamma\rightarrow 0$) the difference between the Bethe solutions $|\delta k|$ is not smaller that $2\sqrt{|\bar{c}|/L}$. Thus, by neglecting the term $i\bar{c}\, \mathrm{sign}(x_j-x_s)$ in  Eq.~(\ref{LLBetheWaveFunction}) one easily obtains 
 \begin{equation}
\displaystyle{
\lim_{c \rightarrow 0}\Phi_N(\{x\}_N,\{k\}_N)\propto\!\sum_{\pi\in\mathcal{S}_N}\!\! \mathrm{exp}\left( i\sum_{j=1}^N k_{\pi(j)}x_j  \right)
}, 
\label{LLSolutionWeakLimit}
\end{equation}
which coincides with the expectations concerning the system of noninteracting bosons.

On the other hand, for very strong repulsion one expands $\theta(\xi)$ around $\bar{c}=\infty$ ($\gamma = \infty$) with the assumption $|\xi|\ll \bar{c}$, i.e. $\theta(\xi) \approx 2\xi/\bar{c}$, getting
\begin{align}
\displaystyle{k_j\stackrel{\gamma \rightarrow \infty}{=}  \frac{2\pi}{L} I_j^\text{p} - \frac{2}{\bar{c}L} \sum_{s=1}^{N}(k_j-k_s)
},
\label{BetheEqsStrongIndividualSolution} 
\end{align}
which for $N=2$ implies $\delta k =  \frac{2 \pi}{L} \Delta  I^\text{p} \left( 1+ \frac{4}{\bar{c}L}\right)^{-1}$.
The above results indicates that for $\gamma\rightarrow \infty$ the solutions of the Bethe Eqs.~(\ref{LLBetheEqsPeriodicLog}) tend to be separated by  $\frac{2\pi}{L}\Delta  I^\text{p} $, where $\Delta  I^\text{p}\in \mathbb{Z}$ is the difference between the distinct numbers $ I_j^\text{p}$ parameterizing the solutions $k_j$. This result is a beautiful example of the so-called \emph{fermionization} phenomenon when the gas of impenetrable bosons can be treated as a system of noninteracting fermions \cite{Girardeau60,Yukalov2005, Paredes2004}.

For strong interparticle attraction the imaginary part of the complex $k_j$ solution is of the order of $\bar{c}$. Indeed, taking $k_j=k_0 +i \eta \,\, (k_0, \eta \in \mathbb{R})$, the $j$-th of Eqs.~(\ref{LLBetheEqsPeriodic}) reads
\begin{align}
\displaystyle{
  \mathrm{e}^{ik_0 L-\eta L} = -\prod_{s=1}^N\frac{k_{j}-k_{s}-i|\bar{c}|}{k_{j}-k_{s}+i|\bar{c}|}.
} 
\label{LLBetheEqsPeriodicImag}
\end{align}
If now the system containing  $N<\infty$ has a very large size, i.e. $L\rightarrow \infty$, then the left-hand side of Eq.~(\ref{LLBetheEqsPeriodicImag}) approaches 0 (or $\infty$) for $\eta>0$ ($\eta<0$). Thus, at least one $k_{s\neq j}$ has to be equal
\begin{align}
\displaystyle{
  k_{s\neq j}=\left\{
\begin{array}{ccccc}
k_j - i |\bar{c}|	+\mathcal{O}\left(\mathrm{e}^{-\eta L}	\right) & \text{for} & \eta>0 \\ 
k_j + i |\bar{c}|	+\mathcal{O}\left(\mathrm{e}^{-\eta L}	\right) & \text{for} & \eta<0 
\end{array}  
  \right.
}. 
\label{LLPeriodicString_VS_ETA}
\end{align}
In a two-body case for $\gamma \rightarrow -\infty$  one gets two paired solutions $k_{\pm} = k_0 \pm i|\bar{c}|/2$. For $N\geq2$, the individual bound state forms a structure called a string consisting of $2\leq r\leq N$ quasimomenta that share the real parts and differ only in values on the imaginary axis (see also Refs.~\cite{Thacker81, Takahashi99, CastinHerzog01, Sykes2007}). We stress that the imaginary parts are always symmetrically distributed around zero, which ensures that the corresponding energy is always a real number. In the $N$-particle system in question the eigenstate may be constructed with $M_S$ strings (multi-string state) of different lengths $r$ and $f$ free (unpaired) real-valued quasimomenta.  Technically speaking, the string of length $r$  consists of $r$ quasimomenta that are given by the following formula
\begin{align}
\displaystyle{
  k_{\alpha}^{s,r} =k_0^{s,r}+i\frac{|\bar{c}|}{2}(r+1-2\alpha)+ \delta_{\alpha}^{s,r}, 
}
\label{LLPeriodicString}
\end{align}
where $k_0^{s,r}\in \mathbb{R}$, $\delta_{\alpha}^{s,r}\in \mathbb{C}$, $\alpha = 1,2, \ldots, r, $  and the deviations $\delta _{\alpha}^{s,r}$ vanish exponentially with $|\gamma|$, i.e. $|\delta _{\alpha}^{s,r}|\sim \mathrm{e}^{-\zeta |\gamma|}, \, \zeta >0$. In our notation, the upper index $s$ enumerates the consecutive strings consisting of $r$ quasimomenta, i.e. Eq.~(\ref{LLPeriodicString})  represents the quasimomenta corresponding to the $s$-th string of length $r$. In general, one can deal with $M_r$ strings built up by the same number $r$ of quasimomenta so $s=1,\ldots, M_r$ and  $N=\sum_{r}r M_r  +f$, $M_{S}=\sum_{r}M_r$.

It is noteworthy that for $N<\infty$ and $\bar{c}<0$ the above-mentioned limit $L\rightarrow\infty$  corresponds to $\gamma \rightarrow -\infty$. Indeed, if $\eta\rightarrow \pm \infty$ but $L<\infty$ the same analysis leads to similar solutions of Eqs.~(\ref{LLBetheEqsPeriodic}) as in  Eqs.~(\ref{LLPeriodicString}). In both cases we reach $\gamma\rightarrow -\infty$, but one should remember that this limit is obtained in two distinctly different ways -- while in the first case we deal with an infinite ring circumference, the second approach requires an infinite value of the  coupling constant $\bar{c}$.
The approximate ground state solutions obtained for $\gamma \rightarrow \pm \infty$ are compared with the exact solutions of Eqs.~(\ref{LLBetheEqsPeriodic})\&(\ref{LLBetheEqsPeriodicLog}) in the upper panels of Fig.~\ref{f2}. For illustration, the scheme of multi-string eigenstate  is shown in Fig.~\ref{f2}(c).

The Bethe solution $\Phi_N$, Eq.~(\ref{LLBetheWaveFunction}),  significantly differs between the strongly repulsive and the strongly attractive regimes. While in the former case $\Phi_N$ reduces to $\det[\mathrm{e}^{ik_jx_s}]\prod_{j>s} \mathrm{sign}(x_j-x_s)$,  in the latter case we deal with the so-called string solutions, for which the imaginary parts of the quasimomenta $k_j$ behave proportionally to $\bar{c}$. It turns out that for the ground state, i.e. the single string eigenstate with zero total momentum, the corresponding wave function can be cast into the following form \cite{Thacker81,CastinHerzog01,Sykes2007}
\begin{eqnarray}  
\Phi_N(\{x\}_N,&\{k\}_N)
 \stackrel{\gamma\rightarrow -\infty}{\propto} \mathrm{exp} \left(\!\frac{\bar{c}}{2}\sum_{j>s}|x_j-x_s| \!\right).
\label{LLGSSolutinString}
\end{eqnarray}

\subsection{Open boundary conditions}
\label{openc_boundary_conditions}

By imposing open boundary conditions we in fact consider a new Hamiltonian with additional infinite square well trapping potential (hard walls), namely
\begin{equation}
\hat{\mathcal{H}}_N^{U}
=-\frac{\hbar^2}{2m} \sum_{j=1}^N \frac{\partial^2}{\partial x_j^2}  + \sum_{j=1}^N U(x_j)+2c\sum_{ j>s}\delta(x_j-x_s),
\label{LLHardWallHamiltonian}
\end{equation} 
\begin{equation} 
 \qquad\text{with} \quad U(x)=
 \left\{
 \begin{array}{lll}
 0, & & 0<x<L \\
 \infty, & & \text{otherwise}
 \end{array}
 \right.		.														
\label{LLHardWallHamiltonianUU}
\end{equation}
In general, by switching the trapping potential $U(x)$ on, we can break the integrability of the system, but here this is not the case and our problem still has an analytical solution constructed in 1971 by M. Gaudin \cite{Gaudin_BetheWF,Batchelor2006,Gaudin71,Batchelor2005}. The derivation requires two steps that we briefly sketch. Assuming that we operate in the domain  $\mathcal{T}:0< x_1 < x_2 < \ldots < x_N< L$,
the solution $\Psi_N$ has to vanish at $x_1=0$ and $x_N=L$.
Firstly, we look for the solution on the semi-infinite axis $x\geq0$ that vanishes at $x_1=0$. For this purpose M. Gaudin used the McGuire's optical analogy for the general problem of particles interacting via $\delta$-potential \cite{mcguire64}. That is, the proper solution $\Psi_N$ can be constructed by a superposition of the elementary solutions of the Lieb-Liniger model $\Phi_N(\{x\}_N,\{k\}_N)$ corresponding to the same energy $E_N$. Therefore, it is sufficient to consider $2^N$ states parameterized by $ \{k\}_N=\{k_1,k_2, \ldots, k_N\} $ where
$k_j=\epsilon_j|k_j|$, $\epsilon_j=\pm1$ and if $k_j\in\mathbb{R}$, we can choose $0<|k_1|<|k_2|<\ldots<|k_N|$ without loss of generality.
Thus, the solution we look for can be expressed as the following superposition
\begin{eqnarray}
\begin{array}{ll}
 \Psi_N(\{x\}_N,\{|k|\}_N)
 \vspace{0.2cm}\\
\qquad \displaystyle{=\sum_{\epsilon_1,\ldots, \epsilon_N}A(\epsilon_1,\ldots,\epsilon_N)\,\Phi_N(\{x\}_N,\{k\}_N).
}
\end{array}
\label{LLHWsuperposition}
\end{eqnarray}
Taking the elementary solution $\Phi_N$ restricted to the $\mathcal{T}$ domain and assuming that $\Psi_N$ vanishes at $x_1=0$ one obtains (see~\ref{appendixLL_open_1})
\begin{eqnarray}
\begin{array}{l}
\Psi_N(\{x\},\{|k|\}_N)\\
\quad\displaystyle{\propto \sum_{\{\epsilon\}} \! \sum_{\pi\in\mathcal{S}_N}}  \displaystyle{\!\! \epsilon_1 \epsilon_2 \, \dotsb \, \epsilon_N \, \mathrm{exp}\left(\! i\sum_{n=1}^N k_{\pi(n)}x_n \! \right)}
 \\
\qquad\qquad\displaystyle{\times\prod_{j>s} \! \left( \! 1-\frac{i\bar{c}}{k_{j}+k_{s}} \! \right)\!\!\left(\!1-\frac{i\bar{c}}{k_{\pi(j)}-k_{\pi(s)}} \! \right),
} 
\end{array}
\label{LLHWPsi}
\end{eqnarray}
where  $k_j=\epsilon_j |k_j|$ and we need to remember that the above solution is valid only for $0\leq x_1 <\ldots<x_N$.

Secondly, by imposing the vanishing at $x_N=L$, 
one gets  the so-called \emph{\textbf{Gaudin's equations}} \cite{Gaudin_BetheWF,Gaudin71}
\begin{eqnarray}
\begin{array}{llll}
\mathrm{e}^{i 2k_j L}&=
\displaystyle{\prod_{\substack{ s=1 \\ s\neq j}  }^{N} \frac{\left( k_s+k_j+i\bar{c} \right) \left( k_s-k_j-i\bar{c} \right) }{\left( k_s+k_j-i\bar{c} \right) \left( k_s-k_j+i\bar{c} \right) }} \\
&=
\displaystyle{\prod_{\substack{ s=1 \\ s\neq j}  }^{N} \frac{ \left( k_j +i\bar{c} \right)^2 -k_s^2}{\left( k_j -i\bar{c} \right)^2 -k_s^2}, \qquad j=1,2,\ldots,N.}
\end{array}
\label{GaudinEqs}
\end{eqnarray}
The above product contains only nonzero numbers $k_j$, such that $|k_j|\neq |k_s|$ if $j\neq s$, and does not depend on the sign of $k_j$, i.e. the Gaudin's equations are invariant under reflections $k_j \rightarrow - k_j$. By construction, the solutions $\{k_j\}$ determine the energy of $\Psi_N$
 \begin{align}
\displaystyle{
E_N=\frac{\hbar^2}{2m}\sum_{j=1}^N k_j^2
}.   
\label{PsiEnergy}
\end{align}

Let us focus on the repulsively interacting case ($\bar{c}>0$), when we expect to deal with real solutions $k_j\in \mathbb{R}$. Due to the fact that the reflection of any quasimomentum, i.e $k_\alpha \rightarrow -k_\alpha$, does not change the Gaudin's Eqs.~(\ref{GaudinEqs}), it is sufficient to restrict to the positive values of $k_j>0$ only. Rewriting the Gaudin's Eqs.~(\ref{GaudinEqs}) to the logarithmic form
\begin{align}
\displaystyle{
k_j L =\pi m_j -\frac{1}{2}\sum_{\substack{s=1 \\ s\neq j}}^N\left[\theta(k_j+k_s)+\theta(k_j-k_s)\right], 
} 
\label{GaudinEqsLog}
\end{align} 
with  $m_j\in \mathbb{Z}$ and employing the relations in Eq.~(\ref{ThetaFunction}) 
or equivalently
\begin{align}
\displaystyle{
\mathrm{arctan}(\xi)=\frac{\pi}{2}\mathrm{sign}(\xi)-\mathrm{arctan}\left(\frac{1}{\xi}\right),  
} 
\label{equalitiesGaudin2}
\end{align}
together with the equality in Eq.~(\ref{thetaToarctan}) one reads \cite{ Gaudin_BetheWF,Batchelor2006,Gaudin71,Batchelor2005,Tomchenko15,Tomchenko17,Tomchenko17_2}
 \begin{eqnarray}
k\displaystyle{_j L = \pi  n_j^\text{o}\!+\!\sum_{\substack{s=1 \\ s\neq j}}^N\!\Bigg[\mathrm{arctan}\bigg( \frac{\bar{c}}{k_{js}^{+}} \bigg)\!+\!\mathrm{arctan}\bigg( \frac{\bar{c}}{ k_{js}^{-}} \bigg)\Bigg]}, 
\label{GaudinEqsArctg}
\end{eqnarray}
where $k_{js}^\pm =k_j\pm k_s$ and the parameterizing numbers $ n_{j=1,,\ldots,N}^\text{o}\in\mathbb{Z}$. Basing on the continuity of solutions principle, Gaudin has deduced that to deal with admissible physically different solutions, it is enough to consider $1\leq n_1^\text{o}\leq n_2^\text{o} \leq \ldots \leq n_N^\text{o}$ only \cite{Gaudin_BetheWF, Gaudin71, Tomchenko15, Tomchenko17, Tomchenko17_2}. Indeed, let us assume that the set $\{n_j^\text{o}\}$ of positive integers parameterizes the set of positive solutions $\{k_j\}$ of the logarithmic Eqs.~(\ref{GaudinEqsArctg}). It is easy to show that by changing the sign of a single parameterizing number, e.g. $n_l^\text{o}\rightarrow \widetilde{n}_l^\text{o}=-n_l^\text{o}$, we simultaneously change the sign of the corresponding solution, i.e. $k_l\rightarrow\widetilde{k}_l=-k_l$, leaving the other solutions $k_{j\neq l}$ unchanged.
Such an operation leads to exactly the same wave function in Eq.~(\ref{LLHWPsi}), which means that both solutions are physically equivalent. It is also easy to note that 
 for $\widetilde{n}_l^\text{o}=0$  the corresponding $l$-th Gaudin's equation is satisfied when $k_l=0$. 

The fact that the Gaudin's Eqs.~(\ref{GaudinEqsArctg})  possess unique real solutions  for each set of the parameterizing numbers $n_j^\text{o}$ has been proven recently by M.~Tomchenko  \cite{Tomchenko17}. Moreover, it has been shown that the set  $\{n_j^\text{o}\}$ consisting of $p$ identical parameterizing numbers, i.e. $n_{s+1}^\text{o}\hspace{-0.02cm}= \hspace{-0.02cm} \ldots  \hspace{-0.02cm} = \hspace{-0.02cm}n_{s+p}^\text{o}$,  corresponds to $p!$  physically equivalent solutions of Eqs.~(\ref{GaudinEqsArctg})~\cite{Tomchenko17}.

There is a beautiful relationship between the repulsive systems with periodic and open boundary conditions. Note that using  the relation (\ref{equalitiesGaudin2}) one can rewrite the Bethe Eqs.~(\ref{LLBetheEqsPeriodicArcTan}) to the Gaudin-like form
 \begin{align}
\displaystyle{
 k_j L= 2\pi n_j^\text{p}+ 2\sum_{\substack{s=1  \\ s\neq j}}^N \mathrm{arctan} \bigg(\frac{\bar{c}}{k_j-k_s}\bigg),
 } 
\label{LLBetheEqsPeriodicArcTan3} 
\end{align} 
where $n_j^\text{p}$ is an integer number independently on the parity of $N$. Since the relation (\ref{equalitiesGaudin2}) holds true for $\xi\neq 0$, both formulations in Eqs. (\ref{LLBetheEqsPeriodicArcTan}) and (\ref{LLBetheEqsPeriodicArcTan3}) are mathematically equivalent. The case $\xi=0$ is excluded from considerations by the fact that physically relevant solutions have to be distinct, i.e. $\forall_{j,s}:k_j\neq k_{s\neq j}$. Note that the solutions of the new form of the Bethe Eqs.~(\ref{LLBetheEqsPeriodicArcTan3}) are still uniquely parameterized by the new parameterizing numbers $n_j^\text{p}$ that do not have to be distinct and satisfy the relation $n_1^\text{p}\leq n_2^\text{p} \leq \ldots \leq n_N^\text{p}$ known from the open boundary conditions case \cite{Tomchenko17_2}.

On the other hand, by employing the same tricky substitution, Eq.~(\ref{equalitiesGaudin2}), and assuming that $1\leq n^\text{o}_1\leq n^\text{o}_2\leq\ldots \leq n^\text{o}_N$, the Gaudin's equations (\ref{GaudinEqsArctg}) can be cast into the Bethe-like form
 \begin{eqnarray}
k_j L = \pi I^\text{o}_j-\sum_{\substack{s=1 \\ s\neq j}}^N\!\Bigg[\mathrm{arctan}\bigg( \frac{k_{js}^+}{\bar{c}} \bigg)\!+\!\mathrm{arctan}\bigg( \frac{k_{js}^-}{\bar{c}} \bigg)\Bigg],
\label{LLGaudinArcTanToBetheForm2} 
\end{eqnarray} 
where the parameterizing numbers $I_j^\text{o}=n_j^\text{o}+j-1$ 
are integers ordered so that $1\leq I_1^\text{o}< I_2^\text{o}< \ldots < I_N^\text{o}$.

Additionally, we can find an analogy between the $2N$-particle system in a ring of size $2L$  and $N$-particle system confined in a hard wall box of length $L$ (both systems are assumed to have the same densities equal to $N/L$). Indeed, by looking for solutions in the form
  \begin{eqnarray} 
\{k\}_{2N}=\{ \{ -k \}_N, \{ k\}_N \} , &\quad k_{j=1, \ldots,N}>0, \label{LLHWCircBox-1} 
\\ 
\{n^\text{p}\}_{2N}=\{ \{ -n^\text{o} \}_N, \{ n^\text{o}\}_N \} , &\quad n_{j=1,\ldots,N}>0, 
\label{LLHWCircBox} 
\end{eqnarray} 
for the periodic system within the Gaudin-like formulation, Eq.~(\ref{LLBetheEqsPeriodicArcTan3}), we obtain \cite{Tomchenko15}
   \begin{eqnarray} 
k_j 2L&-2\pi n_j^\text{p}=2\sum_{\substack{s=1  \\ s\neq j}}^{2N} \mathrm{arctan} \bigg(\frac{\bar{c}}{k_{j}-k_s}\bigg)  \label{LLHWCircBox2b} 
\\ \nonumber
&=2\sum_{\substack{s=1  \\ s\neq j}}^{N}\Bigg[ \mathrm{arctan} \bigg(\frac{\bar{c}}{k_{js}^+}\bigg) + \mathrm{arctan} \bigg(\frac{\bar{c}}{k_{js}^-}\bigg) \Bigg]+f_j,
\end{eqnarray} 
which coincides with Eqs.~(\ref{GaudinEqsArctg}) except the term $f_j=2\,\mathrm{arctan} \left(\bar{c}/2k_j\right)$. Due to the uniqueness of solutions, Gaudin concluded that the energy of the $N$-particle system of size $L$ with open boundary conditions parameterized by  the set $\{n^\text{o}\}_N$, Eq.~(\ref{LLHWCircBox-1}), is twice smaller than the energy of the $2N$-particle periodic system determined by the set $\{n^\text{p}\}_{2N}$, Eq.~(\ref{LLHWCircBox}) \cite{Gaudin_BetheWF,Gaudin71}.
As shown in Ref.~\cite{Tomchenko15}, such correspondence is not entirely exact. This analogy will be also discussed in the context of the ground state energy in the thermodynamic limit
(see   Sec.~\ref{Gs_LL}).

 Let us also briefly study the behaviour of solutions  of the Gaudin's equations in the limiting cases of the interparticle interaction strengths (see also Refs.~\cite{Batchelor2006,Batchelor2005,Tomchenko15}).  First of all we analyze a two-body problem getting 
  \begin{align} 
  \begin{array}{lll}
 \delta k L-\pi \Delta n^\text{o}& \stackrel{ \bar{c}\rightarrow 0_+}{\approx}  & 2 \bar{c}/\delta k, 
 \\
 \delta k L-\pi \Delta n^\text{o} &\stackrel{\bar{c}\rightarrow \infty}{\approx} & \pi \,\mathrm{sign}(\delta k)-2 \delta k/\bar{c} ,
\label{LLHWDiff} 
\end{array}
\end{align} 
where $\delta k = k_{2} - k_1$ and $\Delta n^\text{o}=n_2^\text{o}-n_1^\text{o}$. Now, it is clear that in the strongly repulsive regime  ($\gamma \rightarrow \infty$) $\delta k\approx  \frac{\pi}{L}(\Delta n^\text{o}+\mathrm{sign}(\delta k)) \left(1-\frac{2}{\bar{c} L}\right)$.
Thus, two solutions that are parameterized by equal numbers $n_1^\text{o}=n_2^\text{o}$, i.e. $\Delta n^\text{o} = 0$, are separated by $\delta k\approx\frac{\pi}{L}\left(1-\frac{2}{\bar{c} L}\right)$, which coincides with the expectations for the Tonks-Girardeau regime ($\gamma \rightarrow \infty$), for which the system can be mapped onto the gas of noninteracting fermions \cite{Girardeau60,Yukalov2005, Paredes2004}. On the other hand, in the weakly repulsive limit ($\gamma \rightarrow 0_+$) for $\Delta n^\text{o} = 0$, one obtains $\delta k  = \pm \sqrt{2 \bar{c}/L}$. The result also holds true in the weakly attractive case ($\gamma \rightarrow 0_-$), where $\delta k  = \pm i\sqrt{2 |\bar{c}|/L}$ indicating the existence of a bound state.

By expanding $\theta$ or $\mathrm{arctan}$ function in a chosen regime of interparticle interactions, one easily finds the following approximate equations
 \begin{eqnarray} 
  \begin{array}{lll} 
\displaystyle{k_j \stackrel{\gamma \rightarrow \infty}{=}\frac{\pi}{L}m_j \left(1+\frac{2(N-1)}{\bar{c}L} \right)^{-1}} ,  \vspace{0.2cm}\\
\displaystyle{k_j \stackrel{\gamma \rightarrow 0_\pm}{=}\frac{\pi}{L}d_j+ \frac{\bar{c}}{L}\sum_{\substack{s=1\\ s\neq j}}^{N}\left(\frac{1}{k_j+k_s}+\frac{1}{k_j-k_s}\right)},
\end{array}
\label{LLApproxeqsHWStrong__WEAK} 
\end{eqnarray} 
where, as in the periodic boundary conditions case, cf. Eqs.~(\ref{BetheSolsWeaklyInteracting}), parameters $d_j=0,\pm1,\pm2,\ldots$. Thanks to the compact form of the Gaudin's equations' solutions for $\gamma\rightarrow \infty$, one can easily determine the corresponding ground state energy. In such a limit we deal with a fermionized system, so the $m_j$ numbers should satisfy the relation $m_j=j$ and thus \cite{Batchelor2005}
 \begin{align}
\begin{array}{llll}
\displaystyle{E_{GS}^\text{o} \stackrel{\gamma \rightarrow \infty}{=} 
\frac{\hbar^2 \pi^2N (N+1)(2N+1)}{12m L^2\left[ 1+2(N-1)/\bar{c}L \right]^{2}}. 
} 
\end{array} 
\label{LLHWEnergy_Strong}  
\end{align}

 \begin{figure}[h!] 
\begin{center}\includegraphics[scale=0.19]{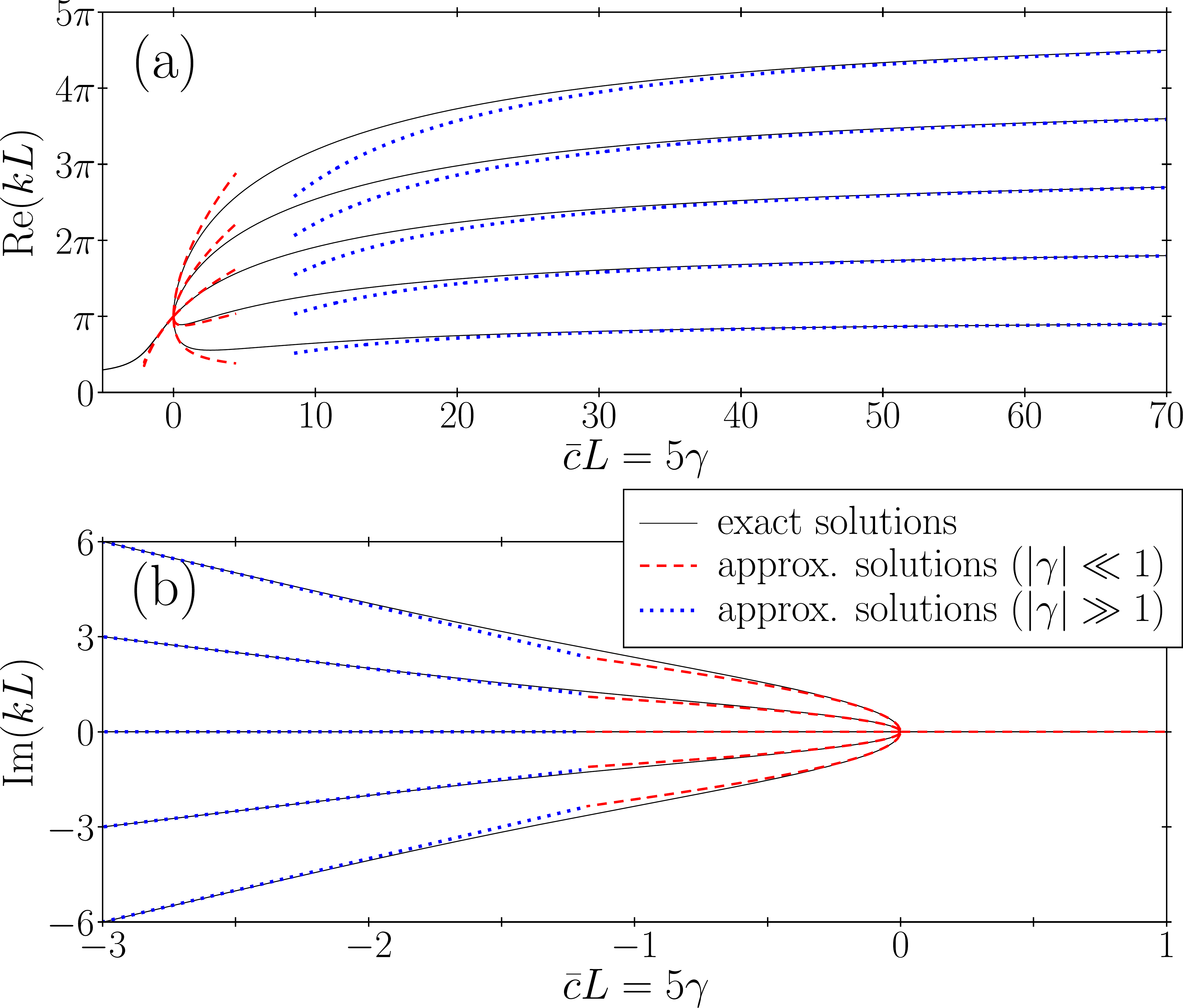} \end{center}
\vspace{-0.5cm}
\caption{  
Solutions of the Gaudin's Eqs.~(\ref{GaudinEqs}) for $N=5$ and for different values of the interaction strength $\gamma$. Panels (a) and (b) refer to the real and the imaginary parts of the solutions, respectively. Starting from the ground state quasimomenta in the noninteracting case, i.e. $k_{j=1,\ldots,N}\rightarrow \frac{\pi}{L}$, we have obtained continuous curves (solid black lines) representing the exact solutions of Eqs.~(\ref{GaudinEqs}), both in the attractive and the repulsive regime. Such results are compared with the solutions of the approximate equations for weakly (dashed red lines) and strongly (dotted blue lines) interacting limits, which were obtained for the numbers $d_{j=1,\ldots,N}=1$ and $\{m_j\}=\{1,2,3,4,5\}$, respectively. The imaginary parts of the approximate solutions in the strongly attractive case are given by $|\bar{c}|(3-\alpha)$, where $\alpha=1,2,\ldots,5$, cf. Eqs.~(\ref{LLPeriodicString}).    }
\label{f3}
\end{figure}

Similarly to the periodic case, one can anticipate that the solutions of Eqs.~(\ref{GaudinEqs}) in the strongly attractive limit ($\gamma \rightarrow -\infty$) may form a bound structures like strings described in Sec.~\ref{periodic_boundary_conditions}.  Indeed, the substitution $k_j=k_0+i \eta$ ($k_0,\eta\in \mathbb{R}$) in the large system size limit $L\rightarrow \infty$ reveals the necessity of existence of a solution of the form $k_{s\neq j}=\pm k_0 \mp i|\bar{c}|$ for $\eta >0$ and $k_{s\neq j}=\pm k_0 \pm i|\bar{c}|$ for $\eta <0$. It is enough to choose $+k_0\pm i|\bar{c}|$. On the other hand, dealing with the ground state in the weakly attractive limit all the quasimomenta should have real parts close to $\pi/L$ and imaginary parts behaving like $\sqrt{|\bar{c}|/2L}$. 
In Fig.~\ref{f3} we compare exact solutions of the Gaudin's Eqs.~(\ref{GaudinEqs}) obtained for the ground state with the solutions of the corresponding approximate Eqs.~(\ref{LLApproxeqsHWStrong__WEAK}).

At the end  we would like to point out  that in the noninteracting limit ($\gamma\rightarrow 0$) the solution $\Psi_N$ constructed by M. Gaudin, Eq.~(\ref{LLHWPsi}), reduces to the solution of the problem of noninteracting bosons confined in an infinite square well potential \cite{Syrwid2017HW}, namely (for proof see~\ref{appendixLL_open_1})
 \begin{eqnarray}
\Psi_N(\{k\}_N,\{x\}_N) \stackrel{\bar{c}\rightarrow 0}{\propto}
 \sum_{\pi\in \mathcal{S}_N} \prod_{s=1}^N\sin\left(|k_{\pi(s)}|x_s\right),
  \label{LLHWWeaklyInteractingLimit3}  
\end{eqnarray} 
where $k_j\stackrel{\bar{c}\rightarrow 0}{\longrightarrow} \frac{\pi}{L}n_j^\text{o} $.

\subsection{Ground state}
\label{Gs_LL}  

We have already discussed analytical solutions of the time-independent Schr\"{o}dinger  equation for the Lieb-Linger model and the stipulations that have to be satisfied in the presence of periodic and open boundary conditions. Here, we are going to analyze the parameterization and properties of the ground state in both cases. 
All the considerations will be restricted to the repulsively interacting case ($\bar{c}>0$).

\subsubsection{Periodic boundary conditions}
\label{LLGroundState_PBC}

Let us start with the investigations of the ground state parameterization. Due to antisymmetry of the general solution $\Phi_N(\{x\}_N,\{k\}_N)$, Eq.~(\ref{LLBetheWaveFunction}), in quasimomenta we know that all the parameterizing numbers $ I_j^\text{p}$ have to be distinct. Hence, applying  \emph{Theorem 3} one can always write $\frac{2\pi}{L}  I^\text{p}_1 \leq k_1< k_2<\ldots < k_N \leq \frac{2\pi}{L}  I^\text{p}_N$, 
which corresponds to $ I^\text{p}_1< I^\text{p}_2<\ldots< I^\text{p}_N$. By the continuity of the Bethe solutions for a given parameterization $\{ I_j^\text{p}\}$, we notice that to determine the ground state parameterization, it is enough to minimize the energy in the case of  infinitely strong interactions ($\gamma\rightarrow \infty$) when  the quasimomenta  $k_j \rightarrow \frac{2\pi}{L} I_j^\text{p}$, i.e. one needs to find the collection of distinct numbers $ I_j^\text{p}$ that minimizes
 \begin{align}
\displaystyle{
\lim_{\gamma \rightarrow \infty}E_N=\frac{4\pi^2\hbar^2}{2mL}\sum_{j=1}^{N} I_j^{\text{p}\, 2}.
}
\label{GSLL2}   
\end{align}
It is clear  that the condition is satisfied by numbers 
 \begin{align}
\displaystyle{
 I_j^\text{p}=j-1-\frac{N-1}{2}, \qquad j=1,2,\ldots,N.
} 
\label{GSLL3}  
\end{align}
Thus, the set $\{ I_j^\text{p} \}_{\text{GS}}$ corresponding to the ground state (GS) has the following symmetric structure
 \begin{equation}
\displaystyle{
\begin{array}{lccc}
\left\{-\frac{N-1}{2},-\frac{N-3}{2}, \ldots,-1, 0,1, \ldots ,\frac{N-1}{2}\right\}\Big|_{N \text{ odd}}
\\
\left\{-\frac{N-1}{2},-\frac{N-3}{2}, \ldots,-\frac{1}{2},\frac{1}{2}, \ldots,\frac{N-1}{2}\right\}\Big|_{N \text{ even}}

\end{array}
}.  
\label{GSLLISet}  
\end{equation}
It can be shown that since we keep the chosen order of $I_j^\text{p}$ numbers ($I_{j+1}^\text{p}>I_j^\text{p}$)
the relation between the Bethe-like and the Gaudin-like parameterizations of eigenstates of the periodic system reads
 \begin{align}
\displaystyle{
 I_j^\text{p}=n_j^\text{p}+j-\frac{N+1}{2}, \qquad j=1,2,\ldots,N.
} 
\label{PBC_BetheVSGaudinParam}  
\end{align}
If so, the ground state in the Gaudin-like formulation of the Bethe Eqs.~(\ref{LLBetheEqsPeriodicArcTan3}) is parameterized as follows
 \begin{align}
\displaystyle{
\{n_j^\text{p} \}_{\text{GS}}=\{0,0,\ldots,0\} \qquad \text{for any } N  
}. 
\label{GSLLWSet}  
\end{align}

There is one more very beautiful feature of the ground state solutions. That is, by considering weakly interacting case ($\gamma\rightarrow 0$) and the ground state parameterization, i.e. $n_{j=1,2,\ldots,N}^\text{p}=0$, we easily obtain 
 \begin{align}
\displaystyle{
k_jL=\sum_{\substack{s=1\\ s\neq j}}^N \frac{2\bar{c}}{k_j-k_s} \Longrightarrow q_j=\sum_{\substack{s=1\\ s\neq j}}^N \frac{1}{q_j-q_s}, 
}  
\label{GSLLWSet0Hermite1}  
\end{align}
where the numbers $q_j=\sqrt{L/2\bar{c}}k_j$, as the solutions of the above equations, turn out to represent zeroes of the $N$-th Hermite polynomial $H_N(q)$, i.e. $H_N(q_j)=0$ for $j=1,2,\ldots,N$.
Thus, the ground state solutions of the Bethe equations for weakly interacting Lieb-Liniger model with periodic boundary conditions are given in terms of the roots of the Hermite polynomial $H_N(q)$, i.e. $k_j=\sqrt{2\bar{c}/L}\, q_j+\mathcal{O}(\bar{c})$ \cite{Gaudin_BetheWF,Jiang15}.

We will also briefly discuss the thermodynamic limit, i.e. $N,L\rightarrow \infty, \, \rho=\frac{N}{L}=\mathrm{const}$. The lowest energy eigenstate in the presence of periodic boundary conditions is given by the parameterization in Eqs.~(\ref{GSLLISet}). It can be shown that the resulting quasimomenta fill the symmetric interval $[-\mathcal{Q},\mathcal{Q}]$ and   $\Delta k_j =k_{j+1}-k_j=\mathcal{O}(L^{-1})$.
By the fact that $\Delta k_j>0$, one denotes the density of states as \cite{LiebLiniger1,BogoliubovKorepinInverseScattering,Gaudin_BetheWF,pethick}
  \begin{align}
\displaystyle{
D_\text{p}(k_j)=\lim \frac{1}{L(k_{j+1}-k_j)}>0
}.
\label{LLPeriodicDensOfStates}  
\end{align}
 While the above limit ($\lim$) refers to the thermodynamic limit, the lower index "p" corresponds to the periodic boundary conditions imposed on the system. 
  As it was shown by Elliott H. Lieb and Werner Liniger (see Ref.~\cite{LiebLiniger1}), the density of states has to satisfy the relation
  \begin{align} 
\displaystyle{ 
D_\text{p}(k)-\frac{1}{2\pi}\int\limits_{-\mathcal{Q}}^{\mathcal{Q}}\frac{2\bar{c} \, D_\text{p}(\mu)}{\bar{c}^2+(k-\mu)^2}d\mu=\frac{1}{2\pi}
},
\label{LLPeriodicTherm3}  
\end{align}
and according to the definition in Eq.~(\ref{LLPeriodicDensOfStates}) 
 \begin{align} 
\displaystyle{ 
\rho=\frac{N}{L}=\int\limits_{-\mathcal{Q}}^{\mathcal{Q}}D_\text{p}(\mu)\mathrm{d}\mu
}.
\label{LLPeriodicTherm2}  
\end{align}
Moreover, it was also shown that Eq.~(\ref{LLPeriodicTherm3}) has a unique solution \cite{LiebLiniger1,BogoliubovKorepinInverseScattering}, which for $\gamma\rightarrow \infty$ takes the following form
\begin{align} 
\displaystyle{ 
D_\text{p}(k)=\left\{
\begin{array}{cll}
\displaystyle{\frac{1}{2\pi}} & \text{for} & |k|\leq \mathcal{Q} 
\vspace{0.2cm}
\\ 
\displaystyle{0}  &\text{for} & |k|> \mathcal{Q}  
\end{array}\right.
},
\label{LLPeriodicTherm4}  
\end{align}
indicating that in the regime of impenetrable bosons $\lim \Delta k_j=\frac{2\pi}{L}$.

Additionally, we can determine the ground state energy, i.e.
  \begin{align} 
\displaystyle{ 
\lim E_{GS}^\text{p}=\frac{\hbar^2L}{2m}\int\limits_{-\mathcal{Q}}^{\mathcal{Q}}k^2 D_\text{p}(k)\mathrm{d}k. 
}
\label{LLPeriodicGSThLim}  
\end{align}
For very strong repulsion (see also Refs.~\cite{LiebLiniger1,Gaudin_BetheWF,pethick,Jiang15,Ristivojevic2014}) one finds $\frac{2\bar{c}}{\bar{c}^2+(k-\mu)^2}\approx 2/\bar{c}$ which leads to $D_\text{p}(k)=\frac{1}{2\pi}\left(1+2/\gamma\right)$ and $\mathcal{Q}=N\frac{\pi}{L}/\left(1+2/\gamma\right)$.   In consequence,
  \begin{eqnarray} 
  \begin{array}{ll}
 \displaystyle{ \lim E_{GS}^\text{p} \stackrel{\gamma \rightarrow \infty}{\approx} \frac{\hbar^2}{2m}\frac{N \pi^2\rho^2}{3(1+2/\gamma )^2}}
 \vspace{0.2cm}\\
\qquad\qquad\displaystyle{=\frac{\hbar^2}{2m}\frac{N \pi^2\rho^2}{3}\left(1-\frac{4}{\gamma}+\frac{12}{\gamma^2}+\ldots\right)}.
\end{array}
\label{LLPeriodicGSThLim}  
\end{eqnarray}

The studies of the opposite limit, i.e. $\gamma\rightarrow 0_+$, turned out to be very illuminating. M. Gaudin noticed \cite{Gaudin_BetheWF,Gaudin71}  that the Fredholm equation of the second kind, Eq.~(\ref{LLPeriodicTherm3}), is identical to the \emph{Love's equation} appearing in the problem of a circular capacitor	\cite{Love49}. With the help of Hutson's method \cite{Hutson63}, Gaudin showed that for $\gamma\rightarrow 0_+$ the ground state energy in the thermodynamic limit reads \cite{Gaudin_BetheWF,Gaudin71}
  \begin{align} 
\displaystyle{ 
\lim E_{GS}^\text{p} \stackrel{ \gamma \rightarrow 0_+}{=}  \frac{N\hbar^2}{2m}\rho^{2}\left( \gamma-\frac{4}{3\pi}\gamma^{3/2}+ \ldots\right) 
}, 
\label{LLPeriodicWeakGSThermLim}  
\end{align}
and coincides with the result that was previously obtained by means of the perturbation theory of Bogoliubov \cite{LiebLiniger1}. Note that by taking a simple limit $2\bar{c}/(\bar{c}^2+(k-\mu)^2)\stackrel{\bar{c} \rightarrow 0_+}{\longrightarrow} 2 \pi \delta (k-\mu)$ one obtains diverging density of states $D_\text{p}(k)$. An asymptotic analysis carried out by Gaudin showed that
  \begin{eqnarray} 
&D_\text{p}(k)  \stackrel{\gamma \rightarrow 0_+}{=}
 \frac{\sqrt{\mathcal{Q}^2-k^2}}{2\pi \bar{c}}   \label{LLPeriodicDensyWeak}  
 \\ \nonumber
&+\frac{\mathcal{Q}}{4\pi^2 \sqrt{\mathcal{Q}^2-k^2}}\left( \!\frac{k}{\mathcal{Q}}\ln \left[  \hspace{-0.05cm}\frac{\mathcal{Q}-k}{\mathcal{Q}+k} \hspace{-0.03cm}\right]\! +\!\ln \left[  \hspace{-0.05cm}\frac{16\pi e \mathcal{Q}}{\bar{c}}\hspace{-0.03cm}\right] \right)
\!+\!\scalebox{0.7}{$\mathcal{O}$} (1),
\end{eqnarray}
where the dominating term resembles a \emph{semi-circle law}, typically appearing in Random Matrix Theory in the context of Gaussian ensembles analysis \cite{Mehta}. 
This result can be also predicted by noting that for $N\gg1$, the distribution of the zeroes of the $N$-th Hermite polynomial  is given by $\frac{1}{\pi}\sqrt{2N-q^2}$ \cite{Gaudin_BetheWF,Szego39}. Employing the result that in the weakly interacting limit the ground state quasimomenta can be expressed in terms of the roots $q_j$ of the $N$-th Hermite polynomial, i.e. $k_j=q_j\sqrt{2\bar{c}/L}$, one finds  $D_\text{p}(k)\propto \sqrt{\mathcal{Q}^2-k^2}$, where $\mathcal{Q}=\sqrt{4 \bar{c}\rho}$. The proportionality factor $(2\pi \bar{c})^{-1}$ appears due to Eq.~(\ref{LLPeriodicTherm2}).

\subsubsection{Open boundary conditions}
\label{LLGroundState_OBC}

 In the weakly interacting limit one expects that all the quasimomenta corresponding to the ground state should approach  $\frac{\pi}{L}$. Simultaneously, we know that for $\bar{c}\rightarrow 0_+$ the solutions of the Gaudin's Eqs.~(\ref{GaudinEqsArctg})  $k_j\longrightarrow \frac{\pi}{L}n_j^\text{o} $, what suggests the following ground state  parameterization 
  \begin{align}
\displaystyle{
\{n_j^\text{o}\}_{GS} = \{1,1,\ldots,1\} \hspace{0.8cm} \text{for any } N.  
} 
\label{GSHWLL1}  
\end{align}
Although, the strict proof of such a ground state parameterization for arbitrary $\bar{c}>0$ is absent \cite{Tomchenko15}, we can look also at the strongly repulsive limit ($\gamma\rightarrow \infty$), where the solutions of the Gaudin's Eqs.~(\ref{GaudinEqsArctg}) with the anticipated parameterization, Eq.~(\ref{GSHWLL1}), tend to  $k_{j}=\frac{\pi}{L}j$, c.f. Fig.~\ref{f3}(a), matching the expectations for the Tonks-Girardeau gas \cite{Girardeau60,Yukalov2005, Paredes2004}.

Note that the ground state parameterization in Eq.~(\ref{GSHWLL1}) has exactly the same structure as in the case of  periodic boundary conditions. Indeed, in both systems the set of parameterizing numbers corresponding to the ground state consists of equal numbers, i.e. $n_{j=1,\ldots,N}^\text{p}=0$ and $n_{j=1,\ldots,N}^\text{o}=1$. In addition, similarly to the periodic case, the ground state quasimomenta of the weakly interacting ($\bar{c}\rightarrow 0$) $N$-particle Lieb-Liniger model with open boundary conditions can be also expressed in terms of the roots of the $N$-th Hermite polynomial (see Ref.~\cite{Tomchenko15}), namely
   \begin{align}
\displaystyle{
k_j=\frac{\pi}{L} +\sqrt{\frac{\bar{c}}{L}}q_j+\frac{(N-1)\bar{c}}{2\pi} +\mathcal{O}(\bar{c}^{\,3/2}).
}
\label{LLHWHermite4} 
\end{align}

To shortly analyze the thermodynamic limit one may carry out the same reasoning as for the periodic boundary conditions. The density of states can be defined in the same way, i.e by Eq.~(\ref{LLPeriodicDensOfStates}), but with the restriction  $0<k_1<k_2<\ldots< k_N$. Hence, 
   \begin{align}
\displaystyle{
\rho=\frac{N}{L}=\int\limits_{k_1}^{k_N}D_\text{o}(k)\mathrm{d}k,
}
\label{LLHWDensityofstatesNorm} 
\end{align}
where the lower index "$\text{o}$" refers to the open boundary conditions. The density of states $D_\text{o}(k)$ has to satisfy the following integral equation \cite{Tomchenko15} 
\begin{eqnarray}
 \pi D_\text{o}(k)+g_k
=\int\limits_{k_1}^{k_N}\! \frac{\bar{c}\,D_\text{o}(q)\,\mathrm{d}q}{\bar{c}^2\!+\!(k\!+\!q)^2}+\int\limits_{k_1}^{k_N}\!\frac{\bar{c}\,D_\text{o}(q)\,\mathrm{d}q}{\bar{c}^2\!+\!(k\!-\!q)^2},
\label{LLHWsubtract3} 
\end{eqnarray}
where $g_k=\frac{1}{L}\frac{2\bar{c}}{\bar{c}^2+4k^2}-1$ and for $\gamma\rightarrow \infty$
  \begin{align} 
\displaystyle{ 
D_\text{o}(k)=\left\{
\begin{array}{cll}
\displaystyle{\frac{1}{\pi}} & \text{for} & k_1\leq k\leq k_N 
\vspace{0.2cm}
\\ 
\displaystyle{0}  &\text{for} & k<k_1, \,\, k> k_N 
\end{array}\right.
},
\label{LLHWsubtract4}  
\end{align}
reflecting the fact that  $\lim \Delta k_j\stackrel{\gamma\rightarrow \infty}{\longrightarrow}\frac{\pi}{L}$.

Now, we get back to the problem of the relationship between the periodic system containing of $2N$ bosons living in a ring of size $2L$  and the system of $N$ bosons confined in an infinite square well potential of size $L$.  In the ground state case half of the corresponding quasimomenta of the periodic system are positive and can be ordered as follows $0<\widetilde{k}_1<\widetilde{k}_2<\ldots < \widetilde{k}_N$. The second half $\widetilde{k}_{j=-1,-2,\ldots,-N}$  satisfy the relation $\widetilde{k}_{-j}=-\widetilde{k}_{j}$.  Note that with the ground state parameterization in Eqs.~(\ref{GSLLISet}) and~(\ref{GSLLWSet}), such a case fulfills the previously discussed conditions in Eqs.~(\ref{LLHWCircBox-1})--(\ref{LLHWCircBox}). Dealing with the Bethe-like formulation of the periodic problem, Eqs.~(\ref{LLBetheEqsPeriodicArcTan}), one finds the following relations \cite{Tomchenko15}
\begin{eqnarray}
\begin{array}{c}
\displaystyle{\int\limits_{\widetilde{k}_1}^{\widetilde{k}_N}D_\text{p}(q)\mathrm{d}q=\frac{N}{2L}},
\vspace{0.1cm}\\ 
\displaystyle{
2\pi D_\text{p}(\widetilde{k})-1
=\int\limits_{\widetilde{k}_1}^{\widetilde{k}_N} \frac{2\bar{c}\, D_\text{p}(q)\, \mathrm{d}q}{\bar{c}^2\!+\!(\widetilde{k}\!+\!q)^2}+\int\limits_{\widetilde{k}_1}^{\widetilde{k}_N}\frac{2\bar{c}\, D_\text{p}(q)\, \mathrm{d}q}{\bar{c}^2\!+\!(\widetilde{k}\!-\!q)^2}
},
\end{array}
\label{LLHWvsPBC_2}
\end{eqnarray}
resembling Eqs.~(\ref{LLHWDensityofstatesNorm})--(\ref{LLHWsubtract3}), except $g_k+1$. The  analysis presented in Ref.~\cite{Tomchenko15} shows that there is no significant distinction between the considered systems with periodic and open boundary conditions when $N\gamma\gg 1$. This fact can be understood in a very easy way. Indeed, in the limit $L\rightarrow \infty$ Eqs.~(\ref{LLHWDensityofstatesNorm})--(\ref{LLHWsubtract3}) are identical with Eqs.~(\ref{LLHWvsPBC_2}), if only $D_\text{o}(k)=2D_\text{p}(k)$. Note that, when keeping the particle density constant and finite one rewrites $g_k+1=\frac{1}{N\gamma} \frac{2\gamma^2 \rho^2}{\gamma^2 \rho^2+4k^2}$
 and thus if $N \gamma \gg 1$, the contribution in question is negligible.   If so, the positive quasimomenta of the periodic system are equal to the quasimomenta of the system with open boundary conditions, i.e. $k_j=\widetilde{k}_{j\geq1}$, and $D_o(k)=2D_p(k)$. Now it is clear that in such a case
\begin{eqnarray}
\begin{array}{l}
\displaystyle{
\lim E_{GS}^\text{p}(2N,2L)=2\frac{\hbar^2 (2L)}{2m}\int\limits_{\widetilde{k}_1}^{\widetilde{k}_N}q^2 D_\text{p}(q)\mathrm{d}q}
\\
\qquad\displaystyle{
\stackrel{N\gamma\gg 1}{=} 2\frac{\hbar^2 L}{2m}\int\limits_{k_1}^{k_N}q^2 D_\text{o}(q)\mathrm{d}q=2\lim E_{GS}^\text{o}(N,L).
}
\end{array}
\label{LLHWvsPBC_3} 
\end{eqnarray}

Let us now determine the thermodynamic ground state energy in the presence of open boundary conditions. For $\bar{c}\rightarrow \infty$ ($\gamma\rightarrow \infty$ for finite $\rho$), the integrand  and the contribution $\frac{1}{L}\frac{2\bar{c}}{\bar{c}^2+4k^2}$ reduce to $\frac{2}{\bar{c}}D_\text{o}(q)$ and $\frac{2}{\bar{c}L}$, respectively. According to Eq.~(\ref{LLHWDensityofstatesNorm}) one gets $D_\text{o}(k)=\frac{1}{\pi}\left(1+2/\gamma -2/N\gamma  \right)$, which coincides with $2D_\text{p}(k)$ obtained in the same limit, because in such a case $N\gamma\gg 1$. Hence, for strong interparticle repulsion, the resulting thermodynamic ground state energy is identical to this obtained for the periodic system, cf. Eq.~(\ref{LLPeriodicGSThLim}), 
  \begin{eqnarray} 
  \begin{array}{l}
 \lim E_{GS}^\text{o} \stackrel{ \gamma \rightarrow \infty}{\approx}  \displaystyle{\frac{\hbar^2}{2m}\frac{N \pi^2\rho^2}{3(1+2/\gamma )^2}}
\vspace{0.1cm} \\
  \qquad\qquad=\displaystyle{\frac{\hbar^2}{2m}\frac{N \pi^2\rho^2}{3}\left(1-\frac{4}{\gamma}+\frac{12}{\gamma^2}+\ldots\right).
}
\end{array}
\label{LLHWGSThLim}  
\end{eqnarray}

The very weakly repulsive case ($\bar{c},\gamma\rightarrow 0_+$, $\rho<\infty$) can be investigated using the distribution of the zeroes of the Hermite polynomials, i.e. $\propto\sqrt{2N-q^2}$ for $N\gg 1$, and the ground state energy in such a limit is given by the following formula (see also Refs.~\cite{Batchelor2005,Tomchenko15})
  \begin{align} 
\displaystyle{ 
 \lim E_{GS}^\text{o} \stackrel{\gamma \rightarrow 0_+}{\approx} \frac{\hbar^2 N}{2m}\left(\frac{\pi^2}{L^2}+\frac{3}{2}\rho^2\gamma\right).
}
\label{LLHWGSThLim2}  
\end{align}

\subsection{Elementary excitations}
\label{LLElementaryExcitations} 

\begin{figure}[h]
\begin{center}\includegraphics[scale=0.22]{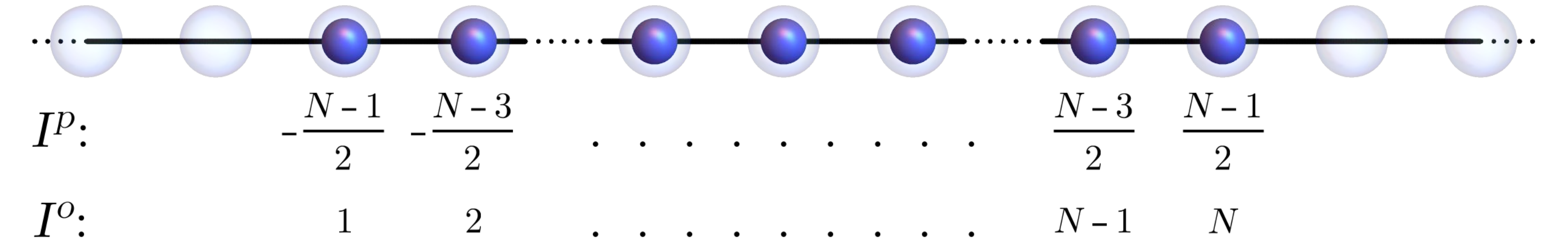} \end{center}
\vspace{-0.55cm}
\caption{ 
Graphical representation of the ground state parameterization of the Lieb-Liniger model within the Bethe-like formulation. Blue filled spheres correspond to the configuration of the parameterizing numbers in the case of periodic $I^\text{p}$ and open $I^\text{o}$ boundary conditions that are indicated below.}
\label{f4}
\end{figure}

 The above discussion shows that there are many similarities between the Lieb-Liniger systems with periodic and open boundary conditions.
 Surprisingly, the parameterizations of the ground state have almost the same structure. Here, we show that this is also the case of the elementary excitation scenarios. That is, in both cases one can distinguish two branches of the elementary excitations. It turns out that while the first one corresponds to  sound waves \cite{Lieb2, Tomchenko17_2}, the second one is related to dark solitons \cite{Syrwid2017HW,Kulish76, Ishikawa80, Syrwid15, Syrwid16}. For the sake of convenience, we restrict our analysis to the Bethe-like formulation only. The parameterizations corresponding to the Gaudin-like description can be easily reproduced from the relations $I_j^\text{p}=n_j^\text{p}+j-\frac{N+1}{2}$ and $I_{j}^\text{o}=n_j^\text{o}+j-1$. In Fig.~\ref{f4}, we present the graphical representation of the ground state parameterization in the both considered cases.  

\subsubsection{Periodic boundary conditions}
\label{LLElementaryExcitations_PBC}

In the ground state case one deals with two limiting values of quasimomenta $\pm \mathcal{Q}$, which resemble Fermi surface in the Fermi system. Since we analyze the interacting Bose system, such a name has to be treated as a mental shortcut  and will be adopted by us only for convenience. Our considerations are based on the quasimomenta instead of the momenta of particles.  Hence, from now $\mathcal{Q}$ will be dubbed Fermi quasimomentum and denoted as $k_F$. The quasimomenta $k$ corresponding to the ground state have to satisfy $|k| \leq k_F$.  
As mentioned above, the elementary excitations in the Lieb-Liniger model can be divided into the two following types:

\vspace{0.15cm}
\underline{\emph{\textbf{Type I (particle excitations)}}}

 It relies on the excitation from the Fermi surface, i.e. $k_\pm=\pm k_F$, to $q_+>k_F$ or $q_-<-k_F$. In the Tonks-Girardeau limit, for which $k_j=\frac{2\pi}{L}I_j^\text{p}$ and $k_F=\frac{\pi(N-1)}{L}$, such an excitation corresponds to the following energy $\epsilon_I$ and momentum $p$  
  \begin{align} 
&\epsilon_I = \frac{\hbar^2}{2m}\left(q_\pm^2-k_F^2 \right),   &
p=\hbar \left( q_\pm \mp  k_F \right), 
\label{LL_PBC_Type1_1}  
\\
&\epsilon_I(p)=\frac{p^2 +2\hbar k_F|p|}{2m}.
\label{LL_PBC_Type1_2}  
\end{align}

In the limit $N\rightarrow \infty$ the type--I excitation can be realized by inserting an additional particle into the system and that is why the type--I excitations are often dubbed {\it \bf particle excitations} \cite{Lieb2,Franchini}. Namely, starting with the $N$-particle ground state parameterized by  $\{I_j^\text{p}\}_\text{GS}=\left\{-\frac{N-1}{2},-\frac{N-3}{2},\ldots, \frac{N-1}{2} \right\}$, after the type--I excitation one considers the $N+1$-particle system parameterized by $\{\widetilde{I}_j^\text{p}\}_I=\left\{-\frac{N}{2}, -\frac{N-2}{2}, \ldots, \frac{N-2}{2}, \frac{N}{2}+M \right\}$. Note that here we have chosen the positive momentum of the excitation equal to $p=\frac{2\pi}{L}M$. The particle insertion changes the values of all the initial quasimomenta $\{k_1,k_2,\ldots, k_N\}$, so that after the particle excitation one deals with new quasimomenta  $\{k'_1, k'_2, \ldots, k'_N, q\}$, where in general $q\neq p$.

Analyzing the difference $\Delta k_j=k'_j-k_j$, that tells how the excitation affects the values of quasimomenta, one finds  \cite{Lieb2,Franchini}
\vspace{-0.3cm}
  \begin{align} 
\displaystyle{
2\pi \mathcal{J}_I^\text{p}(k)=-\pi -\theta(k-q)+\int\limits_{-k_F}^{k_F}\frac{2\bar{c}\, \mathcal{J}_I^\text{p}(\mu)\mathrm{d}\mu}{\bar{c}^2\!+\!(k\!-\!\mu)^2}
}, 
\label{LL_PBC_Type1_Therm_7}  
\\
\begin{array}{l}
p=\displaystyle{\hbar\left[ q+\int\limits_{-k_F}^{k_F}\mathcal{J}_I^\text{p}(\mu) \mathrm{d}\mu\right]},
\vspace{0.1cm}\\ 
\epsilon_I=
\displaystyle{\frac{\hbar^2}{2m}\left[q^2+2\int\limits_{-k_F}^{k_F}\mu \mathcal{J}_I^\text{p}(\mu)\mathrm{d}\mu\right]-\mu_\text{ch}},
\end{array}
\label{LL_PBC_Type1_Therm_9}  
\end{align}
which allows us to extract valuable physical information about the excitation. The system response to the particle insertion is described by the response function $\mathcal{J}_I^\text{p}(k)=\omega(k)  D_\text{p}(k)$, where $\omega(k)= L\Delta k$. The chemical potential $\mu_\text{ch}$ appears as a consequence of an addition of a single particle to the system. As it was pointed out in Ref.~\cite{Lieb2}, since $\pi+\theta(k-q)$ is positive definite, Eq.~(\ref{LL_PBC_Type1_Therm_7}) has a unique and negative definite solution.	Hence, $\mathcal{J}_I^\text{p}$ and $\omega$ are negative, which implies $\Delta k_{j} <0$ for all $j=1,2,\ldots,N$. In other words, the insertion of a particle with the momentum $q>k_F$ into the system, prepared initially in the ground state, decreases the values of all the other quasimomenta.

To calculate the dispersion relation $\epsilon_I(p)$ in the thermodynamic limit for arbitrary $\gamma\geq0$, the Fredholm integral equation of the second kind, Eq.~(\ref{LL_PBC_Type1_Therm_7}), has to be solved. Such a problem can be attacked numerically with the method of Neumann series \cite{Rahman}. In the Tonks-Girardeau limit ($\gamma=\infty$) the spectrum is given by Eq.~(\ref{LL_PBC_Type1_2}) and shifted by $-\mu_\text{ch}$, namely 
 \begin{align} 
\displaystyle{
\lim_{\gamma\rightarrow \infty}\epsilon_I(p) =-\mu_\text{ch} +\frac{p^2 +2\hbar \pi \rho |p|}{2m}
},   
\label{LL_PBC_Type1_Therm_12}  
\end{align}
where we used the fact that $k_F\stackrel{\gamma\rightarrow \infty}{=}\pi \rho$.
For very weak repulsion ($\gamma \rightarrow 0_+$) one reproduces the free particles limit $\epsilon_I(p)=-\mu_\text{ch}+ p^2/2m$. For  $q\gg k_F$,  the response function can be reduced to $\mathcal{J}_I^\text{p}(k)\approx-\frac{2\rho\gamma}{q}D_\text{p}(k)$ \cite{Lieb2}. Therefore, if $p/\hbar\rho\gg \gamma <\infty$,
 \begin{align} 
 \begin{array}{l}
\displaystyle{p\approx \hbar q - \frac{2\hbar\rho^2 \gamma }{q},}
 \\
 \displaystyle{\epsilon_{I}(p)\approx-\mu_\text{ch}+\frac{p^2}{2m}+\frac{2\hbar^2\rho^2 \gamma }{m}.}
 \end{array}
\label{LL_PBC_Type1_Therm_13}  
\end{align}

 For small $\gamma$ the first branch of the elementary excitations  reproduces the Bogoliubov spectrum (sound waves) \cite{Lieb2,Ishikawa80,Franchini}. Indeed, the detailed analysis presented in Ref.~\cite{Lieb2} showes that for small $\gamma$ and $p$ the dispersion relation becomes linear, i.e.~$\epsilon_I(p)\approx p v_s$, where  $v_s\equiv v_s^\text{mic}=\lim_{p\rightarrow 0}\frac{\partial\epsilon_I(p)}{\partial p}=\frac{\hbar}{m}\rho \sqrt{\gamma}$ is the (microscopic) velocity of sound -- speed of the long wave ($p\rightarrow 0$) excitations. In the macroscopic (thermodynamic) description the velocity of sound can be calculated as $v_s^\text{th}=\sqrt{\frac{L}{\rho m} (\partial^2 E_{GS} / \partial L^2)}$, where for $E_{GS}^\text{p}$ in Eq.~(\ref{LLPeriodicWeakGSThermLim}) one finds $v_s^\text{th}(\gamma\rightarrow 0_+)\approx\frac{\hbar}{m}\rho\sqrt{\gamma(1-\gamma^{1/2}/2\pi)}$. It has been also shown that there is no energy gap between the ground state and the lowest possible particle excitation when $N\hspace{-0.02cm}\rightarrow \infty$, i.e. the lower edge of the spectrum corresponds to $p=0$ \cite{Lieb2}.

\vspace{0.15cm}
\underline{\emph{\textbf{Type II (hole excitations)}}}

 Such an excitation takes place when one of the quasimomenta lying below the Fermi surface $0<q<k_F$ ($0>q>-k_F$) is raised just above $k_F$ ($-k_F$). In the Tonks-Girardeau limit the new value of the  quasimomentum in question is equal to $\pm k_F\pm \frac{2\pi}{L}=\pm \frac{\pi(N+1)}{L}$. Thus, the corresponding energy and momentum of the type--II excitation in such a regime are given by 
  \begin{align} 
  \begin{array}{l}
\displaystyle{
\epsilon_{II} = \frac{\hbar^2}{2m}\left(k_F+\frac{2\pi}{L} \right)^2\!-\! \frac{\hbar^2 q^2}{2m},}   
\vspace{0.2cm}\\
\displaystyle{p=\hbar \left( \pm k_F\pm \frac{2\pi}{L} -  q \right), }
\end{array}
\label{LL_PBC_Type2_1}  
\\
\displaystyle{
\epsilon_{II}(p)=\frac{-p^2+2\hbar k_F |p| +4 \hbar |p|\pi/L }{2m}
}, 
\label{LL_PBC_Type2_2}  
\end{align}
where the momentum of the excitation $|p| \leq \rho \pi$.

It turns out that in order to describe this kind of excitations in the limit $N\rightarrow \infty$, it is enough to remove a single particle from the $N$-particle ground state. In this way we create a hole in the initial set of the parameterizing numbers $\{I_j^\text{p}\}_\text{GS}=\left\{-\frac{N-1}{2},-\frac{N-3}{2},\ldots, \frac{N-1}{2}\right\}$, so that the final parameterization of the $N-1$-particle excited state reads $\{\widetilde{I}_j^\text{p}\}_{II}\!=\!\left\{-\frac{N-2}{2},\ldots, \frac{N}{2}\!-\!M\!-\!1,\times ,\frac{N}{2}\!-\!M\!+\!1,\ldots, \frac{N}{2}\right\}$, where $\times$ denotes the missing parameterizing number and the corresponding change of momentum is equal to $p=\frac{2\pi}{L}M$. This is the reason why the type--II excitations are often called \textbf{hole excitations} \cite{Lieb2,Franchini}. Proceeding similarly as in the case of the type--I excitations 
we can find \cite{Lieb2,Franchini}
\vspace{-0.2cm}
\begin{align} 
\displaystyle{
2\pi \mathcal{J}_{II}^\text{p}(k)=\pi +\theta(k-q)+\int\limits_{-k_F}^{k_F}\frac{2\bar{c}\,\mathcal{J}_{II}^\text{p}(\mu)\mathrm{d}\mu}{\bar{c}^2\!+\!(k\!-\!\mu)^2}
},
\label{LL_PBC_Type2_Therm_1}  
\\
\begin{array}{l}
\displaystyle{
p=\hbar\left[-q+\int\limits_{-k_F}^{k_F}\mathcal{J}_{II}^\text{p}(\mu)\mathrm{d}\mu\right] 
},\vspace{0.1cm}\\
\displaystyle{
\epsilon_{II}=\frac{\hbar^2}{2m}\left[-q^2+2\int\limits_{-k_F}^{k_F}\mu\mathcal{J}_{II}^\text{p}(\mu)\mathrm{d}\mu  \right]+\mu_\text{ch}
},
\end{array}
\label{LL_PBC_Type2_Therm_2}  
\end{align}
where $q<k_F$ is the hole quasimomentum. In contrast to the case of the type--I excitation, here the response function $\mathcal{J}_{II}^\text{p}(k)=L\Delta k D_\text{p}(k)=\omega(k)D_\text{p}(k)$ is positive definite. It means that after a type--II excitation the quantity $\Delta k=k'-k$  is positive. Therefore, the hole excitation corresponding to a positive momentum increases the values of quasimomenta. A detailed analysis leads to an observation that in the two limiting cases of the repulsive interactions the type--II spectrum reduces to $\epsilon_{II}(p)|_{\gamma=0}=0$ and $\epsilon_{II}(p)|_{\gamma=\infty}=-\frac{p^2}{2m}+\frac{\hbar \rho \pi |p|}{m}$. In general, as in the type--I excitation case, the response function $\mathcal{J}_{II}^\text{p}(k)$ in Eq.~(\ref{LL_PBC_Type2_Therm_1}) and then the type--II spectrum, Eq.~(\ref{LL_PBC_Type2_Therm_2}), have to be found numerically (see also Ref.~\cite{Lieb2}).



The type--II excitations in the periodic Lieb-Liniger model turn out to correspond to the lowest energy eigenstates for a given nonzero total momentum \cite{Kanamoto2008,Kanamoto2010}. In literature such states are dubbed {\it yrast} states \cite{Siemens1987,Bertsch1999}. The name comes from the Swedish word "yr", which means "dizzy". Hence, literally, the {\it yrast} state is the "dizziest" one \cite{Siemens1987}.

The nature of the hole ({\it yrast}) excitations was a subject of a long discussion, triggered 40 years ago, when it was noticed that in the weak interaction regime the type--II excitation spectrum coincides with the mean-field dark soliton dispersion relation \cite{Kulish76,Ishikawa80}. Such an observation provoked the investigations aimed at understanding the relationship between the type--II eigenstates and dark solitons \cite{Kanamoto2008,Kanamoto2010, Komineas2002,Jackson2002,Karpiuk2012,Karpiuk2015,Sato2012,
Sato2012arxiv,Sato2016,Gawryluk2017}. It turns out that dark soliton signatures emerge in the course of the measurement of particle's positions, if the system is initially prepared in the type--II eigenstate \cite{Syrwid15,Syrwid16}. This unequivocal connection between the {\it yrast} eigenstates and dark solitons will be discussed in details in Sec.~\ref{QuantumSolitonsInMBstates}.

\vspace{0.15cm}
\underline{\emph{\textbf{Umklapp excitation}}}

\begin{figure}[h!]
\begin{center}\includegraphics[scale=0.218]{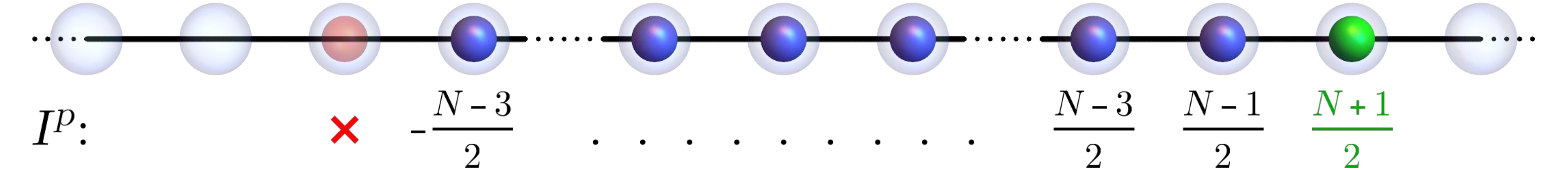} \end{center}
\vspace{-0.7cm}
\begin{center}\includegraphics[scale=0.218]{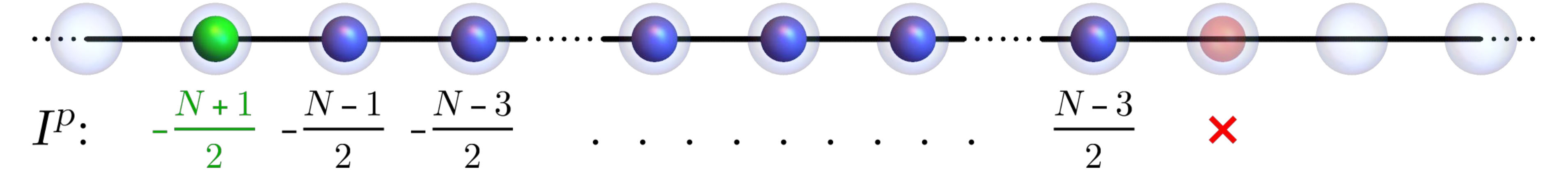} \end{center}
\vspace{-0.3cm}
\caption{
Graphical representation of the Umklapp excitation. The upper (lower) scheme presents the excitation with positive (negative) momentum. The parameterizing numbers are given in the Bethe-like formulation. 
}
\label{f7}
\end{figure}

Let us now imagine that one takes the quasimomentum occupying one of the Fermi edges, i.e. \linebreak $k=\pm k_F$, and creates an excitation by expelling this quasimomentum just above the opposite Fermi edge, i.e. $\mp k_F$ \cite{Lieb2}. The so-called Umklapp process on the level of the state parameterization in the Bethe-like formulation is illustrated in Fig.~\ref{f7}.

The Umklapp excitation is the third elementary excitation we consider here. Note that within the type--II excitation expelling a quasimomentum $q>0$ just above $k_F$ and afterwards the Umklapp excitation to $-k_F$ we effectively realize the  excitation from $q>0$ to $-k_F$. Therefore, it is clear that the three presented types of elementary excitations  allow us to describe all the possible excitations in the system.  For convenience we extend the definition of the type--II excitation, so that the hole quasimomentum is not divided into the positive and negative regions, i.e. from now we assume that the hole excitation may take place from $|q|<k_F$ to (just above) $\pm k_F$.

 We conclude presenting the graphical representation of particle and hole excitations for $N=8$ particle Lieb-Liniger system with periodic boundary conditions. Due to the fact that there are no physical differences between excitations with $p$ and $-p$, we show the case $p>0$ only.

\begin{figure}[h!]
\vspace{-0.3cm}
\begin{center}\includegraphics[scale=0.29]{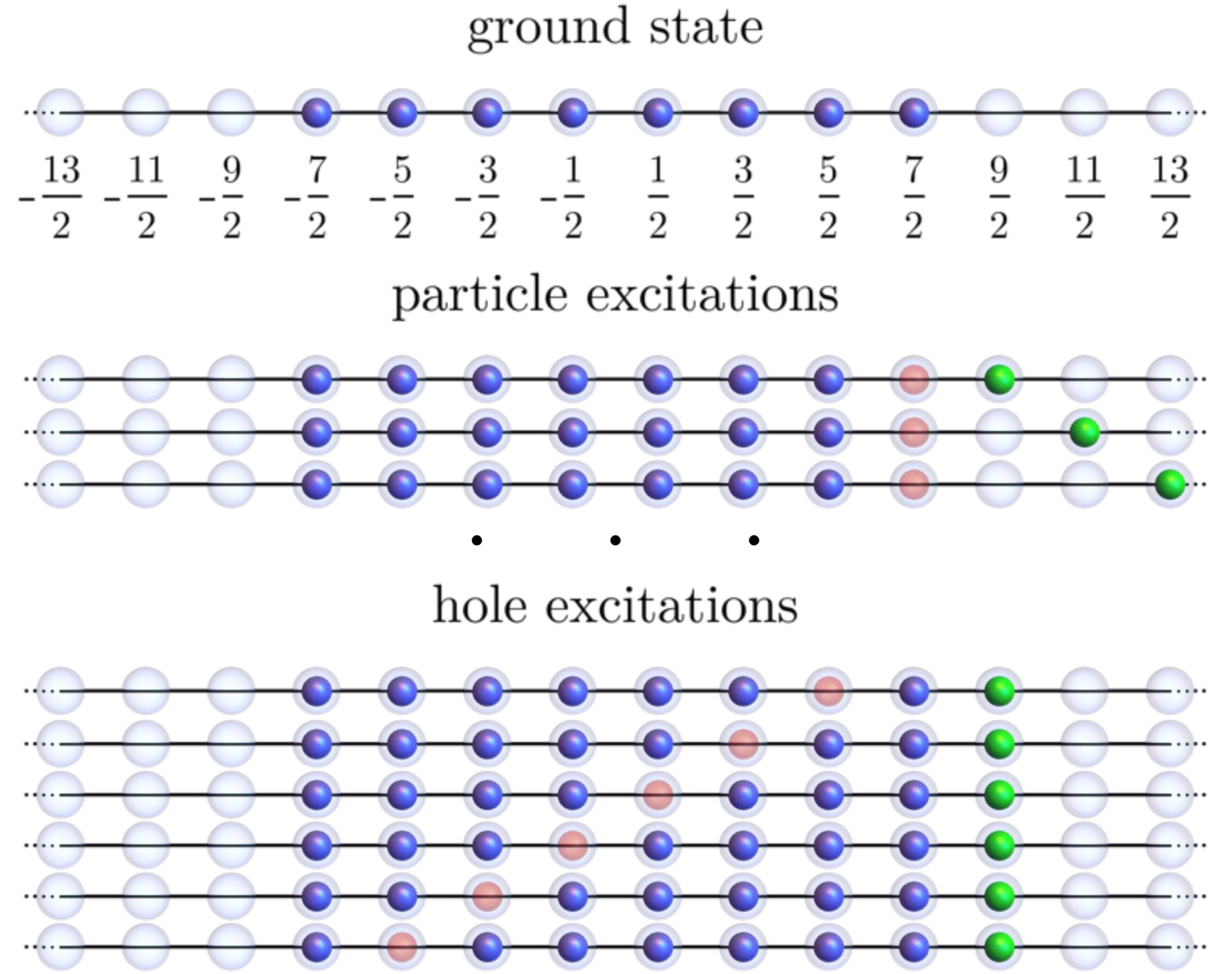}\end{center}
\vspace{-0.3cm}
\caption{ 
Graphical representations of the ground state and both the type--I (particle) and type--II (hole) excitations for the $N=8$ particle system with periodic boundary conditions in the Bethe-like parameterizing numbers formulation. 
} 
\label{f8}
\end{figure}

\subsubsection{Open boundary conditions}
\label{LLElementaryExcitations_OBC}

In contrast to the periodic case in the presence of an infinite square well potential the Fermi surface corresponds to a single positive quasimomentum $k_F$, where we choose the positive solutions of the Gaudin's equations only. In the Tonks-Girardeau limit the Fermi quasimomentum is equal to $k_F=\rho\pi$. As discussed before, there are many similarities between the systems with periodic and open boundary conditions. Here, we show that also the spectra have almost identical structure. The discussion is mainly based on Refs.~\cite{Tomchenko15,Tomchenko17_2}.   

\vspace{0.15cm}
\underline{\emph{\textbf{Type I (particle excitations)}}}

 As in the periodic case, in order to consider the type--I excitation in the problem with open boundary conditions, one excites the quasiparticle from the Fermi surface, i.e. while the Bethe-like parameterization of the ground state is given by $I_{j=1,2,\ldots,N}^\text{o}=j$, for the type--I excited eigenstate we take $\widetilde{I}_{j=1,2,\ldots,N-1}^\text{o}=j$ and $\widetilde{I}_N^\text{o}=N+r\,\, (r\geq 1)$. Hence, starting from the ground state  quasimomenta $\{k_1,k_2,\ldots, k_N=k_F\}$ we end up with $\{k'_1,k'_2,\ldots,k'_{N-1},k'_N=q\}$, with the assumption that in both cases $k_j<k_{j+1}, \, k'_j<k'_{j+1}$. A scrupulous analysis of the relevant quantity $\Delta k_j=k'_j-k_j=\omega(k_j)/L$, describing the system response to the excitation, can be found in Ref.~\cite{Tomchenko15}. The resulting energy of the type--I excitation reads
\begin{align}
\displaystyle{\epsilon_I\approx \frac{\hbar^2}{2m}\Bigg[\frac{\omega^2(k_F)}{L^2}\!+\!2 k_F \frac{\omega(k_F)}{L}\!+\!2\!\!\int\limits_{k_1}^{k_{N-1}}\!\!\! D_\text{o}(k) \omega(k) k \mathrm{d}k\Bigg],}
\label{Open_type_10_therm} 
\end{align}

In the presence of open boundary conditions the wave function is a superposition of the elementary solutions corresponding to different eigenvalues of the total momentum operator, Eq.~(\ref{LLMomentum1stQuantized}). Thus, the total momentum is not defined. Let us now try to define a quantity resembling the total momentum using only physical arguments. A free particle in the infinite square well potential can be described by a wave function $\propto \mathrm{sin}\left(\frac{\pi}{L} n^\text{o} x\right)$, where $n^\text{o}=1,2,3\ldots$ (we consider only $n^\text{o}>0$), with the corresponding absolute value of "momentum" $\frac{\hbar\pi}{L}n^\text{o}$. In the noninteracting many-body case the "total momentum" is proportional to the sum of the corresponding $n^\text{o}$ parameters. Therefore, in the ground state we expect to deal with $\hbar\pi\rho$ "momentum". According to the Gaudin's Eqs.~(\ref{GaudinEqsArctg}) one obtains
 \begin{equation}
\displaystyle{
\sum_{j=1}^{N}k_j=\frac{\pi}{L}\sum_{j=1}^N n_j^\text{o}+\frac{1}{L}\sum_{\substack{j,s=1 \\ s\neq j}}^N \mathrm{arctan} \left( \frac{\bar{c}}{k_j+k_s} \right)
}.
\label{Open_New_momentum_Def_0} 
\end{equation}
Hence, the natural definition of the new "total momentum" in the case of open boundary conditions that comes to mind reads \cite{Tomchenko17_2}
 \begin{eqnarray}
 \begin{array}{lll}
\wp&=&\displaystyle{\hbar \Bigg[\sum_{j=1}^{N}k_j-\frac{1}{L}\sum_{\substack{j,s=1 \\ s\neq j}}^N \mathrm{arctan} \left( \frac{\bar{c}}{k_j+k_s} \right)\Bigg]}
\\
&=&\displaystyle{\hbar\frac{\pi}{L}\sum_{j=1}^N n_j^\text{o}=\hbar\frac{\pi}{L}\sum_{j=1}^N (I^\text{o}_j-j+1).
}
\end{array}
\label{Open_New_momentum_Def_1} 
\end{eqnarray}
Note that $\wp$ satisfies the required quantization rule and for the ground state gives $\hbar \pi \rho$. Within the definition  in Eq.~(\ref{Open_New_momentum_Def_1}), the "momentum" of the type--I excitation is  equal to 
 \begin{align}
\displaystyle{
p=\wp\left(\{\widetilde{I}^\text{o}\}_I\right)-\wp\Big(\{I^\text{o}\}_\text{GS}\Big)=\frac{\hbar\pi}{L}r
},
\label{Open_New_momentum_Of_Type_I} 
\end{align}
where $r$ is a positive integer.

	Recently, M. Tomchenko pointed out (see Ref.~\cite{Tomchenko17_2}) that the equality between velocities of sound calculated basing on the microscopic and macroscopic approaches, i.e. $v_s^\text{mic}=v_s^\text{th}$, can be reached when one employs the "total momentum" in Eq.~(\ref{Open_New_momentum_Def_1}) and is strongly violated when one takes into account the definition in Eq.~(\ref{LLMomentum1stQuantized}). Moreover, for $\gamma \leq 1$ the dispersion relation of the type--I excitation closely follows the corresponding Bogoliubov law  $\epsilon_B(p)=\frac{p}{2m}\sqrt{p^2+4\hbar^2\rho^2\gamma}$ \cite{Tomchenko17_2,Bogolyubov47}, when one uses the "momentum" of the excitation defined as in Eq.~(\ref{Open_New_momentum_Def_1}).
	
	In the Tonks-Girardeau regime for the type--I excitation one gets $\omega(k<k_N)=0$ and $\omega(k_N)=\pi r=p L/\hbar$. Thus, according to Eq.~(\ref{Open_type_10_therm}), the dispersion relation reads $\epsilon_I=\frac{p^2}{2m}+\frac{\hbar k_F p}{m}$ (where for $\gamma\rightarrow \infty$: $k_F=k_N=\pi\rho$), which coincides with the type--I spectrum of the periodic Lieb-Liniger gas with  infinitely strong interparticle repulsion.

\vspace{0.15cm}
\underline{\emph{\textbf{Type II (hole excitations)}}}	
	
 Such an excitation relies on the transfer of a single quasimomentum from the Fermi sea, i.e. $k<k_F$, just above the Fermi surface $k_F$. After the excitation the corresponding eigenstate is parameterized as follows $\widetilde{I}_{j\leq r}^\text{o}=j, \, \widetilde{I}_{j> r}^\text{o}=j+1$. Therefore, we have a hole in the parameterizing collection, which should be somehow reflected in the values of $\Delta k_j=k'_j-k_j=\omega(k_j)/L$, in particular for $j> r$. Thus,  the key equations that have to be investigated, read 
 \begin{eqnarray}
 \begin{array}{ll}
\Delta k_j L - \pi h_{j, r} \\
\displaystyle{= \sum_{\substack{s=1 \\ s\neq j}}^N\bigg[\mathrm{arctan}\bigg( \frac{k_j+k_s}{\bar{c}} \bigg)+\mathrm{arctan}\bigg( \frac{k_j-k_s}{\bar{c}} \bigg)\bigg]}
\\
\displaystyle{-\sum_{\substack{s=1 \\ s\neq j}}^N\bigg[\mathrm{arctan}\bigg( \frac{k'_j+k'_s}{\bar{c}} \bigg)+\mathrm{arctan}\bigg( \frac{k'_j-k'_s}{\bar{c}} \bigg)\bigg],
}
\end{array}
\label{Open_type_2_therm_1} 
\end{eqnarray} 
where $h_{j,r} = 1$ for $j>r$ and 0 otherwise. 
 We expect that $\omega(k_{j> r})$ are significantly larger than $\omega(k_{j\leq r})$. For large $\gamma$ the parameters $\omega$ should be almost equal in distinct ranges of $j$, i.e. for $j\leq r$ and for $j> r$.  Having in mind the definition in Eq.~(\ref{Open_New_momentum_Def_1}), one notices that the "momentum" of such an excitation is equal to $p=\hbar\frac{\pi}{L}(N-r)$. 

It is easy to find that in the Tonks-Girardeau regime ($\gamma\rightarrow \infty$) $\omega(k_{j\leq r})=0$ and $\omega(k_{j>r})=\pi$.
 Thus, the energy of the type--II excitation reads ($k_F=\pi\rho$)
 \begin{align}
\displaystyle{
\lim\limits_{\gamma\rightarrow \infty}\epsilon_{II}= -\frac{p^2}{2m}+\frac{\hbar k_F p}{m}
},
\label{Open_type_2_therm_2} 
\end{align}
 and is identical to the type--II dispersion relation obtained for the strongly interacting ($\gamma\rightarrow \infty$) periodic system. This coincidence is not accidental and, similarly to the periodic case, the type--II excitations in the presence of open boundary conditions are strictly related to quantum dark solitons \cite{Syrwid2017HW} (see  Sec.~\ref{QuantumSolitonsInMBstates}).

\begin{figure}[h]
\begin{center}\includegraphics[scale=0.28]{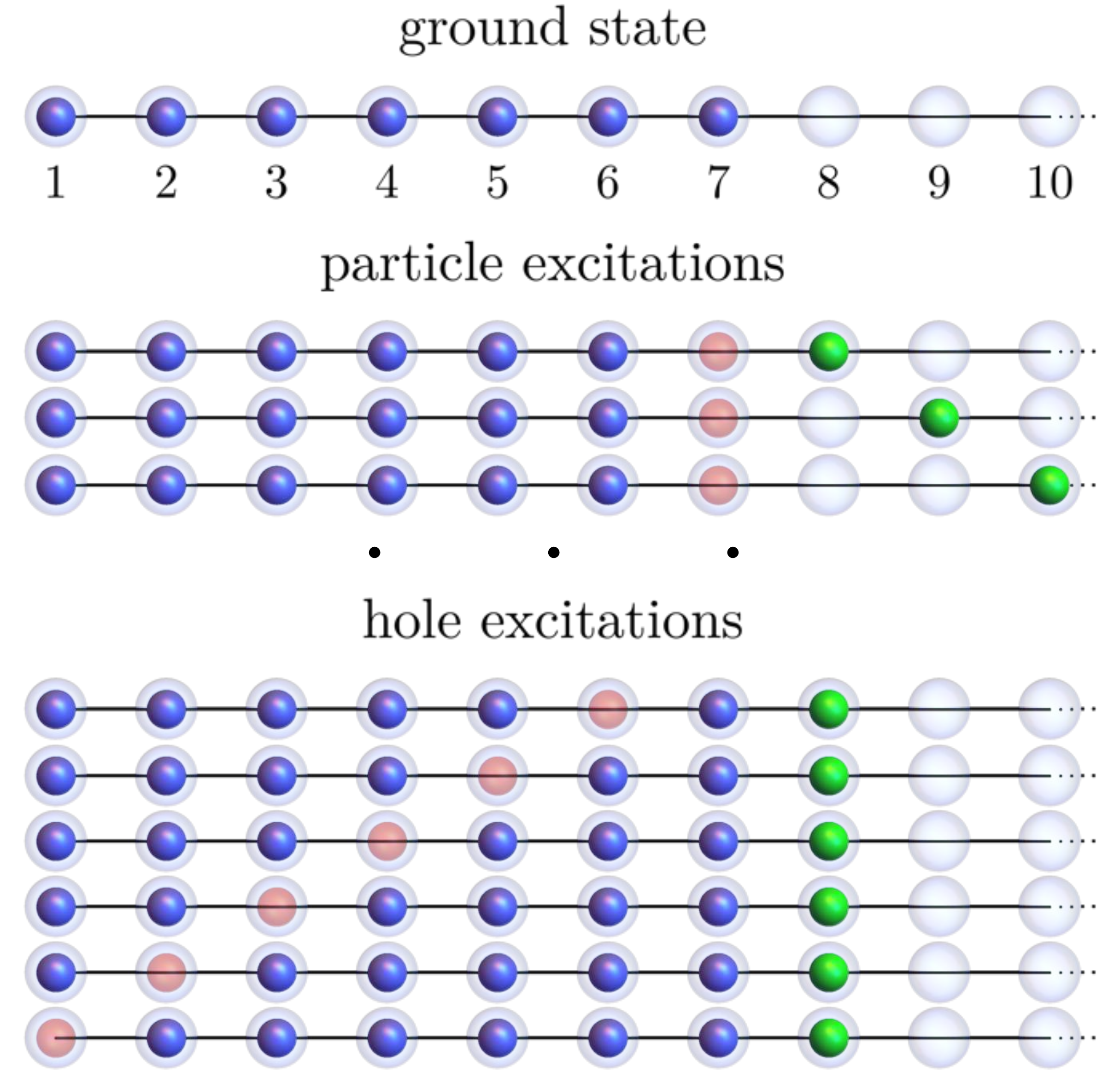}\end{center}
\vspace{-0.3cm}
\caption{ 
Graphical representations of the ground state and the two types of the elementary excitations for the $N=7$ particle Lieb-Liniger system with open boundary conditions in the Bethe-like parameterization. 
}\vspace{-0.cm}
\label{f9}
\end{figure}

 As before, we conclude with the graphical representation (Fig.~\ref{f9}) of both types of elementary excitations in the considered system.

\section{Yang-Gaudin model: Ultracold two-component Fermi gas in 1D  }
\label{Gaudin-Yang_model}

Let us now consider a one-dimensional nonrelativistic ultracold Fermi gas consisting of $N$ fermions of equal mass $m$ with two internal degrees of freedom, denoted as  two distinct spin projections $\sigma=\downarrow,\uparrow$. In general, we can assume that we deal with $N_\downarrow$ ($N_\uparrow$) particles belonging to spin $\downarrow$ ($\uparrow$) component, where $N=N_\downarrow +N_\uparrow$. Moreover, if the interparticle interaction can be restricted to $s$--wave scattering only, then it can be modeled by a point-like $ \delta$--potential as in the case of the Lieb-Liniger model. Due to the Pauli exclusion rule such scattering may occur only between fermions with different internal degrees of freedom (spin projections). Such a system is described by the Yang-Gaudin Hamiltonian \cite{Gaudin_BetheWF,Batchelor2006,Takahashi99,Yang67,Gaudin67,Guan2013,Recher2013,
Recher2013a, ShamailovPhD, SyrwidFermi2018,Shamailov2016}
\begin{align}
\begin{array}{l}
\displaystyle{\mathcal{H}^{Y-G}_{N_\downarrow,N_\uparrow}=-\frac{\hbar^2}{2m}\Bigg[ \sum_{j=1}^{N_{\downarrow}}\frac{\partial^2}{\partial x_j^{\downarrow 2}} + \sum_{s=1}^{N_{\uparrow}}\frac{\partial^2}{\partial x_s^{\uparrow 2}} \Bigg]} 
\\ 
\qquad\qquad\qquad\quad\displaystyle{+ 2c\sum_{j=1}^{N_\downarrow}\sum_{s=1}^{N_\uparrow} \delta(x_j^\downarrow -x_s^\uparrow).}
\end{array}
\label{YG_hamiltonian}
\end{align}
There is no spin-flipping term in the Hamiltonian so the particles belonging to different spin components can be distinguished. $N_\downarrow$ and $N_\uparrow$  are conserved quantities as well as the total momentum defined as follows
\begin{align}
\displaystyle{
\mathcal{P}_{N_\downarrow,N_\uparrow}=-i\hbar \sum_{j=1}^{N_\downarrow}\frac{\partial}{\partial x_j^\downarrow} -i\hbar \sum_{j=1}^{N_\uparrow}\frac{\partial}{\partial x_j^\uparrow}
}. 
\label{YG_totalmomentum}
\end{align}

 Similarly to the case of the Lieb-Liniger model, it is convenient to use the notation with $\bar{c}$ parameter defined in Eq.~(\ref{cbar}). The key quantity that measures the strength of interparticle interactions again reads 
\begin{align}
\displaystyle{
\gamma=\frac{\bar{c}}{\rho}
},
\label{YG_gamma}
\end{align}
where $\rho=\frac{N_\downarrow+N_\uparrow}{L}$ denotes the particle density for a system of size $L$.

The Yang-Gaudin model describes another many-body system, for which the Bethe ansatz method \cite{Bethe1} can be successfully applied. Nevertheless, in comparison to the case of the Lieb-Liniger model (identical indistinguishable bosons), in the presence of internal degrees of freedom the Bethe ansatz procedure has to be generalized (nested Bethe ansatz). The model we are going to analyze in details was initially solved for one and two fermionic impurities in a Fermi gas, i.e. for $N_\downarrow=N-1$, $N_\uparrow =1$ and for $N_\downarrow=N-2$, $N_\uparrow =2$ \cite{mcguire1965,mcguire1966,FlickerLieb1967}. In 1967 C. N. Yang and M. Gaudin constructed the solution that is valid for arbitrary  numbers $N_\downarrow$ and $N_\uparrow$ \cite{Yang67,Gaudin67}.  It should be stressed that the problem of a multicomponent ($M\geq2$ internal degrees of freedom) Fermi gas with contact interactions can also be solved analytically with the help of the nested Bethe ansatz. The idea of the construction of a general solution is presented in Ref.~\cite{Batchelor2006}. 

Here we focus on the case of the spin--$\frac{1}{2}$ Fermi system only. 
In order to construct the eigenvalues of the Yang-Gaudin Hamiltonian, Eq.~(\ref{YG_hamiltonian}),  one considers different sectors of particles' positions in which any two particles cannot meet. When dealing with $N=N_\downarrow+N_\uparrow$ fermions we have $N!$ possibilities of the particle ordering which can be assigned to the permutations $Q$ belonging to the permutation group $\mathcal{S}_N$, i.e.  
\begin{align}
\displaystyle{
0\leq X_{Q(1)}<X_{Q(2)}<\ldots < X_{Q(N-1)}\leq L   
},
\label{YG_ordering}
\end{align}
where $X_i =x_i^{\sigma_i}, \,\, \sigma_i=\downarrow,\uparrow, \,\, i=1,2,\ldots,N$.  Due to the fact that the particle interaction takes place only when $x_j^\downarrow=x_s^\uparrow$, one can use the key idea of the Bethe ansatz: the eigenfunction of the Hamiltonian in Eq.~(\ref{YG_hamiltonian}) can be represented as a superposition of plane waves with the coordinate independent amplitudes, chosen so that the boundary conditions related to the contact potential are satisfied. Thus, one may start with the following ansatz wave function
\begin{eqnarray}
\Psi_Q(\{x^\downarrow\}_{N_\downarrow},\{x^\uparrow\}_{N_\uparrow},\{k\}_N,\{\Lambda\}_{N_\uparrow})
\label{YG_Wf_Old_1}
\\ \nonumber
\propto \hspace{-0.2cm}
\sum_{\substack{\pi\in\mathcal{S}_N \\ \,\,\, \tau\in\mathcal{S}_{N_\uparrow}}} \hspace{-0.2cm}
A_Q(\pi,\tau,\{k\}_N,\{\Lambda\}_{N_\uparrow})\,\mathrm{exp}\left(i\sum_{j=1}^N k_{\pi(j)}X_{Q(j)}\right),
\end{eqnarray}
where without loss of generality we  assumed that $N_\downarrow\geq N_\uparrow$. The Yang-Gaudin eigenstates are uniquely parameterized by the set of the \emph{quasimomenta} $\{k_j\}_{j=1,\ldots,N}$ and auxiliary parameters $\{\Lambda_s\}_{s=1,\ldots,N_\uparrow}$ called \emph{spin-roots}, which are required to describe a system with two internal degrees of freedom. Note that in order to find the wave function explicitly, one needs to determine $N!^2N_\uparrow!$ amplitudes $A_Q(\pi,\tau)$  for all possible permutations $Q,\pi$ and $\tau$. Although these amplitudes are not independent, the task is a considerable challenge. The arduous details of the derivation can be found in Refs.~\cite{Takahashi99,Yang67,Gaudin67,Recher2013}. The resulting wave function is assumed to be antisymmetric in $\{k\}$, $\{\Lambda\}$ and also in $\{x^{\sigma=\downarrow, \uparrow}\}$. The last of the antisymmetry properties entails vanishing of the wave function when $x^\sigma_j=x^\sigma_s$ for $j\neq s$, which underpins the statement that the contact potential is invisible for particles belonging to the same spin component.

It is clear that the level of complication dramatically increases when we deal with a multicomponent gas instead of a single component one. Note that in order to use the full wave function, it is required to perform the summation of very cumbersome mathematical expressions over all elements of the permutation groups $\mathcal{S}_N$ and $\mathcal{S}_{N_\uparrow}$, for every single $Q\in\mathcal{S}_N$. Hence, although the concept of the Bethe ansatz is rather easy, the resulting form of solution is immensely intricate and almost useless in real calculations. 

The cumbersome structure of the eigenstates was a kind of an impediment to scientific progress in the investigations of the Yang-Gaudin system. Fortunately, in 2013 C. Recher and H. Kohler showed that the eigenstates in question can be cast into a much more compact determinant form \cite{Recher2013,Recher2013a}. That is, in the single component case, i.e. $N_\downarrow=N, \, N_\uparrow =0$, the interparticle potential does not play any role and the wave function is given by a Slater determinant. Considering the general case, in which  $N_{\downarrow}$ and $N_{\uparrow}$ are arbitrary and the interparticle interactions are present, Recher and Kohler realized that instead of the very complicated superposition of plane waves, one can look for the solution, being a superposition of the determinants of the plane waves.
 This brilliant idea turned out to be very fruitful and led to  reformulation of the eigenfunctions of the Yang-Gaudin Hamiltonian, Eq.~(\ref{YG_hamiltonian}). Indeed, the eigenstates in question can be written as follows (see Refs.~\cite{Recher2013,Recher2013a})
\begin{eqnarray}
\displaystyle{
\begin{split}
\Psi(\{x^\downarrow\}_{N_\downarrow},\{x^\uparrow\}_{N_\uparrow},&\{k\}_N,\{\Lambda\}_{N_\uparrow})\\
&\propto \hspace{-0.1cm}
\sum_{\pi\in\mathcal{S}_{N_\uparrow}} \mathrm{sgn}(\pi)\, \mathcal{W}_{\pi}^\uparrow \,\, \mathrm{det} \, \Phi_{N\times N},
\end{split}
}
\label{YG_Wf_New_1}
\end{eqnarray}
where 
\begin{align}
\displaystyle{
 \mathcal{W}_{\pi}^\uparrow =\prod_{j<s}^{N_\uparrow}\left[i\big(\Lambda_{\pi(j)}-\Lambda_{\pi(s)}\big)
 +\bar{c}\, \mathrm{sign}\big(x^\uparrow_s -x^\uparrow_j\big)
 \right]
}.
\label{YG_Wf_New_2}
\end{align}
The $N\times N$ matrix $\Phi_{N\times N}$ is constructed from two rectangular matrices ($N\times N_\downarrow$ and $N\times N_\uparrow$) that, in our notation, are separated by the vertical bar
\begin{eqnarray}
\displaystyle{
\begin{split}
 \Phi&_{N\times N}=
 \Bigg( \bigg[ \prod_{s=1}^{N_\uparrow}\mathcal{A}_j\big(\Lambda_{\pi(s)},x_l^\downarrow-x_s^\uparrow\big)\mathrm{e}^{i k_j x_l^\downarrow}\bigg]
 \\ 
 &\,\Bigg|
\bigg[ \prod_{s\neq m}^{N_\uparrow}\mathcal{A}_j\big(\Lambda_{\pi(s)},x_m^\uparrow-x_s^\uparrow\big)\mathrm{e}^{i k_j x_m^\uparrow}\bigg]
 \Bigg)_{\substack{j=1,\ldots, N=N_\uparrow+N_\downarrow\\
                  l=1,\ldots, N_\downarrow\\
                  m=1,\ldots, N_\uparrow}},
\end{split}
}
\label{YG_Wf_New_3}
\end{eqnarray}
with
\begin{align}
\displaystyle{
 \mathcal{A}_j(\Lambda, x)=i(k_j-\Lambda)+\frac{\bar{c}}{2}\,\mathrm{sign}(x)
}.
\label{YG_Wf_New_4}
\end{align}
It is easy to show that the wave function $\Psi$ satisfies the required symmetry properties, namely
\begin{eqnarray}
 \Psi(&\rho_\downarrow\{x^\downarrow\}_{N_\downarrow},\{x^\uparrow\}_{N_\uparrow},\{k\}_N,\{\Lambda\}_{N_\uparrow})
 \label{YG_eigs_sym_1}
 \\ \nonumber
 &=\mathrm{sgn}(\rho_\downarrow)\Psi(\{x^\downarrow\}_{N_\downarrow},\{x^\uparrow\}_{N_\uparrow},\{k\}_N,\{\Lambda\}_{N_\uparrow}),
\\  
 \Psi(&\{x^\downarrow\}_{N_\downarrow},\rho_\uparrow\{x^\uparrow\}_{N_\uparrow},\{k\}_N,\{\Lambda\}_{N_\uparrow})
\label{YG_eigs_sym_2} 
 \\ \nonumber
 &=\mathrm{sgn}(\rho_\uparrow)\Psi(\{x^\downarrow\}_{N_\downarrow},\{x^\uparrow\}_{N_\uparrow},\{k\}_N,\{\Lambda\}_{N_\uparrow}),
\\
 \Psi(&\{x^\downarrow\}_{N_\downarrow},\{x^\uparrow\}_{N_\uparrow},\nu\{k\}_N,\{\Lambda\}_{N_\uparrow})
 \label{YG_eigs_sym_3}
 \\ \nonumber
 &=\mathrm{sgn}(\nu)\Psi(\{x^\downarrow\}_{N_\downarrow},\{x^\uparrow\}_{N_\uparrow},\{k\}_N,\{\Lambda\}_{N_\uparrow}),
\\
 \Psi(&\{x^\downarrow\}_{N_\downarrow},\{x^\uparrow\}_{N_\uparrow},\{k\}_N,\eta\{\Lambda\}_{N_\uparrow})
 \label{YG_eigs_sym_4}
 \\ \nonumber
 &=\mathrm{sgn}(\eta)\Psi(\{x^\downarrow\}_{N_\downarrow},\{x^\uparrow\}_{N_\uparrow},\{k\}_N,\{\Lambda\}_{N_\uparrow}).
\end{eqnarray}
where $\rho_\sigma\in  \mathcal{S}_{N_\sigma} \, (\sigma=\downarrow,\uparrow), \, \nu \in \mathcal{S}_N, \, \eta \in \mathcal{S}_{N_\uparrow}$. Thanks to the appearance of $\mathrm{sign}\big(x_l^{\sigma_l}-x_s^{\sigma_s}\big)$ functions, the new expression for the eigenstates, Eq.~(\ref{YG_Wf_New_1}), is valid for any particle order, Eq.~(\ref{YG_ordering}). Hence, we do not have to be concerned this problem anymore. Note that when $N_\uparrow =0$ the solution, as expected, turns into the Slater determinant describing a noninteracting Fermi gas, because the summation and the right submatrix disappear. In this case $\Lambda$ parameters are not present and  $\mathcal{A}_j(\Lambda, x)\rightarrow\mathrm{const}$.
The fact that it is possible to incorporate the very cumbersome summations over $\mathcal{S}_N$ that are present in Eq.~(\ref{YG_Wf_Old_1}) into the determinants, allows us to investigate the eigenstates of the Yang-Gaudin model much more effectively. Nevertheless, the physical meaning of the determinants of the $\Phi_{N\times N}$ matrices is still intractable to be fully unscrambled.

Since the Yang-Gaudin Hamiltonian in Eq.~(\ref{YG_hamiltonian}) commutes with the total momentum operator in Eq.~(\ref{YG_totalmomentum}), the following eigenequations have to be simultaneously satisfied 
\begin{eqnarray}
 \mathcal{H}^{YG}_{N_\downarrow,N_\uparrow}\Psi(\{x^\downarrow\}_{N_\downarrow},\{x^\uparrow\}_{N_\uparrow},\{k\}_N,\{\Lambda\}_{N_\uparrow})  \label{YG_eigs_sim1}
 \\ \nonumber
\qquad =E_{N_\downarrow,N_\uparrow}\Psi(\{x^\downarrow\}_{N_\downarrow},\{x^\uparrow\}_{N_\uparrow},\{k\}_N,\{\Lambda\}_{N_\uparrow}),
\\
 \mathcal{P}^{YG}_{N_\downarrow,N_\uparrow}\Psi(\{x^\downarrow\}_{N_\downarrow},\{x^\uparrow\}_{N_\uparrow},\{k\}_N,\{\Lambda\}_{N_\uparrow})  \label{YG_eigs_sim2}
 \\ \nonumber
 \qquad=P_{N_\downarrow,N_\uparrow}\Psi(\{x^\downarrow\}_{N_\downarrow},\{x^\uparrow\}_{N_\uparrow},\{k\}_N,\{\Lambda\}_{N_\uparrow}),
\end{eqnarray}
with the eigenvalues
\begin{align}
\displaystyle{
 E_{N_\downarrow,N_\uparrow}=\frac{\hbar^2}{2m}\sum_{j=1}^{N}k_j^2, \qquad P_{N_\downarrow,N_\uparrow}=\hbar\sum_{j=1}^{N}k_j
}. 
\label{YG_eigvals}
\end{align}
Note that the auxiliary parameters $\{\Lambda\}_{N_\uparrow}$ that are present due to the existence of two possible spin projections, do not contribute to the physical quantities like energy and total momentum. The structure of the eigenvalues in Eqs.~(\ref{YG_eigvals}) is very similar to the case of the Lieb-Liniger model, which is due to the Bethe ansatz method. That is, the key element of the construction of solutions is an observation that all particles behave like noninteracting ones, except when at least two of them sit on top of each other. 

The problem is still analytically solvable when we switch on the trap in the form of the infinite square well potential (hard walls).
The final solution can be found in a similar manner to the case of the Lieb-Liniger model with open boundary conditions.
Details of the derivation are neatly presented in Ref.~\cite{Batchelor2006}.

\subsection{Periodic boundary conditions}
\label{FermiBoundaries}

The energy levels are determined by the quasimomenta $k_j$, whose permitted values depend on the boundary conditions imposed on the system. Here we focus on periodic boundary conditions, for which the sets of the quasimomenta $\{k\}_N$ and spin-roots $\{\Lambda\}_{N_\uparrow}$ have to satisfy the following Bethe ansatz equations \cite{Batchelor2006,Guan2013}
\begin{eqnarray}
\displaystyle{ \mathrm{e}^{i k_j L}=\prod_{n=1}^{N_\uparrow}\frac{k_j-\Lambda_n+i\frac{\bar{c}}{2}}{k_j-\Lambda_n-i\frac{\bar{c}}{2}}}, 
\label{YG_PBC_conds_1}
\\
\displaystyle{
\prod_{j=1}^{N}\frac{\Lambda_{m}-k_j+i\frac{\bar{c}}{2}}{\Lambda_{m}-k_j-i\frac{\bar{c}}{2}}=\prod_{\substack{n=1 \\ n\neq m}}^{N_\uparrow}\frac{\Lambda_m-\Lambda_n+i \bar{c}}{\Lambda_m-\Lambda_n-i \bar{c}}},
\label{YG_PBC_conds_2}
\end{eqnarray}
where  $j=1,2,\ldots,N$, $m=1,2,\ldots,N_\uparrow$ and we assumed that $N_\downarrow \geq N_\uparrow$.
At first sight one observes that the structure of the above equations recalls the Bethe equations discussed in Sec.~\ref{periodic_boundary_conditions}. Nonetheless, in the present case we need to also deal with the second level Bethe ansatz Eqs.~(\ref{YG_PBC_conds_2}) related to the spin-roots parameters. Note that when there is no spin-up particles, i.e. $N_\uparrow=0$, the Bethe ansatz equations transit to the well-known stipulations for the case of free fermions, i.e. $\mathrm{e}^{i  k_j L}=1$. 

Let us now rewrite the Bethe ansatz equations to the logarithmic form
\begin{eqnarray}
\displaystyle{
k_j L+\sum_{n=1}^{N_\uparrow}\theta(2(k_j-\Lambda_n))
=2\pi \mathcal{Y}^\text{p}_j,
}
\label{YG_PBC_conds_log_1}
\\
\sum_{j=1}^N \theta(2(\Lambda_m-k_j))-\sum_{\substack{n=1 \\ n\neq m}}^{N_\uparrow}\theta(\Lambda_m-\Lambda_n)  
=2\pi J_m^\text{p}.
\label{YG_PBC_conds_log_2}
\end{eqnarray}
 Again, we need to deal with the scattering phase function $\theta$, Eq.~(\ref{ThetaFunction}), but with the quasimomenta and spin-roots mixed as arguments. The parameterizing numbers $\mathcal{Y}^\text{p}_j$ ($J_m^\text{p}$) are integers when $N_\uparrow$ ($N_\downarrow$) is even (odd) and half-odd integers when $N_\uparrow$ ($N_\downarrow$) is odd (even). Due to the antisymmetry relations (\ref{YG_eigs_sym_3}) and (\ref{YG_eigs_sym_4}) all the parameterizing numbers have to be distinct. Otherwise, two (or more) solutions $k_j$ or $\Lambda_m$ may be equal implying unphysical vanishing of the wave function in the entire accessible space. 
 	
Now, we briefly analyze the regimes of very weak and very strong interactions. Formally, we expand $\theta$ in $\bar{c}$ keeping $N,L$ finite and fixed. For $\bar{c}\rightarrow 0$, $\theta(\xi)\approx -\pi -2\bar{c}/\xi+2\pi r$ ($r\in\mathbb{Z}$), where it is assumed that $|\xi|\gg|\bar{c}|$. In the strongly repulsive limit ($\bar{c}, \gamma \rightarrow \infty$) one expects that $|\xi|\ll \bar{c}$, which leads to $\theta(\xi)\approx 2\xi/\bar{c}$. Such an expansion is no longer correct in the presence of strong attraction, when we anticipate the existence of bound state pairs with $\Im(k)\propto |\bar{c}|$, i.e. the string solutions analyzed for the Bose case in Sec.~\ref{periodic_boundary_conditions}.
 Extended and more rigorous discussion can be found in Ref.~\cite{Batchelor2006}.

\begin{figure}[h!] 
\includegraphics[scale=0.195]{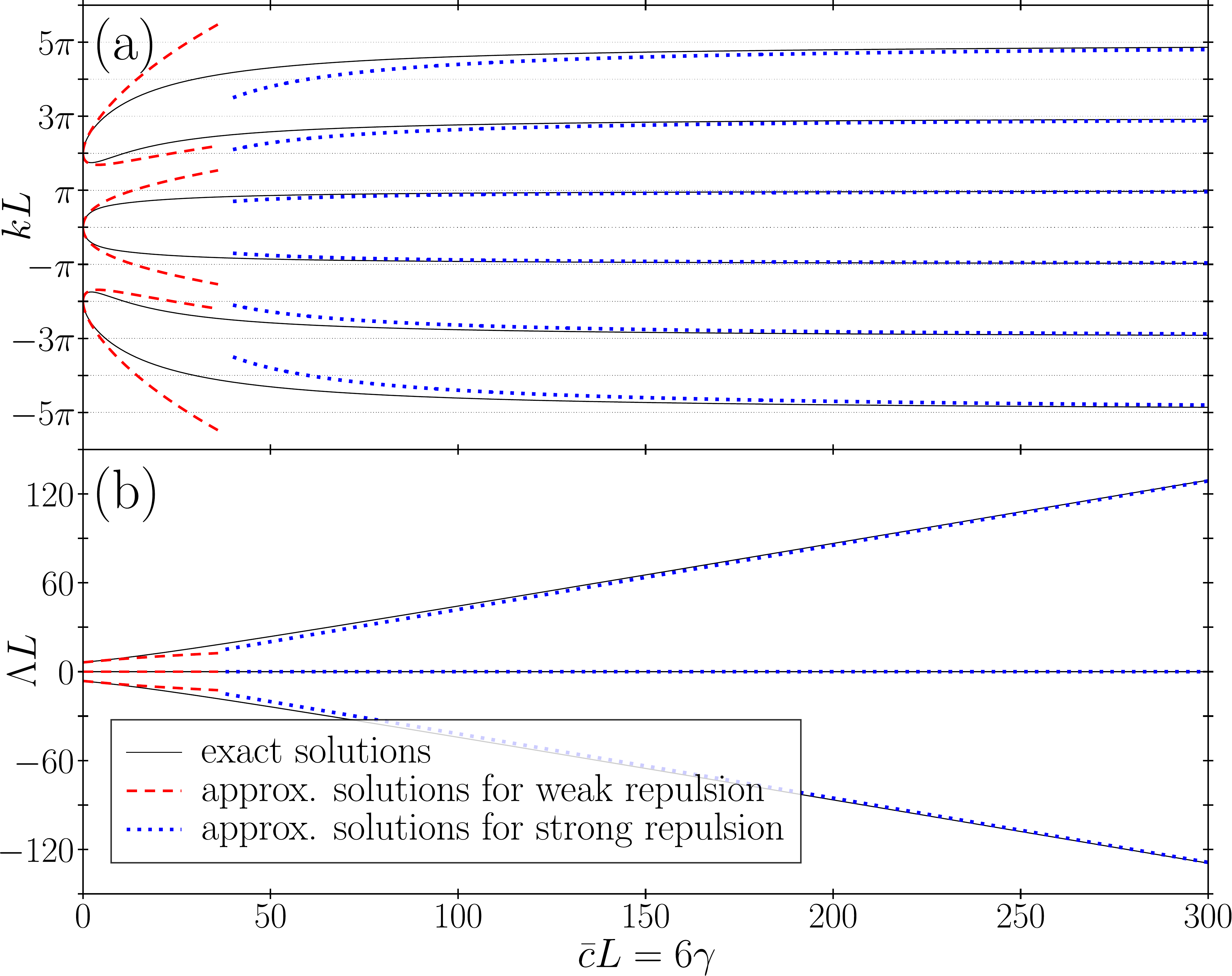}
\vspace{-0.6cm}
\caption{
Numerical solutions of the Bethe ansatz Eqs.~(\ref{YG_PBC_conds_1})--(\ref{YG_PBC_conds_2}) for $N=6$ fermions, prepared in a balanced ($N_{\downarrow,\uparrow}=3$) ground state for a wide range of $\gamma\geq0$. The exact solutions (thin black lines) are compared with the solutions obtained from the approximate equations for the weak, Eqs.~(\ref{YG_PBC_conds_log_weak_1})--(\ref{YG_PBC_conds_weak_log_2}), and strong, Eqs.~(\ref{YG_PBC_conds_log_strongatr_1}), interactions. The spin-root solutions illustrated in panel (b) follow the predictions in Eqs.~(\ref{YG_PBC_conds_log_strongatr_2}) and behave linearly with $\bar{c}$ for fixed $L$.  All the results are purely real as pointed out in the main text.
}
\label{f10}
\end{figure}

For $\gamma\rightarrow 0_{\pm}$ one finds
\begin{eqnarray}
\displaystyle{
k_j\approx\frac{2\pi}{L}m_j^\text{p}+\frac{\bar{c}}{L}\sum_{n=1}^{N_\uparrow}\frac{1}{k_j-\Lambda_n}
},
\label{YG_PBC_conds_log_weak_1}
\\
\displaystyle{
\sum_{j=1}^N\frac{\bar{c}}{k_j-\Lambda_m}+\sum_{\substack{n=1 \\ n\neq m}}^{N_\uparrow}\frac{2\bar{c}}{\Lambda_m-\Lambda_n}\approx2\pi u_m^\text{p}},
\label{YG_PBC_conds_weak_log_2} 
\end{eqnarray}
where $j=1,2,\ldots,N$, $m=1,2,\ldots,N_\uparrow$ and $m_j^\text{p},u_m^\text{p}\in \mathbb{Z}$.
In the easiest nontrivial case, i.e. for $N_{\downarrow,\uparrow} =1$ and $m_1^\text{p}=m_2^\text{p}=u_1^\text{p}=0$, one obtains  $\Lambda_1=0,\, k_1=-k_2=\pm\sqrt{\bar{c}/L}$. Note that for $\bar{c}<0$ the resulting quasimomenta become conjugate, i.e. $k_1=-k_2=\pm i\sqrt{|\bar{c}|/L}$, and correspond to a bound state  of $\downarrow$-$\uparrow$ pair of fermions. Solutions of the approximate Eqs.~(\ref{YG_PBC_conds_log_weak_1})--(\ref{YG_PBC_conds_weak_log_2}) are compared with the exact ones in Fig.~\ref{f10}.


In the strongly repulsive regime ($\gamma\rightarrow \infty$) we get
\begin{eqnarray}
\displaystyle{k_j\approx\frac{2\pi}{L}\mathcal{Y}_j^\text{p}\left( 1 - \frac{4 N_\uparrow}{\bar{c}L} \right)=\frac{2\pi}{L}\mathcal{Y}_j^\text{p}\left( 1 - \frac{4 N_\uparrow}{N\gamma} \right),
}
\label{YG_PBC_conds_log_strongatr_1}
\\
\displaystyle{
\Lambda_m=\lambda_m\bar{c}  \qquad \text{with} \qquad\lambda_m=\mathrm{const},
}
\label{YG_PBC_conds_log_strongatr_2}
\end{eqnarray}
where  the sum over $\Lambda_n/\bar{c}L$ in Eq.~(\ref{YG_PBC_conds_log_strongatr_1}) was neglected. When $N_{\downarrow,\uparrow}=1$ the ground state should correspond to zero total momentum and can be obtained for  $\mathcal{Y}^\text{p}_\pm = \pm \frac{1}{2}$ and $J_1^\text{p}=0$, where the indices $j=1,2$ are replaced by $\pm$. Thus, in such a case, the solutions read $k_\pm =\pm\frac{\pi}{L}(1-2/\gamma)$ and $\Lambda_1=0$. In general, for $\bar{c}\rightarrow \infty$ ($N,L=\mathrm{const}<\infty$) spin-roots $\Lambda_m \propto \bar{c}$, which causes that the expansion we used may not be always valid. The coefficients $\lambda_{m=1,2,\ldots,N_\uparrow}$ 
 can be determined basing on the discussion presented in Ref.~\cite{Batchelor2006}.
 Note that the resulting quasimomenta typically tend to $k_j=\frac{2\pi}{L}\mathcal{Y}_j^\text{p}$ coinciding with expectations for the problem of $N$ free fermions confined in a ring of size $L$.

\subsubsection{Strong attraction ($\gamma\rightarrow -\infty$)}
\label{YG_StrongAttr}

In the presence of the attractive interparticle interactions ($\gamma<0$),  one can anticipate  formation of bound pairs of two conjugate quasimomenta  \cite{Batchelor2006,Takahashi99,Guan2013}. Indeed, it turns out that for strong attraction one deals with string states similar to those discussed in Sec.~\ref{periodic_boundary_conditions}, i.e. bound states built up by $r=2$ quasimomenta $k_{m,\pm}$ possessing identical real parts and differing only in imaginary parts, 
which are interspaced by a corresponding spin-root parameter $\Lambda_m \in \mathbb{R}$ \cite{Batchelor2006,Takahashi99,Guan2013}. 
By substituting $k_{m,\pm}\stackrel{\bar{c}\rightarrow - \infty}{\longrightarrow} \Lambda_m\pm i \alpha\bar{c}$, with $\alpha>0$, into the Bethe ansatz equations, one finds that $\alpha=\frac{1}{2}$ and  the paired quasimomenta in the strongly attractive limit read
\begin{align}
\displaystyle{
k_{m,\pm}=\Lambda_m\pm i\frac{\bar{c}}{2}, \quad\text{where} \quad 1\leq m \leq N_\uparrow
}. 
\label{YG_P_O_string}
\end{align}
 It is assumed that any deviations from the string solutions, in Eq.~(\ref{YG_P_O_string}), vanish in the limit $\gamma\rightarrow -\infty$.
Such an approach is an essence of the so-called string hypothesis \cite{Takahashi99} and will be always applied by us in the upcoming discussion. Note that in the polarized case ($N_\downarrow>N_\uparrow$) or for some highly excited states in the balanced system ($N_\downarrow=N_\uparrow$), one has to also deal with the unpaired quasimomenta.  Suppose now that we consider a state for which we have $M\leq N_\uparrow$ pairs of conjugate quasimomenta. Additionally, let us assume that $k_{j=1,\ldots,M}=k_{j,+}, \, k_{j=M+1,\ldots,2M}=k_{j,-}$.  Thus, in the case of periodic boundary conditions the Bethe ansatz equations for paired quasimomenta ($j=1,2,\ldots,M\leq N_\uparrow$) can be reduced to the form consisting of real parameters only (see~\ref{appendixYang_Gaudin_String})
\begin{eqnarray}
\mathrm{e}^{i 2\Lambda_j L}=\prod_{\substack{l=M+1}}^{N_\uparrow}\frac{\Lambda_l-\Lambda_j+i\bar{c}}{\Lambda_l-\Lambda_j-i\bar{c}}  
\label{YG_string_4}
 \\ \nonumber
 \qquad\qquad\times \prod_{r=2M+1}^N\frac{k_r-\Lambda_j-i\frac{\bar{c}}{2}}{k_r-\Lambda_j+i\frac{\bar{c}}{2}}
\prod_{\substack{n=1 \\ n\neq j}}^{N_\uparrow}\frac{\Lambda_j-\Lambda_n+i\bar{c}}{\Lambda_j-\Lambda_n-i\bar{c}}.
\end{eqnarray}
The spin-roots $\Lambda_{j=1,\ldots,M}$ are in one-to-one correspondence with the paired quasimomenta $k_{j,+}=\Lambda_j+i\frac{\bar{c}}{2}$ and  $k_{j,-}=\Lambda_{j-M}-i\frac{\bar{c}}{2}$. Nevertheless,  $N+N_\uparrow-3M$ equations are still required to determine the values of the remaining unpaired quasimomenta. To do so one can use Eqs.~(\ref{YG_PBC_conds_1}) for $j\geq2M+1$ and  Eqs.~(\ref{YG_PBC_conds_2}) for $m\geq M+1$.

For $\gamma\rightarrow -\infty$ the energy associated with a pair  $k_{j,\pm}$ is proportional  to $\Lambda_j^2-\bar{c}^2/2$. Therefore, any excitation that tears such a pair apart requires an enormous amount of energy $\propto \bar{c}^2$, which leads us to the conclusion that when $\gamma\rightarrow -\infty$ ($N,L=\mathrm{const}<\infty$) none of the pairs is torn. If so, $M=N_\uparrow$ and we get the following system of logarithmic equations \cite{Yang67,Sutherland2004}
\begin{eqnarray}
\begin{array}{l}
\displaystyle{
 k_j L+\sum_{n=1}^{N_\uparrow}\theta(2(k_j-\Lambda_n))=2\pi \mathcal{Y}^\text{p}_j,   
}
\label{YG_string_hypo_1}
\end{array}
\\
\begin{array}{l}
\displaystyle{ 
2\Lambda_m L+  \sum_{r=2N_\uparrow +1}^{N}  \theta(2(\Lambda_m-k_r))}
\vspace{-0.3cm}
\\
\qquad\qquad\qquad\displaystyle{+\sum_{\substack{n=1 }}^{N_\uparrow}\theta(\Lambda_m-\Lambda_n) =2\pi \ell^\text{p}_m,}
\end{array}
\label{YG_string_hypo_2}
\end{eqnarray}
where $j=2N_\uparrow+1,2N_\uparrow+2,\ldots,N$, $m=1,2,\ldots,N_\uparrow$, $ n_m\in \mathbb{Z}$ and the parameterizing numbers $\ell_j^\text{p}$ are integers (half-odd integers) when $N-N_\uparrow +1$ is even (odd). Moreover, by the assumption, all the parameters $k_j$ and $\Lambda_m$ which appear in the above equations are real numbers. Note that within the string hypothesis we reduce the number of equations by $2N_\uparrow$.

\begin{figure}[h!] 
\includegraphics[scale=0.187]{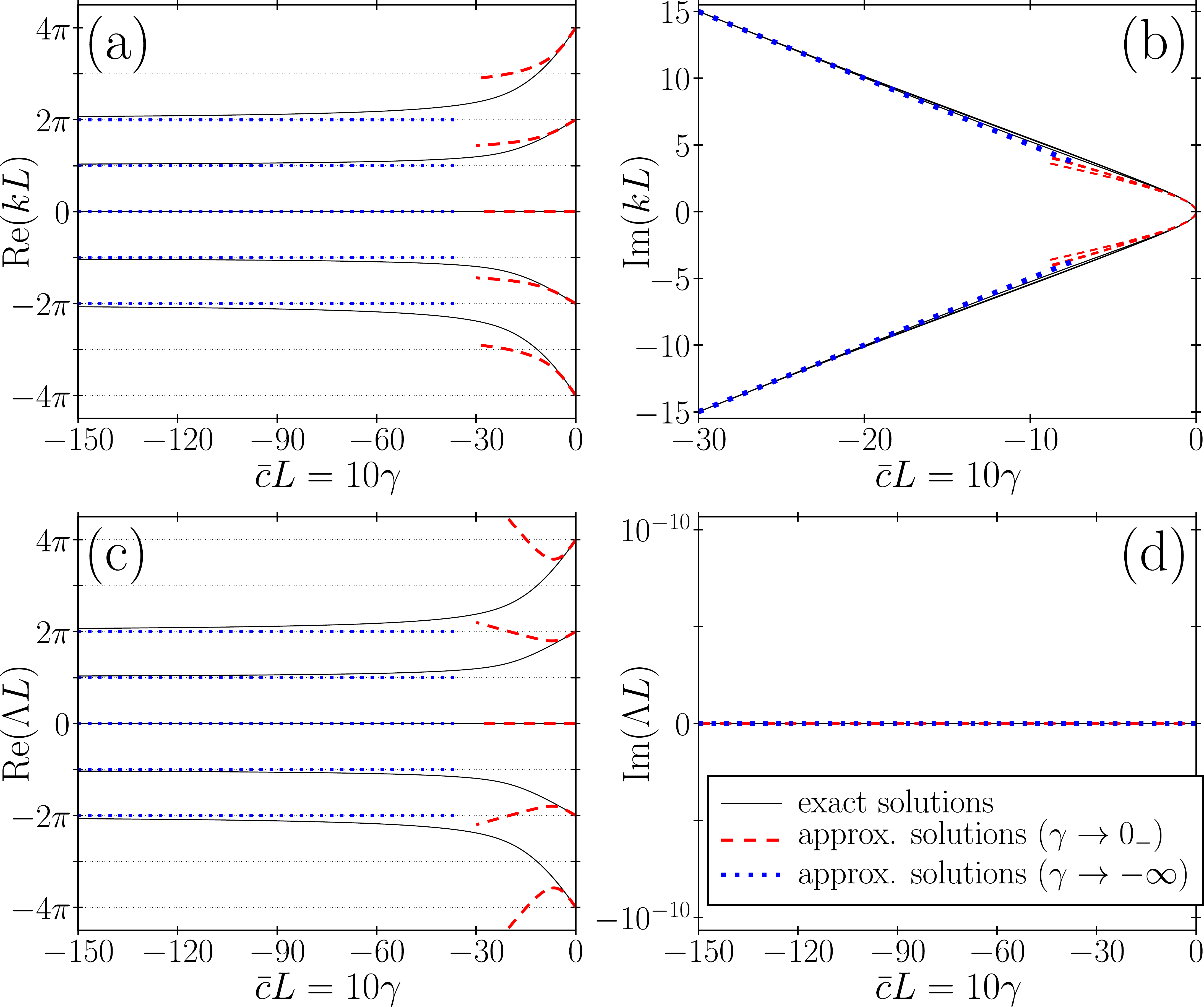} 
\vspace{-0.6cm}
\caption{  
Solutions of the Bethe ansatz Eqs.~(\ref{YG_PBC_conds_1})--(\ref{YG_PBC_conds_2}) for the attractively interacting unpolarized periodic system with $N=10$ particles ($N_{\downarrow,\uparrow}=5$) prepared in the ground state.  While panels (a) and (b) present the real and the imaginary parts of the resulting quasimomenta, the corresponding spin-root solutions are depicted in panels (c) and (d). Exact results (thin black lines) are compared with the approximate ones, obtained from Eqs.~(\ref{YG_PBC_conds_log_weak_1})--(\ref{YG_PBC_conds_weak_log_2}) and~(\ref{YG_P_O_string}) for weak (dashed red lines) and strong (dotted blue lines) attraction, respectively.  
 The spin-roots are purely real in the entire examined regime of $\gamma$.
}
\label{f11}
\end{figure}

As already mentioned, $\theta(\xi)\approx 2\xi/\bar{c}$ when $|\xi|\ll|\bar{c}|$, so the solutions of Eqs.~(\ref{YG_string_hypo_1}) and~(\ref{YG_string_hypo_2})  reduce to 
$k_j= \frac{2\pi}{L}\mathcal{Y}_j^\text{p}+\mathcal{O}\left(1/\bar{c}\right)$ and $ \Lambda_m=\frac{\pi}{L}\ell_m^\text{p}+\mathcal{O}\left(1/\bar{c}\right)$.
Thus, in the strongly attracting limit ($\gamma\rightarrow -\infty$) the formulas for the total momentum and energy read
\begin{eqnarray}
\displaystyle{
P_{N_\downarrow,N_\uparrow}=\hbar\left[2\sum_{m=1}^{N_\uparrow}\Lambda_m+\sum_{j=2N_\uparrow+1}^N k_j\right]
},
\label{YG_string_hypo_momentum}
\\
\displaystyle{
E_{N_\downarrow,N_\uparrow}=\frac{\hbar^2}{2m}\left[2\sum_{m=1}^{N_\uparrow}\left(\Lambda_m^2-\frac{\bar{c}^2}{4}\right)+\!\sum_{j=2N_\uparrow+1}^N\! k_j^2\right]
}.
\label{YG_string_hypo_energy}
\end{eqnarray}
Note that the binding energy per a conjugate pair is equal to $-\frac{\hbar^2 }{4m}\bar{c}^2$. In Fig.~\ref{f11}, we compare the exact numerical solutions of the Bethe ansatz Eqs.~(\ref{YG_PBC_conds_1})--(\ref{YG_PBC_conds_2})  for  $N_{\downarrow,\uparrow} =5$ particle ground state with the corresponding solutions of the approximate equations discussed above.

\subsection{Ground state}
\label{YG_Gs}

In this section we study and characterize the ground state of the unpolarized ($N_\downarrow=N_\uparrow$) Yang-Gaudin system in the presence of periodic boundary conditions (see also Refs.~\cite{Batchelor2006,Guan2013,Recher2013,ShamailovPhD,Shamailov2016,GuanZhong2012a,GuanZhong2012b,IidaWadati2005,IidaWadati2008}.

\begin{figure}[h!] 
\includegraphics[scale=0.185]{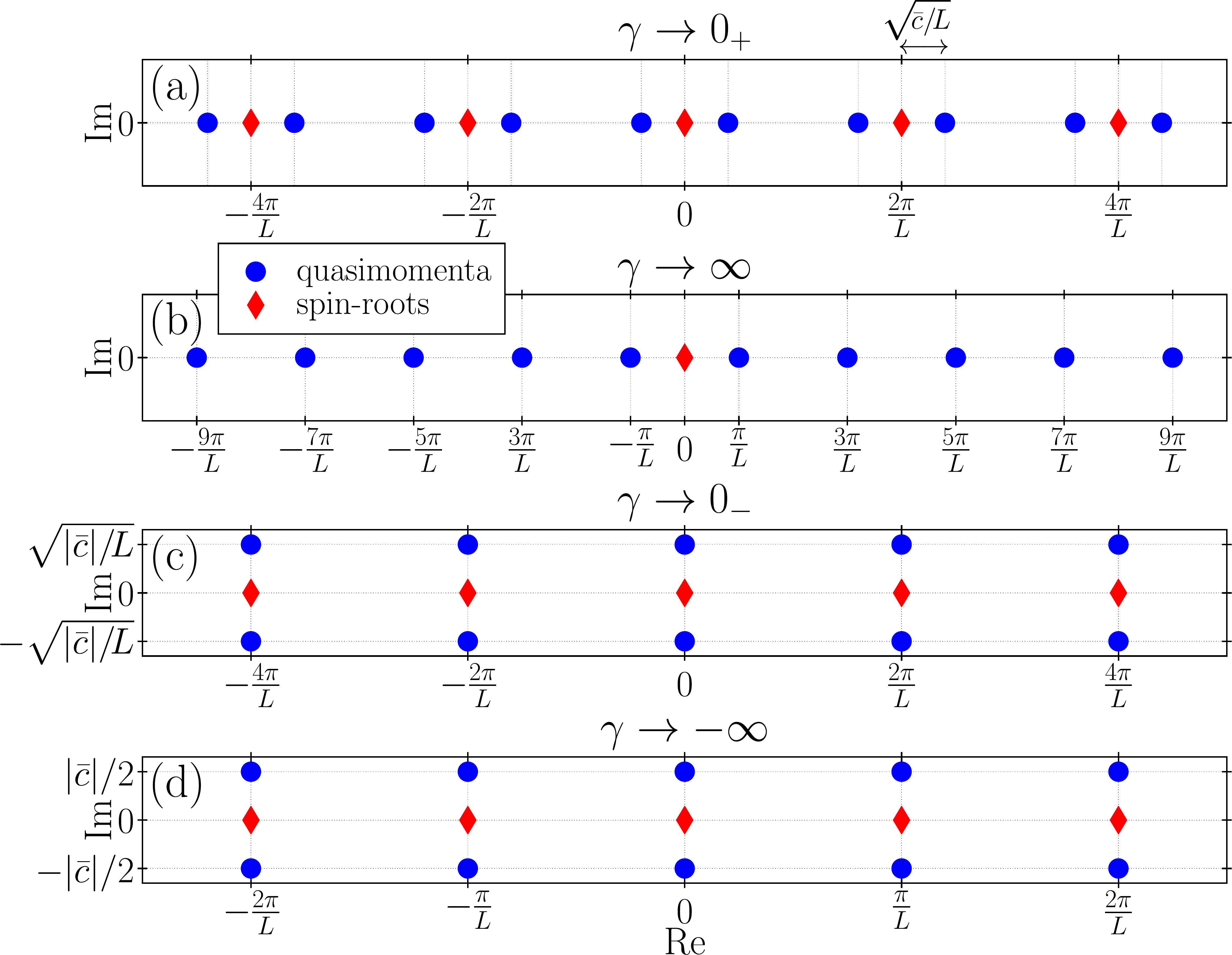} 
\vspace{-0.5cm}
\caption{Graphical representation of the quasimomenta and spin-roots in the complex plane for  the balanced ($N_{\downarrow,\uparrow}=5$) ground state of the Yang-Gaudin system with periodic boundary conditions. While panels (a) and (b) present very weakly ($\gamma\rightarrow 0_{+}$) and strongly ($\gamma\rightarrow \infty$) repulsive limits, the cases of weak ($\gamma\rightarrow 0_{-}$) and strong  ($\gamma\rightarrow -\infty$) attraction are illustrated in panels (c) and (d), respectively. For $\gamma\rightarrow \infty$ the spin-roots wander out proportionally to $\bar{c}$. Hence, except a single one equal to 0, they are not visible in scheme (b). In the repulsively interacting case ($\gamma>0$) the ground state solutions are purely real and the imaginary axis was left only for a direct comparison with the case of the attractive interactions ($\gamma<0$).
}
\label{f12}
\end{figure}

In order to find a structure of the $\{k\}$ and $\{\Lambda\}$ solutions corresponding to the ground state,  one needs to remember about the antisymmetry properties possessed by the Yang-Gaudin eigenstates, Eqs.~(\ref{YG_eigs_sym_1})--(\ref{YG_eigs_sym_4}). Due to antisymmetry in $k$'s, we conclude that for the periodic and balanced system, the energy, Eq.~(\ref{YG_eigvals}), can be minimized when the quasimomenta are symmetrically distributed around 0.
For weak interactions ($\gamma\rightarrow 0_\pm$) the interparticle coupling plays a very little role and fermions with opposite spins can form pairs of different nature that depends on the sign of $\bar{c}$, namely \cite{Batchelor2006,Shamailov2016}
\begin{equation}
\displaystyle{k_{j,\pm} \stackrel{\gamma \rightarrow 0_\pm}{=} \Lambda_j \pm \sqrt{\frac{\bar{c}}{L}}+\mathcal{O}(\bar{c}),\quad
\Lambda_j\stackrel{\gamma \rightarrow 0_\pm}{=}\frac{2\pi}{L}m_j^\text{p}},
\label{YG_GS_weak1}
\end{equation}
where for the ground state
\begin{align}
\displaystyle{
m_j^\text{p}=-\frac{N_\uparrow+1}{2}+j, \qquad  j=1,2,\ldots, N_\uparrow
}. 
\label{YG_GS_weak2}
\end{align}
Note that while for the repulsive case ($\bar{c}>0$) all the quasimomenta are real numbers, in the presence of the attractive interactions ($\bar{c}<0$) the pairs $k_{j,\pm}$ are complex conjugate and the corresponding binding energy is equal to $-\frac{\hbar^2|\bar{c}|}{mL}$.  The schemes of the quasimomenta and spin-roots distributions for the weakly interacting ground state are presented in Fig.~\ref{f12}(a)\&(c).
 The ground state energy for the weakly interacting unpolarized system is given by  
\begin{align}
\displaystyle{
  E_{GS}^\text{p} \stackrel{\gamma \rightarrow 0_\pm}{=} \frac{\hbar^2 \rho^2}{2m}\left(\frac{\pi^2}{3N}(N_\uparrow^2-1)+N_\uparrow \gamma \right)
}.
\label{YG_GS_weak_energy}
\end{align}

In the strongly repulsive case ($\gamma\rightarrow \infty$) one deals with a system resembling $N$ noninteracting spinless fermions living in a ring of length $L$ and the eigenstates should be given by the quasimomenta  in Eqs.~(\ref{YG_PBC_conds_log_strongatr_1}).
 According to the previous discussion (see Sec.~\ref{FermiBoundaries}), the parameterizing numbers $\mathcal{Y}_j^\text{p}$ are integers (half-odd integers) for $N$ odd (even) and the corresponding spin-root parameters $\Lambda\propto \bar{c}$.
In the repulsive regime, the ground state parameterization of the logarithmic Eqs.~(\ref{YG_PBC_conds_log_1})--(\ref{YG_PBC_conds_log_2}) reads
\begin{equation}
\begin{array}{lllll}
\displaystyle{\mathcal{Y}_j^\text{p}=-\frac{N+1}{2}+j}, &&& \hspace{-0.1cm} j=1,2,\ldots, N, 
\\
\displaystyle{J_m^\text{p}=-\frac{N_\uparrow+1}{2}+m}, &&& \hspace{-0.1cm}m=1,2,\ldots, N_\uparrow.
\end{array}	
\label{YG_GS_strong1_11}
\end{equation}
Taking the approximate solutions in Eqs.~(\ref{YG_PBC_conds_log_strongatr_1}), we can estimate the energy of the balanced ground state in the strongly repulsive limit (see also Refs.~\cite{Batchelor2006,Guan2013,GuanZhong2012a,GuanZhong2012b})
\begin{align}
\displaystyle{
 E_{GS}^\text{p} \stackrel{\gamma \rightarrow \infty}{=} \frac{\hbar^2 }{2m}\frac{\pi^2\rho^2(N^2-1)}{3N}\left(1-\frac{4}{\gamma}\right)
},
\label{YG_GS_strong_energy}
\end{align}
which is not entirely accurate, due to the not rigorous enough analysis presented in the previous section.
 The true value of the first order correction in $\gamma^{-1}$ is reduced by the factor $\ln (2)\approx0.693$ \cite{Guan2013,GuanZhong2012a}, i.e. one should replace $4/\gamma$ by $4\ln (2) /\gamma$.

On the other hand, in the presence of very strong attraction the quasimomenta form tightly bound two-component string states. The energy minimization requires  the symmetric distribution of the real parts of the quasimomenta, which coincide with the spin-roots, see Eqs.~(\ref{YG_P_O_string}). If so, the parameterizing numbers for the unpolarized ground state read
\begin{align}
\displaystyle{
\ell_m^\text{p}=-\frac{N_\uparrow+1}{2}+m, \qquad m=1,2,\ldots,N_\uparrow
}. 
\label{YG_GS_strong_attr1}
\end{align}
For strong attraction the spin-roots take the form  
\begin{align}
\displaystyle{
\Lambda_m \stackrel{\gamma \rightarrow -\infty}{=} \frac{\pi}{L}\ell_m^\text{p}\left(1-\frac{N_\uparrow}{L|\bar{c}|}\right)^{-1}, 
}
\label{YG_GS_weak1_corrs}
\end{align}
and the quasimomenta are given by Eq.~(\ref{YG_P_O_string}). Thus, the corresponding ground state energy reads
\begin{align}
\displaystyle{
E_{GS}^\text{p} \stackrel{\gamma \rightarrow -\infty}{=}   \frac{\hbar^2 \rho^2}{2m}\left( \frac{\pi^2}{12N}(N_\uparrow^2-1)-\frac{N \gamma^2 }{4}  \right)
}.
\label{YG_GS_strong_energy}
\end{align}
The ground state quasimomenta and spin-roots for both strongly repulsive and strongly attractive interactions are illustrated in Fig.~\ref{f12}(b)\&(d).

For the purpose of the thermodynamic description, we introduce the following quasimomenta and, if it is applicable, spin-roots densities
\begin{align}
\begin{array}{l}
\displaystyle{
D^\text{p}_F(k_j)=\lim \frac{1}{L(k_{j+1}-k_j)}},
\vspace{0.2cm}\\ 
 \displaystyle{\sigma^\text{p}_F(\Lambda_m)=\lim \frac{1}{L(\Lambda_{m+1}-\Lambda_m)}
},
\end{array}
\label{YG_densities}
\end{align}
where the order of $k_j$ and $\Lambda_m$ is chosen so that both $D_F^\text{p}(k)$ and $\sigma_F^\text{p}(\Lambda)$ are positive definite. The thermodynamic limit denoted by "$\lim$" means that we consider $N,N_\uparrow,L\rightarrow \infty$ with fixed and finite densities $\rho=N/L$ and $\rho_\uparrow=N_\uparrow/L$. In such a case the system is governed by the following integral equations  \cite{Batchelor2006,Guan2013,GuanZhong2012a} 
\begin{eqnarray}
\displaystyle{
\frac{1}{2\pi}=D_{F}^\text{P}(k)-\frac{2\bar{c}}{\pi}\int \limits_{-Q_\Lambda}^{Q_\Lambda}\frac{\sigma_F^\text{p}(\Lambda)\mathrm{d}\Lambda}{\bar{c}^2+4(k-\Lambda)^2}
},
\label{YG_IntegralEqs_1}
\\
\displaystyle{
\sigma_F^\text{p}(\Lambda)=\frac{2\bar{c}}{\pi}\!\!\int\limits_{-Q_k}^{Q_k}\!\!\!\frac{D_F^\text{P}(k)\mathrm{d}k}{\bar{c}^2\!+\!4(\Lambda\!-\!k)^2}
\!-\!\frac{\bar{c}}{\pi}\!\!\int \limits_{-Q_\Lambda}^{Q_\Lambda}\!\!\!\frac{\sigma_F^\text{p}(\Lambda')\mathrm{d}\Lambda'}{\bar{c}^2\!+\!(\Lambda \!-\!\Lambda')^2},
}
\label{YG_IntegralEqs_2}
\end{eqnarray}
where $Q_\Lambda$ and $Q_k$ are proper limiting values for $\Lambda$'s and $k$'s distributions for the ground state in the thermodynamic limit. The above Fredholm equations describe properly the case of the repulsive interactions, for which the parameterization of the logarithmic equations is given by the prescription (\ref{YG_GS_strong1_11}). The densities defined in Eqs.~(\ref{YG_densities}) satisfy
\begin{equation}
\displaystyle{
\int\limits_{-Q_k}^{Q_k}D_F^\text{p}(k)\mathrm{d}k =\frac{N}{L}, 
\qquad
 \int\limits_{-Q_\Lambda}^{Q_\Lambda}\sigma_F^\text{p}(\Lambda)\mathrm{d}\Lambda =\frac{N_\uparrow}{L}
}.
\label{YG_denses_rep}
\end{equation}

To consider the attractive regime, we apply the aforementioned string hypothesis. Namely, we start with Eqs.~(\ref{YG_string_hypo_1})--(\ref{YG_string_hypo_2}) getting \cite{Batchelor2006,Guan2013,GuanZhong2012a} 
\begin{align}
\displaystyle{
\frac{1}{2\pi}=D_{F}^\text{P}(k)-\frac{2\bar{c}}{\pi}\int \limits_{-Q_\Lambda}^{Q_\Lambda}\frac{\sigma_F^\text{p}(\Lambda)\mathrm{d}\Lambda}{\bar{c}^2+4(k-\Lambda)^2}
},
\label{YG_IntegralEqs_3}
\\
\begin{array}{l}
\displaystyle{
\frac{1}{\pi}=\sigma_F^\text{p}(\Lambda)-\frac{2\bar{c}}{\pi}\int\limits_{-\widetilde{Q}_k}^{\widetilde{Q}_k}\frac{D_F^\text{P}(k)\mathrm{d}k}{\bar{c}^2+4(\Lambda-k)^2}  }
\\
\qquad\qquad\qquad\quad\displaystyle{
-\frac{\bar{c}}{\pi}\int \limits_{-Q_\Lambda}^{Q_\Lambda}\frac{\sigma_F^\text{p}(\Lambda')\mathrm{d}\Lambda'}{\bar{c}^2+(\Lambda -\Lambda')^2},
}
\end{array}
\label{YG_IntegralEqs_4}
\end{align}
where the quantity $\widetilde{Q}_k$ refers to the limiting values of the free (unpaired) quasimomenta in the considered thermodynamic limit. Note that while Eq.~(\ref{YG_IntegralEqs_3}) describes only the unpaired quasimomenta, the description of the paired ones is incorporated in the analysis of $\sigma_F^\text{p}(\Lambda)$, because due to the string hypothesis $k_{m,\pm}=\Lambda_m\pm i\bar{c}/2$. In such a case $D_F^\text{P}(k)$ and $\sigma_F^\text{p}(\Lambda)$ fulfill the following normalization conditions 
\begin{equation}
\displaystyle{
\int\limits_{-\widetilde{Q}_k}^{\widetilde{Q}_k}\!D_F^\text{p}(k)\mathrm{d}k =\frac{N\!-\!2N_\uparrow}{L}, 
\quad
 \int\limits_{-Q_\Lambda}^{Q_\Lambda}\!\sigma_F^\text{p}(\Lambda)\mathrm{d}\Lambda =\frac{N_\uparrow}{L}
}.
\label{YG_denses_attr}
\end{equation}
The Fredholm equations for the attractively interacting ground state have been meticulously analyzed in Ref.~\cite{IidaWadati2005}, as well as in the presence of an external magnetic field in Ref.~\cite{IidaWadati2008}. 
Here we restrict to the unpolarized case only. It means that $N =2N_{\downarrow,\uparrow}$ and there are no unpaired quasimomenta in the system. Thus,  Eq.~(\ref{YG_IntegralEqs_3}) drops out and Eq.~(\ref{YG_IntegralEqs_4}) reduces to
\begin{align}
\displaystyle{
\frac{1}{\pi}=\sigma_F^\text{p}(\Lambda)-\frac{\bar{c}}{\pi}\int \limits_{-Q_\Lambda}^{Q_\Lambda}\frac{\sigma_F^\text{p}(\Lambda')\mathrm{d}\Lambda'}{\bar{c}^2+(\Lambda -\Lambda')^2}
},
\label{YG_IntegralEqs_5}
\end{align}

Let us start our analysis with the strongly interacting regime, i.e. $|\gamma|\rightarrow \infty$ (here $\rho=\mathrm{const}<\infty$). In such a case we assume that the following expansion $1/(\bar{c}^2+\Delta^2)\approx 1/\bar{c}^{2}$, where $\Delta =(k-\Lambda)$ or  $\Delta =(\Lambda-\Lambda')$, is valid. Hence, for the strongly repulsive 
\begin{align}
\begin{array}{lclcl}
\displaystyle{D_F^\text{p}(k)}  & \displaystyle{\stackrel{\gamma \rightarrow \infty}{=}} &
\displaystyle{\frac{1}{2\pi}\left( 1+\frac{2}{\gamma} \right)}, && |k|\leq Q_k,
\vspace{0.2cm}\\
 \displaystyle{\sigma_F^\text{p}(\Lambda)} & \displaystyle{\stackrel{\gamma \rightarrow \infty}{=}}
 &  \displaystyle{\frac{3}{2\pi \gamma} }, &&   |\Lambda|\leq Q_\Lambda,
\end{array}
\label{YG_GS_strong_rep1}
\end{align}
and for the strongly attractive unpolarized system
\begin{align}
	\displaystyle{\sigma_F^\text{p}(\Lambda)\stackrel{\gamma \rightarrow -\infty}{=} \frac{1}{\pi}\left(1+\frac{1}{2\gamma}  \right),} \qquad |\Lambda|\leq Q_\Lambda.
 \label{YG_GS_strong_attr1}
\end{align}
By the normalization conditions we find that for strong repulsion $Q_k=\pi \rho\,(1+2/\gamma)^{-1}$ and $Q_\Lambda=\pi \bar{c}/6$ and for strong attraction $Q_\Lambda=\pi \rho\,(4+2/\gamma)^{-1}$. The corresponding ground state energies in the considered limits can be estimated as follows
\begin{eqnarray}
 \mathop{\lim E_{GS}^\text{p}}_{\gamma \rightarrow \infty} = \frac{\hbar^2 L}{2m}&\int\limits_{-Q_k}^{Q_k}k^2 D_F^\text{p} (k) \mathrm{d}k 
 \label{YG_GS_strong_rep2}
 \\ \nonumber
 &
 =\frac{\hbar^2}{2m}\frac{N\pi^2 \rho^2}{3}\left(1-\frac{4}{\gamma}+\frac{12}{\gamma^2}+\ldots\right),
\end{eqnarray}
\begin{eqnarray}
\mathop{\lim E_{GS}^\text{p}}_{\gamma \rightarrow -\infty}&=  \frac{\hbar^2 L}{m}\left[\int\limits_{-Q_\Lambda}^{Q_\Lambda}\Lambda^2 \sigma_F^\text{p} (\Lambda) \mathrm{d}\Lambda- \frac{\rho\bar{c}^2}{4} \right] 
\label{YG_GS_strong_attr2}
\\ \nonumber
&
=\frac{\hbar^2\rho^2N}{2m}\left(\frac{\pi^2}{48}-\frac{\gamma^2}{4}-\frac{\pi^2}{48\gamma}+\frac{\pi^2}{64\gamma^2} +\ldots\right).
\end{eqnarray}
Although the approach we used is very simple,  the higher order corrections found by us in the strongly attractive limit agree (up to $\gamma^{-2}$) with the result of more rigorous analysis reported in Ref.~\cite{IidaWadati2005}.  On the other hand, the strongly repulsive regime turns out to be more cumbersome and the energy in Eq.~(\ref{YG_GS_strong_rep2}) is not entirely accurate. The correct values of the higher order corrections were calculated in Ref.~\cite{GuanZhong2012a}.

Let us now pay more attention to the strongly attractive regime, where we expect that fermions belonging to different spin components form tightly bound $\downarrow$--$\uparrow$ pairs of size much smaller than the interparticle separation $L/N$. Such pairs should have bosonic properties with the exception that two pairs cannot be located at the same point in space, i.e. two identical fermions cannot meet. Hence, one may deduce that we in fact deal with the system being very similar to a Bose gas containing $N/2$ bosons of mass $2m$, which implies $\gamma\rightarrow 4\gamma$. Indeed, by such a substitution in Eq.~(\ref{LLPeriodicGSThLim}) we reproduce the energy given by Eq.~(\ref{YG_GS_strong_attr2}), where the binding energy $\propto -N\gamma^2$ can be neglected	 \cite{IidaWadati2005}. Nonetheless, the sign of $\gamma$ is opposite in both cases. The observation suggests that the ground state energies of both systems coincide for $|\gamma|=\infty$, but there is no strict equivalence between the strongly attractive ground state of $N$-particle Yang-Gaudin system with particle mass $m$ and the strongly repulsive Lieb-Linger gas consisting of $N/2$ bosons of mass $2m$. Recently, S. Chen et al. \cite{Chen2010} pointed out that for the strongly attractive periodic Lieb-Linger model, it is possible to find solutions of the Bethe Eqs.~(\ref{LLBetheEqsPeriodic}) that correspond to a highly excited metastable phase of a Bose gas called the super Tonks-Girardeau phase \cite{Guan2013,Shamailov2016,Chen2010,Astrakharchik2004sTg, Astrakharchik2005sTg, Batchelor2005sTg, Haller2009sTg}. The energy of such an excited state is identical to the energy in Eq.~(\ref{YG_GS_strong_attr2}) excluding the term $\propto-N\gamma^2$.
 Furthermore, the coincidence is exact and does not depend on the strong interaction expansion. Namely, taking the Bethe ansatz Eqs.~(\ref{YG_string_4}), with $M=N_\uparrow=N/2$ and substituting $\bar{c}\rightarrow \bar{c}/2$ as well as $2\Lambda_j=k_j$, one recovers the Bethe Eqs.~(\ref{LLBetheEqsPeriodic}). The relation between $\bar{c}$ in the Bose ($B$) and dimerized Fermi ($F$) system, i.e. $\bar{c}_B=2\bar{c}_F$, has been already noted in Ref.~\cite{Mora2005}, where the authors calculated the dimer-dimer scattering length.
in a quasi one-dimensional balanced four-body gas of ultracold fermions throughout the BEC-BCS crossover. 	

The weakly interacting regime has been meticulously analyzed in Ref.~\cite{GuanZhong2012a}.  Due to the fact that the calculations are quite cumbersome,  we just recall the resulting ground state energy for the unpolarized Yang-Gaudin gas with periodic boundary conditions 
\begin{align}
\displaystyle{
 \lim E_{GS}^\text{p}\stackrel{\gamma \rightarrow 0_\pm}{=}\frac{\hbar^2 \rho^2 N}{2m}\left(\frac{\pi^2}{12}\pm\frac{|\gamma|}{2}+\mathcal{O}(\gamma^{2})\right)
}.
\label{YG_GS_weak_rep_attr}
\end{align}

\subsection{Elementary excitations}
\label{YGElementExcitations}

In this section we analyze the hole-like excitations in the Yang-Gaudin model with periodic boundary conditions. In a conceptual sense, they are similar to those discussed in the Lieb-Liniger case (see Sec.~\ref{LLElementaryExcitations_PBC}). The extended discussion concerning the particle-like excitations is intentionally omitted. Nonetheless, it is very important to understand that in the weakly interacting limit ($\gamma\rightarrow 0_\pm$) such an excitation (for positive momentum) tears a pair of the quasimomenta $k_{N/2,\pm}=\frac{\pi(N/2-1)}{L}\pm \sqrt{\bar{c}/L}$  apart, leaving one of the partners at $\frac{\pi(N/2-1)}{L}$ and exciting the second one above the Fermi surface, i.e. to the value $\frac{\pi(N/2-1)}{L}+\frac{2\pi}{L}n$, with $n=1,2,\ldots$, for the unpolarized case ($N_{\downarrow,\uparrow}=N/2$). The energy associated with this kind of excitation can be easily estimated and takes the value 
\begin{align}
\displaystyle{
 \epsilon\stackrel{\gamma \rightarrow 0_\pm}{\approx}\frac{P^2+\frac{\pi}{L}\hbar P(N-2)}{2m}
}, 
\label{YG_GS_weak_rep_attr_type1}
\end{align}
where $P$ denotes the total momentum. 
 For $\gamma\rightarrow\infty$ the particle-like excitation relies on wrenching a quasimomentum out of the Fermi surface, i.e. for a positive momentum  $k_N=\frac{\pi(N-1)}{L}\rightarrow\frac{\pi(N-1)}{L}+\frac{2\pi}{L}n$, with $n=1,2,\ldots$. In such a case 
\begin{align}
\displaystyle{
 \epsilon \stackrel{\gamma \rightarrow \infty}{\approx} \frac{P^2 +\frac{2\pi}{L}\hbar P (N-1)}{2m}
}. 
\label{YG_GS_strong_rep_attr_type1}
\end{align}
 For strong attraction the mechanism is very similar, with the exception that a whole tightly bound pair of quasimomenta is a subject to an excitation. Thus, the pair $k_{N/2,\pm}=\frac{\pi(N/2-1)}{2L}\pm i \frac{\bar{c}}{2}$ occupying the Fermi surface is thrown above to  $k_{N/2,\pm} +\frac{\pi}{L}n$, where $N_{\downarrow,\uparrow}=N/2$, $n=1,2,\ldots$, and $P=\hbar\frac{2\pi}{L}n>0$, giving 
\begin{align}
\displaystyle{
\epsilon \stackrel{\gamma \rightarrow -\infty}{\approx} \frac{P^2 +\frac{\pi}{L}\hbar P (N-2)}{4m}
}.
\label{YG_GS_strong_attr_attr_type1}
\end{align}
Note that the effective mass is equal to $2m$. In all the cases we only estimated the energy of the particle-like excitations with the assumption that the other values of the quasimomenta remain unchanged. In fact, the excitation affects their values, what have to be taken into account especially for the intermediate strength of the interparticle interactions, i.e. when the value of $|\gamma|$ is far from both considered limits $|\gamma|\rightarrow 0$ and $|\gamma|\rightarrow\infty$.

\subsubsection{Hole excited eigenstates}
\label{YGElementExcitationsHoles}

Similarly to the Lieb-Liniger model case (see Sec.~\ref{LLElementaryExcitations}), one can consider the excitation that whips a quasiparticle lying somewhere below the Fermi surface off and put it just above the Fermi surface.  Here we restrict to the unpolarized case, i.e. $N_{\downarrow,\uparrow}=N/2$, and analyze excitations associated with positive momenta only.  The cases with the negative momenta can be  reproduced by changing the signs of all the quasimomenta and spin-roots.

\vspace{0.15cm} 
$\bullet$\emph{\textbf{ repulsive interactions}}

In order to create a \emph{single hole excitation} in the presence of weak interparticle repulsion we tear the pair of quasimomenta  $k_{j,\pm}=\frac{2\pi}{L}m_j^\text{p}\pm \sqrt{\bar{c}/L}$ with $-\frac{N_\uparrow-1}{2}<m_j^\text{p}<\frac{N_\uparrow-1}{2}$, so that one of the conjugate partners takes the value $k_{j}=\frac{2\pi}{L}m_j^\text{p}$ and the other one $k'_j=\frac{\pi(N_\uparrow+1)}{L}$. We still need to answer the question of how it affects the corresponding spin-root parameter $\Lambda_j$. Assuming that all the other quasimomenta and spin-roots remain unchanged it can be shown \cite{ShamailovPhD,Shamailov2016} that, in order to satisfy the Bethe ansatz equations, $\Lambda_j$ has to move to a position located between $k_j$ and $k'_j$, namely $\Lambda_j\rightarrow\Lambda'_j=\frac{1}{2}\big(k_j+k'_j \big)=\frac{\pi}{L}j$, where $j=2,\ldots,N_\uparrow-1$. When $j$ can be expressed as $j=2s-N_\uparrow -1$ for some $s=1,2,\ldots,N_\uparrow$, the new $\Lambda_j'$ tends to occupy the same position as $\Lambda_s$. In such a case $\Lambda_j'$ and $\Lambda_s$ split away by $\pm i \sqrt{\bar{c}/L}$ forming a conjugate pair  $\Lambda_{j,\pm}=\frac{\pi}{L}j\pm i \sqrt{\bar{c}/L}$, i.e. the real splitting in $k_{j,\pm}$ becomes the imaginary splitting of spin-roots.

\begin{figure}[t!] 
\begin{center}\includegraphics[scale=0.195]{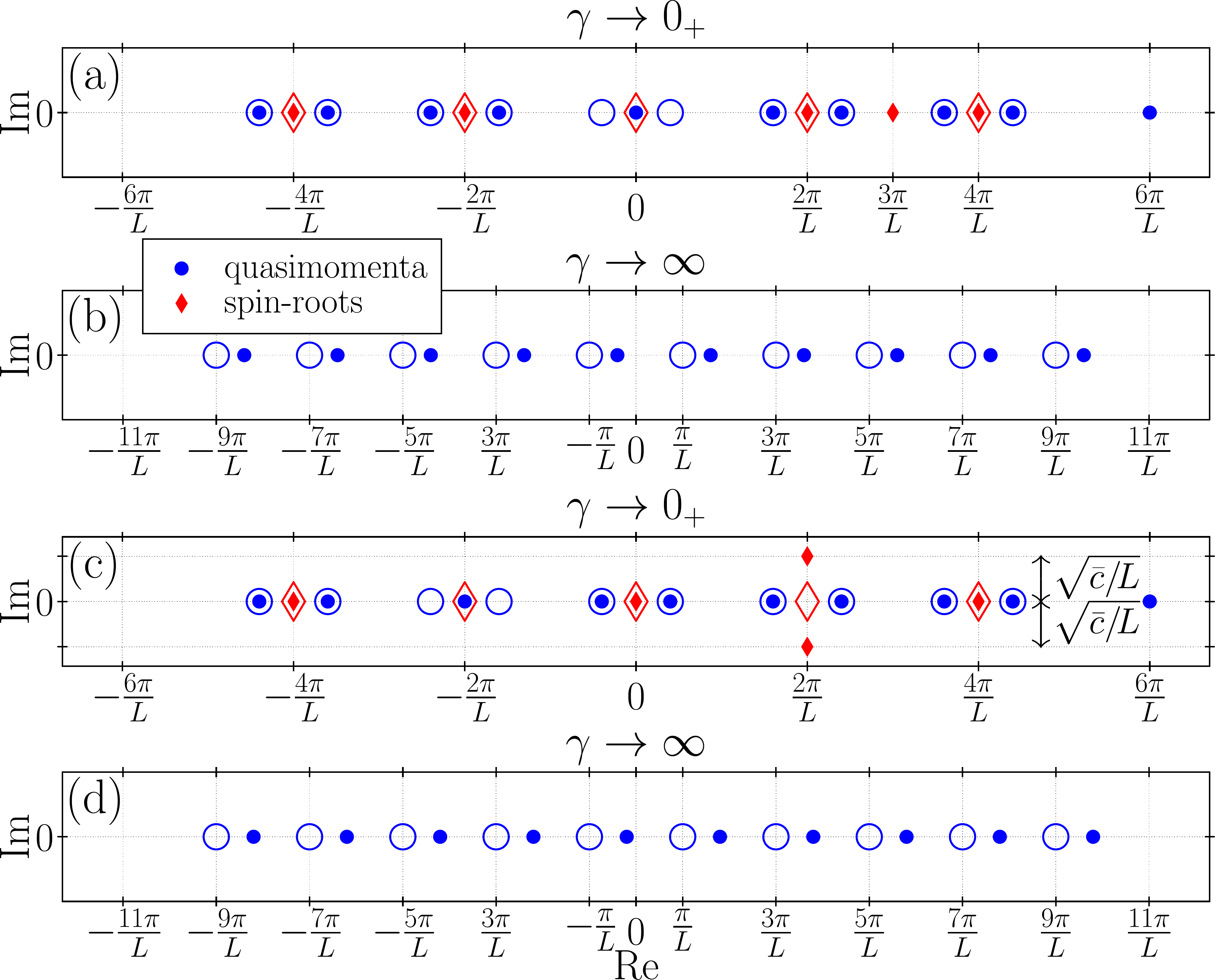} \end{center}
\vspace{-0.5cm}
\caption{
Schemes of a single hole excitation in the unpolarized ($N_{\downarrow,\uparrow} = 5$) Yang-Gaudin model with periodic boundary conditions in the weakly and strongly repulsive limits. The quasimomenta and spin-roots are depicted in the complex planes by blue filled circles and red filled diamonds, respectively. Empty symbols refer to the ground state solutions. Panels (a) and (b) correspond to the weakly and strongly repulsive cases of the single hole excitation with $P=\hbar\frac{6\pi}{L}$. Panels (c) and (d) illustrate the parameterization of the single hole excited eigenstate with  $P=\hbar\frac{8\pi}{L}$ for weak and strong repulsion, respectively. For $\gamma\rightarrow 0_+$ one observes the imaginary splitting of spin-roots (see the main text). All the solutions presented in panels (a), (b) and (d) are purely real.
}
\label{f13}
\end{figure}

For $\gamma\rightarrow \infty$ a single hole excitation of momentum $P$ shifts all the quasimomenta so that they can be written as $k_j=\frac{2\pi}{L}\left(\mathcal{Y}_j^\text{p}+\frac{P}{\hbar N} \right)$, where $\mathcal{Y}_j^\text{p}=-\frac{N+1}{2}+j$ \cite{ShamailovPhD}. Note that here we do not deal with an excitation of a single quasiparticle, but with a translation of all of them by the same value. Moreover, we have no simple analytical predictions on the values of spin-roots that for $\gamma\rightarrow \infty$ behave proportionally to $\bar{c}$.  The single hole excitation scenarios in both weakly and strongly repulsive limits are schematically presented in Fig.~\ref{f13}.

The energies of the single hole excited eigenstates in the weakly and strongly repulsive limits read
\begin{equation}
\displaystyle{
 E_{H}^\text{p} \! \stackrel{\gamma \rightarrow 0_+}{=}\! E_{GS}^\text{p}\!+\!\frac{\hbar^2 \rho^2}{2m}\!\!\left(\!-\frac{P^2}{\hbar^2 \rho^2}\!+\!\frac{\pi P(N\!+\!2)}{\hbar \rho N}\!-\!\frac{2\gamma}{ N} \right)
}, 
\label{YG_Yrast_weak_rep_energy_lim}
\end{equation}
\begin{equation}
\displaystyle{
  E_{H}^\text{p}  \stackrel{\gamma \rightarrow \infty}{=} E_{GS}^\text{p}+\frac{ P^2}{2mN}
},
\label{YG_Yrast_strong_rep_energy_lim}
\end{equation}
where $P$ denotes the total momentum and the lower index $H$ refers to a single hole excitation. Note that both dispersion relations reveal completely different momentum dependencies. Such a dramatic discrepancy can be explained by noting that for $\gamma>0$ the character of single hole excitations changes from the weakly to the strongly interacting regime. Such fact is reflected by bifurcations of solutions of the Bethe ansatz Eqs.~(\ref{YG_PBC_conds_1})--(\ref{YG_PBC_conds_2}) (see Ref.~\cite{ShamailovPhD}).

One can imagine that for $\gamma\rightarrow 0_+$ the excitation kicks the whole pair of quasimomenta, that lays below the Fermi surface, instead of tearing it apart. It means that $k_{j,\pm}=\frac{2\pi}{L}m_j^\text{p}\pm \sqrt{\bar{c}/L}$, with $-\frac{N_\uparrow-1}{2}<m_j^\text{p}<\frac{N_\uparrow-1}{2}$, is shifted to $k'_{j,\pm}=\frac{\pi (N_\uparrow +1)}{L}\pm \sqrt{\bar{c}/L}$. In such a case the corresponding spin-root parameter $\Lambda_j$ is also expelled just above the Fermi surface and takes the value $\Lambda'_j=\frac{\pi(N_\uparrow +1)}{L}$. This kind of excitation will be dubbed a \emph{double hole excitation} \cite{ShamailovPhD}.

\begin{figure}[t!] 
\begin{center}\includegraphics[scale=0.195]{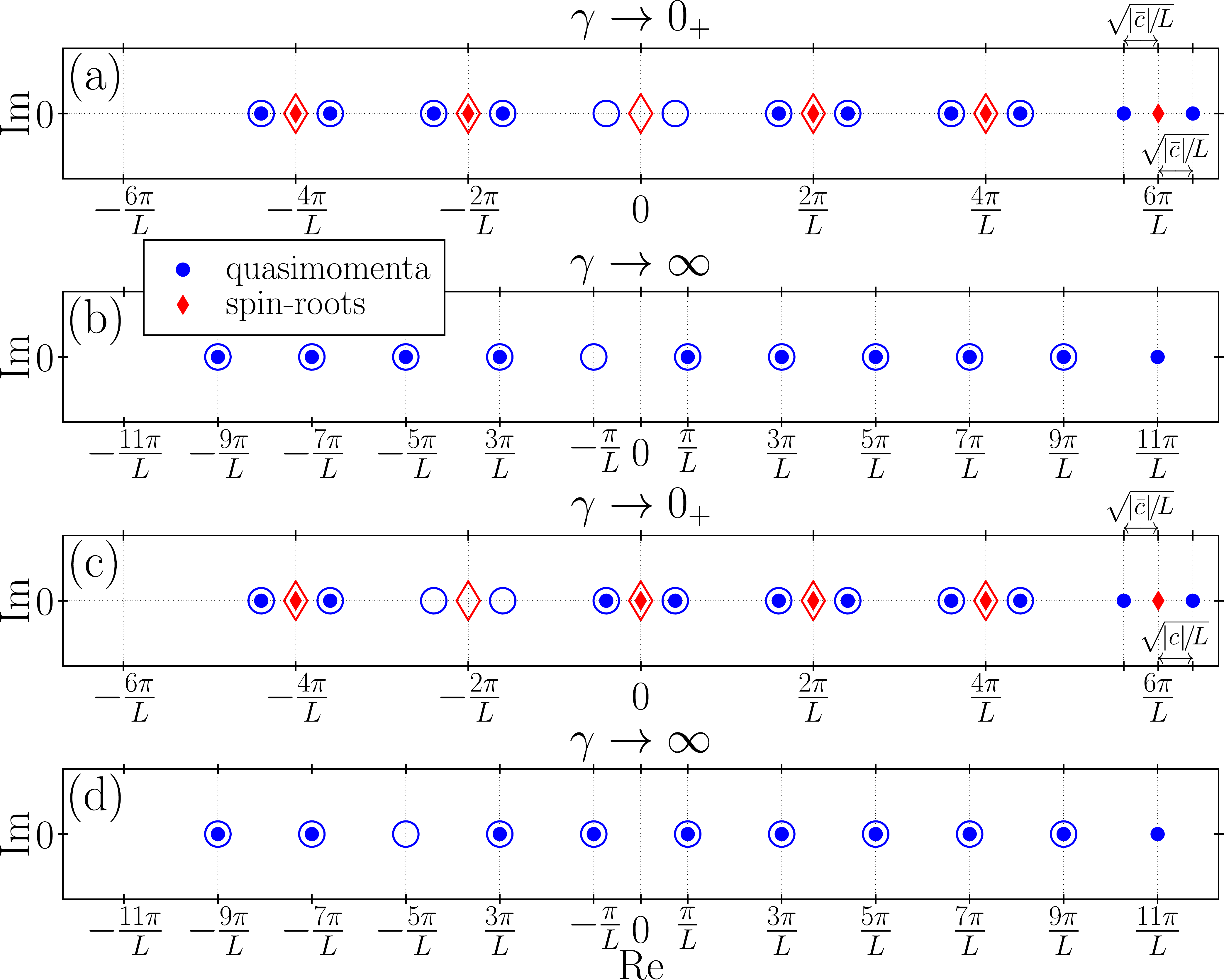} \end{center}
\vspace{-0.5cm}
\caption{
Graphical representation of the quasimomenta (blue filled circles) and spin-roots (red filled diamonds) corresponding to the double hole excited eigenstates for the balanced ($N_{\downarrow ,\uparrow} =5$) Yang-Gaudin system with periodic boundary conditions in the presence of weak and strong repulsion. For comparison, the ground state solutions are depicted by empty symbols. While panels (a) and (b) refer to weakly and strongly repulsive cases of the double hole excitation with  $P=\hbar\frac{12\pi}{L}$, the excitations with $P=\hbar\frac{16\pi}{L}$ for weak and strong repulsion are illustrated in panels (c) and (d), respectively. All the solutions presented in panels (a)--(d) are purely real.
} 
\label{f14}
\end{figure}

For strong repulsion it is clear that the double hole excitation corresponds to the shift of a single quasimomentum $k_{1<j<N}=\frac{2\pi}{L}\mathcal{Y}_j^\text{p}$ with $\mathcal{Y}_j^\text{p}=-\frac{N+1}{2}+j$, just above the Fermi surface, i.e. to $k_j'=\frac{\pi(N+1)}{L}$. Note that such excitations resemble single hole excitations in the Lieb-Liniger gas.  Double hole excitations are schematically presented in Fig.~\ref{f14}.

For the balanced Yang-Gaudin system, the energies of double hole excited eigenstates $E_{2H}^\text{p}$, in both considered limits, take the following form
\begin{equation}
\displaystyle{
 E_{2H}^\text{p} \! \stackrel{\gamma \rightarrow 0_+}{=}  \! E_{GS}^\text{p}+\frac{\hbar^2 \rho^2}{2m}\!\left(\!-\frac{P^2}{2\hbar^2 \rho^2}+\frac{\pi P(N\!+\!2)}{\hbar \rho N}\right)
},
\label{YG_Yrast_weak_rep_energy_lim2}
\end{equation}
\begin{equation}
\displaystyle{
 E_{2H}^\text{p}  \!\stackrel{\gamma \rightarrow \infty}{=}\! E_{GS}^\text{p}+\frac{\hbar^2 \rho^2}{2m}\!\left(\!-\frac{P^2}{\hbar^2 \rho^2}+\frac{2\pi P(N\!+\!1)}{\hbar \rho N}\right)
},
\label{YG_Yrast_strong_rep_energy_lim2}
\end{equation}
where  $P$  is the total momentum.
In contrast to the single hole excitation case, there is no significant difference between the spectra presented above. This may suggest that in such an excitation scenario the quasimomenta being the solutions of the Bethe ansatz equations change smoothly with $\gamma$.

The thermodynamic description of both single and double hole excitations have been neatly analyzed in a comprehensive thesis of S. Shamailov \cite{ShamailovPhD}.

\vspace{0.2cm} 
$\bullet$\emph{\textbf{ attractive interactions}}

For $\gamma\rightarrow 0_-$ a single hole excitation relies on tearing of a pair of quasimomenta $k_{j,\pm}=\frac{2\pi}{L}m_j^\text{p}\pm \sqrt{\bar{c}/L}$ with $-\frac{N_\uparrow-1}{2}<m_j^\text{p}<\frac{N_\uparrow-1}{2}$, as in the previously described weakly repulsive case. The only difference appears when $\Lambda_j$ is expelled to a position coinciding with $\Lambda_s$, where for attractive interactions both spin-roots also form a pair, but splitted on the real axis only, i.e. $\Lambda_j,\Lambda_s\rightarrow \Lambda_{j,\pm}=\frac{\pi}{L}j\pm\sqrt{|\bar{c}|/L}$ \cite{ShamailovPhD,Shamailov2016}. Hence, we can say that the imaginary splitting of  the quasimomenta $k_{j,\pm}$ may be replaced by the  real splitting of the spin-roots.

\begin{figure}[h!] 
\begin{center}\includegraphics[scale=0.195]{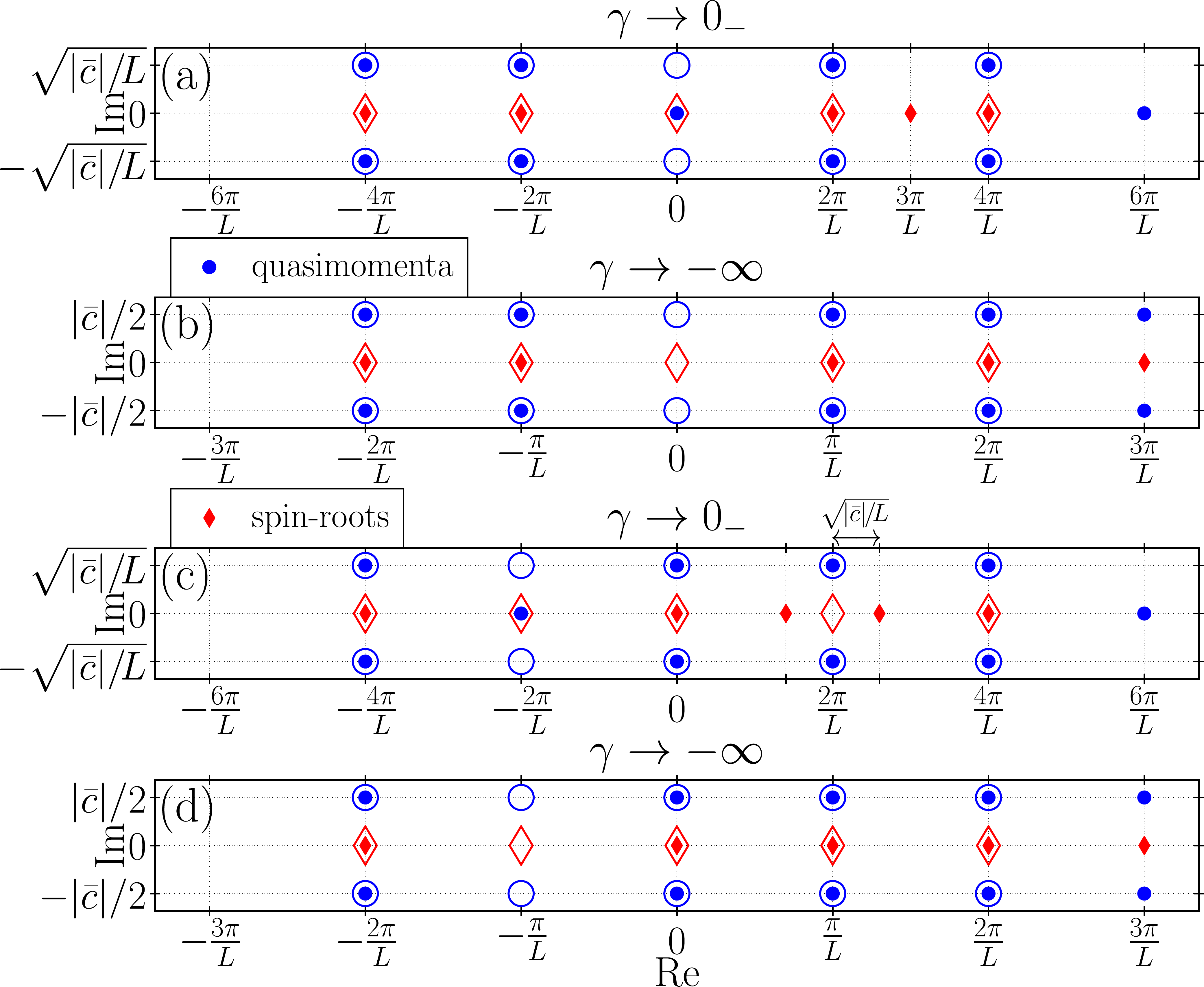} \end{center}
\vspace{-0.5cm}
\caption{
Schematic representation of the quasimomenta (blue filled circles) and spin-roots (red filled diamonds), that are related to two chosen single hole excited eigenstates ({\it yrast} eigenstates) of the periodic Yang-Gaudin system consisting of $N=2N_{\downarrow,\uparrow}=10$ particles in the presence of weak and strong interparticle attraction. While the panels (a) and (b) present the single hole ({\it yrast}) excitation characterized by $P=\hbar\frac{6\pi}{L}$, panels (c) and (d) correspond to the single hole ({\it yrast}) excited eigenstate with $P=\hbar\frac{8\pi}{L}$. As we  pointed out in the main text, in the weakly attractive limit the single hole excitation tears a pair of quasimomenta apart and shifts one of the conjugate partners just above the Fermi surface.  The corresponding spin-root parameter is shifted between the elements of the pair that was torn. If the shifted position coincides with the value of the other spin-root, we observe splitting along the real axis. Such a scenario is illustrated in panel (c). On the other hand, for $\gamma\rightarrow -\infty$, the {\it yrast} eigenstate is a result of the excitation of the whole conjugate pair of quasimomenta.
Empty symbols refer to the corresponding ground state solutions.
}
\label{f15}
\end{figure}

In contrast to the repulsive case, the structure of the single hole excitation is maintained in the presence of strong attraction. That is, for $\gamma\rightarrow -\infty$ it is energetically favourable to kick the whole bound pair of quasimomenta just above the Fermi surface, instead of tearing it apart. Therefore, in order to create a single hole excited eigenstate in this limit, one needs to shift a pair lying below the Fermi surface $k_{1<j<N_\uparrow,\pm}=\frac{\pi(2j -N_\uparrow -1)}{2L}\pm i\frac{\bar{c}}{2}$, just above to  $\frac{\pi (N_\uparrow+1)}{2L} \pm i\frac{\bar{c}}{2}$. Together with the pair $k_{j,\pm}$, one expels the corresponding spin-root $\Lambda_j=\frac{\pi(2j -N_\uparrow -1)}{2L}\rightarrow \Lambda'_j=\frac{\pi (N_\uparrow+1)}{2L}$. To realize a transition from the weakly to the strongly attractive regime one needs to cross the region where the two totally different excitation scenarios meet. It has been shown that there is no smooth connection between them and one deals with the bifurcation of the Bethe ansatz solutions \cite{ShamailovPhD, SyrwidFermi2018,Shamailov2016}. Furthermore, we stress that both presented single hole excitation schemes correspond to the lowest energy eigenstate for a given total momentum. Thus, in such a case one deals with the so-called {\it yrast} eigenstates \cite{ShamailovPhD, SyrwidFermi2018,Shamailov2016}. In Fig.~\ref{f15} we present a scheme of the Bethe ansatz solutions related to the {\it yrast} excitations in the Yang-Gaudin system.

The energy of the unpolarized Yang-Gaudin system after a single hole ({\it yrast}) excitation  characterized by the total momentum $P$ in the presence of weak and strong interparticle attraction reads
\begin{equation}
\displaystyle{
  E_{H}^\text{p}  \!\stackrel{\gamma \rightarrow 0_-}{=}\! E_{GS}^\text{p}\!+\!\frac{\hbar^2 \rho^2}{2m}\!\left(\!-\frac{P^2}{\hbar^2 \rho^2}\!+\!\frac{\pi P(N\!+\!2)}{\hbar \rho N}\!-\!\frac{2\gamma}{ N}\right)
},
\label{YG_Yrast_weak_attr_energy_lim}
\end{equation}
\begin{equation}
\displaystyle{
 E_{H}^\text{p} \!\stackrel{\gamma \rightarrow -\infty}{=}\!  E_{GS}^\text{p}\!+\!\frac{\hbar^2 \rho^2}{4m}\!\left(\!-\frac{P^2}{\hbar^2 \rho^2}\!+\!\frac{\pi P(N\!+\!2)}{\hbar \rho N}\right)
}.
\label{YG_Yrast_strong_attr_energy_lim}
\end{equation}
In spite of the fact that we deal with the bifurcation of the solutions of the Bethe ansatz equations, the character of the {\it yrast} dispersion relation remains almost unchanged.

For the purpose of the thermodynamic description of the balanced attractively interacting Yang-Gaudin system, it is sufficient to consider the string hypothesis, Eqs.~(\ref{YG_string_hypo_2}), only. To realize a single hole ({\it yrast}) excitation in the thermodynamic limit one can remove a pair of conjugate quasimomenta lying below the Fermi surface. Thus, we start with the ground state given by the sets of the quasimomenta $\{k_{1,\pm},k_{2,\pm},\ldots, k_{N_\uparrow,\pm}\}$ and the spin-roots $\{\Lambda_1,\Lambda_2,\ldots,\Lambda_{N_\uparrow}\}$ that are related with ${\{\ell_j^{p}\}\!=\!\big\{-\frac{N_\uparrow-1}{2}, -\frac{N_\uparrow-3}{2}, \ldots, \frac{N_\uparrow-1}{2} \big\} }$. After the excitation we deal with the system consisting of $N_\uparrow -1$ pairs of $\downarrow$--$\uparrow$ fermions, which is characterized by $\{k_{1,\pm}',k_{2,\pm}',\ldots, k_{N_\uparrow-1,\pm}'\}$ and $\{\Lambda_1',\Lambda_2',\ldots,\Lambda_{N_\uparrow-1}'\}$ with parameterization $\{\widetilde{\ell}_j^{p}\}=\big\{-\frac{N_\uparrow-2}{2}, -\frac{N_\uparrow-4}{2} ,\ldots, \frac{N_\uparrow-2M-2}{2}, \times ,\frac{N_\uparrow-2M+2}{2},\ldots \frac{N_\uparrow}{2} \big\}$, where $1<M<N_\uparrow-1$.
 Note that for an unpolarized system  Eqs.~(\ref{YG_string_hypo_2}) are almost identical to the Bethe Eqs.~(\ref{LLBetheEqsPeriodicLog}). Moreover, from the spin-roots point of view, the scheme of a single hole excitation is also very similar. Therefore, it is clear that we should obtain similar integral equations as for the hole excited eigenstates in the periodic Lieb-Liniger model, i.e. Eqs.~(\ref{LL_PBC_Type2_Therm_1})--(\ref{LL_PBC_Type2_Therm_2}), but for the spin-roots $\Lambda'$s, namely (see also Refs.~\cite{ShamailovPhD,Shamailov2016})
\begin{align} 
2\pi \mathcal{\sigma}_{F,H}^\text{p}(\Lambda)\!=\!\pi \!+\!\theta(\Lambda\!-\!q)\!
+\!2\bar{c}\!\!\int\limits_{-Q_\Lambda}^{Q_\Lambda}\!\!\frac{\mathcal{\sigma}_{F,H}^\text{p}(\Lambda')\mathrm{d}\Lambda'}{\bar{c}^2\!+\!(\Lambda\!-\!\Lambda')^2},
 \label{YG_PBC_SHole_Therm_1}  
\end{align}\vspace{-0.5cm}
\begin{align} 
P=\hbar\left[-q+2\int\limits_{-Q_\Lambda}^{Q_\Lambda}\sigma_{F,H}^\text{p}(\Lambda)\mathrm{d}\Lambda\right],
\label{YG_PBC_SHole_Therm_2}  
\end{align}\vspace{-0.3cm}
\begin{eqnarray} 
\begin{array}{lll}
\epsilon_{H}^\text{p}&=&E_{H}^\text{p}-E_{GS}^\text{p} 
\\
&=&
\displaystyle{\frac{\hbar^2}{2m}\left[-2q^2\!+\!4\!\!\!\int\limits_{-Q_\Lambda}^{Q_\Lambda}\!\!\!\Lambda\mathcal{\sigma}_{F,H}^\text{p}(\Lambda)\mathrm{d}\Lambda \right]\!+\!2\mu_\text{ch}
},
\end{array}
\label{YG_PBC_SHole_Therm_3}  
\end{eqnarray}
where $|q|<Q_\Lambda$ is the quasimomentum of the hole. Additional factor 2  in the energy in Eq.~(\ref{YG_PBC_SHole_Therm_3}) as well as before the integral in Eq.~(\ref{YG_PBC_SHole_Therm_2}) appears due to the fact that we remove a pair of particles instead of a single particle.
The {\it yrast} dispersion relation in the case of strong interparticle attraction reads \cite{Shamailov2016}
\begin{eqnarray} 
\mathop{\lim E_{H}^\text{p}}_{\gamma \rightarrow -\infty}&=\mathop{\lim E_{GS}^\text{p}}_{\gamma \rightarrow -\infty} \label{YG_PBC_SHole_Therm_4}  
\\ \nonumber
&+\frac{\hbar^2\rho^2}{4m}\!\left(-\frac{P^2}{\hbar^2\rho^2}\!+\!\frac{\pi P}{\hbar \rho}\right)\!\left(\frac{2\gamma}{1\!+\!2\gamma}\right)^2\!+\mathcal{O}\left(\gamma^{-3}\right),
\end{eqnarray}
and coincides with Eq.~(\ref{YG_Yrast_strong_attr_energy_lim}) for $N\gg 1$ and $\gamma\rightarrow -\infty$.

It turns out that the single hole excited eigenstates of the unpolarized attractively interacting Yang-Gaudin system with periodic boundary conditions share a few features with their bosonic brothers (type--II eigenstates), see Sec.~\ref{LLElementaryExcitations}. As it was pointed out in the literature \cite{Kanamoto2008,Kanamoto2010,Shamailov2016}, such eigenstates are the so-called {\it yrast} states, i.e. the lowest energy eigenstates for a given total momentum. In Secs.~\ref{QuantumSolitonsInMBstates} \&~\ref{QuantumSolitonsInMBstatesYG} we show that the {\it yrast} eigenstates of both Lieb-Liniger and Yang-Gaudin models are unequivocally connected with quantum dark solitons.

\section{Solitons in mean-field approximation}
\label{DarkSolitons}


When analyzing ultracold Bose gases one often deals with the Bose-Einstein condensation phenomenon \cite{pethick,PitaevskiiStringariBEC,CastinArxiv}. The simplest way to proceed in this case is to assume that all particles occupy the same single particle state and behave collectively. Thanks to this, one can easily derive the so-called Gross-Pitaevskii equation (GPE), for which the many-body problem is reduced to a description of the macroscopically occupied single particle state \cite{pethick,PitaevskiiStringariBEC,CastinArxiv}. One can say that each particle lives in the same averaged potential (proportional to the density of the atomic cloud), which is induced by the milieu of other particles. It turns out that the nonlinearity, present in the GPE, can kill dispersive effects so that the corresponding solution maintains its shape during time evolution. The resulting structures, called solitons, are ubiquitous phenomena and appear in a wide range of physical systems, i.e. they appear, among others, in nonlinear optics, fluid dynamics as well as ultracold atomic gases \cite{Kivshar2003Book}. The latter systems turned out to be an excellent playground for both theoretical and experimental investigations of matter-wave solitons \cite{Denschlag2000, Strecker2002, Khaykovich2002, Becker2008,Stellmer2008, Weller2008, Theocharis2010, Gawryluk2006,Burger1999, Pawlowski2015}.

	In a one-dimensional space the GPE possesses two types of stable soliton solutions: bright solitons that can appear for the attractive interparticle interactions and dark solitons that are present when one deals with the interparticle repulsion. In contrast to the attractive case, when the bright soliton can be observed even in the ground state, the dark soliton solution is always a manifestation of a collective excitation in the system. 
	In a Bose-Einstein condensate (BEC) they can be realized experimentally thanks to the so-called phase imprinting method \cite{Denschlag2000, Becker2008, Stellmer2008, Burger1999, Dobrek1999, Andrelczyk2001, CarrBrand2001}. That is, a short laser pulse of intensity varying over the atomic cloud can modify the spatial distribution of the phase of the condensate wave function. In particular, the experimental setup can be prepared in such a way that only half of the condensate cloud is modified by the laser radiation and acquires an additional $\pi$ phase. The procedure carves a dark soliton notch at a position where the phase of the wave function abruptly changes. Both bright and dark solitons were successfully generated and observed in  ultracold atoms laboratories. An experimental examination of their general properties confirmed  theoretical predictions  provided by the GPE \cite{Denschlag2000, Strecker2002, Khaykovich2002, Becker2008,Stellmer2008, Weller2008, Theocharis2010, Burger1999, Boisse2017}. 

Here we discuss  bright and dark soliton solutions of the one-dimensional GPE in the infinite space and after imposing periodic and open boundary conditions. We also briefly describe the long-standing conjecture concerning  
the relationship between dark solitons the type--II eigenstates of the periodic Lieb-Liniger model \cite{Kulish76,Ishikawa80}.


\subsection{Nonlinear Schr\"{o}dinger equation}
\label{Solitons_Inf} 

Let us start with the Lieb-Liniger model that is described by the Hamiltonian in Eq.~(\ref{LiebLinigerHamiltonian}). By employing the evolution equation  $i\hbar\, \partial_{t}\hat{A}(t)=[\hat{A}(t),\hat{\mathrm{H}}]$, and the canonical commutation relations, Eqs.~(\ref{BoseFieldsCummutators01})--(\ref{BoseFieldsCummutators}), one finds the time-dependent nonlinear Schr\"{o}dinger equation \cite{BogoliubovKorepinInverseScattering,Ishikawa80}
\begin{eqnarray} 
i\hbar\, \partial_t \hat{\Psi}_H(x,t)=&-\displaystyle{\frac{\hbar^2}{2m}\partial_x^2\hat{\Psi}_H(x,t)}  \label{NLS_1}
\\ \nonumber
&+2c \,\hat{\Psi}_H^\dagger(x,t) \hat{\Psi}_H(x,t)\hat{\Psi}_H(x,t), 
\end{eqnarray}
It is known that there is no Bose-Einstein condensation in a one-dimensional space. Nevertheless, for finite ($L<\infty$) and weakly interacting systems maintained at very low temperatures the number of particles occupying nearly zero momentum states may become a significant fraction of the total number of particles $N$ (see also Refs.~\cite{SachaBook,Dalfovo1999, VanDruten1996,VanDruten1997}). Under such assumptions, during the analysis of low-lying excitations we can approximate the quantum Bose field operator $\hat{\Psi}(x,t)$ by  a classical field $\Psi(x,t)$, i.e. $\hat{\Psi}(x,t)\longrightarrow \sqrt{N}\Psi(x,t)$, which leads to the one-dimensional time-dependent Gross-Pitaevskii equation (GPE)~\cite{pethick,Pitaevskii61,Gross61,Tsuzuki71,PitaevskiiStringariBEC}
\begin{eqnarray} 
i\hbar\, \partial_t \Psi(x,t)= &- \frac{\hbar^2}{2m}\partial_x^2\Psi(x,t)  \label{NLS_2}
\\ \nonumber
&+2c N|\Psi(x,t)|^2 \Psi(x,t)- \mu_\text{ch}\Psi(x,t).
\end{eqnarray}

The nonlinear contribution, $\propto|\Psi|^2$, plays a crucial role in the formation of stable nontrivial structures. That is, by an interplay with the dispersive spreading, caused by the kinetic term $\propto p^2$, the solution may be stabilized so that it can propagate without any change of its shape. This kind of solutions of the nonlinear wave equations are called solitons and can be observed in many different physical systems \cite{Kivshar2003Book}. Here we focus on the one-dimensional matter-wave solitons in an ultracold Bose gas described by the GPE in Eq.~(\ref{NLS_2}).
For this purpose, one assumes the following form of a solution
\begin{align} 
\displaystyle{
\Psi(x,t)=\Psi_0(x-vt)\, \mathrm{e}^{-\frac{i}{\hbar} \mu_\text{ch}t}
 },
 \label{NLS_3}
\end{align}
that propagates with a constant velocity $v$ maintaining its shape. Moreover, the following boundary conditions have to be imposed
\begin{align} 
\displaystyle{
\Psi_0\bigg|_{\substack{ x\rightarrow \pm \infty   \\  |t|<\infty } }=\mathrm{const}, \qquad \frac{\mathrm{d} \Psi_0}{\mathrm{d}x}\bigg|_{ \substack{x\rightarrow \pm \infty   \\  |t|<\infty } }=0
 }.
 \label{NLS_4}
\end{align}
Proceeding as in Refs.~\cite{pethick,PitaevskiiStringariBEC} one can obtain nontrivial analytical bright and dark soliton solutions.

\vspace{0.2cm}
\underline{\emph{\textbf{Bright soliton solution ($c<0$)}}}

\begin{figure}[t!] 
\begin{center}\includegraphics[scale=0.265]{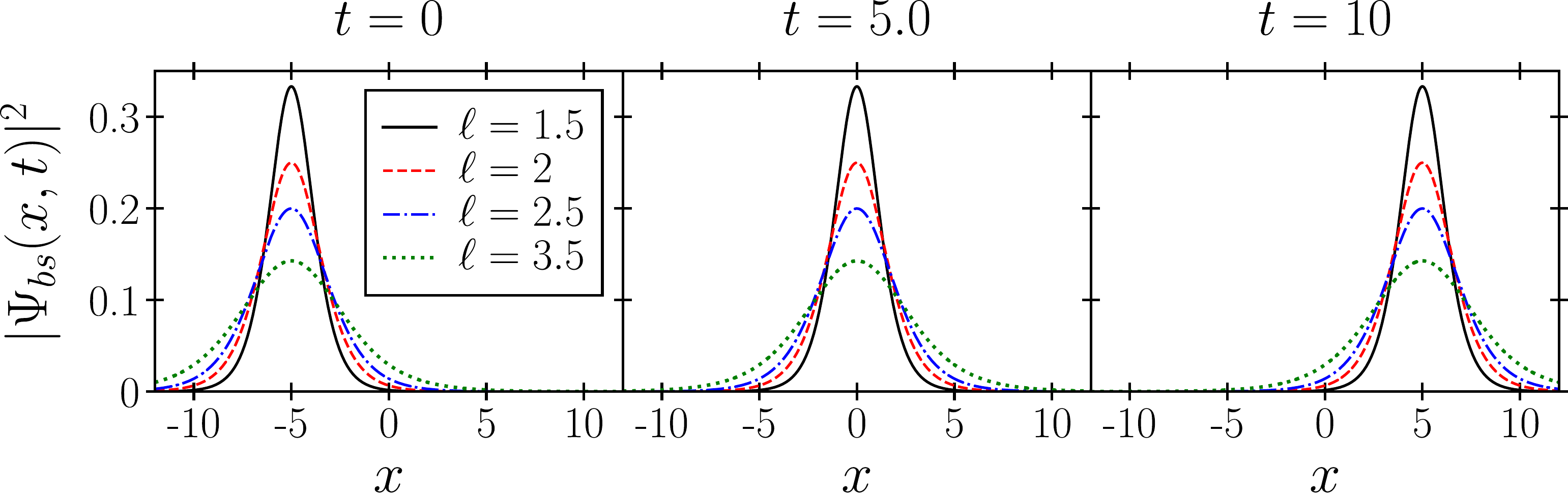} \end{center}
\vspace{-0.5cm}
\caption{
Time evolution of a bright soliton probability density for different soliton widths $\ell$ indicated in the legend. The results were obtained for $\hbar=m=v=1$ and $x_0=-5$.   
} \vspace{-0.cm}
\label{figBrightSol}
\end{figure}

 Let us start with bright solitons that can form in the system ground state, when $c<0$. In such a case, it is energetically favourable to form a clump of particles instead of spread them uniformly in space. 
 In 1D the time-dependent GPE possesses an analytical bright soliton solution of form  \cite{pethick,PitaevskiiStringariBEC,CastinArxiv,SachaBook} 
  \begin{align} 
\displaystyle{
\Psi_{bs}(x,t)=\frac{1}{\sqrt{2\ell}}\frac{\mathrm{e}^{i \alpha x}\,\mathrm{e}^{-i \kappa t}}{\mathrm{cosh}\left[(x-x_0-v t)/\ell\right]}
 },
 \label{BrightSol1}
\end{align}
 where $\alpha= m v/\hbar$, $\kappa = \frac{1}{2}\left[m v^2/\hbar-\hbar/ (\ell^2m)\right]$, $x_0$ is the bright soliton position at $t=0$ and $\ell=\hbar^2/(m |c|N)$ is the soliton width. The nonspreading wave packet structure moves with an arbitrary velocity $v$. 
In Fig.~\ref{figBrightSol}, we show the bright soliton probability density for several values of $\ell$ and in different time moments.

The energy in a bright soliton state, Eq.~(\ref{BrightSol1}),
  \begin{eqnarray} 
 E_{bs}	&= N \int\limits_{-\infty}^{\infty}\left[\frac{\hbar^2}{2m}\left|\frac{\partial}{\partial x } \Psi_{bs}\right|^2 - |c| N |\Psi_{bs}|^4\right]\mathrm{d}x
 \label{BrightSolEnergy}
 \\ \nonumber
&=N\frac{ m v^2}{2}+N\frac{\hbar^2}{6 \ell^2m }-N^2\frac{|c|}{3\ell}=N\frac{ m v^2}{2}	-N^2\frac{|c|}{6\ell},
\end{eqnarray}
reveals a particle-like behaviour of the analyzed structure.
By a direct calculation we immediately obtain the chemical potential $\mu_\text{ch}= -N|c|/2\ell$ telling us that each particle inserted to the system interacts with $N$ bosons that form the soliton  and the corresponding interaction energy is on average  equal to  $-N|c|/2\ell$.

\vspace{0.2cm}
\underline{\emph{\textbf{Dark soliton solution ($c>0$)}}}

In contrast to the bright soliton case, the dark soliton state is always a result of a collective excitation in the repulsively interacting ($c>0$) system and corresponds to a breach in the atomic cloud, i.e. a density notch in the probability distribution. Moreover, such a density dip is always associated with a phase flip. The corresponding wave function, being an analytical solution of Eq.~(\ref{NLS_2}) reads \cite{pethick,PitaevskiiStringariBEC,SachaBook} (see also a meticulous analysis presented in Ref.~\cite{Tsuzuki71})
\begin{eqnarray} 
\Psi_{ds}(x,t)&=\mathrm{e}^{-\frac{i}{\hbar}\mu_\text{ch} t}\sqrt{\rho_0}  \label{DarkSol}
\\ \nonumber
& \qquad\times\Bigg\{i\frac{v}{v_s}\!+\! \beta\,\mathrm{tanh}\left[\frac{\beta}{\xi}(x\!-\!x_0\!-\!vt)\right]\!\Bigg\}, 
\end{eqnarray}
where $\beta=\sqrt{1-v^2/v_s^2}$.
The above wave function describes a dark soliton propagating with a velocity $|v|\leq v_s$, where $v_s=\sqrt{2\rho_0  c N  /m}$ is the velocity of Bogoliubov phonons (propagation velocity of long wavelengths disturbances -- speed of sound). 
The quantity $x_0$ is a soliton position at $t=0$. 
 Note that far away from the soliton notch the probability density is equal to $|\Psi|^2=\rho_0$ and thus $\rho=\rho_0N$. The corresponding chemical potential reads $\mu_\text{ch}=2 \rho c$.

The soliton width is given by $\xi/\beta$, where the so-called \emph{healing length} $\xi=\sqrt{\hbar^2/2c\rho m}$ is the distance over which the disturbances in a condensate disappear. The dark soliton perishes when $v\longrightarrow v_s$. Indeed, in such a limit  the soliton depth $\rho_0\beta^2\stackrel{v\rightarrow v_s}{\longrightarrow} 0$ and the soliton width $\xi/\beta \stackrel{v\rightarrow v_s}{\longrightarrow} \infty$.
The density notch is strictly associated with the phase flip and they are located at the same position. If $\varphi(x,t)$ denotes a phase of $\Psi_{ds}(x,t)$, Eq.~(\ref{DarkSol}), then according to the result
   \begin{equation} 
\displaystyle{   
\lim_{x\rightarrow \pm\infty}\Psi_{ds}(x,|t|<\infty)= \sqrt{\rho_0}\left[i\frac{v}{v_s}\pm\sqrt{1-\frac{v^2}{v_s^2}} \,\right]
 },
 \label{LimValsDarkSol}
\end{equation}
one finds the change of phase along the soliton
\begin{eqnarray} 
\Delta\varphi&=\varphi(\infty,t)-\varphi(-\infty,t)  \label{DarkSolPhaseChange}
\\ \nonumber
&=2\, \mathrm{arctan}\left( \frac{v}{\sqrt{v_s^2-v^2}} \right)-\pi =-2\,\mathrm{arccos}\left(\frac{v}{v_s}\right).
\end{eqnarray}

\begin{figure}[h!] 
\begin{center}\includegraphics[scale=0.265]{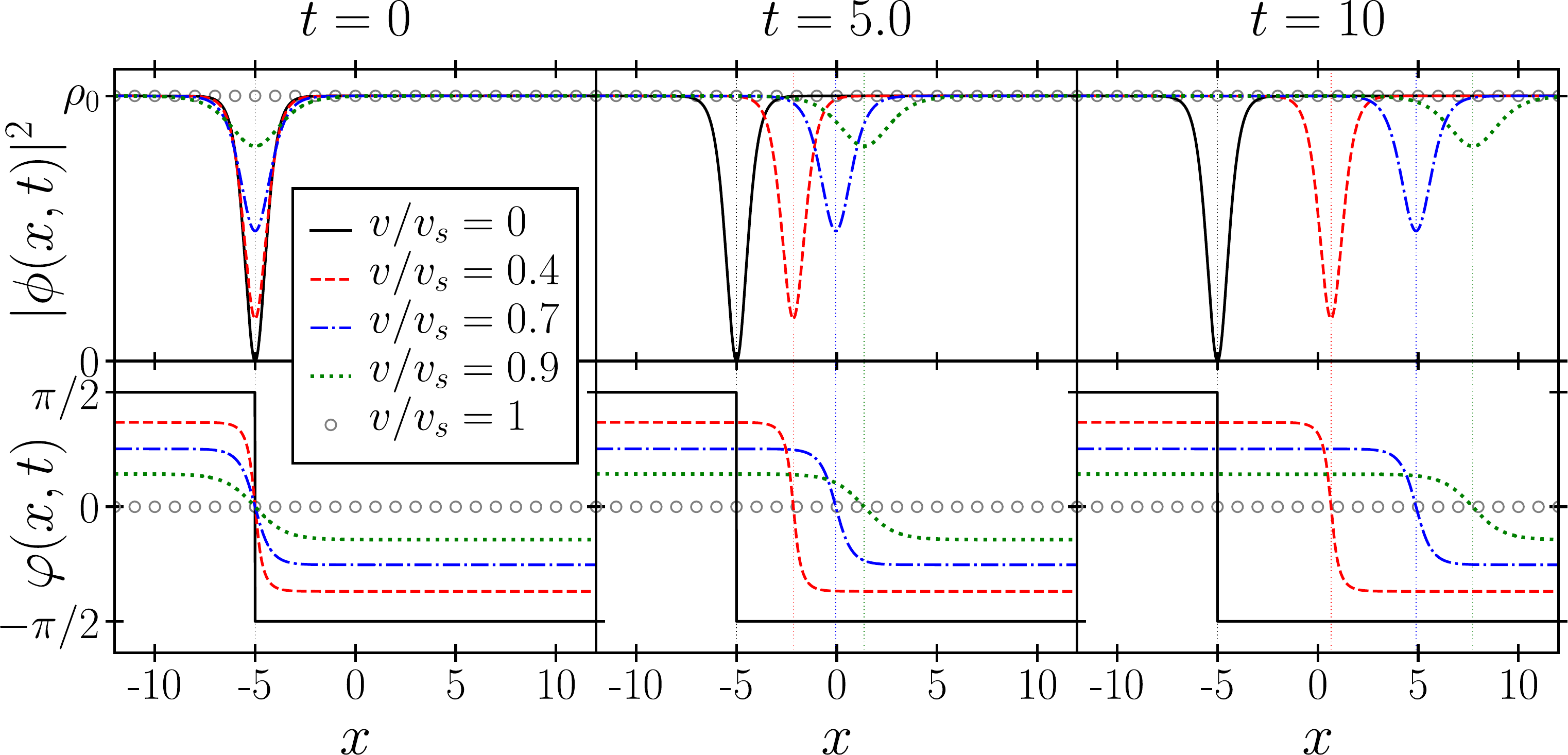} \end{center}
\vspace{-0.6cm}
\caption{
Time evolution of a dark soliton probability density and the corresponding phase distribution for various propagation velocities indicated in the legend. Note that the higher the speed, the shallower the soliton notch is. For $v<v_s$, each phase distribution reveals the phase flip which coincides with the density notch position for all time moments. The results were obtained for $m=\hbar=1$, $x_0=-5$ and $\rho c=1$.
} 
\label{figDarkSol}
\end{figure}

A stationary dark soliton, for which $v=0$ and $\beta=1$, has  depth $\rho_0$, width $\xi$ and is sometimes called a {\it black soliton} due to a zero value of the corresponding probability density at the bottom of the density notch.  In such a case one  observes a single point $\pi$--phase flip at a density dip position.
In Fig.~\ref{figDarkSol} we present the evolution of dark soliton wave function, Eq.~(\ref{DarkSol}), for different propagation velocities $v$.

To calculate the energy associated with the dark soliton solution $\Psi_{ds}$, it is convenient to switch to the grand canonical ensemble, where the corresponding energy can be cast into the following form \cite{pethick, PitaevskiiStringariBEC} 
\begin{equation} 
\displaystyle{
E_{ds}=\!\int\limits_{-\infty}^\infty\!\left[ N\frac{\hbar^2}{2m}\left|\frac{\partial \Psi_{ds}}{\partial x} \right|^2\!+\!c\left(N|\Psi_{ds}|^2\!-\!\rho\right)^2\right]\mathrm{d}x
 }.
 \label{DarkSolEnergy}
\end{equation}
Noting that $N\frac{\hbar^2}{2m}\left|\frac{\partial}{\partial x} \Psi_{ds}\right|^2 =c\left(N|\Psi_{ds}|^2\!-\!\rho\right)^2$ one  gets
\begin{align} 
\displaystyle{
E_{ds}=\frac{4}{3}\hbar \rho v_s \left(1-\frac{v}{v_s}\right)^{3/2} 
 }.
 \label{DarkSolEnergy2}
\end{align}
For small velocities $E_{ds}\stackrel{v\ll v_s}{\approx} \frac{4}{3}\hbar \rho v_s -2\frac{\hbar \rho}{v_s} v^2$ 
revealing a particle-like behaviour with a negative effective mass $m_{ds}=-4 \hbar \rho /v_s$. This result is strictly related to the density reduction inside a soliton notch.

To find the soliton momentum one needs to note that there are two important contributions. Namely, apart from the \emph{local momentum} generated around the density dip in a range of the order of $\xi$ \cite{PitaevskiiStringariBEC}
\begin{align} 
\displaystyle{
p_1=-i \hbar\int\limits_{-\infty}^\infty \Psi_{ds}\frac{\partial \Psi_{ds}}{\partial x} \,\mathrm{d}x=-2\hbar\rho \frac{v}{v_s}\sqrt{1-\frac{v^2}{v_s^2}}
 }, 
 \label{DarkSolMomentum1}
\end{align}
we need to take into account the asymptotic change of phase that does not affect the energy. Indeed, if we assume that the system is confined in a very large ring, i.e. with a circumference much larger than the soliton width $\xi/\beta$, then it is clear that the periodic boundary conditions and consequently the requirement of a single valued wave function enforces the appearance of an additional counterflow related to the phase difference given by Eq.~(\ref{DarkSolPhaseChange}). Such a contribution is produced very far from the soliton and can be written as  (see Refs.~\cite{PitaevskiiStringariBEC,Ishikawa80})
	\begin{align} 
p_2=-\hbar \rho  \int \limits_{-\infty}^\infty\frac{\partial\Delta \varphi}{\partial x}\mathrm{d}x
=2 \hbar \rho \, \mathrm{arccos}\left( \frac{v}{v_s} \right).
 \label{DarkSolMomentum2} 
\end{align}
Thus, the total momentum of the dark soliton is equal to $p=p_1+p_2$ and reads
	\begin{align} 
p=2\hbar\rho\Bigg[\mathrm{arccos}\left( \frac{v}{v_s} \right) -\frac{v}{v_s}&\sqrt{1-\frac{v^2}{v_s^2}}\,\Bigg],   
\label{DarkSolMomentum3}
\end{align}
which for $v\ll v_s$ reduces to $p\approx\hbar \pi \rho - 4 \hbar \rho \frac{v}{v_s}$.

Let us come back to the fully quantum many-body problem described within the Lieb-Liniger model. As discussed in Sec.~\ref{Lieb-Liniger_model}, the type--II spectrum for a very weak repulsion ($\gamma \ll 1$) has been extensively studied in the context of the relation to dark solitons. In 1976, P. Kulish, S. V. Manakov, and L. D. Faddeev noted that the semiclassical dark soliton spectrum, Eqs.~(\ref{DarkSolEnergy2}) and (\ref{DarkSolMomentum3}), closely follows the type--II dispersion relation of the Bose gas with weakly repulsive contact interactions \cite{Kulish76}. A few years later, in 1980, M. Ishikawa and H. Takayama  (see Ref.~\cite{Ishikawa80}) pointed out that such a relation becomes exact in a very weak repulsion limit. Indeed, by using the result of M. Kac and H. Pollard, presented in Ref.~\cite{Kac50}  (see also Ref.~\cite{Hutson63}), they found the energy and momentum of the hole excitations in the Lieb-Liniger model with periodic boundary conditions for $\gamma \rightarrow 0_+$. It turned out that the resulting type--II spectrum  coincides with this obtained for the mean-field dark soliton, i.e. within the semiclassical treatment of the nonlinear Schr\"{o}dinger equation, Eq.~(\ref{NLS_2}). 

\begin{figure}[h!] 
\includegraphics[scale=0.33]{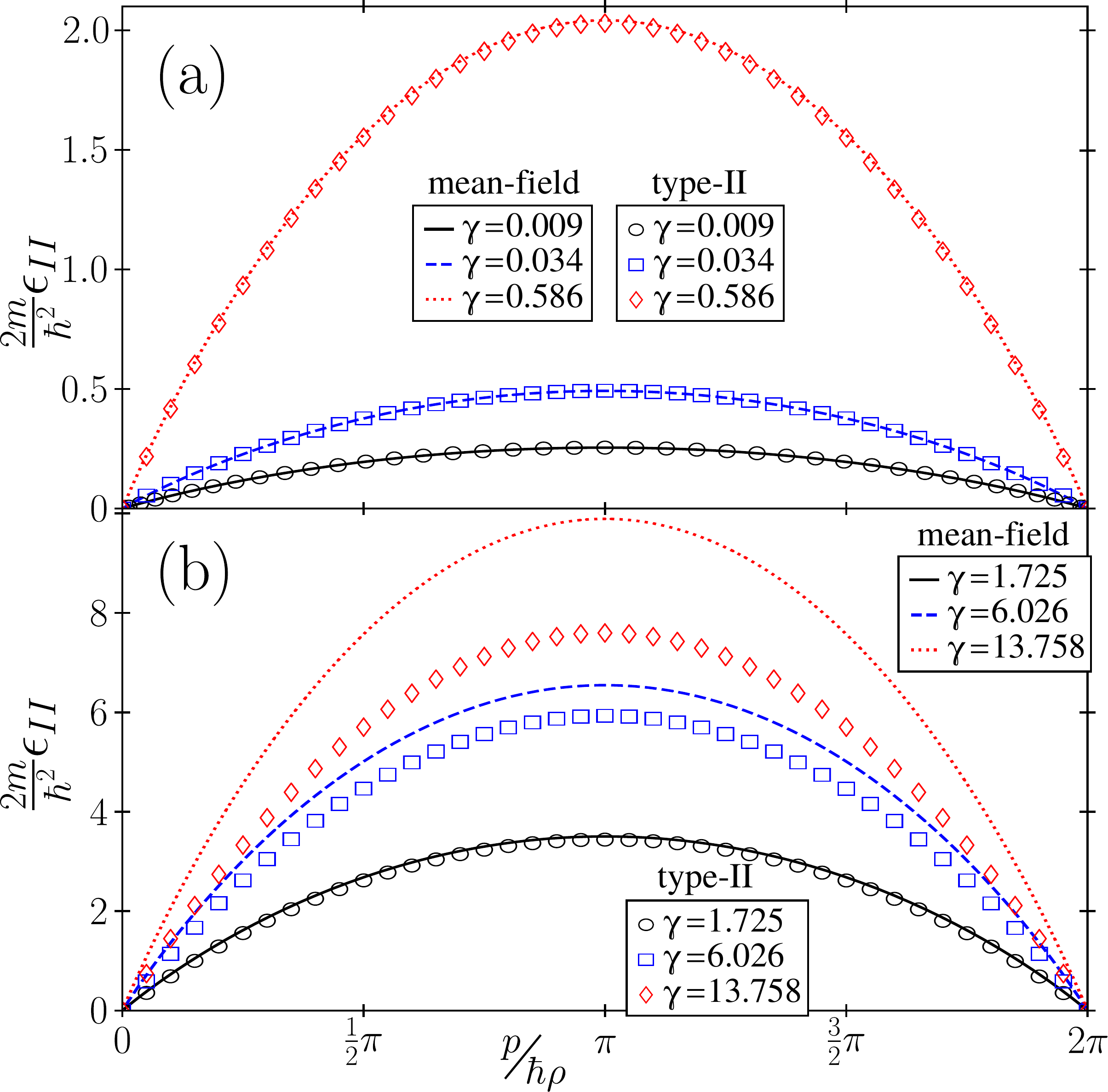} 
\vspace{-0.2cm}
\caption{
Panels (a) and (b) present the mean-field dark soliton dispersion relation compared with the type--II spectrum of the periodic Lieb-Liniger model for weak and strong repulsion, respectively. The interaction strength $\gamma$ is indicated in legends.  
} 
\label{DarkSolConjDispRel}
\end{figure}

In Fig.~\ref{DarkSolConjDispRel}, we present a comparison between the mean-field dark soliton spectrum and the type--II dispersion relation obtained by the numerical solutions of the corresponding integral Eqs.~(\ref{LL_PBC_Type2_Therm_1})--(\ref{LL_PBC_Type2_Therm_2}). Note that both relations coincide in the weak interaction regime. This result, being the main argument in favour, provoked the discussion about the solitonic nature of the type--II eigenstates.

\subsection{Periodic boundary conditions}
\label{Solitons_PBC} 

As for now we discussed the soliton solutions of the Gross-Pitaevskii equation in an infinite space. It is clear that the profiles of the solutions, analyzed in the previous section, are valid when tails of the soliton structure do not feel the boundaries, i.e. when $\xi\ll L<\infty$ and when $x_0$ is far from the system edges. Now, we impose the periodic boundary conditions that entails the following requirements
\begin{align} 
\displaystyle{
\Psi(0)=\Psi(L), \qquad \frac{\mathrm{d}}{\mathrm{d}x}\Psi(0)=\frac{\mathrm{d}}{\mathrm{d}x}\Psi(L)
 }.
 \label{LL2Sec3_Per_1}
\end{align}
Note that when merging the ends of the dark soliton wave function we deal with an additional momentum, as in Eq.~(\ref{DarkSolMomentum2}). Nevertheless, for finite systems ($L<\infty$) the corresponding momentum per particle does not vanish even for a totally black soliton. Thus, the stationary soliton solution in the system of a ring geometry should be investigated in a rotating frame of reference. Otherwise, one needs to deal with the time-dependent nonlinearity $|\Psi(x,t)|^2$.

In 1D the stationary Gross-Pitaevskii equation in a frame rotating with angular velocity $\omega$ reads
 \begin{equation}
\displaystyle{
\left[
\left(-i\frac{\partial}{\partial \theta} - \Omega \right)^2 \!+2\bar{c}_\theta N |\Psi(\theta)|^2 -\bar{\mu}_{\text{ch},\theta}\right]\!\Psi(\theta)=0
},
 \label{LL2Sec3_Per_2}
\end{equation}
where the angle $\theta=2\pi \frac{x}{L}$ indicates the position in a ring and dimensionless parameters are given by
 \begin{align}
 \begin{array}{ll}
&\displaystyle{\Omega=\frac{m\omega L^2}{4\pi^2\hbar}},
\qquad
\displaystyle{\bar{c}_\theta =\frac{L}{2\pi}\frac{2m}{\hbar^2}c=\frac{L}{2\pi}\bar{c}}, 
\vspace{0.2cm}\\ 
&
\displaystyle{\bar{\mu}_{\text{ch},\theta}=\left(\frac{L}{2\pi}\right)^2\frac{2m}{\hbar^2}\mu_\text{ch}=\left(\frac{L}{2\pi}\right)^2\bar{\mu}_\text{ch}}. 
\end{array}
 \label{LL2Sec3_Per_2_params}
\end{align}
A trivial solution is simply given by the plane wave $\Psi_{J}(x)=\mathrm{e}^{i J\theta}/\sqrt{2\pi}$
with $J\in \mathbb{Z}$ being the so-called phase winding number. In order to find the stationary soliton-train (ST) solutions one applies the ansatz $\Psi_{J,j}^{(\text{ST})}(x)=\sqrt{\rho_j(\theta)} \, \mathrm{e}^{i\varphi_{J,j}(\theta)}$,
where the density $\rho_j(\theta)$ and the phase $\varphi_{J,j}(\theta)$ are real-valued functions that have to satisfy the following cyclicity conditions 
 \begin{align}
\begin{array}{ll}
&\rho_{j}(\theta+2\pi)-\rho_j(\theta)=0, \\ 
&\varphi_{J,j}(\theta+2\pi)-\varphi_{J,j}(\theta)=2\pi J.
\end{array}
 \label{LL2Sec3_Per_5}
\end{align}
 The parameter $j=1,2,3,\ldots$ indicates the number of solitons related to the solution $\Psi_{J,j}^{(\text{ST})}$. Within the ansatz for $\Psi_{J,j}^{(\text{ST})}$, Eq.~(\ref{LL2Sec3_Per_2})  can be reduced to the following system of differential equations \cite{WuZaremba2013}
 \begin{align}
\displaystyle{
\frac{1}{4}\left(\frac{\mathrm{d}\rho}{\mathrm{d} \theta}\right)^2-\bar{c}_\theta N \rho^3+ \bar{\mu}_{\text{ch},\theta} \rho^2+V\rho+W^2=0	
},
 \label{LL2Sec3_Per_13}
\\
\frac{\mathrm{d}\varphi}{\mathrm{d}\theta}= \Omega +\frac{W}{\rho}, \qquad V,\,\,W=\mathrm{const}.
 \label{LL2Sec3_Per_9}
\end{align}
Equation~(\ref{LL2Sec3_Per_13}) turns out to have analytical solutions that can be cast into the following form \cite{ KanamotoCarrUeda2009, WuZaremba2013}
\begin{eqnarray}
\displaystyle{
\sqrt{\rho_{j}} }
= \left\{ 
\begin{array}{llllll}
\displaystyle{\!\!\mathcal{N}\sqrt{1\!+\!\eta \, \mathrm{dn}^2\!\left(\kappa_{j}\widetilde{\theta}\, \big| m\right)}   } & \text{for} & \bar{c}_\theta>0 \vspace{0.2cm}
\\ 
\displaystyle{\!\!\mathcal{N}\sqrt{\mathrm{dn}^2\!\left(\kappa_{j}\widetilde{\theta}\, \big| m\right)\!-\!\eta m'}  }
& \text{for} & \bar{c}_\theta<0
\end{array}
\right. ,
 \label{LL2Sec3_Per_DENSTITY}
\end{eqnarray}
where $\widetilde{\theta}=\theta-\theta_0$, $\kappa_j=j K(m)/\pi$ and the Jacobi elliptic function
 \begin{equation}
\displaystyle{
\mathrm{dn}(u|m)=\sqrt{1\!-\!m \,\mathrm{sin} \phi}, \quad u=\int \limits_{0}^\phi \frac{\mathrm{d}\theta}{\sqrt{1\!-\!m\, \mathrm{sin}^2\theta}}
}.
 \label{LL2Sec3_Per_JacobiEllipticDN}
\end{equation}
is parameterized by the so-called elliptic parameter $m$ that together with the complementary elliptic parameter $m'=1-m$ are real numbers and belong to the unit interval, i.e. $m,m'\in [0,1]$ \cite{KanamotoCarrUeda2009, WuZaremba2013,Abramovitz}. The normalization condition, $\int_{0}^{2\pi}\rho_j(\theta)\mathrm{d}\theta=1$, leads to 
 \begin{equation}
 \displaystyle{\mathcal{N}=\left\{ 
\begin{array}{lcccc}
\displaystyle{ \!\sqrt{\frac{K(m)}{2\pi\big[K(m)\!+\!\eta E(m)\big]}}   }
& \text{for} & \bar{c}_\theta>0 
\vspace{0.2cm}
\\ 
\displaystyle{ \!\sqrt{\frac{K(m)}{2\pi\big[E(m)\!-\!\eta m' K(m)\big]}}  }
& \text{for} & \bar{c}_\theta<0
\end{array}
\right. 
}.
 \label{LL2Sec3_Per_Norms}
\end{equation}
The complete elliptic integrals of the first $K(m)$ and the second kind $E(m)$ are defined as follows \cite{Abramovitz}
 \begin{align}
 \begin{array}{l}
\displaystyle{
K(m)=\int\limits_{0}^{\pi/2}\frac{\mathrm{d}\theta}{\sqrt{1-m \,\mathrm{sin}^2\theta}},}
\\
\displaystyle{E(m)=\int\limits_{0}^{\pi/2}\sqrt{1-m \,\mathrm{sin}^2\theta}\, \mathrm{d}\theta.
}
\end{array}
 \label{LL2Sec3_Per_ComplIntEqs}
\end{align}
Note that the soliton position $0<\theta_0\leq 2\pi$ is arbitrary due to the translational invariance of the considered system and indicates that the soliton solutions are symmetry broken states. For the sake of convenience, one introduces the following functions \cite{ KanamotoCarrUeda2009, WuZaremba2013}
 \begin{equation}
\displaystyle{
f\equiv\left\{ 
\begin{array}{ccccc}
\!+\left[\pi \bar{c}_\theta N - \mathcal{A}_j+\mathcal{B}_j \right] 
& \text{for} & \bar{c}_\theta>0 
\vspace{0.2cm}
\\ 
 \!-\left[\pi \bar{c}_\theta N - \mathcal{A}_j+\mathcal{B}_j \right]  
& \text{for} & \bar{c}_\theta<0
\end{array}
\right.
},
 \label{LL2Sec3_Per_Function_f}
\end{equation}
 \begin{equation}
\displaystyle{
h\equiv\left\{ 
\begin{array}{ccccc}
\!+\left[\pi \bar{c}_\theta N \!-\! \mathcal{A}_j \!+\! \mathcal{B}_j \!+\! m\mathcal{A}_j \right] 
& \text{for} & \bar{c}_\theta>0 
\vspace{0.2cm}
\\ 
 \!-\left[\pi \bar{c}_\theta N \!-\! \mathcal{A}_j \!+\! \mathcal{B}_j \!+\! m \mathcal{A}_j \right]  
&\text{for} & \bar{c}_\theta<0
\end{array}
\right.
},
 \label{LL2Sec3_Per_Function_h}
\end{equation}
 \begin{align}
 \begin{array}{l}
\displaystyle{
g\equiv \pi \bar{c}_\theta N +\mathcal{B}_j}, \\
\displaystyle{
\mathcal{S}\equiv \mathrm{sign}(J-\Omega)=}
\left\{ 
\begin{array}{cc}
+1, & J>  \Omega  \\
-1, & J<  \Omega  
\end{array}
\right.,
\end{array}
 \label{LL2Sec3_Per_Function_g}
\end{align}
where $\mathcal{A}_j=2 j^2 K^2(m)$ and $\mathcal{B}_j=2j^2 K(m) E(m)$.
Additionally, the parameters $\eta$, describing the soliton properties as its depth ($c>0$) or height ($c<0$),  $W$ and the chemical potential $\mu_\text{ch}$ can be obtained by substituting the solution in Eq.~(\ref{LL2Sec3_Per_DENSTITY}) into the relation~(\ref{LL2Sec3_Per_13}) \cite{ KanamotoCarrUeda2009, WuZaremba2013}
 \begin{eqnarray}
 \begin{array}{l}
\eta = \left\{ 
\begin{array}{clcccc}
-\mathcal{A}_j/g & \in [-1,0] 
&& \text{for} && \bar{c}_\theta>0 
\vspace{0.2cm}
\\ 
 g/( m' \mathcal{A}_j) & \in [0,1]
&& \text{for} && \bar{c}_\theta<0
\end{array}
\right., 
\vspace{0.1cm}\\
\displaystyle{W=\frac{\mathcal{S}}{2\pi^3|\bar{c}_\theta|N}\sqrt{\frac{fgh}{2}},}
 \vspace{0.1cm}\\
\displaystyle{
\bar{\mu}_{\text{ch},\theta}=\frac{3\, \bar{c}_\theta N }{2\pi}\! +\! \left(\!\frac{j}{\pi}\!\right)^2 \!\! \left[ 3 K(m) E(m)-(2\!-\!m)K^2(m) \right]
}.
\end{array}
 \label{LL2Sec3_Per_Function_etaAndW}
\end{eqnarray}

\begin{figure}[h!] 
\begin{center}\includegraphics[scale=0.8]{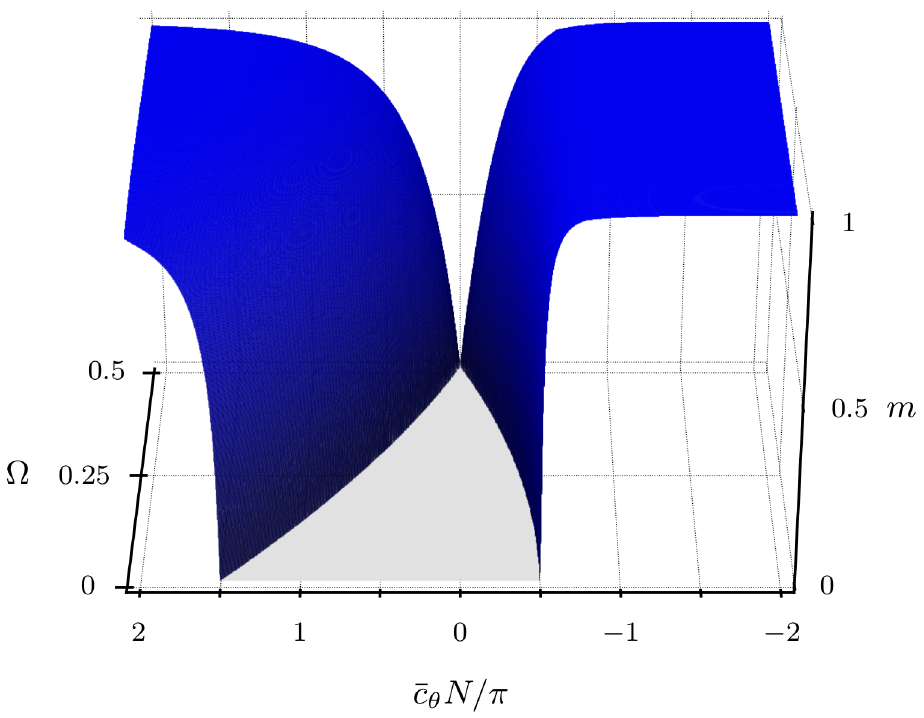} \end{center}
\vspace{-0.5cm}
\caption{ The elliptic parameter $m$ versus  $\Omega$ and  the effective strength of interactions $\bar{c}_\theta N$ for a single soliton state ($j=1$). The corresponding phase winding number is equal to $J=1$ for $\bar{c}_\theta>0$ and $J=0$ for $\bar{c}_\theta<0$.
The gray sheet presented in the plot indicates the range of parameters $\Omega$ and $\bar{c}_\theta N /\pi$ for which the single soliton solutions do not exist.
}
\label{PBC_MF_SolExistence}
\end{figure}

An analysis of the soliton solutions requires the determination of the allowed values of the elliptic parameter $m$. 
Moreover, we also need to deal with the parameter $\Omega$ that, in general, is not free and turns out to be strictly associated with a particular soliton state. Thus, the value of $m$, that corresponds to a concrete solution, is a function of the effective interactions strength  $\bar{c}_\theta N$ and the rescaled angular velocity $\Omega$. The latter quantity plays a key role in the periodicity of the phase $\varphi_{J,j}$. By integrating Eq.~(\ref{LL2Sec3_Per_9}) with $\rho$ in Eq.~(\ref{LL2Sec3_Per_DENSTITY}) one obtains \cite{KanamotoCarrUeda2009, WuZaremba2013}
 \begin{align}
 \begin{array}{l}
\displaystyle{
\varphi_{J,j}(\theta)=\Omega \theta +\frac{\mathcal{S}}{j K(m)}\sqrt{\frac{gh}{2f}}\, \Pi\!\left(\xi;\kappa_j \widetilde{\theta}\, \big |m\right),} \vspace{0.3cm}\\
\xi=\left\{
\begin{array}{ccccc}
\displaystyle{-m \mathcal{A}_j/(2f) } && \text{for} &&  \bar{c}_\theta>0 
\\ 
\displaystyle{+m \mathcal{A}_j/(2f)  } &&  \text{for} &&  \bar{c}_\theta<0 
\end{array}
\right.,
\end{array}
 \label{LL2Sec3_Per_PHASE}
\end{align}
with the elliptic integral of the third kind \cite{Abramovitz}
 \begin{align}
\displaystyle{
\Pi(\xi;u|m)=\int \frac{\mathrm{d}u}{1-\xi \,\mathrm{sn}^2(u|m)}, 
}
 \label{LL2Sec3_Per_EllipticInt}
\end{align}
where $\mathrm{dn}^2(u|m)=1-m\,\mathrm{sn}^2(u|m)$.
The phase $\varphi_{J,j}$ has to satisfy the boundary condition in Eq.~(\ref{LL2Sec3_Per_5}) which, for a given density profile, can be reached by tuning the angular velocity $\Omega$. Now, it is clear that all the parameters: $J,j, m, \Omega$ and $\bar{c}_\theta N $, are interrelated.  In Fig.~\ref{PBC_MF_SolExistence}, we show how the elliptic parameter $m$ depends on $\Omega$ and $\bar{c}_\theta N$ for a single soliton ($j=1$) state.

In contrast to the case of an infinite space, where the soliton can possess an arbitrary width, the presence of periodic boundary conditions imposes some additional restrictions on the soliton solutions. Consequently, a bright soliton appears in the ground state when the attraction, described by $\bar{c}_\theta N<0$ parameter, is strong enough \cite{Kanamoto2003_pt}. Otherwise, the ground state is uniform. To find the critical point $\bar{c}_\theta^{\hspace{0.05cm}\text{cr}}$, for which a quantum phase transition from the uniform to the bright soliton state occurs, we take $j=1$, $J=0$ and $\Omega=0$, which guarantees the lowest energy in the laboratory frame. Additionally, by noting that $\eta|_{\bar{c}_\theta<0}\in[0,1]$ is a monotonically increasing function of $m$, we consider the limit $m\rightarrow 0_+$ that should correspond to $\eta\rightarrow 0_+$. A simple Taylor expansion%
 \begin{equation}
\displaystyle{
\eta  \Big|_{ \substack{\bar{c}_\theta<0   \\  j=1 }} = \left(1\!+\!\frac{2\bar{c}_\theta N}{\pi}\right)\!+\!\left( \frac{1}{2}\!+\!\frac{\bar{c}_\theta N}{\pi}\right)m \!+\!\mathcal{O}\big(m^2\big)
},
 \label{BrightSolitonExistence}
\end{equation}
leads to a conclusion that the quantum phase transition occurs at $\bar{c}_\theta^{\hspace{0.05cm}\text{cr}}=-\frac{\pi}{2N}$ \cite{Kanamoto2003_pt}. Note that the phase, Eq.~(\ref{LL2Sec3_Per_PHASE}), is uniform when $\bar{c}_\theta-\bar{c}_\theta^{\hspace{0.05cm}\text{cr}}\rightarrow 0_+$ 
satisfying the assumption $J=0$. Thus, one can expect a bright soliton existence in the ground state when, cf. Fig.~\ref{PBC_MF_SolExistence},
 \begin{align}
\displaystyle{
\bar{c}_\theta N <-\frac{\pi}{2}
}.
 \label{BrightSolitonExistence2}
\end{align}

\begin{figure}[h!] 
\begin{center}\includegraphics[scale=0.165]{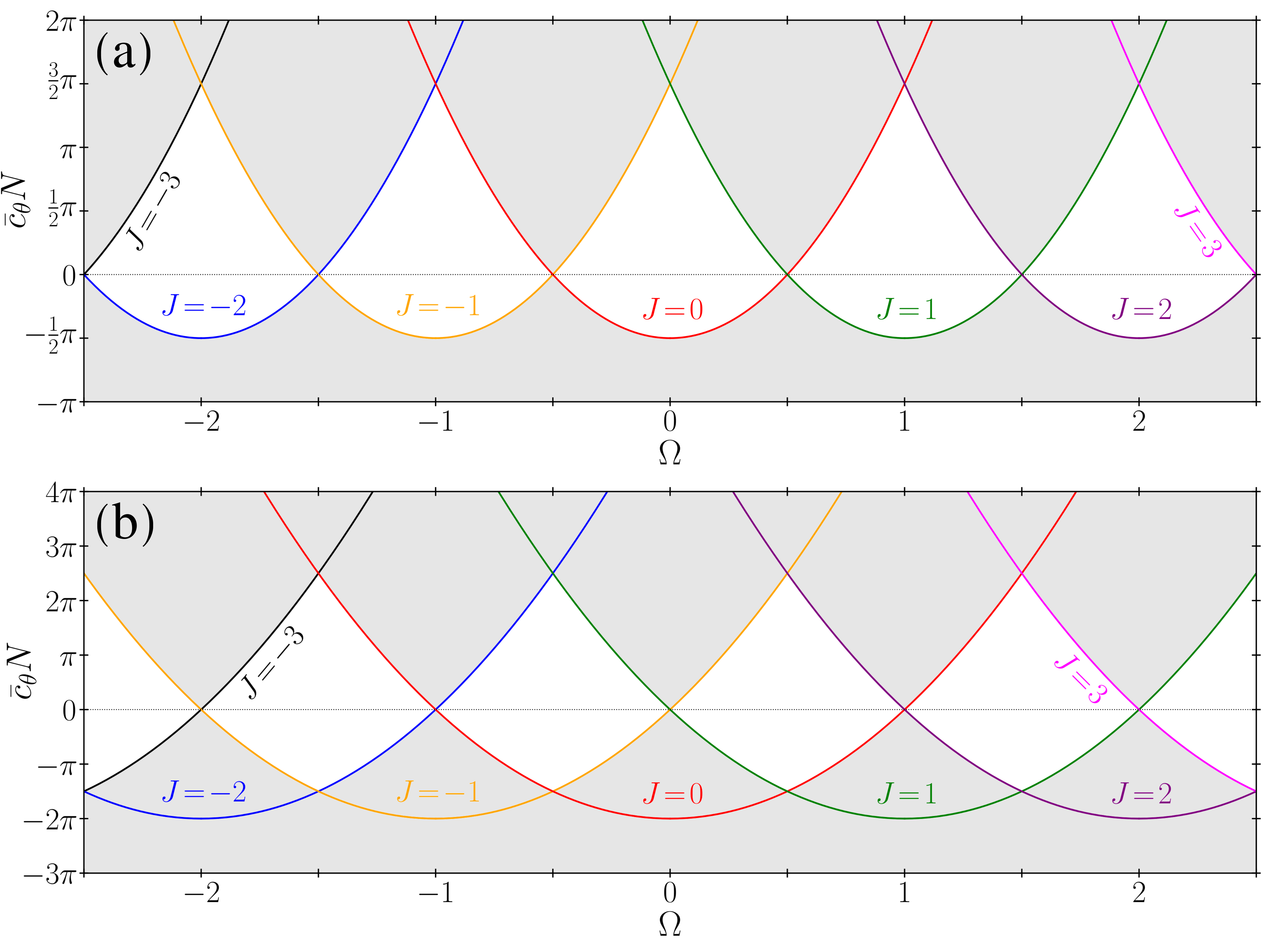} \end{center}
\vspace{-0.5cm}
\caption{ 
Panels (a) and (b) present diagrams of single ($j=1$) and double ($j=2$) soliton solutions existence, respectively. Parabolic curves represent the phase boundaries and correspond to $m=0$.  Different values of the phase winding $J$ are indicated by different colors on the plots.  While the regimes where soliton solutions exist are marked in gray, white areas correspond to regions where only the plane wave solutions can be found. 
}
\label{PBC_MF_SolExistence2}
\end{figure}

In general, in the system with a ring geometry, dark and bright solitons do not exist for arbitrary values of parameters of solutions. It turns out that the hyperplanes of $m=0$ determine boundaries of the soliton solutions existence in the parameter space and separate them from the regions where only the plane wave solutions exist (see Fig.~\ref{PBC_MF_SolExistence2}). 
For $\bar{c}_\theta>0$  dark soliton states are stable and coexist with stable plane wave solutions. On the other hand, in the regime of strong attraction the plane waves become dynamically unstable and the only existing stable solutions correspond to bright solitons.  
A scrupulous discussion concerning a linear stability of the soliton solutions can be found in Ref.~\cite{KanamotoCarrUeda2009}.

\begin{figure}[h!] 
\begin{center}\includegraphics[scale=0.3]{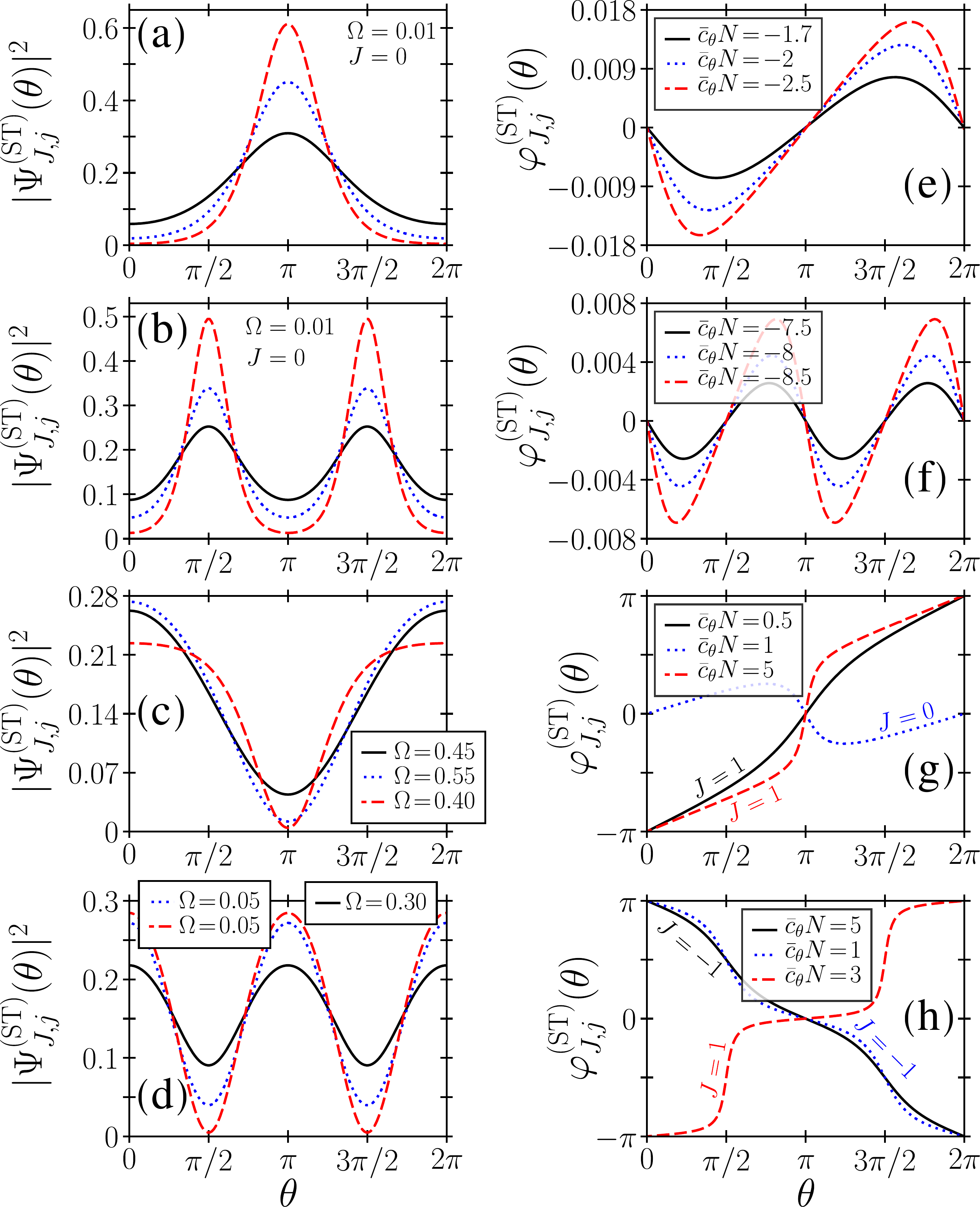} \end{center}
\vspace{-0.5cm}
\caption{
Left column present probability densities for the chosen soliton states. The corresponding phase distributions are illustrated in right column.
While first two rows, i.e. (a), (e) and (b), (f), present single and double bright solitons, the other two, i.e. (c), (g) and (d), (h), show results obtained for single and double dark solitons, respectively. All the curves related to bright solitons were obtained for $\Omega=0.01$, $J=0$ and for different attraction strengths $\bar{c}_\theta N$. For dark solitons we also present the results corresponding to different angular velocities $\Omega$ and phase winding numbers $J$ (see legends).
}\vspace{-0.2cm}
\label{PBC_MF_Sols}
\end{figure}

Let us now focus on the repulsively interacting case ($\bar{c}_\theta>0$) in which dark soliton solutions are expected. It is clear that the corresponding density notch (or density notches, i.e. a dark soliton train) is the shallower the closer to zero the parameter $\eta$ is.  In the limiting case $\eta\rightarrow 0 _-$, the density notch disappears and the solution passes continuously to a plane wave. Such a transition occurs through the parabolic curves indicating the phase boundaries  and corresponding to $m=0$ (see Fig.~\ref{PBC_MF_SolExistence2}). On the other hand, when $\eta\rightarrow -1$ the density dip reaches the zero probability density and we deal with a black soliton (or soliton train). For $-1<\eta<0$, the density notch is partially filled and one can observe gray solitons living between the two parabolic phase boundaries that refer to distinct phase winding numbers $J$ and $J'$. Both $m=0$ curves meet for $\bar{c}_\theta N=0$ at some specific value $\widetilde{\Omega}(j)$, with $j=|J-J'|$. It is worth stressing that two dark soliton trains with $j$ density dips and different phase windings $J$ and $J'$ approach the same soliton train solution at the line defined by $\widetilde{\Omega}(j=|J-J'|)$ \cite{KanamotoCarrUeda2009}. 
In Fig.~\ref{PBC_MF_Sols} we illustrate a few soliton solutions.

At the end we  stress that the average momentum per particle in a soliton state $\Psi_{J,j}$ reads
\begin{align}
\displaystyle{
\frac{\left<P\right>_\Psi}{N}=-i\hbar\int\limits_{0}^{2\pi}\Psi_{J,j}^{*}\frac{\partial \Psi_{J,j}}{\partial\theta}\mathrm{d}\theta
=
\hbar \big(\Omega+2\pi W\big).
}
 \label{L_VS_Omega}
\end{align}

\subsection{Open boundary conditions}
\label{Solitons_OBC}

 It turns out that analytical soliton solutions can be found also in the presence of the infinite square well trapping potential. In such a case it is enough to consider the following stationary GPE 
\begin{align} 
\displaystyle{
-\partial_x^2  \Psi(x)+2\bar{c} N|\Psi(x)|^2 \Psi(x)= \bar{\mu}_\text{ch}\Phi(x)}, 
 \label{NLS_OBC_!}
\end{align} 
 with the boundary conditions $\Psi(0)=\Psi(L)=0$. When $\Psi$ satisfying Eq.~(\ref{NLS_OBC_!}) vanishes at least at a single point in the considered interval one can assume $\Psi$ to be real-valued function and thus Eq.~(\ref{NLS_OBC_!}) reduces to
\begin{align} 
\displaystyle{
-\partial_x^2\Psi(x)+2\bar{c} N\Psi(x)^3 = \bar{\mu}_\text{ch}\Psi(x) 
}, 
 \label{NLS_OBC_!2}
\end{align}
Indeed, by a simple Taylor expansion of $\Psi$ in the vicinity of the hard wall box edges, where the wave function $\Psi$ drops, one can show that the complex phase must be constant \cite{CarrClarkReinhardt2000a}.
 The analytical solutions of Eq.~(\ref{NLS_OBC_!2}) with open boundary conditions take the following form \cite{CarrClarkReinhardt2000a,CarrClarkReinhardt2000b}
\begin{equation} 
\displaystyle{
\Psi_j= \left\{
\begin{array}{lcccc}
\!\!\displaystyle{ \mathcal{N}\, \mathrm{sn}\!\left( K(m)\, \widetilde{x}_j \Big| m \right)   }
& \text{for} & \bar{c}>0 
\vspace{0.1cm}
\\ 
\!\!\displaystyle{ \mathcal{N}\, \mathrm{cn}\!\left( K(m)( \widetilde{x}_j-1) \Big| m \right)  }
& \text{for} & \bar{c}<0
\end{array}
\right.
},
 \label{NLS_OBC_!3}
\end{equation}
with $\widetilde{x}_j=2(j+1)x/L$ and $\mathcal{N}=\beta_j \big[ m/(|\bar{c}|N L^2)\big]^{1/2}$, where $\beta_j=2(j+1)K(m)$.
The Jacobi elliptic functions $\mathrm{sn}(u|m)$ and $\mathrm{cn}(u|m)$ satisfy \cite{Abramovitz}
\begin{align} 
\begin{array}{l}
\displaystyle{
\mathrm{cn}^2(u|m)+\mathrm{sn}^2(u|m)=1,} \\
\displaystyle{m \,\mathrm{sn}^2(u|m)+\mathrm{dn}^2(u|m)=1,}
\end{array}
 \label{NLS_OBC_elliptics}
\end{align}
with $\mathrm{dn}(u|m)$ defined in Eq.~(\ref{LL2Sec3_Per_JacobiEllipticDN}).
The elliptic parameter $m$ can be determined by using the normalization condition
\begin{align} 
\begin{array}{lcccc}
\displaystyle{   \beta_j^2\left(1-\frac{E(m)}{K(m)}\right)=|\bar{c}| N L   }
& \text{for} & \bar{c}>0 
\vspace{0.2cm}
\\ 
\displaystyle{  \beta_j^2\left(\frac{E(m)}{K(m)}-m'\right)=|\bar{c}| N L  }
& \text{for} & \bar{c}<0
\end{array},
 \label{NLS_OBC_!5}
\end{align}
where $j=0,1,2,\ldots$ counts the number of nodes of $\Psi_j$ inside the hard wall box. It should be stressed that we deal with the ground state when $j=0$. In contrast to the repulsively interacting case ($\bar{c}>0$), where $j$ specifies the number of dark solitons being always a manifestation of the system excitations, a single bright soliton state can appear for the attractively interacting ground state ($\bar{c}<0,\,j=0$). Hence, in the latter case, the number of bright solitons is indicated by  $j+1$.  The corresponding chemical potential reads
\begin{equation} 
\displaystyle{
\bar{\mu}_\text{ch}=\frac{2m}{\hbar^2}\mu_\text{ch}=\left\{
\begin{array}{lcccc}
\displaystyle{ +\frac{\beta_j^2}{L^2} (1 + m)   }
& \text{for} & \bar{c}>0 
\vspace{0.2cm}
\\ 
\displaystyle{ -\frac{\beta_j^2}{L^2}  (2m-1) }
& \text{for} & \bar{c}<0
\end{array}
\right.
}.
 \label{NLS_OBC_!6}
\end{equation}

\begin{figure}[h!] 
\begin{center}\includegraphics[scale=0.25]{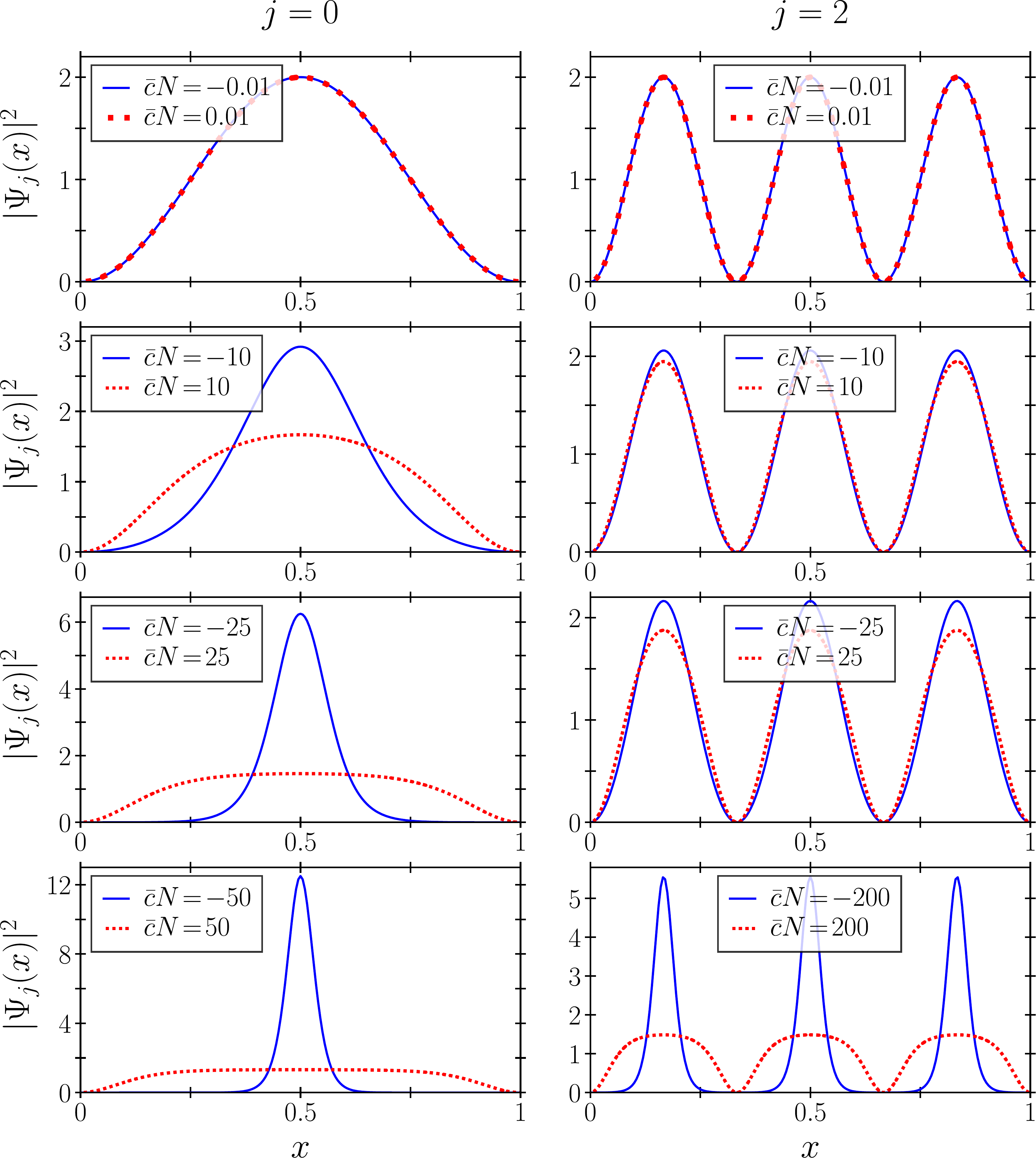} \end{center}
\vspace{-0.5cm}
\caption{ 
Probability densities corresponding to $\Psi_j$ in Eq.~(\ref{NLS_OBC_!3}) for various interaction strengths $\bar{c}N$ (indicated in legends) and for $j=0$ (left column) and $j=2$ (right column). 
}
\label{OBC_MF_Sols}
\end{figure}

It is also worth stressing that when the separation between nodes becomes significantly larger than the soliton width, the analytical expressions given by Eqs.~(\ref{NLS_OBC_!3}) in the neighbourhood of the soliton structures approach the bright and dark soliton solutions obtained for an infinite space \cite{CarrClarkReinhardt2000a,CarrClarkReinhardt2000b}.
In Fig.~\ref{OBC_MF_Sols} we present chosen solutions of the nonlinear Schr\"{o}dinger equation with open boundary conditions.

\section{Quantum dark solitons in many-body eigenstates of Lieb-Liniger gas}
\label{QuantumSolitonsInMBstates}

The mean-field approach does not take into account quantum many-body effects that should be expected as long as we deal with many-body systems. 
Therefore, the Gross-Pitaevskii equation cannot be used to explain the quantum character of solitons, i.e. the beyond mean-field quantum many-body effects \cite{Syrwid2017HW, Syrwid15, Syrwid16, Lai89a, Lai89b, Corney1997, Corney2001, DziarmagaSacha2002, Law2003, DziarmagaKarkuszewski2003, Dziarmaga2004A, DziarmagaSacha2006A, Streltsov2008A, WeissCastin2009, Mishmash2009a, Mishmash2009b, SachaZakrzewski2009A, DziarmagaDeuar2010, Mishmash2010Reply, MartinRuostekoski2010, Gertjerenken2012, Gertjerenken2013, DelandeSachaAvazbaev2013, DelandeSachaMBMatterWave2014, KronkeSmehchel2015, Hans2015A, Marchant2016A, Katsimiga2017a, Katsimiga2017b, Oldziejewski2018, CastinArxiv}. 
Despite the experimental realization of solitons is well established, the observation of their quantum nature constitutes a great challenge and requires innovative laboratory techniques \cite{Boisse2017}.

The conjecture that the type--II ({\it yrast}/hole excited) eigenstates of the weakly interacting Lieb-Liniger gas confined in a ring are related to mean-field dark solitons appeared nearly 40 years ago due to the coincidence between the type--II spectrum and the mean-field dark soliton dispersion relation \cite{Kulish76,Ishikawa80} (see also Sec.~\ref{Solitons_Inf}). Since then, other evidence underpinning the anticipated relation have been found \cite{Kanamoto2008, Kanamoto2010, Komineas2002, Jackson2002, Karpiuk2012, Karpiuk2015, Sato2012, Sato2012arxiv, Sato2016, Gawryluk2017}. Especially, it has been shown by Sato et al.~\cite{Sato2012,Sato2012arxiv,Sato2016} that dark soliton signatures can be visible in the reduced single particle density, calculated for a proper superposition of the type--II eigenstates.
 Note that, in the presence of periodic boundary conditions, the many-body Hamiltonian is invariant under translations of all particles by the same distance. Thus, the system eigenstates also possess the same spatial translation symmetry, which is obviously broken by the mean-field soliton solutions. Therefore, the soliton-like structures cannot be displayed in the reduced single particle density calculated for single eigenstates of the periodic system.

All the arguments concerning the above-mentioned conjecture were indirect and there was no final incontestable answer for the question if a single quantum many-body eigenstate can reveal dark soliton signatures like the density notch and the corresponding phase flip. Here we show that continuous space translation symmetry can be broken by the successive measurement of particles' positions, leading to an emergence of soliton-like structures when the system is initially prepared in a type--II ({\it yrast}/hole excited) eigenstate. We also point out that multiple hole excitations are strictly associated with multiple dark soliton signatures. The system dynamics in the presence of weak and strong repulsion is investigated as well \cite{Syrwid15, Syrwid16}.

In the case of open boundary conditions the hole excitation scenario  and the corresponding dispersion relation is analogous to the periodic case (see Sec.~\ref{LLElementaryExcitations_OBC}). This may suggest dark soliton-like nature of quantum many-body hole excited eigenstates in the system with open boundary conditions. In this section we point out that such an identification is correct and the eigenstates in question are unequivocally connected with quantum dark solitons \cite{Syrwid2017HW}.

\subsection{Periodic case: dark solitons emergence in course of successive measurement of particles' positions}
\label{BoseEmergence}

Eigenstates of periodic systems are invariant under spatial translations, and thus the corresponding reduced single particle density has to be uniform in space revealing no inhomogeneities expected for solitons. This feature effectively precluded the investigations of the type--II eigenstates' solitonic nature. We perceived that the anticipated dark soliton-like structures may emerge after a space translation symmetry breaking, which can be induced by the successive  measurement of positions of particles in the system. 
In order to perform numerical simulations of such a process, one needs to calculate the conditional single particle probability densities for consecutive measurements. That is, to choose the position of the $j$-th particle in the system provided that $j-1$ particles have been already measured at positions $\widetilde{x}_{s=1,2,\ldots,j-1}$,  we have to determine
$\rho_j(x_j)\propto\big<\Psi_{j-1}|\hat{\psi}^\dagger(x_j)\hat{\psi}(x_j)|\Psi_{j-1}\big>$, where $\big|\Psi_s\big>\propto \hat{\psi}(\widetilde{x}_{s})\big|\Psi_{s-1}\big>$ and $\hat{\psi}$ is the Bose field operator. Note that if the initial state $\big|\Psi_0\big>$ is translationally invariant, then $\rho_1(x_1)=\frac{1}{L}$.

In general, the distributions $\rho_j$ are given by very cumbersome multidimensional integrals.
 Fortunately, dealing with the Bethe eigenstates, we can take advantage from the analytical determinant formulas for the so-called Bose field form factors $\mathcal{F}(\{q\}_{M-1},\{\mu\}_{M})=\big<\{q\}_{M-1}\big|\hat{\Psi}(0)\big|\{\mu\}_{M}\big>$ (see Refs.~\cite{Kojima1997,CauxSlavnov2007}). This allows us to express $\rho_j$ as follows \cite{Syrwid15}
\begin{align}
\rho_j(x_j)=\sum_{\{k\}_{N-j}} \frac{\big|\Gamma_j(x_j,\{k\}_{N-j})\big|^2}{N-j+1},
\label{consecutiveMeasurement}
\end{align} 
 where the summation is performed over all sets $\{k\}_{N-j}$ of quasimomenta parameterizing  eigenstates $\big|\{k\}_{N-j}\big>$ of the system containing $N-j$ particles. All eigenstates $\big|\{k\}_{N-j}\big>$ are normalized to unity and 
 \begin{eqnarray}
\Gamma_j(x_j,\{k\}_{N-j})=&\sum_{\{q\}_{N-j+1}}\Gamma_{j-1}(x_{j-1},\{q\}_{N-j+1}) \nonumber
\\ \nonumber
&\times \mathrm{e}^{ \frac{i}{\hbar} [P(\{q\}_{N-j+1})-P(\{k\}_{N-j}) ]x_j} 
\\ 
&\times \mathcal{F}(\{k\}_{N-j},\{q\}_{N-j+1}),
\label{consecutiveMeasurement2}
\end{eqnarray} 
where $P(\{k\}_M)=\hbar\sum_{j=1}^M k_j$ and 
\begin{align}
\Gamma_1(x_1,\{q\}_{N-1})=& \,\mathrm{e}^{\frac{i}{\hbar}[P(\{\kappa\}_N)-P(\{q\}_{N-1})]x_1}
 \nonumber
 \\
&\times\mathcal{F}(\{q\}_{N-1},\{\kappa\}_N).
\label{consecutiveMeasurement3}	
\end{align}
The quasimomenta $\{\kappa\}_N$ correspond to the initial state, i.e. $\big|\Psi_0\big>=\big|\{\kappa\}_N\big>$. In practice, to determine $\rho_j$, one restricts the summations in Eqs.~(\ref{consecutiveMeasurement})--(\ref{consecutiveMeasurement2}) to the relevant sets $\{k\}_{N-j}$ only. Note that we assumed that all measurements are executed at $t=0$.

Employing the prescription in Eqs.~(\ref{consecutiveMeasurement})--(\ref{consecutiveMeasurement3})
 we have computed and analyzed the conditional single particle densities starting with a single type--II eigenstate $\left| \Psi_0 \right>=\left| \{I^\text{p}\}_{II}\right>$  that for $N$-particle system is given by the following parameterizing numbers 
\begin{eqnarray} 
\textstyle{
\{I^\text{p}\}_{II}=\left\{-\frac{N-1}{2},-\frac{N-3}{2},\ldots,\times,\ldots,\frac{N-1}{2},\frac{N+1}{2} \right\},
}
 \label{Syr15_1}
\end{eqnarray}
where $\times$ denotes  $ |I^\text{p}|\leq \frac{N-3}{2}$ replaced by $\frac{N+1}{2}$ (see Sec.~\ref{LLElementaryExcitations_PBC}). The number of states that have to be taken into account in Eqs.~(\ref{consecutiveMeasurement})--(\ref{consecutiveMeasurement2}) dramatically proliferates with $N$. Similar situation takes place when one increases the interparticle repulsion. The accessible computer resources allowed us to investigate the system consisting of $N=8$ particles only. Although we can examine small systems only, we show that it is enough to observe an emergence of dark soliton signatures, i.e. density notch and phase flip, that match the corresponding mean-field dark soliton solutions well.

During the numerical simulations of the one-by-one particle detection process the consecutive particles' positions $\widetilde{x}_j$ are chosen randomly according to the conditional single particle  probability densities $\rho_j(x_j)$, Eqs.~(\ref{consecutiveMeasurement})--(\ref{consecutiveMeasurement3}).
A single run of the procedure corresponds to a single realization of the successive particles' positions measurement and results in $N$ consecutive conditional single particle probability densities and $N$ positions $\widetilde{x}_{j=1,2,\ldots,N}$ of measured particles. 
To analyze an average particle density, we repeat the measurement procedure many times and collect many sets of particles' positions $\{\widetilde{x}_1,\widetilde{x}_2,\ldots,\widetilde{x}_N\}$. Due to the space translation symmetry possessed by the system, a simple single particle density has to be spatially uniform and the position of anticipated soliton signatures is expected to vary from one realization to another.

 We believe that the dark soliton-like structure in each realization should be the more visible, the more particles were measured. 
 Thus, we decided to determine the position $x_0$ of the solitonic signatures by analyzing the wave function for the last $N$-th remaining particle in the system $\phi(x)=\sqrt{\rho_N(x)}\,\mathrm{e}^{i \varphi(x)}$. If the dark soliton structure emerges in the course of the particles' positions measurement, then the last particle density $\rho_N(x)$ should have a minimum and the last particle phase distribution $\varphi(x)$ should exhibit a flip at $x_0$.
  To investigate an average particle density in the context of dark soliton-like nature of the type--II eigenstates, one shifts all particles' positions $\{\widetilde{x}_1,\widetilde{x}_2,\ldots,\widetilde{x}_N\}$, obtained in each realization of the measurement process, by the same distance in a ring so that the corresponding $x_0$ always coincides with $\frac{L}{2}$. The so prepared shifted positions can be used to prepare histograms corresponding to the average particle density in question.

\subsubsection{Weakly interacting regime}
\label{QSolLLWeakInt}
\begin{figure}[h!] 
\begin{center}
\includegraphics[scale=0.33]{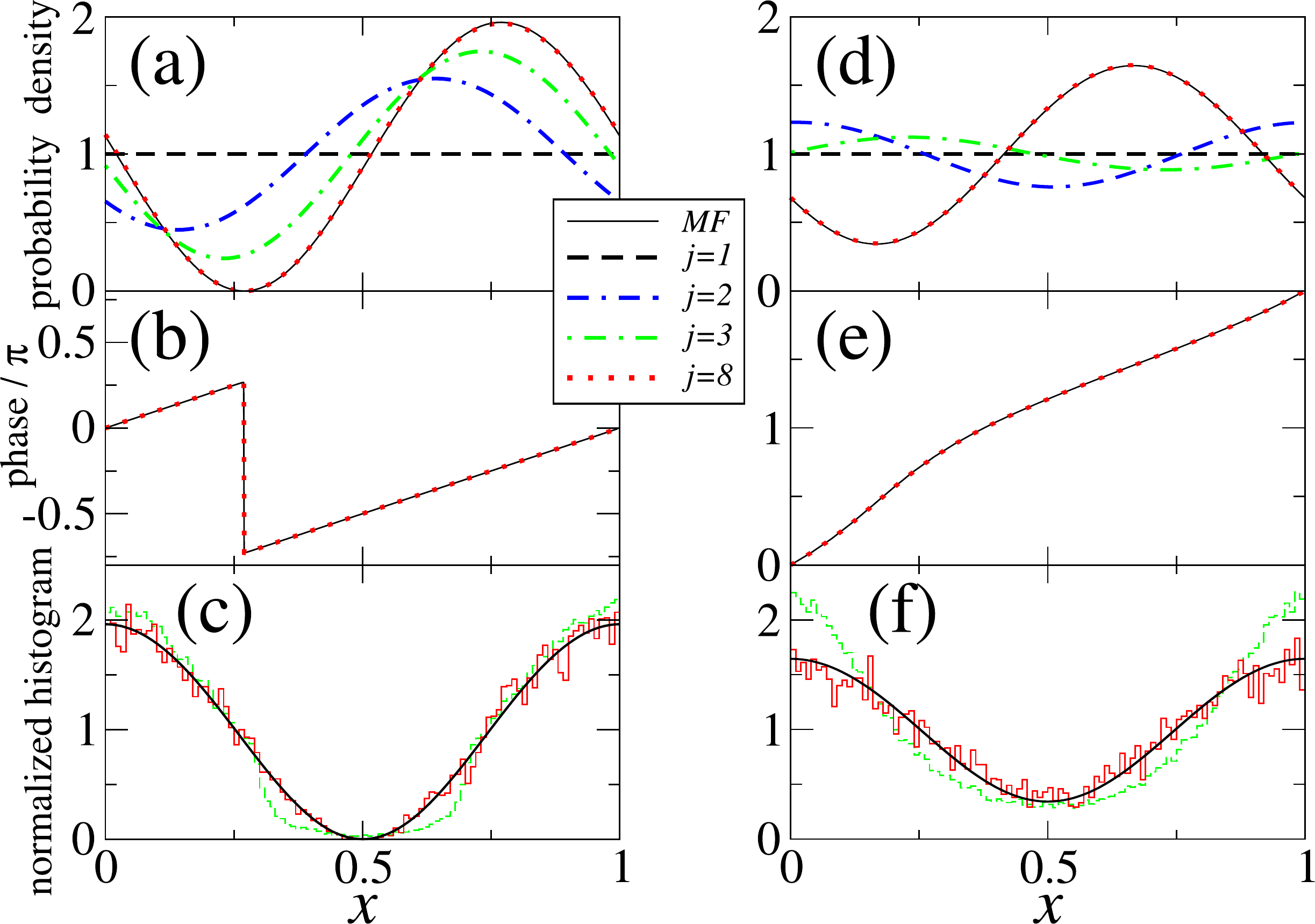}
 \end{center}
\vspace{-0.5cm}
\caption{
Results of particles' positions measurements performed in the $N=8$ particle system of size $L=1$ for weak interparticle interactions $\bar{c}=0.08$ ($\gamma=0.01$). The left and right columns present numerical outcomes obtained for the single hole (type--II) excited eigenstates with $\frac{P}{N}=\hbar\frac{\pi}{L}$ and $\frac{P}{N}=\hbar\frac{7\pi}{4L}$ (parameterizations $\{I^\text{p}\}_{II}^\text{black}$ and $\{I^\text{p}\}_{II}^\text{gray}$), respectively. Panels (a) and (d) illustrate a typical change of the conditional single particle density $\rho_j(x)$ (for $j=1,2,3,8$) in the course of particles' positions measurement. Note that $\rho_{j}(x)$ for $j=8$ matches the corresponding mean-field solution (solid black line named "\emph{MF}"), for details see the main text. A similar agreement can be also observed when looking at the  phase distribution of the last particle wave function obtained in the same realization and shown in panels (b) and (e).  
Histograms representing average particle densities [panels (c) and (f)] are generated from the positions of particles $\{\widetilde{x}_1,\widetilde{x}_2,\ldots,\widetilde{x}_8\}$ measured in $10^4$ realizations and shifted so that the position $x_0$ of the soliton-like structure, visible in the last particle wave function, always coincides with $\frac{L}{2}=0.5$.
While the dashed green histograms are prepared from all the measured positions, the solid red lines represent histograms for which only the last particle positions $\widetilde{x}_8$  were taken into account. The results are compared with the corresponding mean-field predictions for the average momentum equal to the total momentum per particle $\frac{P}{N}$  of the initial many-body eigenstate. Reprinted and adapted from \cite{Syrwid15}
}
\label{Sols_Syr15_1}
\end{figure}

Basing on the mean-field soliton solutions discussed in Sec.~\ref{Solitons_PBC} we can try to predict what kind of solitons should emerge in the type--II eigenstates corresponding to different total momenta per particle $\frac{P}{N}$.  For example, we expect that if the type--II eigenstates possess dark soliton features, then a black soliton should be related to the type--II eigenstate characterized by $\frac{P}{N}=\hbar\frac{\pi}{L}$.
 To see it cf. Fig.~\ref{PBC_MF_SolExistence2}(a) and note that $\frac{P}{N}=\hbar\frac{2\pi }{L}\Omega$. The eigenstates belonging to the same branch of single hole excitations, but corresponding to different $\frac{P}{N}$, are expected to reveal dark (gray) solitons. We investigate both cases by preparing the $N=8$ particle system initially in the two different, "black" and "gray", type--II eigenstates parameterized as follows
\begin{eqnarray}	
\{I^\text{p}\}_{II}^\text{black}&=&\textstyle{\left\{-\frac{7}{2},-\frac{5}{2},-\frac{3}{2},-\frac{1}{2}, \times ,\frac{3}{2},\frac{5}{2},\frac{7}{2},\frac{9}{2} \right\}, }
 \label{Syr15_2}
\\ 
\{I^\text{p}\}_{II}^\text{gray}&=&\textstyle{\left\{-\frac{7}{2},\times,-\frac{3}{2},-\frac{1}{2},\frac{1}{2},\frac{3}{2},\frac{5}{2},\frac{7}{2},\frac{9}{2}\right\},  }
 \label{Syr15_3}
\end{eqnarray}
and corresponding to $\frac{P}{N}=\hbar\frac{\pi}{L}$ and $\frac{P}{N}=\hbar\frac{7\pi}{4L}$, respectively.
In Fig.~\ref{Sols_Syr15_1}, we present the numerical results obtained for the weakly repulsive ($\gamma=0.01$) 8-particle system of size $L=1$. 
Note that starting from a uniform spatial distribution for the first measurement, i.e. $\rho_1(x)=\mathrm{const}$, we observe an emergence of dark soliton-like structures in the course of the particles' positions detection. Indeed, the consecutive density profiles $\rho_j(x)$, visible in panels (a) and (d), approach the corresponding mean-field dark soliton solutions. Moreover, for the last particle both probability density $\rho_N(x)$ and phase distribution $\varphi(x)$ match the mean-field predictions very well, see Fig.~\ref{Sols_Syr15_1}(a)--(e). We stress that the comparison was done for the mean-field dark soliton solutions corresponding to the same $\gamma$  and possessing the same average momenta as the resulting last particle  wave functions. 
 Additionally, the soliton position $x_s$ of the mean-field solution is shifted to coincide with the  minimum of $\rho_N(x)$ (equivalently the phase flip position).

For the type--II eigenstate with $\frac{P}{N}=\hbar\frac{\pi}{L}$ we always observe completely black soliton-like structure in the plot of $\rho_N(x)$ and  a single-point $\pi$-phase flip in $\varphi(x)$. Moreover, the corresponding average momentum for the last particle is equal to $\hbar \frac{\pi}{L}$ in every single realization. 
The situation is different when we consider the eigenstate parameterized by $\{I^\text{p}\}_{II}^\text{gray}$, Eq.~(\ref{Syr15_3}), which can be associated with a gray soliton. In such a case the depth of the density notch visible in $\rho_N(x)$, the shape of $\varphi(x)$ as well as the corresponding average momentum for the last particle in the system vary between subsequent realizations.

 Typically, it is enough to measure a few particles to observe  a clear density notch only slightly different from the corresponding mean-field dark soliton profile. Nonetheless, the first few particles in each realization are measured from more or less uniform distributions, often very different from the expected soliton density.   Due to the fact that in our simulations the first few particles constitute a considerable fraction of $N$, the average density is expected to be significantly distorted. Indeed, this effect is visible in Fig.~\ref{Sols_Syr15_1}(c)\&(f), where dashed green histograms were prepared from all the measured positions. For comparison, the mean-field solutions are obtained assuming the average momentum equal to $\frac{P}{N}$, i.e. the total momentum per particle of the initial many-body eigenstate. On the other hand, the distribution of the last, 8-th, particles' positions, represented by the solid red histogram, in both cases closely follows the corresponding mean-field dark soliton profile \cite{Syrwid15}.

In general, the system can be prepared in single hole eigenstates possessing different total momenta per particle $\frac{|P|}{N}=\hbar\frac{2\pi}{L}\left(1-\frac{j}{N}\right)$,
where $j=1,2\ldots,N-2$ and
a totally dark (black) soliton is expected for $j=\frac{N}{2}$. The larger the difference $\left|j-\frac{N}{2}\right|\leq \frac{N-2}{2}$, the shallower (on average)  density notch structure is visible in the conditional probability density for the last particle in the system. On the other hand, for the type--I eigenstates we do not observe emergence of any soliton-like structures in the process of particle detection and the last particle  density is essentially uniform, revealing no soliton signatures \cite{Syrwid15}.

\begin{figure}[h!] 
\begin{center}
\includegraphics[scale=0.335]{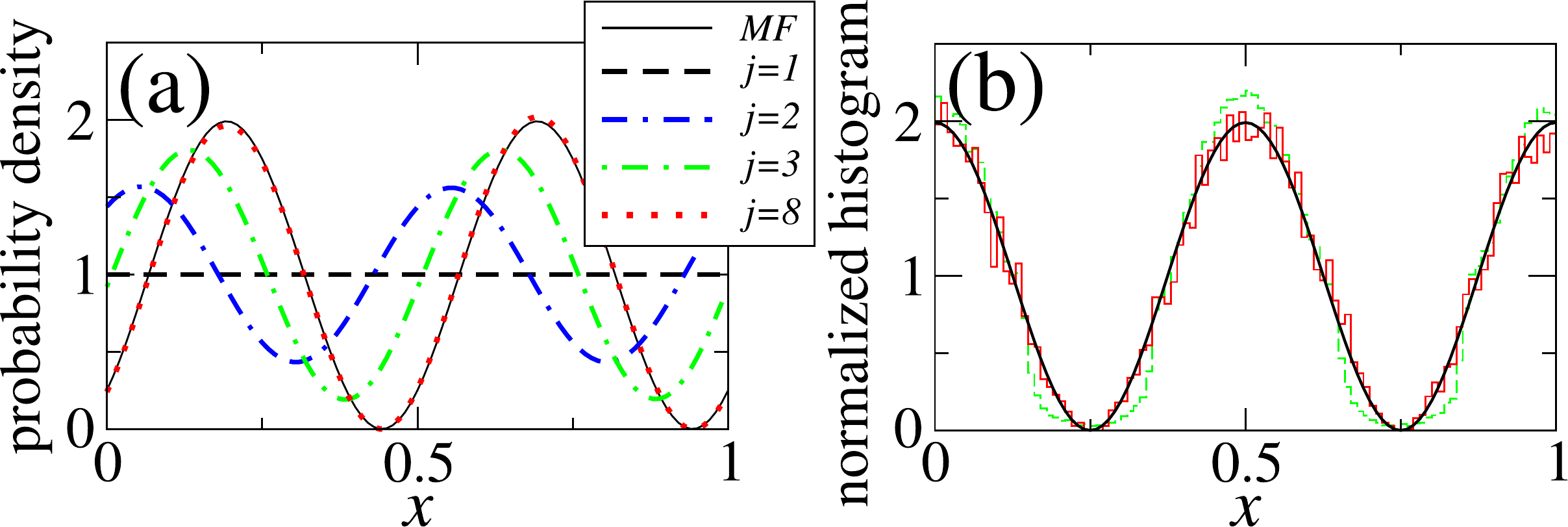}
 \end{center}
\vspace{-0.5cm}
\caption{
Results of successive  particles' positions detections performed in the 8-particle periodic system of size $L=1$. The weakly interacting ($\gamma=0.01$) system was prepared initially in the double hole excited eigenstate $| \{I^\text{p}\}_{II, 2}^\text{black} \!\left.\right>$, Eq.~(\ref{Syr15_5}).  While panel (a) presents  an example of how the consecutive conditional single particle probability density changes in the course of the particles' positions measurement, panel (b) illustrates the comparison between the resulting average particle density and the corresponding mean-field double dark soliton solution. The dashed green and solid red histograms were prepared similarly as in Fig.~\ref{Sols_Syr15_1}, with the exception that the measured particles' positions in each realization have been shifted so that one of the two minima of $\rho_N(x)$ is always located at $\frac{L}{4}$.   
 Reprinted and adapted from \cite{Syrwid15}.
}\vspace{-0.cm}
\label{Sols_Syr15_2}
\end{figure}

One may ask what is going to happen, when we start with the system prepared initially not in a single  but in a double hole excited eigenstate. It turns out that the second hole (type--II) excitation leads to an appearance of another density dip and the corresponding phase flip in the  last particle wave function. Keeping the same system parameters fixed, i.e. $N=8$, $L=1$, $\bar{c}=0.08$ ($\gamma=0.01$), we have performed identical simulations of the successive particles' positions measurements. However, this time we prepared  the system initially in the double hole excited eigenstate parameterized as follows 
\begin{equation} 
\textstyle{
\{I^\text{p}\}_{II, 2}^\text{black}=\left\{-\frac{9}{2},-\frac{7}{2},-\frac{5}{2},-\frac{3}{2},\times, \times ,\frac{3}{2},\frac{5}{2},\frac{7}{2},\frac{9}{2} \right\},
}
 \label{Syr15_5}
\end{equation}
In comparison with $\{I^\text{p}\}_{II}^\text{black}$ in Eq.~(\ref{Syr15_2}), we created a second hole by replacing $I^\text{p}=-\frac{1}{2}\rightarrow -\frac{9}{2}$. Hence, the double hole excited eigenstate $| \{I^\text{p}\}_{II, 2}^\text{black} \!\left.\right>$ is a result of two separate single hole excitations corresponding to single black solitons with opposite total momenta per particle $\frac{P}{N}=\pm\hbar\frac{\pi}{L}$ and then the total momentum of $| \{I^\text{p}\}_{II, 2}^\text{black} \!\left.\right>$ remains zero.
 Therefore, it can be suspected that during the examination of  $| \{I^\text{p}\}_{II, 2}^\text{black} \!\left.\right>$ one should observe an emergence of a double black soliton. 
 Indeed, our simulations confirm this prediction and the resulting conditional single particle probability density $\rho_N(x)$ agrees with the corresponding mean-field double dark soliton solution possessing zero average momentum, see Fig.~\ref{Sols_Syr15_2}(a). Moreover, the phase distribution $\varphi(x)$ of the last particle wave function $\phi(x)$ exhibits two $\pi$-phase flip discontinuities located exactly at the positions of soliton notches visible in the plot of $\rho_N(x)$. The agreement can be also observed when we look at the average particle density obtained from $10^4$ simulations of the measurement process, see Fig.~\ref{Sols_Syr15_2}(b) \cite{Syrwid15}. We stress that for weak interactions ($\gamma=0.01$) the resulting relative distance between the two soliton signatures visible in  $\phi$ is always equal to $\frac{L}{2}$. This is no longer true when one increases the repulsion strength (see next sections).

The results of our numerical simulations obtained so far in the weakly repulsive regime prove that the type--II eigenstates  of the periodic Lieb-Liniger model are strictly related to dark solitons. In the presence of weak interparticle repulsion the comparison between the numerical outcomes and the corresponding mean-field dark soliton solutions is justifiable. In general, such a comparison is legitimate when both $\gamma=\frac{\bar{c}L}{N}\ll 1$ and $\bar{c}N=\mathrm{const}$ for $\bar{c}\rightarrow 0$, $N\rightarrow \infty$.  Note that when dealing with systems consisting of small number of particles like $N=8$, one cannot explore the regime where the corresponding mean-field solitons are much narrower than the system size, because it requires $N^2\gg\bar{c}LN\gg 1 $.

\subsubsection{Beyond mean-field regime of interparticle repulsion}
\label{QSolLLBeyondMF}

\begin{figure}[h!] 
\begin{center}
\includegraphics[scale=0.33]{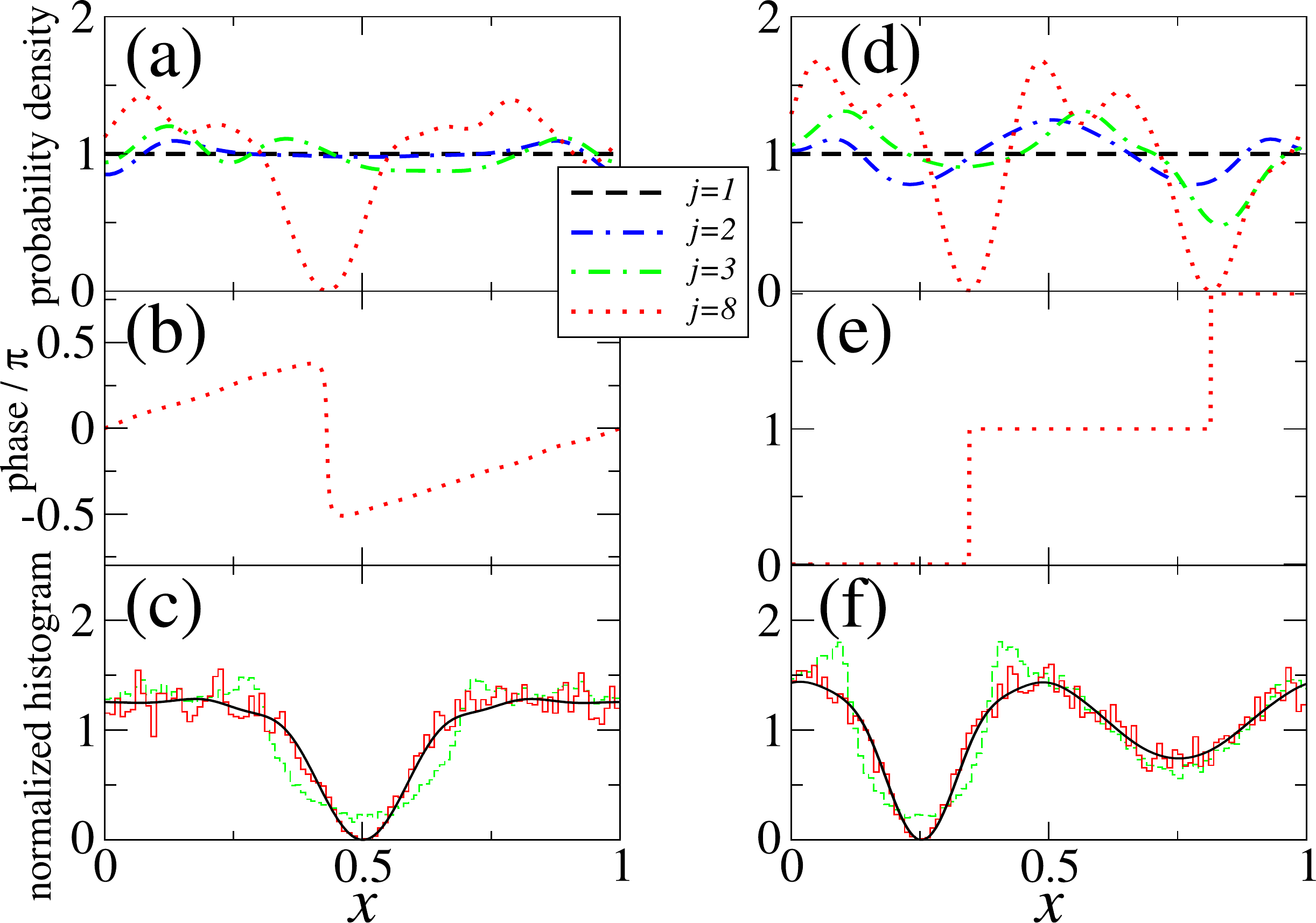}
 \end{center}
\vspace{-0.5cm}
\caption{Results of the one-by-one particles' positions measurement procedure  presented in a similar way as in Figs.~\ref{Sols_Syr15_1} and \ref{Sols_Syr15_2}, but for $\gamma=1$. The left and right columns correspond to the $N=8$ particle system of size $L=1$ prepared initially in the single and double hole excited eigenstates  $| \{I^\text{p}\}_{II}^\text{black} \!\left.\right>$ and $| \{I^\text{p}\}_{II, 2}^\text{black} \!\left.\right>$, Eqs.~(\ref{Syr15_2}) and (\ref{Syr15_5}). Single and double black soliton signatures, i.e a density notch (or notches), panels (a) and (d), and a phase flip (or flips), panels (b) and (e), are clearly visible in the wave function $\sqrt{\rho_N(x)}\, \mathrm{exp}(i\varphi(x))$ of the last particle in the system. The histograms of the average particle densities, panels (c) and (f), are prepared in the same way as in the weakly interacting regime (see the main text). Solid black lines indicate the last particle density $\rho_N(x)$ averaged over $10^4$ realizations of the particles' positions detection.
 Reprinted and adapted from \cite{Syrwid15}.
}\vspace{-0cm}
\label{Sols_Syr15_3}
\end{figure}

Let us now employ the one-by-one particle detection method to the system
with moderate repulsion strength given by $\bar{c}=8$, i.e. $\gamma = 1$  ($N=8$, $L=1$). Typical numerical outcomes obtained for the initial eigenstates $| \{I^\text{p}\}_{II}^\text{black} \!\left.\right>$ and $| \{I^\text{p}\}_{II, 2}^\text{black} \!\left.\right>$, Eqs.~(\ref{Syr15_2}) and(\ref{Syr15_5}),  are presented in panels (a) and (d) of Fig.~\ref{Sols_Syr15_3}, respectively. For $\gamma=1$, where the mean-field approximation is not valid, every single particle measurement is likely to leave a small bend on the conditional single particle probability densities $\rho_j(x)$. This is related to the fact that in the presence of the relatively strong repulsion the consecutive particles do not want to occupy the regions in which other particles have been already measured. Thus, the resulting conditional single particle densities are not as regular as those observed in the case of weak repulsion. Nevertheless, both the probability density $\rho_N(x)$ and phase distribution $\varphi(x)$ of the last particle wave function reveal a clearly visible density notch (or density notches) and the corresponding $\pi$-phase flip (or flips), which can be identified with a single (or double) black soliton-like structure (see panels (a)--(b) and (d)--(e) of Fig.~\ref{Sols_Syr15_3}). 
Note that the soliton notch structures become much narrower when we increase the repulsion strength. This is not the only difference in comparison to the weakly interacting case. It turns out that when the interparticle interactions become significant, the relative distance between the two observed dark soliton signatures fluctuates from one realization to another. Such a behaviour of the resulting spatial separation between the soliton-like structures appears due to quantum many-body effects and is strictly related to the positions of particles measured in every single simulation of the particle detection process \cite{Syrwid15}.

In addition, in both the considered cases we investigate average particle densities given by histograms prepared similarly as in the previous section, see Fig.~\ref{Sols_Syr15_3}(c)\&(f).  For comparison we also show the last particle density $\rho_N(x)$ averaged over $10^4$ realizations (solid black line). Before averaging, for $| \{I^\text{p}\}_{II,2}^\text{black} \!\left.\right>$, each set of measured positions and each density $\rho_N(x)$ were shifted so that the corresponding single soliton signatures (phase flips) are located at the position $x_0=\frac{L}{2}$. For $| \{I^\text{p}\}_{II,2}^\text{black} \!\left.\right>$  each set of measured positions and each resulting $\rho_N(x)$ were shifted in such a way that the position of the one of the phase flips always coincides with $\frac{L}{4}$.
 Note that the other averaged density notch located at $x=\frac{3}{4}L$ is significantly broadened and shallow \cite{Syrwid15}. This is a direct consequence of quantum fluctuations of the relative distances between two dark soliton-like structures. This effect is thoroughly analyzed  in Sec.~\ref{BoseDynamics}.

\subsection{Periodic case: single-shot simulations of hole excited eigenstates and dynamics of quantum dark solitons}
\label{BoseDynamics} 

In Sec.~\ref{BoseEmergence} we analyzed single and double hole excited eigenstates of the periodic Lieb-Liniger system.
 Employing numerical simulations of the successive particles' positions measurements we demonstrated that the measurement process breaks the space translation symmetry possessed by the eigenstates and leads to an emergence of clearly visible dark soliton signatures.
It turns out that in the weakly repulsive regime both the probability density and phase distribution of the last particle wave function match the corresponding mean-field dark soliton solution well.
Additionally, we showed that the soliton-like structures survive also in the beyond mean-field, moderate regime of interparticle interactions.
Due to a dramatic increase of the computation time, the applied method does not allow for an examination of large systems  ($N\gg 1$) as well as the strongly repulsive limit ($\gamma\gg1$).  Nevertheless, the simulations carried out for the $8$-particle system proved that the hole excited eigenstates are unequivocally connected with dark solitons. 

 The one-by-one process of the particle detection is equivalent to taking a single-shot of the atomic cloud density, i.e. a single simultaneous measurement of all particles' positions. Therefore, instead of the successive particle detections, one can employ the Markovian walk in the $N$-dimensional configuration space governed by the Monte Carlo method of Metropolis et al.~\cite{Metropolis}. That is, by a direct sampling of the $N$-particle probability distribution $|\Phi_N(x_1,x_2,\ldots,x_N)|^2$,  Eq.~(\ref{LLBetheWaveFunction}), one generates the so-called Markov chain $\{\mathcal{X}\}$ of sets $\mathcal{X}=\{\widetilde{x}_1,\widetilde{x}_2,\ldots,\widetilde{x}_N\}$ of particles' positions. The routine is based on a simple random walk in the space of $N$ parameters $x_{j=1,2,\ldots,N}\in[0,L]$, where at each step a set of randomly chosen positions $X'=\{x_1',x_2',\ldots,x_N'\}$ is produced. If $\mathcal{X}$ represents the last accepted set of the particles' positions in the procedure (the last element of the Markov chain), then a new candidate $X'$ is accepted (appended to the end of  $\{\mathcal{X}\}$) with the probability $p=\mathrm{min}\left(1, |\Phi_N(X')|^2/|\Phi_N(\mathcal{X})|^2 \right)$.
In the case when $X'$ is not qualified, one again appends $\mathcal{X}$ to the Markov chain. Note that the algorithm takes advantage of the analytical form of the Bethe eigenstates in Eq.~(\ref{LLBetheWaveFunction}), which gives a hope to investigate the regime of strong interparticle repulsion. Despite the fact that we can employ the analytical expressions for the eigenstates, an analysis of a large particle number $N\gg 1$ is still not possible, because the number of terms in the Bethe wave function increases like $N!$, cf. Eq.~(\ref{LLBetheWaveFunction}).

 The elements of the resulting Markov chain $\{\mathcal{X}\}$, i.e.  the accepted sets of particles' positions, can be treated as results of the consecutive single-shot particle detections. In each realization represented by $\mathcal{X}=\{\widetilde{x}_1,\widetilde{x}_2,\ldots,\widetilde{x}_N\}$ one can choose $N-1$ positions (for convenience we  take first $N-1$ positions), fix them, and study the last particle wave function  \cite{Syrwid16}
\begin{equation} 
\displaystyle{
\phi(x)=\Phi_N(\widetilde{x}_1,\widetilde{x}_2,\ldots,\widetilde{x}_{N-1};x)=\sqrt{\rho_N(x)}\,\mathrm{e}^{i\varphi(x)}
},
 \label{Syr16_2}
\end{equation}
where $\Phi_N$ is assumed to be properly normalized.

The idea of single-shot measurements turned out to be very efficient for studying the many-body dynamics of quantum dark solitons in the Lieb-Liniger model. In such investigations, however, a few important details should be kept in mind and the following procedure should be applied. First of all,  eigenstates of the Lieb-Liniger model with periodic boundary conditions are translationally invariant, and thus they exhibit no soliton signatures in the reduced single particle density. Hence, starting with a hole excited eigenstate, an observation of the many-body evolution of dark soliton structures requires breaking of the space translation symmetry.
 For this purpose, at $t=t_i$ one should perform an initial simultaneous measurement of $N_i$ out of $N$ positions of particles. We expect that the larger $N_i$, the better localization of the anticipated soliton structure is. After such a measurement we effectively deal with a $N_r=N-N_i$ particle problem.  The key question that has to be answered, concerns the behaviour of the soliton-like notch in time evolution. Strictly speaking, we want to check if the cloud of the remaining $N_r$ particles, probed at different time moments $t>t_i$, reveals a similar shape of the probability density in the laboratory frame but shifted by $v(t-t_i)$, where $v=\mathrm{const}$, and, if that is so, what is the relation between $v$ and the momentum per particle of the initial many-body eigenstate. To do so, we note that after the initial measurement that fixes $N_i$ positions of particles $\{\widetilde{x}\}_{N_i}$ we are not in the $N$-particle system eigenstate anymore. Instead, as a result of the measurement, the state of the system $\Psi_{N_r}(\{x\}_{N_r})\propto \Phi_N(\{\widetilde{x}\}_{N_i};\{x\}_{N_r})$ belongs to the $N_r$-particle Hilbert space
and $|\Psi_{N_r}(\{x\}_{N_r})|^2$ represents a conditional probability density for the detection of the $N_r$ remaining particles at positions $\{x\}_{N_r}$ and at time $t_i$. Such a state can be always expanded in the basis of $N_r$-particle eigenstates $\psi_{\{k\}_{N_r}}(\{x\}_{N_r})$ of the reduced $N_r$-particle Lieb-Liniger Hamiltonian, namely
 \begin{align} 
\displaystyle{
\Psi_{N_r}(\{x\}_{N_r})=\sum_{\{k\}_{N_r}}C_{\{k\}_{N_r}} \, \psi_{\{k\}_{N_r}}(\{x\}_{N_r})
},
 \label{Syr16_4}
\end{align}
where $\{k\}_{N_r}$ is a set of $N_r$ quasimomenta, which parameterize  $\psi_{\{k\}_{N_r}}\!$ and can be determined by solving the corresponding Bethe Eqs.~(\ref{LLBetheEqsPeriodicLog}). Assuming that the initial measurement takes place at $t_i=0$, the time dependent wave function is given by
  \begin{eqnarray}
  \begin{array}{ll} 
\Psi_{N_r}(\{x\}_{N_r},t)\\
\qquad\displaystyle{=\sum_{\{k\}_{N_r}}  C_{\{k\}_{N_r}} \, \mathrm{e}^{-\frac{i}{\hbar}E_{\{k\}_{N_r} } t } \,\, \psi_{\{k\}_{N_r}}(\{x\}_{N_r}),}
 \end{array}
 \label{Syr16_5}
\end{eqnarray}
where $E_{\{k\}_{N_r}} =\frac{\hbar^2}{2m}\sum_{j=1}^{N_r}k_j$.
The density distribution $|\Psi_{N_r}(\{x\}_{N_r},t)|^2$ reflects the two-time conditional probability density for the measurement of $N_r$ particles at positions $\{x\}_{N_r}$ at time $t$ provided that initially, at $t_i=0$, one found  $N_i$ particles at positions $\{\widetilde{x}\}_{N_i}$. To investigate the time evolution 
we sample this $N_r$-particle distribution with the help of the above-mentioned Monte Carlo method. Every single sequence $\{\widetilde{x}_1,\widetilde{x}_2,\ldots,\widetilde{x}_{N_r}\}$, being a single outcome of the algorithm, corresponds to a single-shot measurement of the remaining cloud of $N_r$-particles. According to the Metropolis routine we collect them into a chain for a given $t$, which allows us to prepare a histogram representing a single particle density related to $\Psi_{N_r}(\{x\}_{N_r},t)$. By repeating the procedure for different moments of time we can monitor the quantum many-body dynamics of the anticipated dark soliton-like structures \cite{Syrwid16}.

The main difficulty is to calculate coefficients $C_{\{k\}_{N_r}}=\big< \psi_{\{k\}_{N_r}} \big|\Psi_{N_r}\big>$, which, in general,  are given by usually very cumbersome $N_r$-dimensional integrals.
Fortunately, we realized that the expansion in Eq.~(\ref{Syr16_5}) can be found in a much simpler way, where one involves the idea of multidimensional linear regression. Indeed, one can sample the  $N_r$-dimensional space of the particles' positions and compute the values of  $\Psi_{N_r}(\{x\}_{N_r})$ and $\psi_{\{k\}_{N_r}}(\{x\}_{N_r})$ for a chosen collection of $\mathcal{M}$ sets of the quasimomenta $\{k\}_{N_r}$. Hence, for every single sequence of positions $\{x\}_{N_r}$  Eq.~(\ref{Syr16_5}) can be treated as a linear equation with $\mathcal{M}$ unknown parameters $C_{\{k\}_{N_r}}$. If so, the required coefficients $C_{\{k\}_{N_r}}$ can be found by a standard fitting procedure, which turned out to be very efficient as long as the chosen collection of sets  $\{k\}_{N_r}$ takes into account all the states with a significant contribution to the expansion in Eq.~(\ref{Syr16_5}) \cite{Syrwid16}. We stress that the procedure is very simple in the weakly interacting limit and becomes numerically expensive when one enters the strongly interacting regime, where a very large number of states $\psi_{\{k\}_{N_r}}(\{x\}_{N_r})$ is required to reproduce $\Psi_{N_r}(\{x\}_{N_r})$ accurately. 

Here, we perform the numerical simulations for the system consisting of $N=8$ particles. To investigate the quantum many-body dynamics of hole excited eigenstates, at $t_i=0$ we measure $N_i=5$ particles, which guarantees a clearly visible soliton structure in $\Psi_{N_r}(\{x\}_{N_r})$. Additionally, the number of the remaining particles $N_r=3$ is not too large thanks to which $C_{\{k\}_{N_r}}$ coefficients can be found with sufficient accuracy.
  Obviously,  $N_i=5$  is a significant fraction of $N=8$ and the state of the system after the initial measurement is very far from the initial eigenstate. Nevertheless, we expect that when $N_i,N\rightarrow \infty $ and  $N\gg N_i$, the initial $N_i$ particles' positions measurement is an infinitesimally weak perturbation that breaks the space translation symmetry and localizes the soliton notch perfectly. In such a case the resulting state $\Psi_{N_r}(\{x\}_{N_r})$ should be almost identical to the initial one $\Phi_N(\{x\}_N)$, but reveal dark soliton features in the single particle density. It turns out that the  soliton-like evolution is still present for hole excited eigenstates even if $N$ is small. In addition, we observe the beyond mean-field effects like a smearing of the density notch and a quantum revival.

\subsubsection{Weakly interacting regime}
\label{QSolLLDynamicsWeakInt}

Let us start with the regime of weak repulsion ($\gamma=0.01$) for which the results of our many-body simulations can be compared with the mean-field predictions. The analysis is  divided into three parts. In the first one we are going to consider the single hole eigenstate with $\frac{P}{N}=\hbar \frac{\pi}{L}$, Eq.~(\ref{Syr15_2}). According to the results of the previous section and the mean-field considerations such a quantum many-body eigenstate is strictly related to a single black soliton and in the laboratory frame is expected to reveal a periodic motion around the ring with the velocity $v=\frac{\pi \hbar}{Lm }$. We also briefly analyze a single gray  soliton visible in the last particle wave function, Eq.~(\ref{Syr16_2}), when the system is prepared initially in a single hole excited (type--II) eigenstate characterized by $\frac{P}{N}\neq\hbar\frac{\pi}{L}$. The second part is devoted to the case of double dark solitons that can be anticipated when one starts with a double hole excited eigenstate. Finally, in the last part we focus on a specific double hole excited eigenstate for which we observe and analyze dark soliton collisions.

\vspace{0.2cm}

$\bullet$\emph{\textbf{ single soliton}}

\begin{figure}[h!] 
\begin{center}
\includegraphics[scale=0.315]{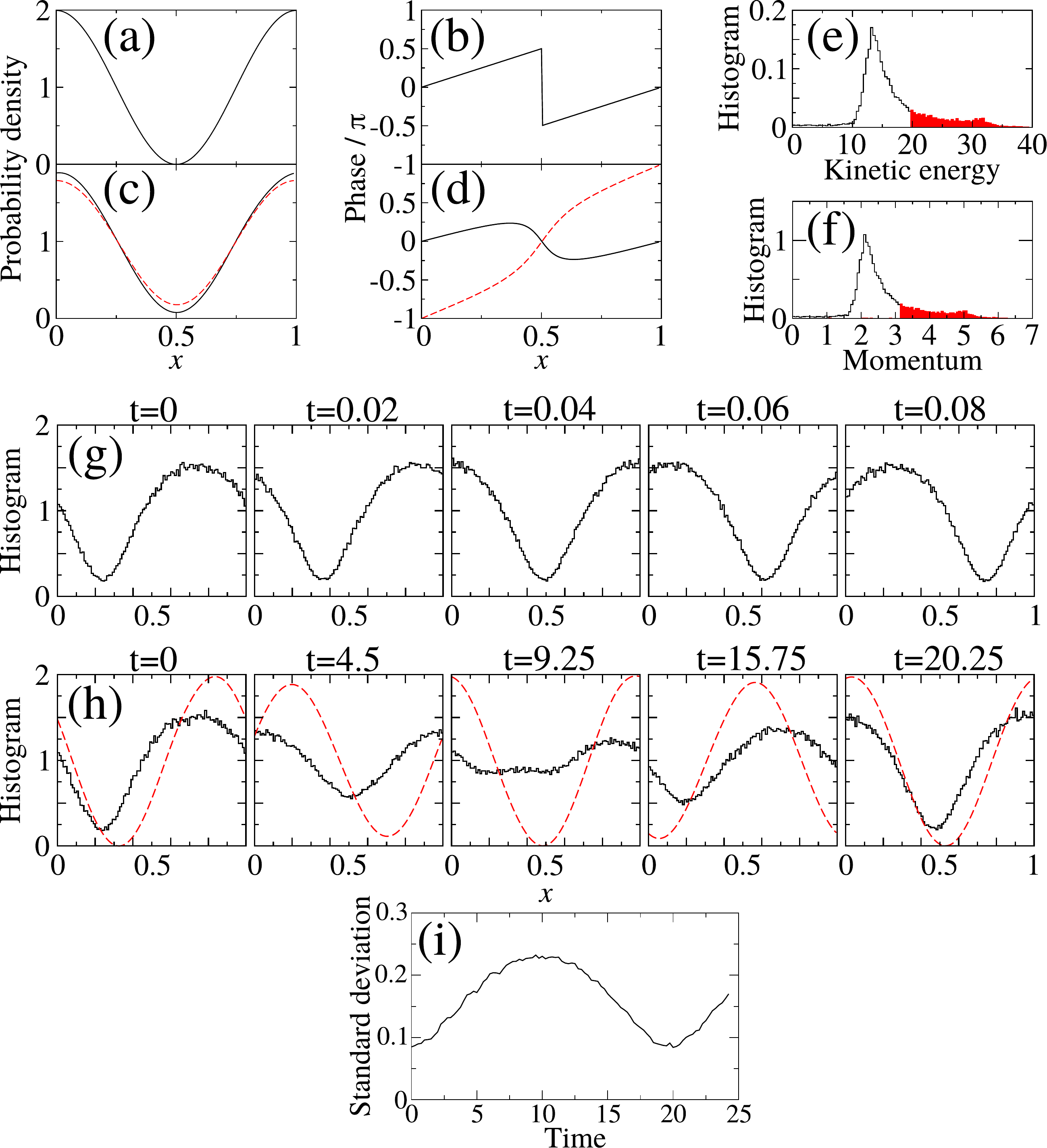}
 \end{center}
\vspace{-0.5cm}
\caption{
Panels (a)--(b) show typical last particle wave function, Eq.~(\ref{Syr16_2}), for the single hole excited eigenstate with $\frac{P}{N}=\hbar\frac{\pi}{L}$, which reveals a clearly visible black soliton notch and a single point $\pi$-phase flip. On the other hand, for the single hole excited eigenstate with $\frac{P}{N}=\hbar\frac{3\pi}{4L}$ (gray soliton eigenstate) we observe many different last particle wave functions that vary between realizations. Two of them, corresponding to different phase winding numbers $J=0$ (solid black line) and $J=1$ (dashed red line), are depicted in panels (c) and (d). For the same initial eigenstate, in panels (e) and (f), we show histograms of the kinetic energies and average momenta calculated for the last particle wave functions obtained in many realizations. Black (Red filled) histograms refer to $J=0$ ($J=1$).  Panels (g)--(i) illustrates a single quantum dark soliton many-body dynamics of the single hole excited eigenstate with $\frac{P}{N}=\hbar\frac{\pi}{L}$ (for details see the main text).  The short time dynamics, panel (g), reveals a periodic motion of the dark soliton-like structure with the velocity $v\approx 2\pi$, which coincides with the mean-field prediction.
 The long time dynamics, panel (h), exhibits quantum many-body effects like a smearing of the density notch structure (visible for $t\approx 9.25$) and a quantum revival ($t\approx 20.25$). For each time we also plotted a typical last particle density (dashed red line). The beyond mean-field effects are related to the time dependence of the soliton position uncertainty (see the main text), whose standard deviation versus time is illustrated in panel (i).    
All the results were obtained for $N=8$, $L=1$, $\gamma=0.01$ and the units chosen so that $\hbar=2m=1$.
Reprinted and adapted from \cite{Syrwid16}.
}
\label{Sols_Syr16_1}
\end{figure}

At the beginning, we are going to analyze the last particle wave function by employing  the above-mentioned Monte Carlo method. We examine the single hole (type--II) eigenstates of the 8-particle system characterized by  $\frac{P}{N}=\hbar\frac{\pi}{L}$ and $\hbar\frac{3\pi}{4L}$, where the corresponding sets $\{I^\mathrm{p}\}$ are related to the excitations $I^\text{p}=\frac{1}{2}\rightarrow \frac{9}{2}$ and $I^\text{p}=\frac{3}{2}\rightarrow \frac{9}{2}$, respectively. The simulations were performed for $\hbar=2m=L=1$ and $\bar{c}=0.08$ ($\gamma=0.01$).  After many steps of the Metropolis algorithm we have collected plenty of configurations of particles' positions corresponding to many realizations of the single-shot measurements.  According to Eq.~(\ref{Syr16_2}), in each realization we can plot the density $\rho_N(x)$ and phase distribution $\varphi(x)$ for the last, $N$-th particle in the system. Typical results of the single-shot experiments obtained for both considered states are presented in Fig.~\ref{Sols_Syr16_1}(a)--(d).
 For the type--II eigenstate possessing $\frac{P}{N}=\hbar\frac{\pi}{L}$, the last particle wave function $\phi$, Eq.~(\ref{Syr16_2}), always reveals a clearly visible density notch of identical shape corresponding to zero probability at the minimum. We also stress that the average momentum related to $\phi$ is always equal to $\hbar\frac{\pi}{L}$ and the corresponding energy does not change between realizations and reproduces the value predicted within the mean-field considerations.
 The density $\rho_N(x)$ drops to zero at a single point $x_0$, where the phase distribution $\varphi(x)$ reveals an abrupt flip by $\pi$, i.e. $\lim_{\varepsilon\rightarrow 0}[\varphi(x_0+\epsilon)-\varphi(x_0-\epsilon)]=\pm \pi  $ modulo $2\pi$, see Fig.~\ref{Sols_Syr16_1}(a)--(b). The cyclicity condition for the phase reads $\varphi(x+L)-\varphi(x)=2\pi J$, where $J\in \mathbb{Z}$ is the so-called phase winding number (see also Sec.~\ref{Solitons_PBC}). It means that in the case of a black soliton, the phase windings  $J$ cannot be distinguished. On the other hand, for a gray soliton $\varphi(x)$ is a smooth function and $J$ can be unambiguously determined. 
The results of our numerical simulations indicate that starting with the type--II eigenstate characterized by $\frac{P}{N}=\hbar\frac{3\pi}{4L}$ the shapes of both the last particle density profile and the corresponding phase distribution $\varphi(x)$ vary from one realization of the single-shot detection to another. Therefore, 
while $J$ takes only two values, $J=1$ (with the probability 0.21) and $J=0$ (with the probability 0.79), which stays in agreement with the mean-field predictions, cf. Fig.~\ref{PBC_MF_SolExistence2}(a),
in a wide array of the realizations one observes the last particle wave functions that correspond to different average momenta and  energies, in Fig.~\ref{Sols_Syr16_1}(e)--(f).
  Despite the fact that the resulting shapes of $\rho_N(x)$ and $\varphi(x)$ depend on the realization, the corresponding  momentum averaged over many realizations is equal to $\frac{P}{N}=\hbar\frac{3\pi}{4L}$. In addition, the average energy of the last particle wave functions coincides with the energy corresponding to the mean-field dark soliton solution  obtained for the same $\gamma=0.01$ and momentum $\frac{P}{N}$ \cite{Syrwid16}.

 During the investigations of the single dark soliton many-body dynamics we focused on the type--II eigenstate  corresponding to a single black soliton, i.e. the single hole eigenstate characterized by  $\hbar\frac{\pi}{L}$. In order to decide at which positions the first $N_i=5$ particles should be measured, we have performed many steps of the Metropolis algorithm and  identified the most probable configuration of particles' positions for a system prepared in the chosen eigenstate. The positions $\{\widetilde{x}\}_{N_i}$, that we fixed in $\Phi_N$, were chosen randomly from this particular most probable set. The remaining $N_r$ particles are described by the wave function in Eq.~(\ref{Syr16_5}). 
In Fig.~\ref{Sols_Syr16_1}(g) we show that in short time dynamics  the resulting dark soliton structure propagates in the laboratory frame with a constant velocity maintining its shape, like a genuine dark soliton.  
The observed velocity $v\approx2\pi$ agrees with the mean-field predictions for a single black soliton with average momentum equal to $\frac{P}{N}$ of the analyzed eigenstate. Additionally, thanks to the application of a full quantum many-body description, we are able to observe beyond mean-field quantum many-body effects. Indeed, in the long time dynamics presented, panel (h), we notice a smearing and re-deepening of the soliton notch. Such quantum phenomena are strictly related to the uncertainty of the soliton position. That is, by analyzing the last particle wave function in many detection processes, we found that the position of dark soliton signatures fluctuates between the realizations. This is also the reason why the density notch visible in panel (g) is not completely black. In panel (h) we also depicted typical last particle densities  (dashed red line). Some of them are also slightly filled, i.e. the corresponding probability density at the minimum is a bit larger than 0, but this contribution is very small in comparison to the smearing observed, for example, at $t=9.25$. The investigations of the standard deviation of the position of soliton signatures visible in the last particle wave functions, Fig.~\ref{Sols_Syr16_1}(i), showed that in the course of time evolution the position of soliton structures becomes more and more blurred. The moment of time corresponding to the maximum value of the uncertainty of the dark soliton position coincides with the time at which the density notch is almost totally smeared, cf. panels (h) and (i) of Fig.~\ref{Sols_Syr16_1} for time $t\approx 10$. Afterwards, for longer times, the location of the soliton structure becomes less uncertain and the shape of the histogram representing the single particle density returns to the initial dark soliton-like profile, cf. panels (h) and (i) of Fig.~\ref{Sols_Syr16_1} for time $t\approx 20$ \cite{Syrwid16}. Such a phenomenon, expected mostly for a few-body systems, is called a quantum revival. It can be expected that with the increase of particle number $N$, but keeping $\bar{c}(N-1)=\mathrm{const}$,  one would observe the increase of time for which the many-body dynamics follows the corresponding mean-field evolution.

\begin{figure}[t!] 
\begin{center}
\includegraphics[scale=0.315]{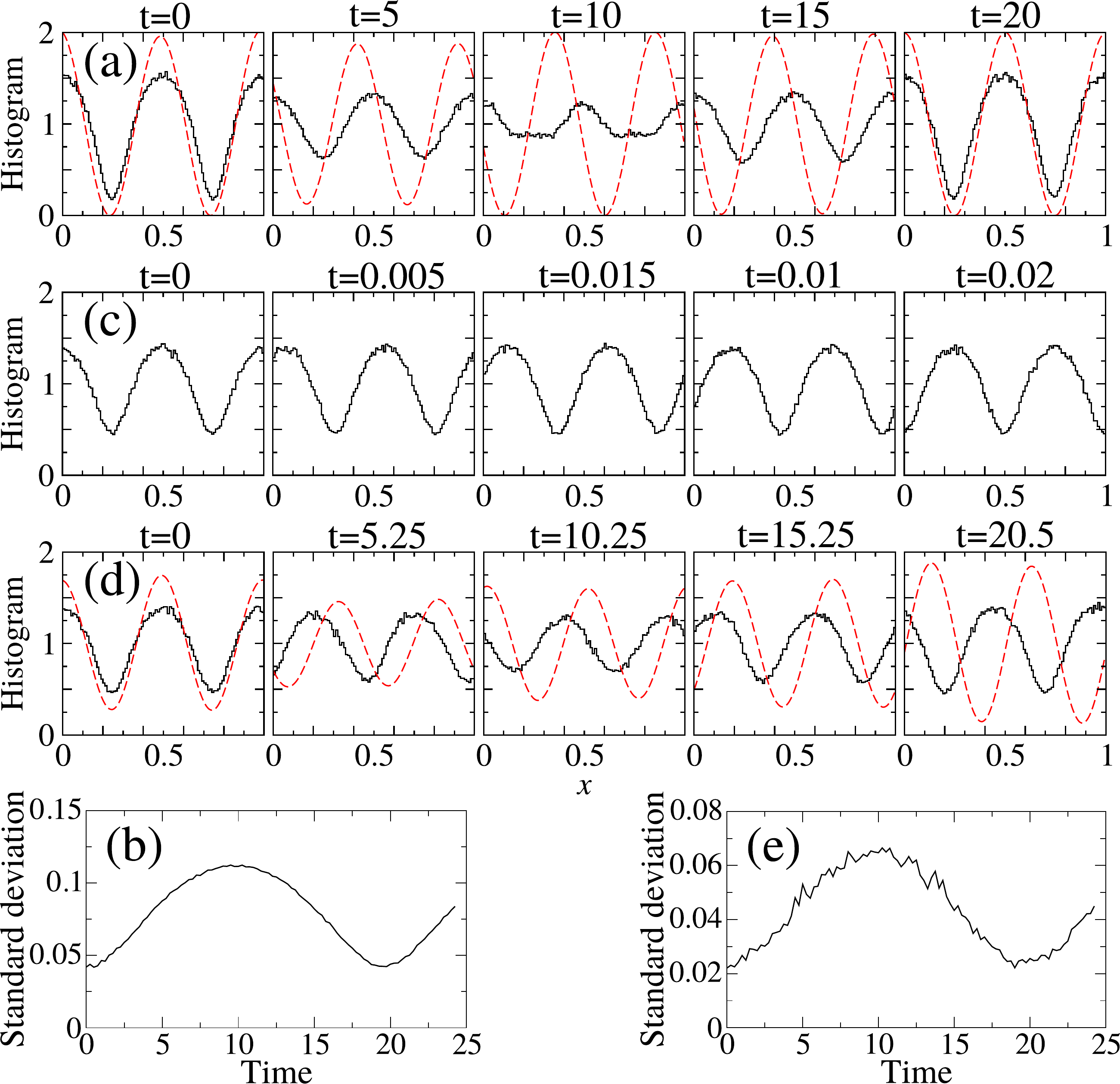}
 \end{center}
\vspace{-0.5cm}
\caption{
Many-body dynamics of double hole excited eigenstates.   The results obtained for   
the initial eigenstate parameterized by $\{I^\text{p}\}_{II, 2}^\text{black}$ are shown in panels (a) and (b). 
In agreement with the mean-field predictions two stationary almost black soliton-like structures are visible in a single particle density after the initial measurement of $N_i=5$ particles. For long times we observe a smearing and a revival of the density notches, which can be attributed to the increasing and decreasing uncertainty of the soliton positions in time. The corresponding standard deviation of the soliton signatures' positions versus time is presented in panel (b). In subsequent histograms, see panel (a), we also show typical last particle densities (dashed red lines). Similar data, but obtained for the system prepared initially in the double hole excited eigenstate with $\frac{P}{N}=\hbar\frac{3\pi}{L}$ (parameterized by $\{I^\text{p}\}_{II, 2}^\text{gray}$), is presented in panels (c)--(e). In such a case one observes two gray soliton structures moving with the predicted velocity $v\approx 4\pi$. The quantum many-body effects are also visible in the long time dynamics. Time dependence of the corresponding standard deviation of the soliton structure position is depicted in panel (e). All the results were obtained for the 8-particle system of size $L=1$ with the weak interparticle repulsion $\gamma=0.01$. The units were chosen so that $\hbar=2m=1$. 
Reprinted and adapted from \cite{Syrwid16}.
}
\label{Sols_Syr16_2}
\end{figure}

\vspace{0.2cm}
$\bullet$\emph{\textbf{ double soliton}}

In order to study the quantum many-body dynamics of double dark soliton structures, we consider two double hole excited eigenstates parametrized by
\begin{eqnarray} 
\{I^\text{p}\}_{II, 2}^\text{black}&=&
\textstyle{\left\{-\frac{9}{2},-\frac{7}{2},-\frac{5}{2},-\frac{3}{2},\times, \times ,\frac{3}{2},\frac{5}{2},\frac{7}{2},\frac{9}{2} \right\},}
 \label{Syr16_1_8} \\
\{I^\text{p}\}_{II, 2}^\text{gray}&=&
\textstyle{\left\{-\frac{7}{2},-\frac{5}{2},\times, \times ,\frac{1}{2},\frac{3}{2},\frac{5}{2},\frac{7}{2},\frac{9}{2},\frac{11}{2} \right\}.}   
 \label{Syr16_1_9}
\end{eqnarray}
While $\big|\{I^\text{p}\}_{II, 2}^\text{black}\big>$, being a result of two hole excitations related to $I^\text{p}=\pm\frac{1}{2}\rightarrow \pm\frac{9}{2}$, corresponds to $P=0$,  the eigenstate $\big|\{I^\text{p}\}_{II, 2}^\text{gray}\big>$ is generated by the exchange $\left\{-\frac{3}{2},-\frac{1}{2}\right\}\rightarrow \left\{\frac{9}{2}, \frac{11}{2}\right\}$ and possesses $\frac{P}{N}=\hbar\frac{3\pi}{L}$. The values of momenta related to the individual type--II excitations suggest that the  eigenstates given by $\{I^\text{p}\}_{II, 2}^\text{black}$ and $\{I^\text{p}\}_{II, 2}^\text{gray}$ should exhibit double black and double gray soliton structures, respectively. 
By analyzing the double dark mean-field soliton solutions corresponding to the average momentum equal to $\hbar\frac{3 \pi}{L}$, one can predict that the anticipated double gray soliton structure should move in the laboratory frame with $v\approx4\pi$ ($\hbar=2m=L=1$).  On the other hand, the eigenstate given by $\{I^\text{p}\}_{II, 2}^\text{black}$ possesses $P=0$ and for weak repulsion there is only one corresponding mean-field double black soliton solution, which is stationary in the laboratory frame \cite{Syrwid16}.

Investigations of quantum many-body dynamics were carried out in exactly the same way as in the previously discussed case of single hole excited eigenstate. 
The obtained results are presented in Fig.~\ref{Sols_Syr16_2}.
Again, the short time many-body evolution closely follows the mean-field predictions and the beyond mean-field effects related to quantum fluctuations are revealed in the long time dynamics.
 That is, the double dark soliton structure corresponding to $P=0$ is stationary in the laboratory frame, see Fig.~\ref{Sols_Syr16_2}(a). Due to the uncertainty of the soliton position that varies in time, cf. Fig.~\ref{Sols_Syr16_2}(b), one can observe a gradual smearing and revival of the density notches. On the other hand, two gray solitons, propagating as a whole structure with the anticipated velocity $v\approx 4\pi$ are visible in the numerical outcomes obtained for the initial eigenstate characterized by $\frac{P}{N}=\hbar\frac{3\pi}{L}$, see histograms in Fig.~\ref{Sols_Syr16_2}(c). We  stress that any differences between both gray solitons, like their depths, velocities as well as the relative distance between them, are unnoticeable in the course of time evolution. Quantum many body effects, i.e. smearing of the density notches and their revival, become visible for long times, see Fig.~\ref{Sols_Syr16_2}(d). The mechanism responsible for such effects is identical as in the former case, however, the uncertainty of the soliton position is smaller and the phenomena are much less evident, cf. panels (b) and (e).

\vspace{0.2cm}
$\bullet$\emph{\textbf{ collisions}}

\begin{figure}[h!] 
\begin{center}
\includegraphics[scale=0.238]{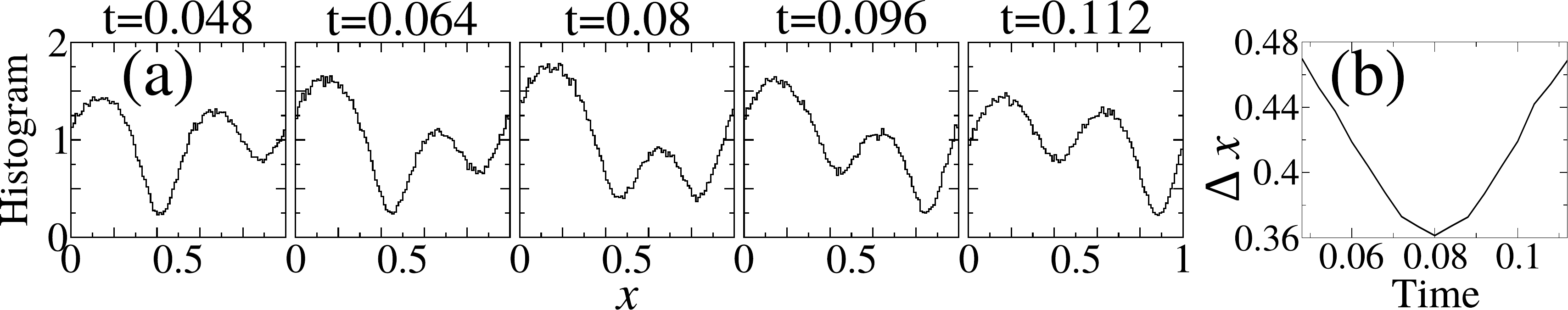}
 \end{center}
\vspace{-0.5cm}
\caption{ 
Panel (a) shows data obtained and presented in similar way as in Figs.~\ref{Sols_Syr16_1} and \ref{Sols_Syr16_2}, but for the system prepared initially in the double hole excited eigenstate parameterized by $\{I^\text{p}\}_{II, 2}^\text{collision}$, Eq.~(\ref{Syr16_1_10}). The dynamics of two distinct density notches is monitored in subsequent histograms. Both deep and shallow dark soliton-like structures move towards each other and collide at $t\approx0.08$. During the collision both density notches become identical and the soliton signatures exchange their positions. The relative distance $\Delta x$ between density minima versus time is illustrated in panel (b). In the simulations the units were chosen so that $\hbar=2m=L=1$ and  $\bar{c}=0.08$ ($\gamma=0.01$).
 Reprinted and adapted from \cite{Syrwid16}.
}
\label{Sols_Syr16_3}
\end{figure}

There are many different ways to generate double hole excitations. As we have shown,  among many possibilities, one can find single eigenstates revealing Gross-Pitaevskii solitons in the course of particles' positions measurement. Moreover, we pointed out that such a dark soliton-like structures closely follow the corresponding mean-field dynamics.  As for now, our investigations of double hole excited eigenstates were focused on cases for which the resulting soliton structures were identical and propagated in the same direction with equal velocities. It is natural to ask if it is possible to prepare the considered system in an eigenstate related to two significantly different dark solitons  that move towards each other and collide. Here we provide a positive answer by analyzing the eigenstate parameterized as follows 
\begin{equation} 
\textstyle{
\{I^\text{p}\}_{II, 2}^\text{collision}=\left\{-\frac{9}{2},-\frac{7}{2},\times, -\frac{3}{2} ,-\frac{1}{2} ,\times,\frac{3}{2},\frac{5}{2},\frac{7}{2},\frac{9}{2} \right\}}.
 \label{Syr16_1_10}
\end{equation}
Such a choice can be interpreted as two individual type--II excitations related to a single black soliton with positive momentum ($I^\text{p}=\frac{1}{2}\rightarrow \frac{9}{2}$) and a single gray soliton with negative momentum ($I^\text{p}=-\frac{5}{2}\rightarrow -\frac{9}{2}$).
The numerical analysis of such an eigenstate we performed as before. By the investigations of the last particle wave function we note that the anticipated black and gray soliton structures are not always visible. Among many realizations of the single-shot measurements one can find both clearly separated and completely blurred soliton notches.  We deduce that the absence of double soliton signatures in the last particle wave function corresponds to the moment of their collision. The solitons are clearly visible in average densities reproduced by histograms of particles' positions. As expected, in the laboratory frame we observe one very deep density notch that moves towards the other, shallow soliton structure. 
In Fig.~\ref{Sols_Syr16_3} we monitor how both density notches propagate and collide, which relies on the exchange of their positions

Let us summarize the results obtained during the investigations of the double hole excited eigenstates of the Lieb-Liniger model with periodic boundary conditions. It turns out that these excitations belong to three different classes, where for convenience we restrict to an even number of particles $N$. Eigenstates strictly related to mean-field double black solitons constitute the first class. They can be generated by two single type--II excitations  characterized by $\frac{P}{N}=+\hbar\frac{\pi}{L}$ and $\frac{P}{N}=-\hbar\frac{\pi}{L}$, which correspond to the replacement $I^\text{p}=\pm\frac{1}{2}\rightarrow \pm (I_F^\text{p}+1)$ (where $I_F^\text{p}=\frac{N-1}{2}$) in the ground state parameterization, cf. Eq.~(\ref{GSLLISet}).	 In such a case one deals with the eigenstate possessing zero momentum per particle and revealing double black soliton-like signatures in the last particle wave function. Identical black soliton signatures, but moving in the laboratory frame, can be obtained by shifting all the parameterizing numbers by some  $s\in\mathbb{Z}$, i.e. $\forall_{j=1,2,\ldots,N}$: $I_j^\text{p}\rightarrow I_j^\text{p}+s$. The resulting double soliton-like structure should propagate in the laboratory frame with $v=\hbar\frac{2\pi s}{Lm}$. In the second class one can find  eigenstates related to the exchange of two neighbouring parameterizing numbers $-I_F^\text{p}< I_j^\text{p},I_{j+1}^\text{p}< I_F^\text{p}$ ($I_{j+1}^\text{p}-I_{j}^\text{p}=1$) by the numbers $\pm(I_F^\text{p}+1)$ and $\pm(I_F^\text{p}+2)$, where $I_j^\text{p}\neq \frac{1}{2}$ for "$+$" and $I_{j+1}^\text{p}\neq-\frac{1}{2}$ for "$-$".  In such a way one creates a double hole excited eigenstate that corresponds to two identical gray solitons moving with the same velocity in the laboratory frame. The process of the particles' positions detection results in many different last particle wave functions that reproduce the corresponding mean-field solutions almost perfectly.
 Finally, the last, third class of double hole excitations is related to the replacement $\{I_r^\text{p},I_u^\text{p}\}\rightarrow \{\pm(I_F^\text{p}+1),\pm(I_F^\text{p}+2)\}$ or $\{I_r^\text{p},I_u^\text{p}\}\rightarrow \{-(I_F^\text{p}+1),+(I_F^\text{p}+1)\}$, where  $-I_F^\text{p}< I_r^\text{p},I_{u}^\text{p}< I_F^\text{p}$ and $|I_{r}^\text{p}-I_{u}^\text{p}|>1$. In such a case the wave function of the last particle in the system may reveal two different dark soliton signatures located at different relative distances. The eigenstates belonging to this class allow us to investigate signatures of soliton collisions, which cannot be described by a single solution of the stationary GPE.

\subsubsection{Strongly interacting regime}
\label{QSolLLDynamicsStrongInt}

It is very nontrivial to answer the question if the soliton-like signatures can be observed when the interparticle interactions are strong so that the mean-field description breaks down. For example, it has been shown that in the system of bosons in optical lattice dark solitons cannot be realized by a standard phase imprinting method  if one enters the Mott insulator regime, for which the phase coherence between the lattice sites is lost \cite{Lewy2010}. Meanwhile, in the case of the Lieb-Liniger model, the existence of the second branch of the elementary excitations that can be associated with dark solitons is completely independent of the repulsion strength. Additionally, we have already shown that the soliton signatures survive in the regime of moderate interactions, where the mean-field approach is not valid. This part of our discussion is devoted to examine what happens with dark soliton structures when $\gamma\gg 1$ and what is their dynamics in the beyond mean-field regime of interactions.
 
\begin{figure}[h!] 
\begin{center}
\includegraphics[scale=0.29]{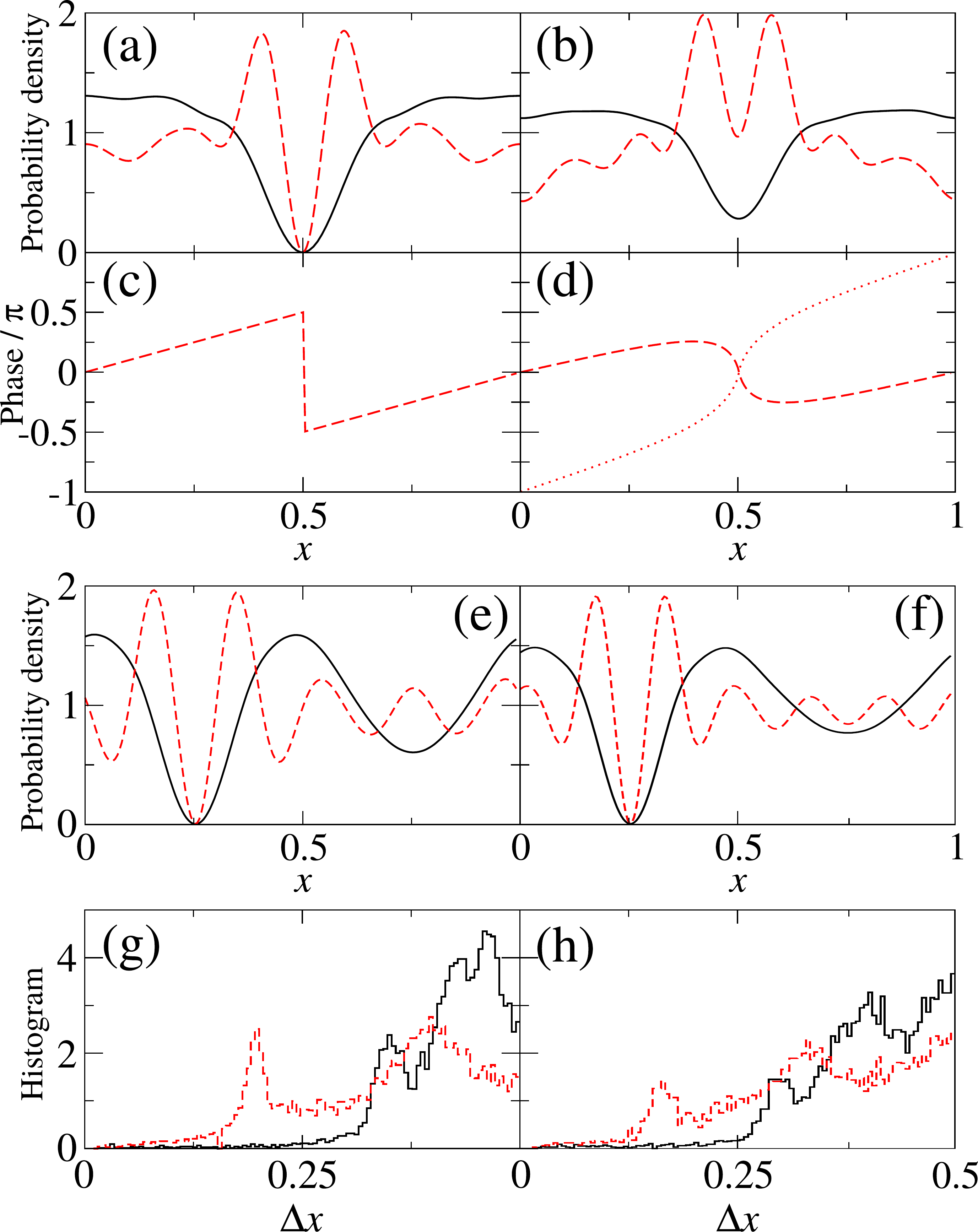}
 \end{center}
\vspace{-0.5cm}
\caption{  Panels (a) and (b) show average last particle densities obtained for the 8-particle system prepared initially in single hole excited eigenstates characterized by $\frac{P}{N}=\hbar\frac{\pi}{L}$ and $\hbar\frac{3\pi}{4L}$, respectively. Solid black (Dashed red) lines correspond to $\gamma=1$ ($\gamma=10^3$). The average phases of the last particle wave functions turned out to be almost independent of the interparticle interaction strength. Hence, we present only these obtained for $\gamma=10^3$. For the eigenstate possessing  $\frac{P}{N}=\hbar\frac{\pi}{L}$ we observe an abrupt phase flip by $\pi$, see panel (c). On the other hand, when the system is initially prepared in the type--II eigenstate with $\frac{P}{N}=\hbar\frac{3\pi}{4L}$, single-shot measurements reveal two types of phase flips related to different phase winding numbers: $J=0$ (dashed red line) and $J=1$ (dotted red line), see panel (d). Double hole excited eigenstates with $\frac{P}{N}=0$ were analyzed in the similar way (see the main text). Average last particle densities for eigenstates consisting of $N=6$ and $N=8$ bosons are illustrated in panels (e) and (f), respectively. While one of the density notch structures is clearly visible, the other one is significantly smeared when $\gamma=1$ (solid black lines) and totally blurred for $\gamma=10^3$ (dashed red lines).  Such an effect is related to the fact that the relative distances between the two phase flip signatures $\Delta x$ vary from one realization to another. The distributions of $\Delta x$ obtained for $N=6$ and $N=8$ are shown in histograms (g) and (h), respectively. The units were chosen so that $\hbar=2m=L=1$. 
 Reprinted and adapted from \cite{Syrwid16}.
}
\label{Sols_Syr16_4}
\end{figure}

Let us start our analysis with single hole excited eigenstates 
of the 8-particle system characterized by the total momenta per particle $\frac{P}{N}=\hbar\frac{\pi}{L}$ and $\frac{P}{N}=\hbar\frac{3\pi}{4L}$, already studied in Sec.~\ref{QSolLLDynamicsWeakInt}.
When investigating the nature of these excitations for $\gamma\gg1$ in every single realization of the particle detection process we identified the position of a phase flip signature visible in the last particle wave function $\phi$, Eq.~(\ref{Syr16_2}). Afterwards, each resulting $\phi$ was shifted so that the phase flip is located at $x=\frac{L}{2}$. The so prepared probability densities $\rho_N(x)$ and phase distributions $\varphi(x)$ were averaged over hundreds of thousands realizations. The results obtained for $\gamma=1$ (solid black lines) and $\gamma=10^3$ (dashed red lines) are shown in Fig.~\ref{Sols_Syr16_4}(a)--(d). The average phase distributions seem not to react to the increase of the repulsion strength and remain almost unchanged in comparison to the weakly interacting case, cf.  Fig.~\ref{Sols_Syr16_4}(c)\&(d) with Fig.~\ref{Sols_Syr16_1}(b)\&(d). On the other hand, average densities  change significantly up to $\gamma\approx100$, after which they become frozen and insensitive on further increase of $\gamma$. The final profiles of the average probability densities reveal a fermionization phenomenon, in which one deals with impenetrable bosons tending to maximize their mutual separation in the configuration space. Indeed, the period of the density modulations and the breadth of the local density notches, visible for $\gamma=10^3$, correspond to the characteristic length scale $\frac{L}{N}$, see dashed red lines in Fig.~\ref{Sols_Syr16_4}(a)\&(b) \cite{Syrwid16}.

We also analyzed the double hole excited eigenstate belonging to the first class discussed before. That is, the systems consisting of $N=6$ and $N=8$ particles were prepared initially in the eigenstate related to an exchange $I^\text{p}=\pm\frac{1}{2}\rightarrow \pm(I_F^\text{p}+1)$. The investigations were based on the same procedure as before with the exception that now, instead of one, we observe two phase flips and the last particle wave functions were shifted so that one of them is always located at  $\frac{L}{4}$. The resulting  average probability densities obtained in the presence of moderate ($\gamma=1$, solid black lines) and strong ($\gamma=10^3$, dashed red lines) interparticle repulsion  for $N=6$ and $N=8$ are presented in panels (e) and (f) of Fig.~\ref{Sols_Syr16_4}, respectively. In addition, in contrast to the weakly interacting case, we observe that the relative distance between the two phase flip signatures $\Delta x$ fluctuates between realizations. The corresponding histograms of $\Delta x$ can be found in Fig.~\ref{Sols_Syr16_4}(g)\&(h). The variation of the soliton signatures separation is responsible for blurring the second density notch when $\gamma=1$ and even its disappearance for strong repulsion $\gamma=10^3$. We stress that when $\gamma\rightarrow \infty$ the first two maxima of $\Delta x$ distributions appear for $\Delta x \approx \frac{L}{N}$ and $\Delta x\approx \frac{2L}{N}$, which is strictly related to the situation in which only one or two bosons were measured between the soliton-like structures \cite{Syrwid16}.

\begin{figure}[h!] 
\begin{center}
\includegraphics[scale=0.32]{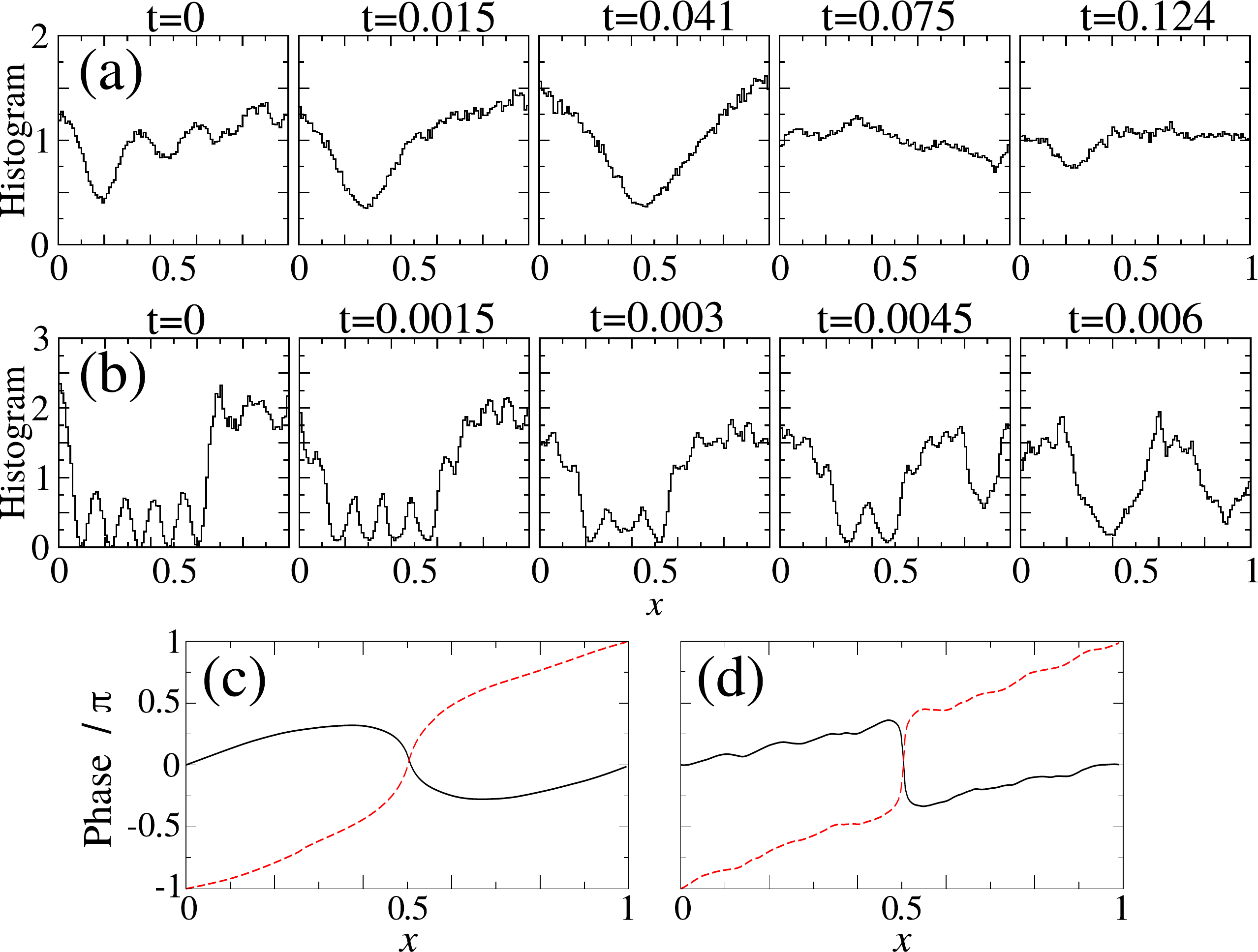}
 \end{center}
\vspace{-0.5cm}
\caption{Time evolution of the 8-particle single hole excited eigenstate characterized by $\frac{P}{N}=\hbar\frac{\pi}{L}$ in the strongly repulsive regime. At $t=0$ we carry out an initial measurement of $N_i =5$ particles at very probable positions and monitor the dynamics of the remaining $N_r=3$ particles, which for $\gamma=1$ and $\gamma=\infty$ is presented in panels (a) and (b), respectively.  
In both cases the phase distributions of the last particle wave functions exhibit phase flip signatures, which are preserved also in the course of time evolution. In panel (c) we show the last particle phase distribution averaged over many realizations for $\gamma=1$  and at $t=0.124$. Similar results but obtained for $\gamma=\infty$ and at $t=0.75 t_c$ ($t_c\approx0.003$) are presented in panel (d). Solid black (Dashed red) lines are related to the phase winding number $J=0$ ($J=1$). The final averaged results were shifted so that the phase flips coincide with $\frac{L}{2}$. In simulations the units were chosen so that $\hbar=2m=L=1$.
 Reprinted and adapted from \cite{Syrwid16}.
}
\label{Sols_Syr16_5}
\end{figure}

During the analysis of the many-body dynamics in the presence of strong interparticle repulsion we focus on the single hole excited eigenstate parameterized by $\{I^\text{p}\}_{II}^\text{black}$, Eq.~(\ref{Syr15_2}). Such an eigenstate is characterized by $\frac{P}{N}=\hbar\frac{\pi}{L}$ and for $\gamma\ll 1$ can be associated with a single black soliton propagating around the ring with the velocity $v=\frac{\hbar\pi}{Lm}$. In order to investigate the time evolution of the anticipated dark-soliton like signatures in the beyond mean-field regime of interactions, we start with the initial measurement of $N_i=5$ particles at very probable positions. For moderate interparticle repulsion strength ($\gamma=1$) we use the idea of expansion in Eq.~(\ref{Syr16_5}). The resulting time evolution of the corresponding single particle probability density is presented in Fig.~\ref{Sols_Syr16_5}(a). While at $t=0$ one can observe a clearly visible density notch located around $x\approx \frac{L}{4}=0.25$, the solitonic structure quickly disappears in time  $t\approx 0.124$.  Nevertheless, the phase flip, being the other dark soliton signature, survives much longer and is still very distinct, even when the density notch is no longer visible. Among many realizations of the single-shot detection, performed for the remaining  $N_r=3$ particles at time $t=0.124$, we identified two types of phase flips (related to $J=0$ and $J=1$) in the last particle wave functions. The phase distributions of different kinds, i.e. with different phase windings $J$, were averaged separately \cite{Syrwid16}. The resulting average last particle phase distributions are depicted in Fig.~\ref{Sols_Syr16_5}(c).

For $\gamma\gg1$ the many-body dynamics cannot be effectively studied basing on the expansion in Eq.~(\ref{Syr16_5}). This is due to extremely large number of 3-particle eigenstates required to reproduce the state after the initial measurement correctly. Fortunately, when $\gamma=\infty$ we deal with impenetrable bosons whose time evolution can be examined with the help of the wave function describing noninteracting fermions, i.e.
\begin{eqnarray}  
\Psi_{N_r=3}(x_1,x_2,x_3)&\propto\Phi_F(\{\widetilde{x}\}_{N_i=5};x_1,x_2,x_3)   \label{Syr16_1_11}
\\ \nonumber 
&\displaystyle{\times S(x_1,x_2,x_3)\prod_{j=1}^{3}\prod_{s=1}^5\mathrm{sign}(x_j-\widetilde{x}_s)
}, 
\end{eqnarray}
where $S(x_1,x_2,x_3)=\prod_{a>b=1}^3\mathrm{sign}(x_a\!-\!x_b)$ and  $\Phi_F$ is the Fermi wave function given by a Slater determinant of plane waves $\propto\mathrm{exp}\left[i k_j x_s\right]$ and $\propto\mathrm{exp}\left[i k_j \widetilde{x}_s\right]$ with $k_j\big|_{\gamma=\infty}=\frac{2\pi}{L}I_j^\text{p}$ ($I_j^\text{p}\in \{I^\text{p}\}_{II}^\text{black}$). Note that  $\psi(x_1,x_2,x_3)=\Psi_{N_r=3}(x_1,x_2,x_3) S(x_1,x_2,x_3)$
represents a 3-particle wave function describing noninteracting fermions, whose time evolution can be simply obtained in the momentum space by employing the Fourier transform. Calculating  $\psi(x_1,x_2,x_3,t)$ at different moments of time one can also obtain $\Psi_{N_r=3}(x_1,x_2,x_3,t)$ \cite{Syrwid16}. The dynamics of the corresponding single particle density is monitored in Fig.~\ref{Sols_Syr16_5}(b). Note that there are no clearly visible dark soliton signatures in the average density even for $t=0$. At the beginning we can observe 5 deep minima located exactly at the positions of the 5 initially measured particles, which is a manifestation of the infinitely strong repulsion in the system. In the course of time evolution the structure of the density dips gradually disappears. Remarkably, the phase flips of windings $J=0$ and $J=1$ can be still recognized in the last particle wave functions for times comparable to the so-called quantum speed limit time $t_c=\mathrm{min}\left(\frac{\hbar \pi}{2\mathcal{E}}, \frac{\pi \hbar}{2\Delta \mathcal{E}}\right) $
describing a typical lifetime of a generic quantum state \cite{Sato2016}. While $\mathcal{E}$ is a mean energy of the considered system with respect to the ground state,   $\Delta\mathcal{E}$ denotes its variance. Thus, as calculated in Ref.~\cite{Syrwid16}, in our 3-particle infinitely interacting problem $t_c\approx0.003$. In Fig.~\ref{Sols_Syr16_5}(d) we show average last particle phase distributions corresponding to $J=0$ (solid black line) and $J=1$ (dashed red line) obtained at time $t=\frac{3}{4}t_c=0.00225$, when almost 25\% realizations revealed windings $J$ and phase distributions for which the identification of phase flips was very unreliable. Nonetheless, the phase distributions averaged over the other 75\% of the realizations possess clearly visible flips. This result allows us to suppose that the investigated type--II eigenstate remains strictly connected with a quantum dark soliton even for a very large $\gamma$.

\subsubsection{Summary}
\label{QSolLLDynamicsSummary}

Hole excited ({\it yrast}) eigenstates of the periodic Lieb-Liniger model were conjectured to be associated with dark solitons due to the coincidence of the type--II spectrum and the mean-field dark soliton dispersion relation. 
Such a supposition cannot be proven by an examination of a reduced single particle density calculated for a single hole eigenstate. 
This is due to the space translation invariance of the periodic system eigenstates, which turned out to be the main impediment on the way to answer the question whether the conjecture is correct or wrong.
Here, we pointed out that the spatial translation symmetry can be broken in the course of the particles' positions measurement resulting in emergence of dark soliton signatures for the hole excited eigenstates. 

Even if one considers a very weakly interacting system, the initial many-body hole excited eigenstate is far from the BEC product state. The situation can be changed during the successive measurement of particles' positions, when the state of the remaining particles is driven the closer the BEC product state the more particles were detected. 
We demonstrated that starting with the weakly interacting system prepared initially in a hole excited eigenstate, the dark soliton signatures localize in the course of particle measurements and 
the resulting last particle wave functions closely follow the corresponding mean-field dark soliton solutions.
Further investigations showed that the mean-field dynamics is well reproduced by the short time many-body evolution of the soliton-like structures encoded in the type--II eigenstates. For longer times, one can observe quantum many-body effects like the smearing and the revival of the density notch structures. Such phenomena are strictly related to the time variability of the soliton position uncertainty. The solitonic nature of the type--II eigenstates in the weakly interacting regime can be deeply understood by analyzing the {\it yrast} states of the noninteracting Bose gas. That is, in such a case the {\it yrast} excitation is associated with a single Fock state $\left|N-K,K\right>$, in which $K$ bosons were shifted from the zero momentum mode to the mode with the momentum equal to $\hbar\frac{2\pi}{L}$. As shown in Ref.~\cite{Oldziejewski2018}, such {\it yrast} states exhibit many features that are typical for solitons. 

 Another type--II (hole) excitation results in the appearance of the second dark soliton-like structure. Depending on the way we excite the second hole, the last particle wave function and many-body time evolution exhibit different behaviour. While some of the double hole excited eigenstates correspond to double black and double gray solitons known from the stationary GPE, there are also eigenstates related to two dark solitons that differ in shape and velocity, which allowed us to study collisions of many-body dark soliton-like structures. 
 
 The emergence of dark soliton signatures can be also observed when we enter the beyond mean-field regime. Single-shot particle detections revealed that the soliton signatures like a density notch and a phase flip can be still recognized in the last particle wave functions and are clearly visible after averaging them over many realizations. Additionally, we pointed out that the phase flip signature survives in dynamics for times comparable to a typical lifetime of a generic quantum states, even for $\gamma=\infty$.
 The quantum nature of the soliton-like structures can be observed  not only in the many-body evolution but also when the time is frozen. Indeed, in the strongly repulsive limit, the relative distance between the two solitonic signatures fluctuates from one realization to another and the resulting last particle phase flips' separation is strictly related to the number of particles measured between them.  

The results of our investigations show that the type--II (hole) excitations in the Lieb-Liniger model with periodic boundary conditions are unequivocally connected with quantum dark solitons. Such a relation is valid not only for weak interactions but also in the presence of strong interparticle repulsion.

\subsection{Open boundary conditions: identification of Gaudin's eigenstates strictly related to single and multiple quantum dark solitons}
\label{OBC_bose_sols}

As in the periodic case, eigenstates of the repulsively interacting Lieb-Liniger system confined in a hard wall box can be parameterized by distinct numbers $I_j^\text{o}$ (see Sec.~\ref{openc_boundary_conditions}). 
There are many other similarities between the Lieb-Liniger model with periodic and open boundary conditions. Here, we focus on
hole excitations whose structure is analogous in both systems (see Sec.~\ref{LLElementaryExcitations}), and show that they are strictly related to quantum dark solitons also in the open boundary conditions case.

Let us begin with infinitely weak repulsion, for which the eigenstates of the considered system reduce to Eq.~(\ref{LLHWWeaklyInteractingLimit3}), 
where $k_{j}\approx\frac{\pi}{L}n_j^\text{o}$ with $n_j^\text{o}=1,2,3,\ldots$. For instance, the noninteracting ground state is reproduced for $n_{j=1,2,\ldots,N}^\text{o}=1$. When we set all $n_j^\text{o}$'s to be equal $s>1$, the wave function $\Psi_N \propto  \prod_{j=1}^N\sin( \pi s x_j/L)$
reveals $s-1$ density notches and $\pi$-phase flips, which resembles $s-1$ dark soliton signatures. 
Obviously, in the noninteracting systems dark solitons cannot be expected. In the presence of the interparticle repulsion, the system eigenstates are given by Eq.~(\ref{LLHWPsi}), where the quasimomenta $\{k\}$ have to be determined from the Gaudin's Eqs.~(\ref{GaudinEqsArctg}) or (\ref{LLGaudinArcTanToBetheForm2}).
Note that, $n_{j=1,2,\ldots,N}^\text{o}=s$ represents $(s-1)$-fold collective excitation,
 which in the Bethe-like parameterization corresponds to $I_j^\text{o}=j+s-1$. Thus, for $s=2$ we deal with single hole excited eigenstate that can be associated with the injection of the "total momentum",  Eq.~(\ref{Open_New_momentum_Def_1}), per particle equal to $\frac{\Delta \wp}{N}=\hbar\frac{\pi}{L}$.
The situation is strikingly similar to the periodic case, for which the type--II eigenstate with $\frac{P}{N}=\hbar\frac{\pi}{L}$ turned out to be unequivocally connected with a single black soliton. Obviously, it could be only an accidental coincidence, but it is not. Employing the idea of single-shot measurements described in Sec.~\ref{BoseDynamics}, we show that single Gaudin's eigenstates parameterized by  $I_{j=1,2,\ldots,N}^\text{o}=j+s-1$ not only resemble $s-1$ dark solitons, but also reveal their clear signatures in many single-shot detections for a wide range of the repulsion strength ($\gamma=0.01,1,100$) \cite{Syrwid2017HW}. 

The numerical simulations were performed for the system consisting of $N=6$ and $N=7$ particles only. 
This is due to a dramatic increase of the computation time with $N$. Indeed, the number of terms that have to be calculated to obtain a single value of the wave function in Eq.~(\ref{LLHWPsi}), proliferates like $2^N N!$, which is $2^N$ times faster than in the periodic case.  For simplicity, the system size is set to be equal to $L=1$.

\subsubsection{First collective excitation}
\label{QSolLLHWFirstCollective}

\begin{figure}[h!] 
\begin{center}
\vspace{0.1cm}
\includegraphics[scale=0.185]{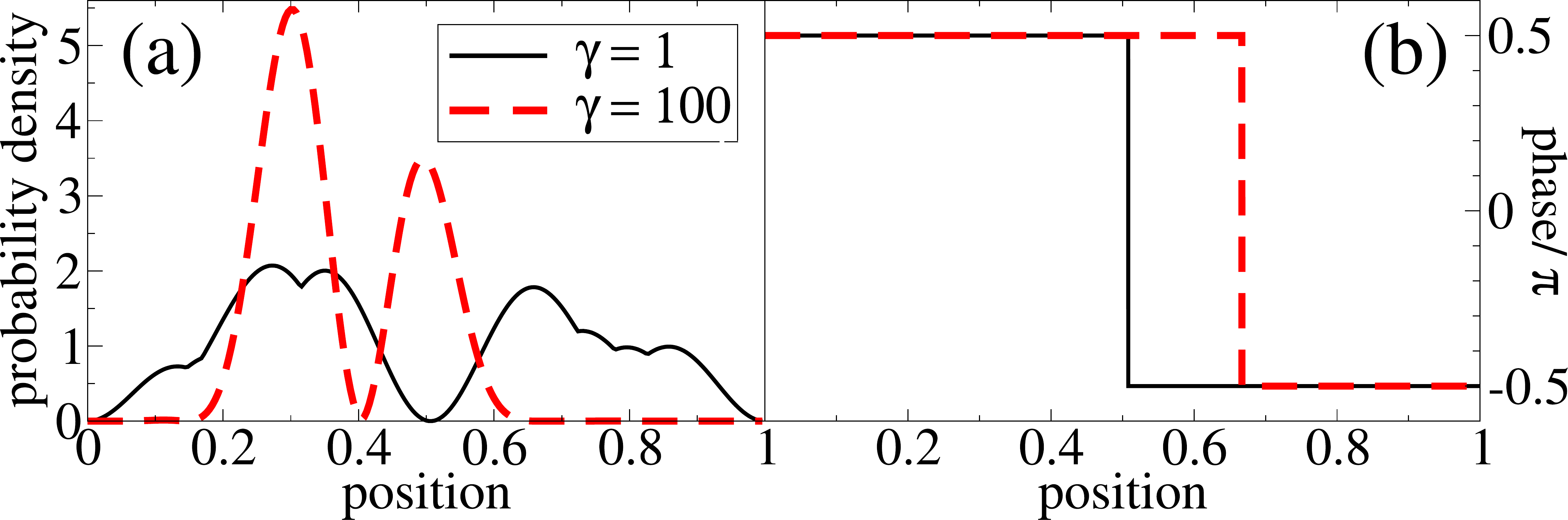}
 \end{center}
\vspace{-0.5cm}
\caption{
Typical last particle wave functions of the 7-particle Lieb-Liniger system with open boundary conditions prepared initially in the single hole excited eigenstate parameterized by $\{I^\text{o}\}_{II,1}^\text{black}$. Panel (a) presents the last particle probability density for the moderate $\gamma=1$ (solid black line) and strong $\gamma=100$ (dashed red line) repulsion. In panel (b) we show the corresponding last particle phase distributions.  
 The case of $\gamma\ll 1$ is intentionally omitted because for very weak repulsion the wave function in questions is $\propto \mathrm{sin} \left(\frac{2\pi}{L}x \right)$. 
 Reprinted and adapted from \cite{Syrwid2017HW}.
}
\label{Sols_Syr17_1}
\end{figure}

Let us start with the single collective excitation, for which the system is initially prepared in the single hole eigenstate parameterized by $\{I^\text{o}\}_{II,1}^\text{black}=\{2,3,\ldots,N,N+1\}$,
that corresponds to $\frac{\Delta \wp}{N}=\hbar\frac{\pi}{L}$.
In Fig.~\ref{Sols_Syr17_1} we show typical last particle wave functions
 \begin{eqnarray}  
 \phi(x)=&\sqrt{\rho_N(x)}\, \mathrm{e}^{i\varphi(x)}    \label{Syr17_4}
 \\ \nonumber
 &\propto \Psi_{N}(\widetilde{x}_1,\widetilde{x}_2,\ldots,\widetilde{x}_{j-1},x,\widetilde{x}_{j+1},\ldots,\widetilde{x}_N, \{k\}),
\end{eqnarray}
obtained for $N=7$ in the presence of moderate $\gamma=1$ (solid black line)  and strong  $\gamma=100$ (dashed red line)  repulsion. Here, $\Psi_{N}$ in Eq.~(\ref{LLHWPsi}) is characterized by the quasimomenta $\{k\}$ parameterized by $\{I^\text{o}\}_{II,1}^\text{black}$. The values $\widetilde{x}$ correspond to positions of $N-1$ initially measured particles where $0\leq\widetilde{x}_1 <  \ldots < \widetilde{x}_{j-1} < x < \widetilde{x}_{j+1} < \ldots < \widetilde{x}_N
\leq L$.  While panel (a) presents the last particle conditional probability densities $\rho_N(x)$, the corresponding last particle phase distributions $\varphi(x)$ are illustrated in panel (b). 
For $\gamma=1$, the last particle density profile reveals not only a density notch signature but also 6 incisions at the positions where 6 particles were initially measured. According to the nature of the interparticle repulsion, the stronger the interactions, the deeper incisions are observed.  The corresponding phase distribution possesses a single point $\pi$-phase flip located exactly at the position of the density notch. On the other hand, for very strong repulsion  ($\gamma=100$), the $\pi$-phase flip signature is still present, but it cannot be associated with any concrete density dip structure in $\rho_N(x)$.  That is, the slight incisions observed for $\gamma=1$ become very deep in the strong interaction regime. This reflects the fact that a very strong repulsion prevents from measurements of consecutive particles near those previously detected. In result the last particle density is often very distorted and that is why we always use positions of the phase flip signatures, that survive even for $\gamma \gg1$, to determine positions of the anticipated dark soliton-like structures \cite{Syrwid16,Syrwid2017HW}.

\begin{figure}[h!]  
\begin{center}
\includegraphics[scale=0.205]{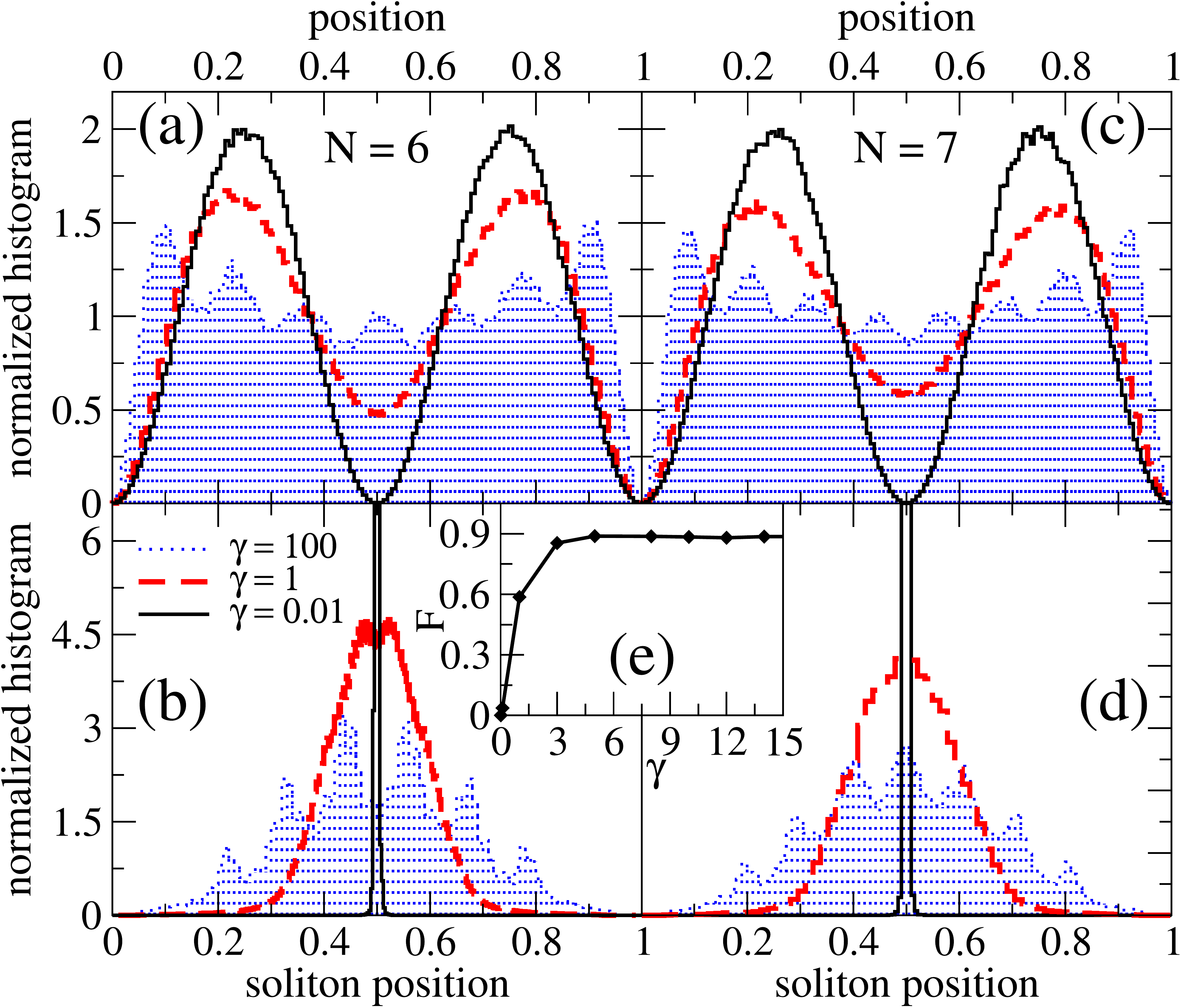}
 \end{center}
\vspace{-0.5cm}
\caption{
Single dark soliton signatures revealed in single-shot measurements of 6 and 7 bosons prepared initially in the first collectively excited eigenstate parameterized by $\{I^\text{o}\}_{II,1}^\text{black}$ for 
weak $\gamma=0.01$ (solid lines), moderate $\gamma=1$ (dashed red lines) and strong $\gamma=100$ (dotted blue shading) repulsion.
  Histograms representing average single particle densities obtained for $N=6$ and $N=7$ are shown in panels (a) and (c), respectively. 
  For $\gamma=0.01$ the density notch is clearly visible. The situation changes with the increase of $\gamma$. While for moderate interactions ($\gamma=1$) the density notch is significantly smeared but still present, for strong interparticle repulsion ($\gamma=100$) one observes a very shallow and wide hollow in the density accompanied by oscillations along its profile (see the main text). Distributions of phase flips visible in the last particle wave functions are depicted in panels (b) and (d) for $N=6$ and $N=7$, respectively. Note that the amplitude of the phase flip position fluctuations increases with the interparticle repulsion strength. In panel (e) we present the degree of filling $F=\rho(L/2)$, calculated for $N=7$. 
 Reprinted and adapted from \cite{Syrwid2017HW}.
}
\label{Sols_Syr17_2}
\end{figure}

Many particle  measurements we study reveal quantum fluctuations of dark soliton-like signatures causing the smearing of the density notch in the average single particle density. When preparing the systems in the eigenstate parameterized by $\{I^\text{o}\}_{II,1}^\text{black}$ we performed hundreds of thousands of single-shot detections realized with the help of the Metropolis algorithm. In such a way we have collected many configurations of particles' positions, which allowed us to generate histograms reproducing average single particle densities. The results obtained for $N=6$ and $N=7$ are presented in the upper panels (a) and (c) of Fig.~\ref{Sols_Syr17_2}, respectively.  In every single realization we have also determined the position of the phase flip indicating the location of the anticipated dark soliton-like structure.  This soliton signature is always visible and its position varies from one realization to another. The amplitude of such fluctuations increases with $\gamma$, which we illustrate in histograms of the phase flip positions, see Fig.~\ref{Sols_Syr17_2}(b)\&(d). Such an  effect is responsible for blurring of the density notch when moderate and strong interparticle repulsion is present, see Fig.~\ref{Sols_Syr17_2}(a)\&(c). In order to quantify the filling of the dark soliton-like notch we introduce a very simple quantity $F=\rho \left(\frac{L}{2}\right)$,
which measures how the average particle density $\rho(x)$ is affected by  quantum fluctuations. The relation between the degree of filling $F$ and $\gamma$ obtained for $N=7$ is plotted in panel (e). Note that the filling quickly increases up to $\gamma\approx 5$, when we observe a saturation. This result is a premise to believe that for $\gamma>5$ the system is strictly dominated by the interparticle repulsion \cite{Syrwid2017HW}.

In the presence of strong repulsion ($\gamma=100$), there are  $N+1$  clearly visible oscillations along the average density profiles. This is  one oscillation more in comparison to the ground state case, for which $N$  local maxima, indicating  average positions of bosons, are expected. Indeed, for  $\gamma\rightarrow \infty$ we enter the Tonks-Girardeau limit and the density should reproduce the result known from the system of noninteracting fermions \cite{Girardeau60,Paredes2004}. The additional maximum, observed when the system is prepared in a single hole excited eigenstate, appears due to the existence of a solitonic structure. Note that the phase flip jostles between  neighboring bosons, modifying their relative distances. It results in one extra local minimum  implying also  one additional local  maximum visible in the average particle density. 
The positions of local density minima, see Fig.~\ref{Sols_Syr17_2}(a)\&(c), roughly coincide with local maxima of the corresponding phase flip distribution, Fig.~\ref{Sols_Syr17_2}(b)\&(d). This is a manifestation of the fact that particles are very unlikely to be detected in the region where the phase flip occurs.

Here, we analyze the system consisting of a small number of particles only. In such a case the oscillations visible in the average particle densities for $\gamma=100$ are clearly visible, but it is expected that their period should  decay like $1/N$. Thus, for a large $N$, due to the vanishing separation between maxima, the density modulations should become negligible leading to an almost flat density profile far the from the system boundaries.

The Hamiltonian in Eq.~(\ref{LLHardWallHamiltonian}) commutes with the many-body parity operator that reflects all the particles' positions $x_j$ with respect to $\frac{L}{2}$. It can be easily seen in the limit $\gamma\rightarrow 0$, for which the considered eigenstate  $\Psi_N\propto  \prod_{j=1}^N\sin( 2\pi x_j/L)$. In such a case all particles occupy the same antisymmetric single particle state $\propto \mathrm{sin}\left(\frac{2\pi x}{L}\right)$, which leads to the vanishing of the single particle density $\rho$ at $x=\frac{L}{2}$. This is not the case in the presence of interparticle repulsion when $\rho\left(\frac{L}{2}\right)>0$, cf. results obtained for $\gamma=1,100$. There is no contradiction with the law of parity conservation and the eigenstate in question still has the same parity as in the noninteracting limit. Indeed, the interparticle interactions induce a coupling between states possessing the same parity, among which one can find also those, where some even number of bosons occupy symmetric modes. Obviously, the superposition of such many-body states is still an eigenstate of the parity operator to the same eigenvalue.

\subsubsection{Higher collective excitations}
\label{QSolLLHWHigherCollective}

\begin{figure}[t] 
\begin{center}
\includegraphics[scale=0.195]{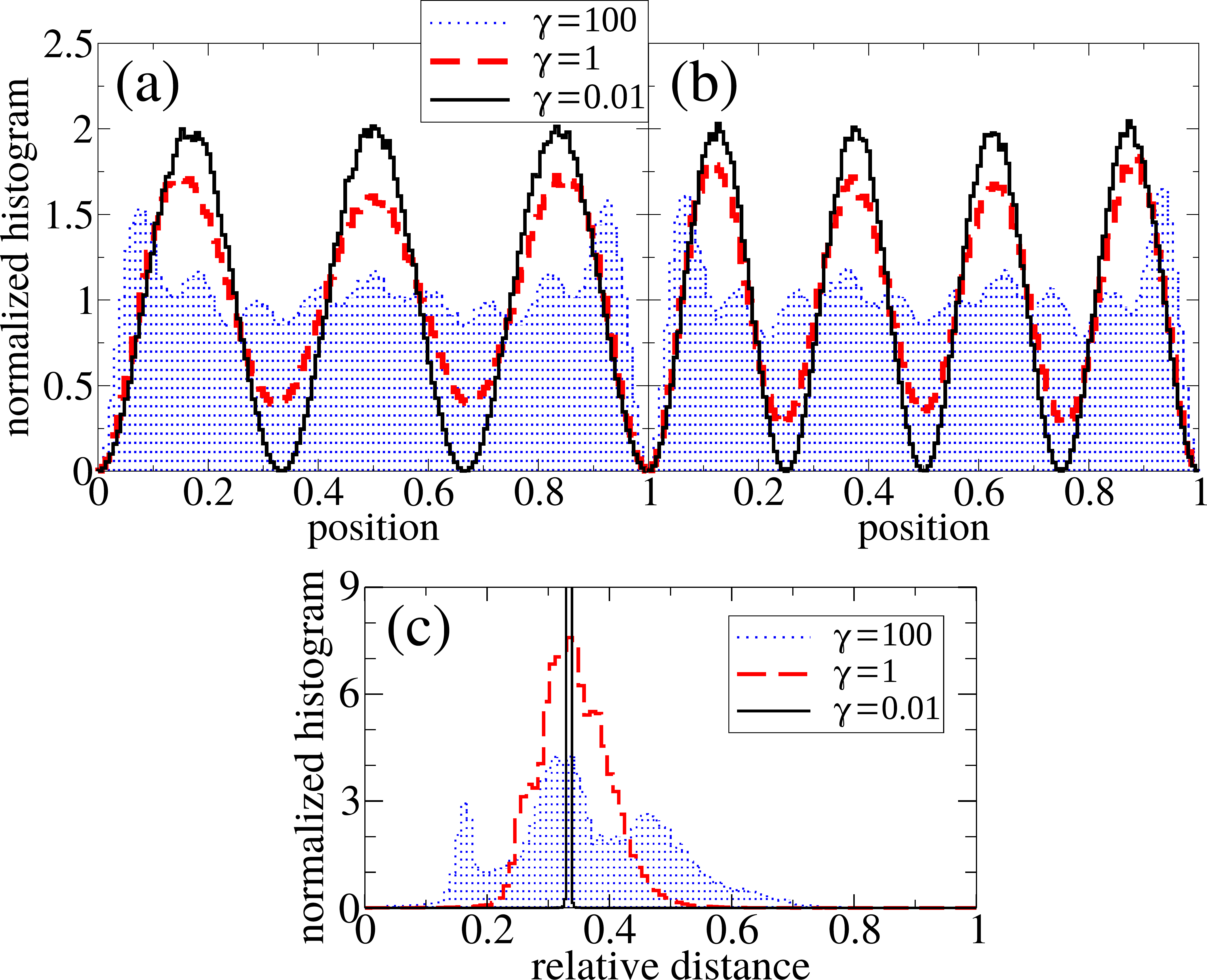}
 \end{center}
\vspace{-0.5cm}
\caption{
 Multiple dark soliton-like signatures observed in many single-shot detections performed for the 7-particle system prepared initially in the eigenstate given by $\{I^\text{o}\}_{II,s-1}^\text{black}$, Eq.~(\ref{Syr17_6}).
 Panels (a) and (b) show the average particle densities obtained in three different interaction regimes: $\gamma=0.01,1,100$, for $s=3$ and $s=4$, respectively. 
In the former case ($s=3$)  we have also investigated the distribution of relative distances between two phase flip soliton signatures, which are  illustrated in panel (c).
 Reprinted and adapted from \cite{Syrwid2017HW}.
}
\label{Sols_Syr17_3}
\end{figure}

We can also expect that through a multiple collective excitation one generates multiple dark soliton-like structures. If so, preparing the system initially in the eigenstate parameterized by  
 \begin{equation}  
\textstyle{
\{I^\text{o}\}_{II,s-1}^\text{black}=\{s,s+1,\ldots,N+s-2,N+s-1\}
 }, 
 \label{Syr17_6}
\end{equation} 
where $\frac{\Delta\wp}{N}=\hbar\frac{\pi(s-1)}{L} $,
one should observe $s-1$ density notches and $\pi$-phase flips. To check this supposition we have performed exactly the same single-shot measurements for the system consisting of $N=7$ bosons as before, but for $s=3$ and $s=4$. The results of single realizations of the detection process and average single particle densities, presented in Fig.~\ref{Sols_Syr17_3}(a)--(b), confirm our predictions. As expected, in each simulation we can always find $s-1$ phase flips by $\pi$ in the last particle wave function, even for very strong repulsion $\gamma=100$.  In the case of weak ($\gamma=0.01$) and moderate ($\gamma=1$)  interactions, the positions of the flips coincide with $s-1$ clearly visible density notches exhibited by the conditional last particle probability density. The average particle densities in different interaction regimes look very similar to those observed for $s=2$, with the exception that instead of one there are $s-1$ density notches. The solitonic notches are partially blurred when $\gamma=1$ and almost completely invisible for $\gamma=100$. In the latter case we again notice the modulations along profiles of the average densities, but this time the number of maxima is equal to $N+s-1$, which coincides with the sum of the number of particles $N$ and the number $s-1$ of the anticipated dark soliton structures (phase flips).  That is, starting with the eigenstate given by Eq.~(\ref{Syr17_6}), one deals with $s-1$ phase flips. They interspace $s-1$ pairs of neighbouring particles modifying their mutual separations and leading to an appearance of $s-1$ additional minima in comparison to the ground state case.

Let us now focus on the $s=3$ case, corresponding to the eigenstate strictly related to a double black soliton. In many single-shot measurements we observe fluctuations of positions of  the two $\pi$-phase flips as well as  variation of the relative distance between them.  As the interparticle repulsion increases, these quantum many-body phenomena intensify, which is responsible for smearing of the density notches visible in Fig.~\ref{Sols_Syr17_3}(a). For $\gamma\ll 1$, a double soliton structure is well approximated by the product of $\mathrm{sin}\left(\frac{3\pi x}{L}\right)$ functions, and thus the soliton signatures are expected to be located at $x=\frac{L}{3}$ and $\frac{2L}{3}$. Hence,  the relative distance between the two $\pi$-phase flips is fixed and equal to $\frac{L}{3}$, see solid black histogram in Fig.~\ref{Sols_Syr17_3}(c). For $\gamma=1$ (dashed red line), the distribution of relative distances between phase flips is significantly broadened but still centred at $\frac{L}{3}$. This result is strictly related to the open boundary conditions imposed on the system. Approaching the system boundaries, the wave function vanishes on a length scale comparable with the half of the density notch width. Let us now imagine that we merge the system edges. In such a case the corresponding probability density resembles not double but triple soliton structure in a ring. Therefore, an average distance between solitons should be equal to $\frac{L}{3}$. One may expect that the visualization we proposed works also for higher collective excitations, i.e. starting with the $(s-1)$-fold collectively excited eigenstate, the mean separation between soliton signatures is approximately given by $\frac{L}{s}$.

In the Tonks-Girardeau limit ($\gamma\rightarrow \infty$), particles tend to be localized at spatial intervals $\frac{L}{N}$ and any phase flip signature should occur halfway between the neighboring bosons. Therefore, one can expect that in such a case two phase flip signatures are most likely distanced by integer multiples of $\frac{L}{N}$.  The numerical results, presented in Fig.~\ref{Sols_Syr17_3}(c), agree with this observation. Indeed, for $\gamma=100$ the distribution of the relative distances between the two phase flips, visible in the last particle wave functions, reveals three peaks located in close vicinity of three first integer multiples of $\frac{L}{N}$. Similar effect was also observed for periodic system, see Fig.~\ref{Sols_Syr16_4}(g)\&(h).

\subsubsection{Summary}
\label{QSolLLHWSummary}

Basing on the analysis of many-body eigenstates in the noninteracting limit  and using the similarities between  systems with periodic and open boundary conditions, we have identified a specific class of many-body eigenstates that are strictly related to quantum dark solitons in the Lieb-Liniger model confined in a hard wall box. 
Genuine dark solitons can be realized only in the presence of nonzero interparticle repulsion, where the parameterizing numbers $I_j^\text{o}$ (or $n_j^\text{o}$) cannot be unambiguously associated with single particle modes. Therefore, it is adequate to ask what is the authentic character of hole excited eigenstates in question. Employing the idea of single-shot measurements we showed that if the eigenstate is parameterized by the set $\{I^\text{o}\}_{II,s-1}^\text{black}$ defined in Eq.~(\ref{Syr17_6}), then $s-1$ dark soliton signatures can be observed in the last particle wave functions obtained in each single realization of the detection process. In the weak and intermediate regime of the interparticle repulsion, $s-1$ density notches are also visible in the resulting average particle densities. In the latter case quantum many-body effects cause fluctuations of phase flip positions, which lead to a partial smearing of the density notch structure. It is worth stressing that in the case of small particle numbers ($N=6,7$) we investigated, the shape of the density notches cannot reproduce a simple hyperbolic tangent function, known from the basic mean-field considerations. This is due to the fact that if $\gamma\ll1$  and $N$ is small, the corresponding healing length $\xi=1/\sqrt{\bar{c}\rho}=L/\sqrt{N^2\gamma}$ is of the same order or even larger than $L$.

 In the strongly repulsive regime ($\gamma\gg 1$), the system is dominated by quantum many-body effects and positions of the last particle phase flips vary widely between realizations.  Consequently, the density notches in the average particle densities  are completely blurred. Studying fluctuations of two phase flips' signatures we showed that distributions of the relative distances between them are the broader, the stronger interactions are present. For very strong interparticle repulsion one observes a peak structure of the distribution indicating the specific distances at which the phase flip  signatures localize more eagerly.

\section{Dark soliton-like many-body yrast eigenstate of Yang-Gaudin gas in ring geometry}
\label{QuantumSolitonsInMBstatesYG}

 At very low temperatures the two-component attractively interacting Fermi gas can undergo a phase transition and form a superfluid state. In the presence of weak interparticle interactions the system is in the Bardeen-Cooper-Schrieffer (BCS) regime  governed by a set of nonlinear Bogoliubov-de Gennes equations \cite{pethick, KetterleZwierlein}. In general, these equations play a particular role in the ground state properties determination. However, they can also be used in order to describe dark soliton solutions corresponding to nearly the same particle densities as in the ground state case, but possessing the dark soliton phase flip signature visible in the BCS pairing function  \cite{DziarmagaSachaArxiv2004, DziarmagaSachaLasPhys2005, Antezza2007A}. Passing through the crossover from the BCS to BEC regime, one can identify the BCS pairing function of a dark soliton solution as the condensate wave function revealing dark soliton signatures in a molecular BEC \cite{Antezza2007A, Pieri2003A}.

 The idea of experimental realization of dark solitons in Fermi systems is based on the identical phase engineering as the one used for bosons. An application of a short laser radiation on a half of the system of noninteracting single-component fermions leads to an emergence of pairs of bright and dark soliton-like structures \cite{Karpiuk2002A,Karpiuk2002B}. The similar method was also applied to a superfluid Fermi system, but the resulting state, announced as a heavy soliton \cite{Yefsah2013a}, turned out to be very unstable and quickly evolves into a vortex, which was predicted numerically \cite{Bulgac2014A,Scherpelz2014A} and confirmed in the laboratory with the help of tomographic imaging \cite{Yefsah2014a}. Therefore, the procedure in which both components of a Fermi gas are subjected to phase imprinting fails. Further studies on the dark soliton BCS pairing function provide a prediction 
 that the experimental realization of the dark soliton excitation in a superfluid Fermi system should rely on the phase engineering, in which the phase of only one of two fermions constituting a Cooper-pair is modified, i.e. the phase should be imprinted on a single component only (see Ref.~\cite{DelandeSachaProper2014}). 
 
An ultracold unpolarized gas of spin-$\frac{1}{2}$ fermions in 1D is described by the Yang-Gaudin Hamiltonian, Eq.~(\ref{YG_hamiltonian}).
The interparticle attraction results in a formation of two-particle bound state pairs built up by fermions belonging to different spin components. The size of such $\downarrow$--$\uparrow$ pairs depends on the attraction strength and can be used to distinguish two regimes of interactions, where the physical properties of the system are significantly different. Namely, one effectively deals with a BCS-like gas of Cooper pairs, when the interparticle attraction is weak so that the average size of $\downarrow$--$\uparrow$ molecules is larger than their mean separation. On the other hand, the very strong attraction leads to a formation of tightly bound pairs resembling impenetrable bosonic dimers (see also  Ref.~\cite{SowaGajda}).  There is a crossover between both the physically different regimes, which can be observed around $\gamma\approx -1$ \cite{Shamailov2016,SyrwidFermi2018}.

In Sec.~\ref{YG_Gs} we discuss similarities and differences between the strongly attractive Yang-Gaudin system and a gas of strongly interacting bosonic dimers which can be described within the Lieb-Liniger model.
While one cannot unambiguously associate the strongly attractive two-component Fermi system with a gas consisting of strongly repelling bosons, we believe that both systems share many properties. 
In Sec.~\ref{QuantumSolitonsInMBstates} we showed that the type--II eigenstates of the Lieb-Liniger model with periodic boundary conditions are unequivocally connected with quantum dark solitons.  Such eigenstates correspond to the lowest energy for a given nonzero total momentum, and thus they are called {\it yrast} states.  
 Recently, S. Shamailov and J. Brand  identified the {\it yrast} eigenstates in the periodic Yang-Gaudin model and showed that, analogously to the Bose case, in the weakly attracting regime the {\it yrast} spectrum reproduces the mean-field dark soliton dispersion relation.

Here, employing the Bethe ansatz method and the idea of single-shot measurements, we investigate the {\it yrast} excitations in the attracting Yang-Gaudin system with periodic boundaries and their possible relationship with dark solitons.
In addition, we study the pairing phenomenon between fermions with opposite spins and determine the size of $\downarrow$--$\uparrow$  pairs of fermions in a wide range of the interparticle attraction strength.
The numerical simulations of the particles' positions detection, confirm the solitonic nature of  {\it yrast} excitations in the periodic Yang-Gaudin model. Due to space translation symmetry possessed by the periodic system eigenstates any soliton-like structures cannot be displayed in the corresponding reduced single particle density. Here, similarly to the Lieb-Liniger case, we show that, if the system  is prepared initially in an {\it yrast} eigenstate, dark soliton signatures emerge in the course of particles' positions measurement and they are clearly visible in the wave function of the last $\downarrow$--$\uparrow$ pair of fermions in the system.

\subsection{Pairing phenomenon in single-shot measurements}
\label{YG_pairing}

We start with an analysis of the pairing phenomenon between spin-$\downarrow$  and spin-$\uparrow$ fermions. For this purpose we consider the periodic attractively interacting Yang-Gaudin model (see Sec.~\ref{Gaudin-Yang_model}) consisting of equal numbers of fermions in each spin component (unpolarized gas), where the following notation $N_\downarrow=N_\uparrow =N$ will be used. Taking advantage of the determinant formulation of the Bethe ansatz wave function, Eqs.~(\ref{YG_Wf_New_1})--(\ref{YG_Wf_New_4}), we perform numerical simulations of the single-shot measurements by means of the Monte Carlo Metropolis routine (see also Sec.~\ref{BoseDynamics}). In such a case, it is convenient to assume that the measured positions $\widetilde{r}_{s}$ in every single sample configuration $\mathcal{R}=\{\widetilde{r}_1,\widetilde{r}_2,\ldots,\widetilde{r}_{2N}\}$ are ordered so that $\widetilde{r}_{s}=\widetilde{x}_s^\downarrow$ and $\widetilde{r}_{N+s}=\widetilde{x}_s^\uparrow$ for $s=1,\ldots,N$. Despite the method is very efficient, the computation time still dramatically increases with $N$. Thus, to collect reliable statistics we are able to study 5+5 ($N=5$) particle system only.   

 We focus on the ground state and the specific {\it yrast} eigenstate characterized by the total momentum equal to $P=\hbar\frac{6\pi}{L}$. 
 Such a choice of the {\it yrast} excitation is dictated by the results obtained for the periodic Lieb-Liniger system, where the {\it yrast} eigenstate with $\frac{P}{N}=\hbar\frac{\pi}{L}$ corresponds to the most distinct, black soliton signatures.
It is expected that in the two-component Fermi system in question dark soliton-like structures can be observed in the wave function of the last pair of $\downarrow$--$\uparrow$ fermions. 
Additionally, in the strongly attractive regime $\big(\gamma=\frac{\bar{c}L}{2N}\rightarrow -\infty\big)$, spin-$\downarrow$ and spin-$\uparrow$ fermions form tightly bound bosonic dimers and the system, to some extent, resembles a strongly repulsive Bose gas. 
Hence, instead of the total momentum per single particle we consider the total momentum per  $\downarrow$--$\uparrow$ pair. That is why we decided to analyze the {\it yrast} eigenstate for which  $\frac{P}{N}=\hbar\frac{6\pi}{5L}$ is as close $\hbar\frac{\pi}{L}$ as it is possible.

 \begin{figure*}[h!] 
\begin{center}
\includegraphics[scale=0.16]{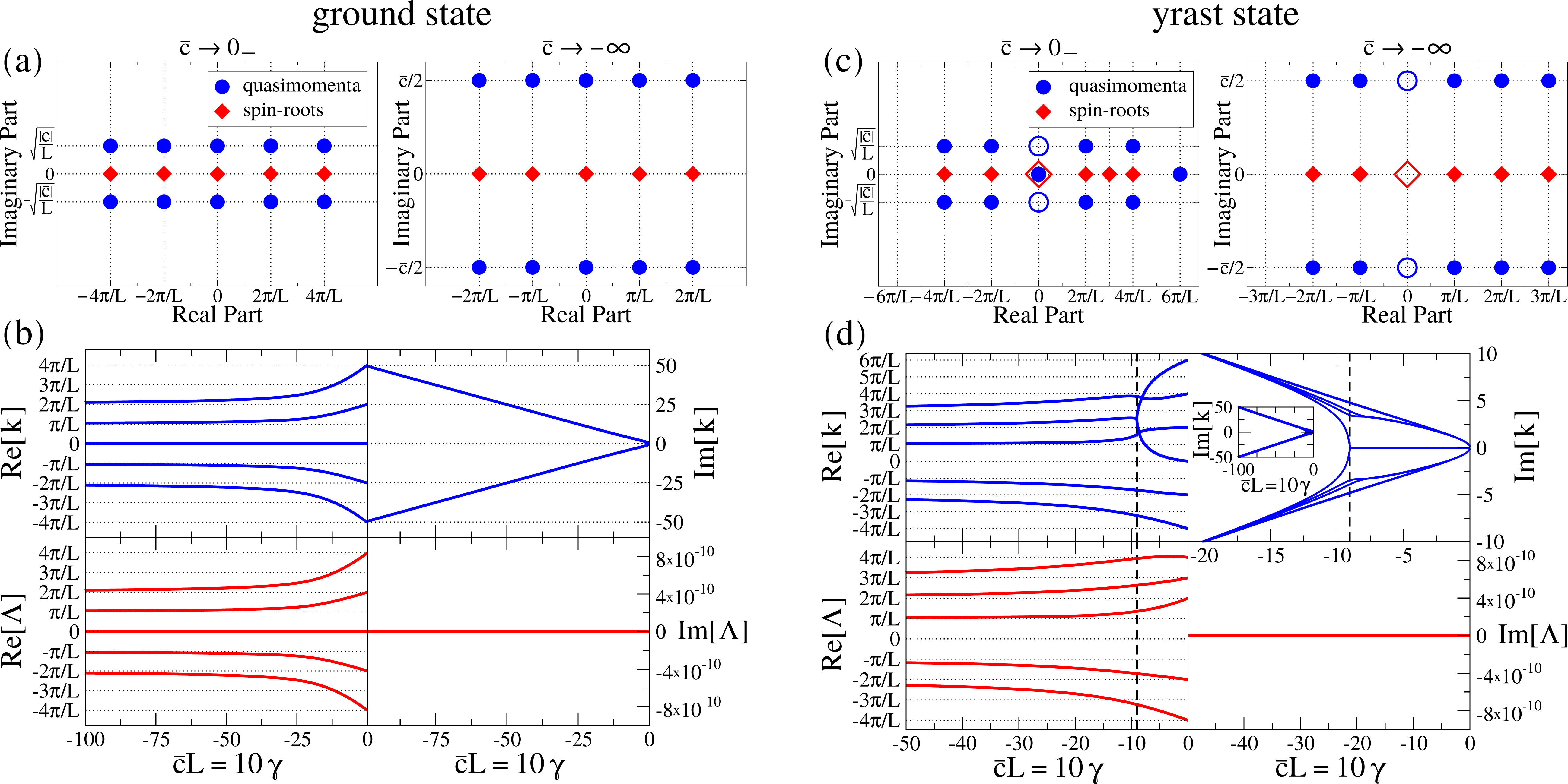}
 \end{center}
\vspace{-0.5cm}
\caption{
Quasimomenta $\{k\}$ and spin-roots $\{\Lambda\}$ solutions of the Bethe ansatz Eqs.~(\ref{YG_PBC_conds_1})--(\ref{YG_PBC_conds_2}) for the unpolarized system consisting of $N_{\downarrow,\uparrow}=N=5$ fermions.
Panels (a) and (c) present  sketches of the solutions in the weakly ($\bar{c}\rightarrow 0_-$) and strongly ($\bar{c}\rightarrow -\infty$) attractive limits for the ground state and the chosen {\it yrast} eigenstate with $P=\hbar\frac{6\pi}{L}$, respectively. The empty symbols are used to indicate which values are changed by the excitation in comparison to the ground state. The numerical solutions of the Bethe ansatz equations versus $\gamma$ are illustrated in panels (b) and (d). 
 Reprinted and adapted from \cite{SyrwidFermi2018}.
}
\label{Sols_Syr18_Fermi_1}
\end{figure*}

In order to study the system eigenstates, one needs to determine $2N$ quasimomenta $k$ and $N$ values of spin-roots $\Lambda$ satisfying the Bethe ansatz Eqs.~(\ref{YG_PBC_conds_1})--(\ref{YG_PBC_conds_2}). 
To do so, we follow the idea of a simple linear extrapolation that allows us to find the solutions $\{k\}_{2N}$ and $\{\Lambda\}_N$ by a consecutive increase or decrease of $\bar{c}$. It is possible, because we have predictions for $\{k\}_{2N}$ and $\{\Lambda\}_N$ in the weak and strong interaction regime (see Secs.~\ref{YG_Gs}\&\ref{YGElementExcitations}). Schemes of the approximate quasimomenta $k$ and spin-roots $\Lambda$ for the 5+5 particle ground state and the chosen {\it yrast} eigenstate for weak ($\bar{c}\rightarrow 0_-$) and strong ($\bar{c}\rightarrow -\infty$) attraction are shown in Fig.~\ref{Sols_Syr18_Fermi_1}(a)\&(c). In Fig.~\ref{Sols_Syr18_Fermi_1}(b)\&(d), we present the numerical solutions of  Eqs.~(\ref{YG_PBC_conds_1})--(\ref{YG_PBC_conds_2}), obtained in a wide range of $\gamma<0$. While for the ground state all the solutions smoothly changes from weak to strong attraction, in the {\it yrast} eigenstate case one observes a bifurcation of the resulting quasimomenta that takes place at $\gamma_b\approx -0.905$. The bifurcation point separates two different {\it yrast} excitation scenarios. For $\gamma\rightarrow 0_-$, the excitation tears the conjugate pair $k_\pm=\pm i\sqrt{|\bar{c}|/L}$ apart and expels one of these quasimomenta just above the Fermi surface, i.e. at  $ k=\frac{6\pi}{L}$. Such an {\it yrast} excitation scheme is very similar to the collective excitation  investigated in a single-component Fermi system, see Refs.~\cite{Karpiuk2002A,Karpiuk2002B,DamskiSachaZakrz2002}. Indeed, for weak attraction, the considered  {\it yrast} eigenstate in the occupation basis reads  $\left|\Psi\right>\approx\frac{\sqrt{2}}{2}\big ( \left|\{y\} \right>_\downarrow \left|\{g\} \right>_\uparrow +  \left|\{g\} \right>_\downarrow \left|\{y\} \right>_\uparrow \big)$, 
where the Fock states ($\sigma=\downarrow,\uparrow$)
 \begin{eqnarray}  
\displaystyle{
\begin{array}{l}
\left|\{g\} \right>_\sigma=\left|\ldots,0_{-3},1_{-2},1_{-1},1_0,1_1,1_2,0_3,\ldots\right>,
\\ 
\left|\{y\} \right>_\sigma=\left|\ldots,0_{-3},1_{-2},1_{-1},0_0,1_1,1_2,1_3,0_4,\ldots\right>,
\end{array}
 }
 \label{Syr18_Fermi_2}
\end{eqnarray}
indicate which single particle states $\propto\mathrm{exp}\left[i 2\pi j x/L \right]$ are occupied ($1_j$) and which are not ($0_j$). Note that it also coincides with the prediction based on the BCS pairing function analysis, where it is expected that the dark soliton-like state corresponds to a collective excitation of a single component only \cite{DelandeSachaProper2014}. We stress that in our simulations the BCS regime is not attainable due to insufficient current computer resources. Indeed, to examine the system in which the size of Cooper pairs exceeds the interparticle separation many times, one needs to deal with $N\gg 5$.

The other scenario of the {\it yrast} excitation takes place for $\gamma<\gamma_b$ and relies on the excitation of a whole pair of conjugate quasimomenta, i.e. for $\gamma\rightarrow -\infty$ the pair $k_\pm=\pm i\frac{|\bar{c}|}{2}$ is expelled just above the Fermi surface at $\frac{3\pi}{L}\pm i\frac{|\bar{c}|}{2}$. Note that the pair is not broken under the excitation, which is a manifestation of the fact that for $\gamma<\gamma_b$ the binding energy per pair exceeds the  energetic cost associated with the excitation of the second conjugate partner (see Sec.~\ref{YGElementExcitations}).

\begin{figure}[h!] 
\begin{center}
\includegraphics[scale=0.35]{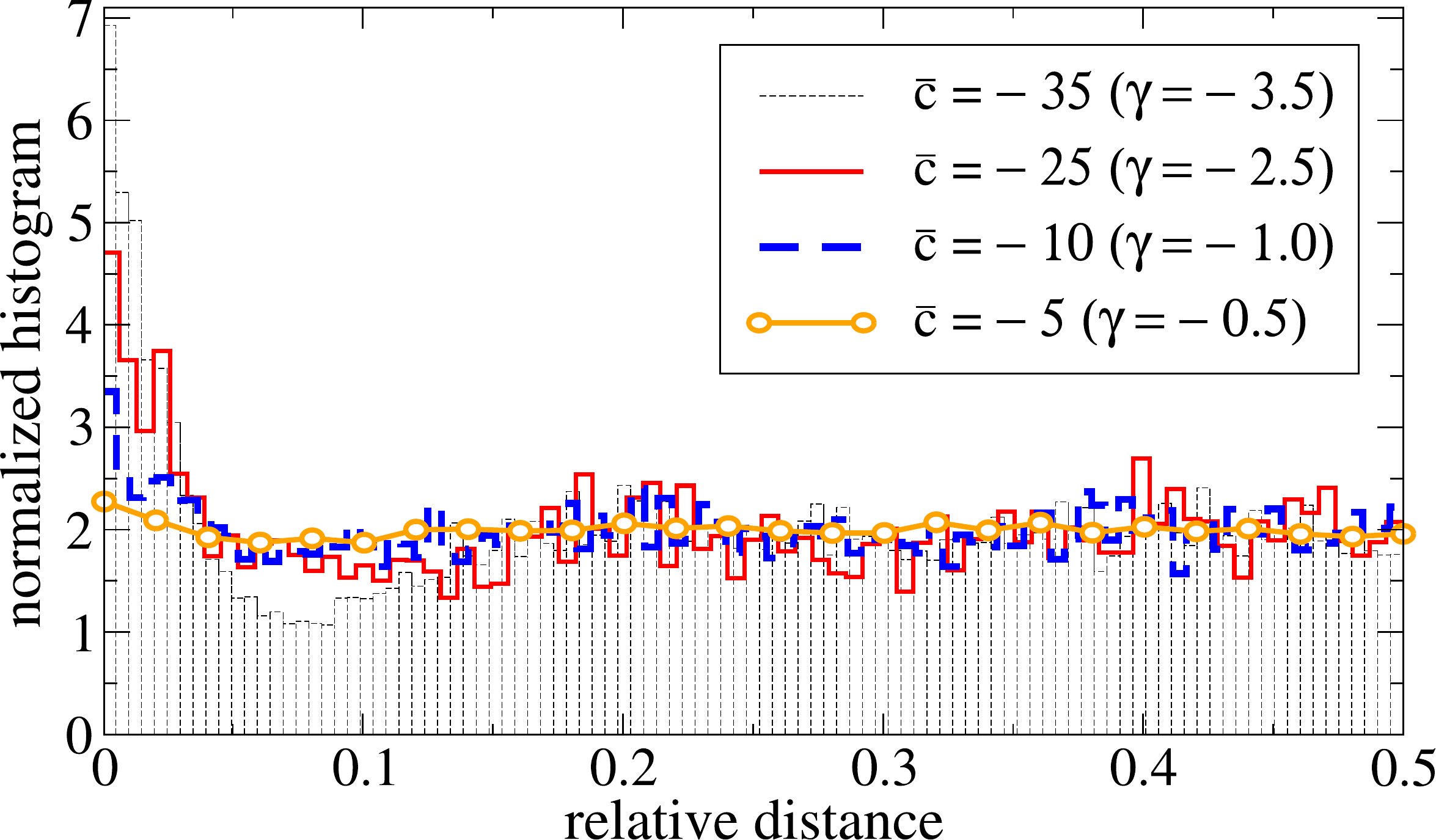}
 \end{center}
\vspace{-0.5cm}
\caption{ Distributions of the relative spatial separations  $\Delta_j^n$,  Eq.~(\ref{Syr18_Fermi_4}), between spin-$\downarrow$ and spin-$\uparrow$ fermions obtained for different attraction strengths in millions of single-shot measurements. The simulations were performed for the ground state of the 5+5 particle periodic system of size $L=1$. 
 Reprinted and adapted from \cite{SyrwidFermi2018}.
}
\label{Sols_Syr18_Fermi_2}
\end{figure}

Having the numerical solutions of the Bethe ansatz equations, i.e. the values $\{k\}_{2N}$ and $\{\Lambda\}_N$, we are ready to analyze the problem of $\downarrow$--$\uparrow$ pairs formation. The entire concept of the analysis we present on the ground state example. Similar study was also performed for the chosen {\it yrast} eigenstate. The investigations are based on single measurements. In each step we generate two sets of particles' positions  $X_j^\sigma=\{x_{j,1}^\sigma, x_{j,2}^\sigma, \ldots, x_{j,5}^\sigma\}$ ($\sigma=\downarrow,\uparrow$),
where $j$ counts the number of realizations. These sets are used to calculate relative distances between spin-$\downarrow$ and spin-$\uparrow$ fermions in a ring of length $L$, where for $n$-th fermions in $j$-th realization such a quantity is defined as  \cite{SyrwidFermi2018}
 \begin{align}  
\displaystyle{
\Delta_j^n=\mathrm{min}\left( \left|x_{j,n}^\downarrow -x_{j,n}^\uparrow  \right|, \left|L-\left|x_{j,n}^\downarrow -x_{j,n}^\uparrow  \right|  \right|	 \right)
 }.
 \label{Syr18_Fermi_4}
\end{align} 
 After many single-shot detections we prepare histograms of spatial relative separations between fermions with opposite spins. In Fig.~\ref{Sols_Syr18_Fermi_2} we show the comparison between results obtained for different $\gamma$. At first sight one observes a background density $\approx 2$ (in the simulations we set $L=1$) visible in each histogram. This is due to the fact that the probability density is not affected by permutations of particles. Consequently,  the positions for which w calculate $\Delta_j^n$'s are in a certain sense random. This result is also independent of the interaction strength. The profiles of the relative distance distributions for different values of $\gamma$ differ mainly in the range from 0 to 0.2. It can be attributed to the fact that when $\downarrow$--$\uparrow$ pairs are larger than their mean separation $\bar{\delta}=\frac{L}{N}=0.2$ ($N=5,\, L=1$), there exists spatial overlap between fermions constituting different pairs. Such an effect blurs the evidences of large-sized Cooper-like pairs in the histograms we investigate. This also means that the method we are going to apply to determine an average size of $\downarrow$-$\uparrow$ pairs, becomes less reliable when it exceeds $\bar{\delta}$. Therefore, the mean separation between pairs $\bar{\delta}$ turns out to be a reference quantity in our analysis. This is also the length scale on which one can observe the pairing phenomenon, while studying  the relative distances between spin-$\downarrow$  and spin-$\uparrow$ fermions \cite{SyrwidFermi2018}.

Let us draw conclusions from the histograms presented in Fig.~\ref{Sols_Syr18_Fermi_2}. 
 For $\gamma\lesssim -1$, one observes an escalation of small-sized $\downarrow$--$\uparrow$ pairs occurrence revealed by a single peak in the relative distance distributions located at $\approx 0$. The appearance of such a peak, which is the higher, the stronger attraction is, indicates that we entered the regime of tightly bound $\downarrow$--$\uparrow$ molecules of an average size smaller than $\bar{\delta}$. On the other hand, when $0>\gamma\gtrsim-1$, the peak is either invisible or hardly recognizable. This is the regime of the BCS-like Cooper pairs formation, where the average size of pair is larger than $\bar{\delta}$. In between, around $\gamma\approx -1$, there is no sharp transition and we observe a crossover.  One can also notice small oscillations along the histogram profiles, whose period $\approx \bar{\delta}$. Additionally, for $\gamma=-3.5$ the distribution reveals a density dip located just after the main peak, at the relative spatial separation $\approx 0.08$. For such a strong attraction, we expect formation of tightly bound bosonic dimers, smaller than $\bar{\delta}$. Nevertheless, the Pauli exclusion principle is still present, and does not allow two $\downarrow$--$\uparrow$ dimers to occupy the same position in the coordinate space. Therefore, the tightly bound $\downarrow$--$\uparrow$ molecules exhibit a natural tendency for a uniform spatial distribution, i.e. they want to be located at equal spatial intervals.  The oscillations of  the relative distance distributions, visible in Fig.~\ref{Sols_Syr18_Fermi_2}, arise through this mechanism. Similar modulations are revealed by the correlation function $\big<   \hat{\Psi}_\downarrow^\dagger (x)   \hat{\Psi}_\uparrow^\dagger (y)\hat{\Psi}_\uparrow (y) \hat{\Psi}_\downarrow (x)  \big>$ calculated in the BCS approximation for the BCS ground state (see also Refs.~\cite{SyrwidFermi2018,KetterleZwierlein}).

For $\gamma\lesssim-1$  fermions with opposite spins form two-particle tightly bound states. Hence, it is adequate to compare the results obtained for the strongly attracting many-body system with the predictions provided by an analysis of a two-body problem, where the probability density of the corresponding two-particle bound state reads $| \psi_{2}(r)|^2\propto \mathrm{e}^{-2|\gamma|r/\bar{\delta}} $ with $r=\big|x^\downarrow - x^\uparrow \big|$ \cite{CastinHerzog01,mcguire64,Guan2013,SyrwidFermi2018}.
Thus, according to the  two-body considerations, the  $\downarrow$--$\uparrow$ dimer  size is given by $\bar{\delta}/|\gamma|$, where for $\gamma=-1$ the size of $\downarrow$--$\uparrow$ pair coincides with $\bar{\delta}$.   

\begin{figure}[h!] 
\begin{center}
\includegraphics[scale=0.34]{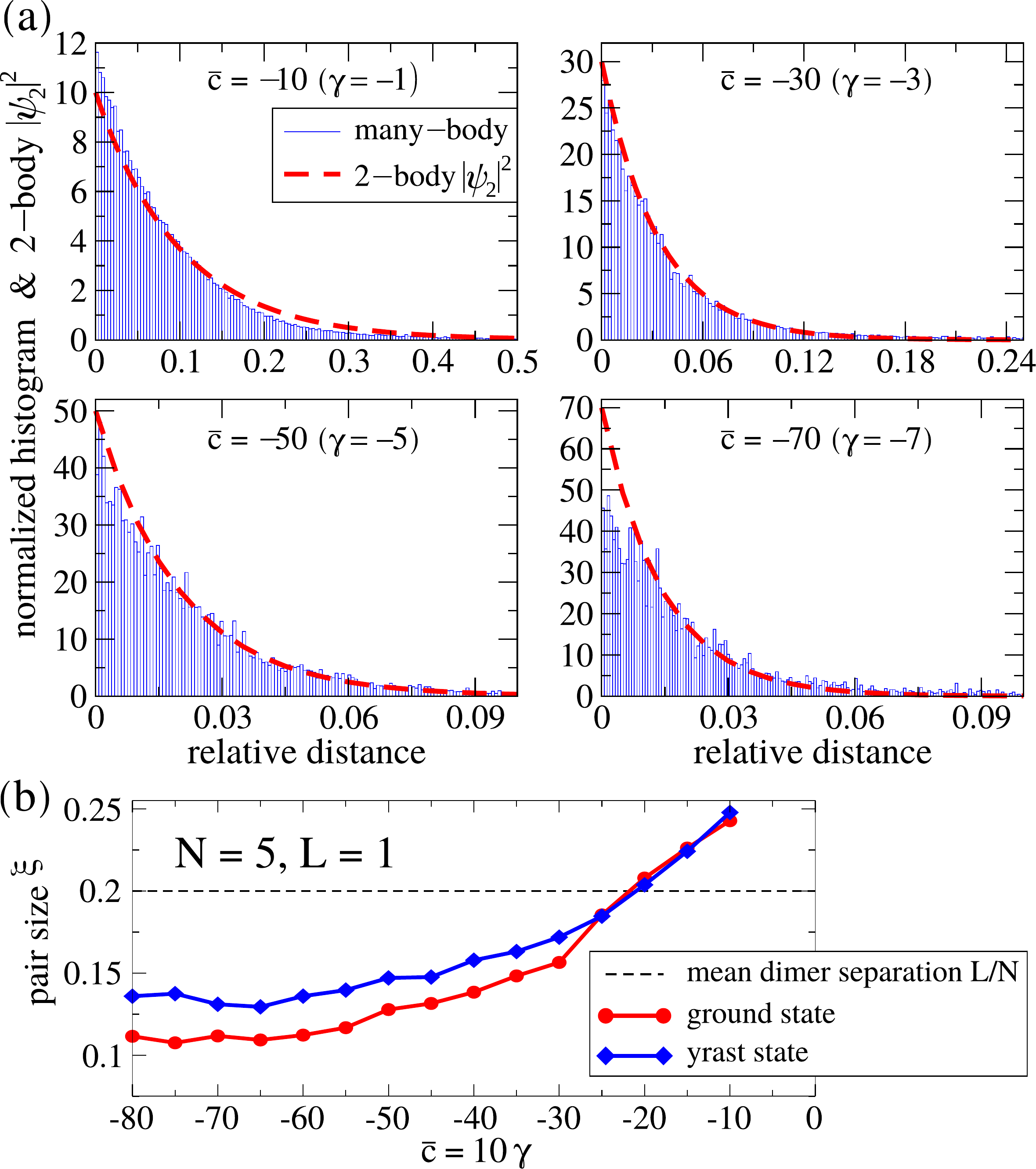}
 \end{center}
\vspace{-0.5cm}
\caption{ Panel (a): histograms of the relative spatial separations, Eq.~(\ref{Syr18_Fermi_6}),  between recognized $\downarrow$--$\uparrow$  pairs of fermions obtained in millions of single-shot simulations for the unpolarized ground state with $N=5$.  Each histogram is compared with the corresponding two-particle probability density $|\psi_2|^2$ (dashed red line), see the main text.
Panel (b): the $\downarrow$--$\uparrow$ pair size $\xi$, Eq.~(\ref{Syr18_Fermi_8}), versus $\bar{c}=10\gamma$ determined for the 5+5 ground state (red line with filled circles) and for the chosen {\it yrast} eigenstate with $P=6\hbar \pi$ (blue line with filled diamonds).  The mean spatial separation between $\downarrow$--$\uparrow$ pairs $\bar{\delta}=0.2$ ($L=1$) is  indicated  by the dashed black line.
 Reprinted and adapted from \cite{SyrwidFermi2018}.
}
\label{Sols_Syr18_Fermi_3}
\end{figure}

 We are going to show that the two-body prediction agrees with the results of our many-body simulations in the regime where $\gamma\lesssim-1$. 
 For this purpose, we need to somehow recognize which fermions from the sets $X_j^\downarrow$ and $X_j^\uparrow$ are actually paired. To do so, in every single $j$-th realization we consider permutations $\tau\in \mathcal{S}_N$ of the order of spin-$\uparrow$ fermions, i.e. $x_{j,n}^\uparrow\rightarrow x_{j,\tau(n)}^\uparrow$. As a result, the relative distances depend on $\tau$ and read
 \begin{equation}  
\displaystyle{
\widetilde{\Delta}_j^n=\mathrm{min}\left( \left|x_{j,n}^\downarrow -x_{j,\tau(n)}^\uparrow  \right|, \left|L-\left|x_{j,n}^\downarrow -x_{j,\tau(n)}^\uparrow  \right|  \right|	 \right)
 }.
 \label{Syr18_Fermi_6}
\end{equation} 
Our idea is based on the assumption that the permutation $\tau$, minimizing the sum $S_j(\tau)=\sum_{n=1}^{N}\widetilde{\Delta}_j^n$,
 corresponds to the relative distances $\widetilde{\Delta}_j^n$ of genuinely paired spin-$\downarrow$ and spin-$\uparrow$ fermions \cite{SyrwidFermi2018}. By proceeding in such a way, we collected all the relative spatial separations $\widetilde{\Delta}_j^n$ in millions of single-shot detections carried out for the 5+5 particle ground state, where in each $j$-th realization the permutation $\tau$ was chosen so that $S_j(\tau)$ is minimal. The resulting histograms of the relative distances between paired fermions are presented in Fig.~\ref{Sols_Syr18_Fermi_3}(a). Note that for $\gamma\lesssim -1$ all the obtained distributions are very similar to the corresponding two-body prediction.  The coincidence between numerical outcomes of the many-body simulations and simple two-body analysis suggests that in the presence of strong  attraction the two-body physics dominates and plays a key role in the process of $\downarrow$--$\uparrow$ pairs formation \cite{SyrwidFermi2018}. 
 Similar agreement can be expected for large $N$, because for $\gamma\lesssim-1$ the spatial breadth of the two-particle bound state is always smaller than $\bar{\delta}$.
 We stress that for $0>\gamma\gtrsim -1$, the method of pair recognition we proposed cannot be used. That is, in such a regime the anticipated $\downarrow$-$\uparrow$ pairs are larger than $\bar{\delta}$,  which makes the idea of minimization of $S_j(\tau)$ physically unjustified.

 It is clear that the width of the two-particle probability distribution is strictly associated with the expectation value of $\widetilde{\Delta}^2$. Therefore, we decided to define the size of $\downarrow$--$\uparrow$ pairs as follows \cite{SyrwidFermi2018}	
 \begin{align}  
\displaystyle{
\xi=2\sqrt{\frac{1}{M}\sum_{j=1}^M \widetilde{\Delta}^2_j} \,\,
 },
 \label{Syr18_Fermi_8}
\end{align}
where $M$ is the number of the  relative spatial separations $\widetilde{\Delta}$, collected in many single-shot measurements. In Fig.~\ref{Sols_Syr18_Fermi_3}(b) we present the resulting $\downarrow$--$\uparrow$ dimer size $\xi$, calculated for the ground state and for the chosen {\it yrast} eigenstate characterized by $P=\hbar\frac{6\pi}{L}$, in a wide range of the attraction strength $\gamma$. Note that while both curves coincide for $\gamma\gtrsim -2.5$, they split when $\gamma\lesssim -2.5$. For such a strong attraction the pairing phenomenon is governed by the two-body physics and in the {\it yrast} eigenstate case the average pair size $\xi$ turns out to be slightly larger than for the ground state.
This can be attributed to an additional kinetic energy associated with the  {\it yrast} excitation \cite{SyrwidFermi2018}.

We stress that the strongly attractive regime is very difficult to be simulated numerically. 
Note that the Yang-Gaudin eigenstate,  Eqs.~(\ref{YG_Wf_New_1})--(\ref{YG_Wf_New_4}), can be expressed as a superposition of exponentials  $\mathrm{e}^{i k_j x_s^\sigma}$, whose correct numerical determination for the complex quasimomenta with $\Im(k_j)\approx\pm i\frac{\bar{c}}{2}$ requires a very high numerical precision when $|\bar{c}|$ is large. Additionally, the Bethe ansatz Eqs.~(\ref{YG_PBC_conds_1})--(\ref{YG_PBC_conds_2}), which have to be solved, are the more unstable, the stronger attraction is. 
 That is why our analysis was carried out with quadruple precision. Nevertheless, it turned out to be insufficient to satisfy the cyclicity conditions of the wave function for $\gamma<-7$. The other problem we have encountered is the increasing number of simulations that have to be performed to reproduce the investigated distributions properly. That is, in the presence of strong attraction the system favourizes bound states, which manifests in extremely high values of the corresponding probability densities for some specific  configurations of particles' positions. Such probability distributions are very narrow, and thus in order to reconstruct them by histograms of particles' positions accurately,  a very large number of steps in the Markovian walk is required.

\subsection{Emergence of dark soliton signatures in many-body yrast eigenstate}
\label{YG_solitons} 

The {\it yrast} spectrum in the weakly attractive  Yang-Gaudin system coincides with the mean-field dark soliton dispersion relation, which makes the many-body {\it yrast} eigenstates promising candidates for the investigations of dark solitons in the Yang-Gaudin model \cite{Shamailov2016}.  Due to the ring system geometry, the eigenstates are invariant under spatial translations, and thus any soliton-like features cannot be displayed in the corresponding reduced single particle  density. We expect that solitonic signatures,  i.e. a density notch and a phase flip,  emerge in the course of particles' positions measurement and can be observed in the wave function describing the last  $\downarrow$-$\uparrow$ pair of fermions.  Hence, we prepare the unpolarized system ($N_\downarrow=N_\uparrow=N$) in the {\it yrast} eigenstate and simulate the process of the measurement of 
$N_\downarrow-1$ positions of the spin-$\downarrow$ fermions and  $N_\uparrow-1$ positions of the spin-$\uparrow$ fermions. As a result $2N-2$ particles' positions $\widetilde{x}^{\downarrow,\uparrow}_{j=1,2,\ldots, N-1}$ are known and fixed. The last two remaining fermions with opposite spins can be described by 
 \begin{align}
\Psi_{2\text{D}}\big(x^\downarrow ,x^\uparrow \big) 
&=\big| \Psi_{2\text{D}}\big(x^\downarrow,x^\uparrow \big)\big|\mathrm{e}^{i \phi_{2\text{D}} (x^\downarrow,x^\uparrow )}
 \label{Syr18_Fermi_9}
\\ \nonumber
&\propto\Psi\big(\big\{  \widetilde{x}^{\downarrow,\uparrow}_{j=1,2,\ldots,N-1} \big\};   x^\downarrow,x^\uparrow  \big),
\end{align}
where $\Psi$ is the corresponding $2N$-particle wave function given by Eqs.~(\ref{YG_Wf_New_1})--(\ref{YG_Wf_New_4}). 
The measurement of $2N-2$ particles' positions we carry out in a two substantially different ways \cite{SyrwidFermi2018}:

\textbf{\emph{I. }} In the first scenario we assume that the particle detection may occur only when two fermions belonging to different spin components sit on top of each other, i.e. $x_s^\downarrow=x_s^\uparrow$, for $s=1,2,\ldots,N-1$. Thus, we measure zero-sized $\downarrow$--$\uparrow$ pairs of fermions only and such a measurement will be dubbed "zero size". Similar process of particle detection takes place in real experiments by means of the rapid ramp technique, in which one measures tightly bound molecules that form thanks to sweep across a Feshbach resonance \cite{Mies2000, VanAbeelen1999, Yurovsky1999, Regal2003, Cubizolles2003, Jochim2003, StreckerPartridge2003, DienerHo2004, Perali2005, AltmanVishwanath2005, Yuzbashyan2005}.

\textbf{\emph{II. }} The second method is based on a direct sampling of the many-body probability distribution, imposing no additional restrictions. In such a case we measure single particles instead of $\downarrow$--$\uparrow$ pairs. In other words, we detect pairs of any size. Hence, such a measurement scenario will be called "any size".
\vspace{0.15cm}

\begin{figure}[t!] 
\begin{center}
\includegraphics[scale=0.205]{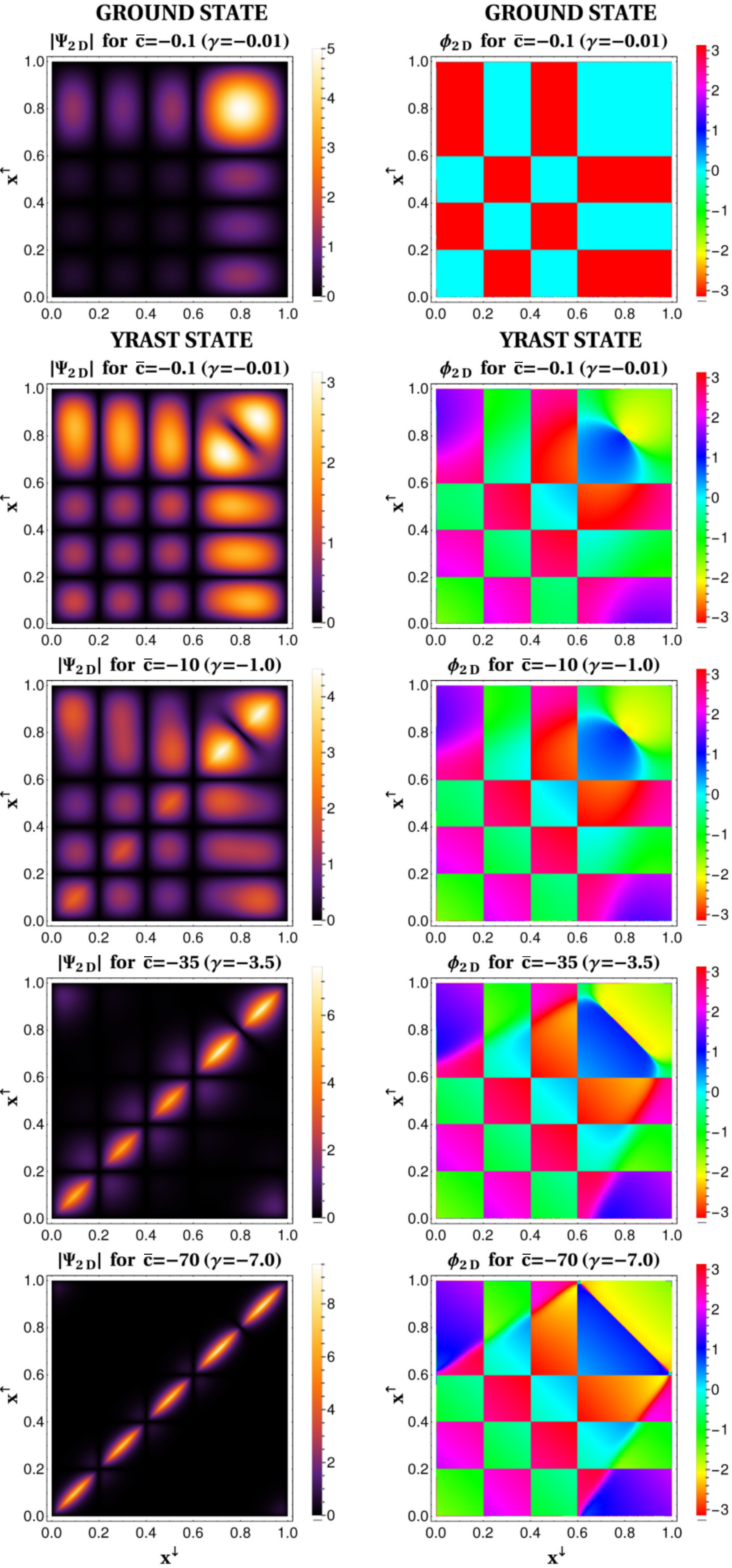}
\end{center}
\vspace{-0.5cm}
\caption{ 
Amplitudes $| \Psi_{2\text{D}}|$ (left column) and phases $\phi_{2\text{D}}$ (right column) of $\Psi_{2\text{D}}$, Eq.~(\ref{Syr18_Fermi_9}), in the 5+5 particle system of size $L=1$. First four zero-sized $\downarrow$-$\uparrow$ pairs were initially measured at positions $\widetilde{x}_j^\downarrow=\widetilde{x}_j^\uparrow=\frac{1}{5}(j-1)$, where $j=1,2,3,4$ (see the main text). In the top panels (first row) we show results obtained for weakly attracting ($\gamma=-0.01$) ground state, where no dark soliton-like structures are visible. Lower panels (rows 2--5) show $\Psi_{2\text{D}}(x^\downarrow,x^\uparrow )$ corresponding to the {\it yrast} eigenstate with $P=6\hbar\pi$ and calculated for different $-0.01 \leq \gamma \leq-7$. The resulting modules $| \Psi_{2\text{D}}|$ and phase distributions $\phi_{2\text{D}}$  reveal very sharp dark soliton signatures located around  $x^\downarrow\approx x^\uparrow \approx0.8$. The nodal structure, visible as a chessboard-like pattern in  $\phi_{2\text{D}}$, appears due to the Pauli exclusion principle, obeyed by the Fermi system (see the main text).   Reprinted and adapted from \cite{SyrwidFermi2018}.
}
\label{Sols_Syr18_Fermi_4}
\end{figure}

At the beginning let us focus on a single realization in the zero-size detection scheme. 
For this purpose we investigate 5+5 ($N=5$)  {\it yrast} eigenstate characterized by  $P=6\hbar\pi \,\,(L=1$), which is expected to reveal the most distinct dark soliton signatures (see Sec.~\ref{YG_pairing}).
 Due to a small number of particles, $N=5$, the resulting last two-particle wave function in a single realization of the detection process may not exhibit clearly visible dark soliton structures. Hence, to get some intuition, we decided to start with optimal positions of the four initially measured pairs of fermions, $\widetilde{x}_j^\downarrow=x_j^\uparrow=\frac{1}{5}(j-1)$ where $j=1,2,3,4$, which should maximize the probability in the ground state case. Note that the Pauli exclusion principle forbids the last remaining $\downarrow$--$\uparrow$ pair of fermions approaching the other, already measured pairs. Consequently, it is expected that such a pair should be most probably located in the largest unoccupied space interval, i.e. between $x^\downarrow=x^\uparrow =0.6$  and 1. This is also the region, where the anticipated dark soliton signatures are supposed to be found. In Fig.~\ref{Sols_Syr18_Fermi_4} we show 
 the modulus $|\Psi_{2\text{D}}|$ (left column) and phase distribution $\phi_{2\text{D}}$ (right column) of  the resulting last two-fermion wave function, Eq.~(\ref{Syr18_Fermi_9}), obtained in a wide range of the attraction strength ($-7\leq \gamma \leq -0.01$). It turns out that for the chosen optimal positions of the initially measured zero-sized $\downarrow$--$\uparrow$ dimers, the wave function  $\Psi_{2\text{D}}(x^\downarrow,x^\uparrow )$ reveals clear dark soliton signatures located around $x^\downarrow \approx x^\uparrow \approx 0.8$. Note that the existence and the position of dark soliton-like structures are independent of $\gamma$ and survive also in the strongly attracting regime. For comparison, we present the results obtained in a similar numerical experiment, but starting with the 5+5 particle ground state, for which no solitonic signatures are visible (see the top panels of Fig.~\ref{Sols_Syr18_Fermi_4}). It is clear that with the increase of the interparticle attraction, the average size of pairs formed by spin-$\downarrow$ and spin-$\uparrow$ fermions decreases. This effect is visible in Fig.~\ref{Sols_Syr18_Fermi_4}, where the diagonal elements of $|\Psi_{2\text{D}}|$ become the more dominant, the stronger attraction is. In addition, both for the ground state and for the chosen {\it yrast} eigenstate the resulting two-dimensional phase distribution $\phi_{2\text{D}}$  exhibits a nodal structure with nodes, visible as phase flips by $\pm \pi$, located along the positions $\widetilde{x}^{\downarrow,\uparrow}_{j=1,2,3,4}$, where four zero-sized $\downarrow$--$\uparrow$ molecules were initially measured. Such a structure resembles a chessboard and appears due to the Pauli exclusion rule completely independent of the interaction strength.  Indeed, a similar behaviour of the wave function can be observed even in the noninteracting case. For example, the system consisting of two identical noninteracting fermions is described by the wave function $\varphi(x,x+\varepsilon)\propto \mathrm{e}^{i \alpha x} \, \mathrm{e}^{i\frac{\alpha}{2}\varepsilon} \, \mathrm{sin} \left(  \beta \epsilon \right)$, where $\alpha\in \mathbb{R}$ and $\beta \in \mathbb{R}_+$, which reveals the $\pi$-phase flip located at $x$ when the relative distance between two fermions $\varepsilon$ passes through zero.

\begin{figure}[h!] 
\begin{center}
\includegraphics[scale=0.355]{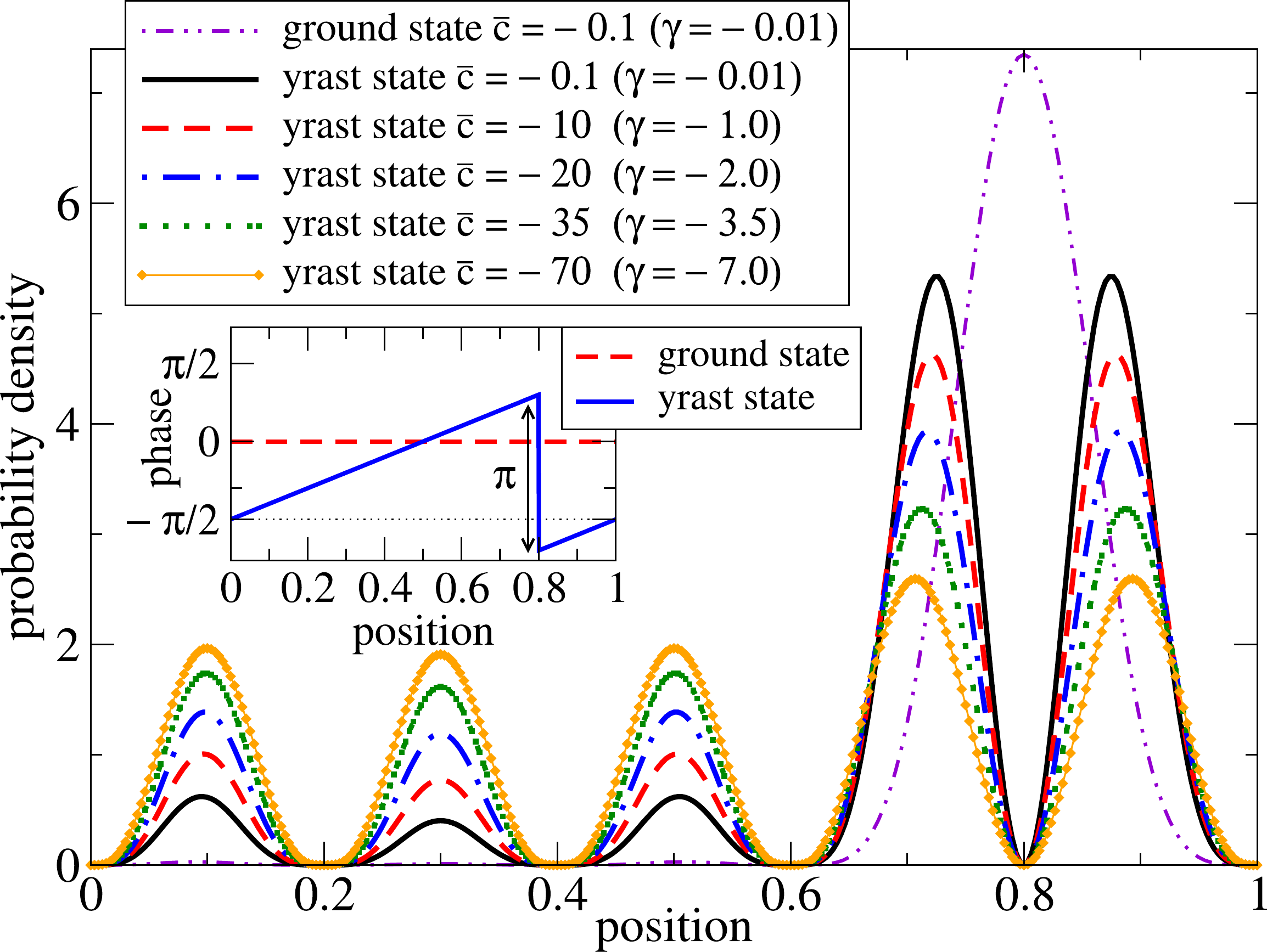}
 \end{center}
\vspace{-0.5cm}
\caption{ Diagonal cuts of  $|\Psi_{2\text{D}}(x^\downarrow,x^\uparrow )|^2$ extracted from the results obtained similarly to those presented in Fig.~\ref{Sols_Syr18_Fermi_4}. The diagonal phase distribution $\phi_{2\text{D}}(x,x)$, presented in the inset, depends only on the initial state and is not affected by changes of $\gamma<0$. While in the ground state case there are no dark soliton signatures, the density notch and $\pi$-phase flip are clearly visible and coincide at $x=0.8$ for the considered  {\it yrast} eigenstate with $P=6\hbar\pi\,\, (L=1)$.
 Reprinted and adapted from \cite{SyrwidFermi2018}. 
}
\label{Sols_Syr18_Fermi_5}
\end{figure}

The density notch and phase flip signatures should be most distinct in the diagonal part of $\Psi_{2\text{D}}(x^\downarrow,x^\uparrow)$. Therefore, we look at the cuts along the diagonals of two-dimensional color maps presented in Fig.~\ref{Sols_Syr18_Fermi_4}. This means that we investigate the wave function for the last $\downarrow$--$\uparrow$ pair of zero size, i.e. $x^\downarrow=x^\uparrow=x$. In Fig.~\ref{Sols_Syr18_Fermi_5} we show the resulting diagonal probability densities $|\Psi_{2\text{D}}(x,x)|^2$ and phase distributions $\phi_{2\text{D}}(x,x)$. The dark soliton-like density notch and accompanying $\pi$-phase flip are always located at $x=0.8$, where the probability density drops to zero. Such a structure has a completely different nature than the zero depth minima located at the positions of four initially measured dimers,  which are expected as long as we deal with the Fermi system. It is hardly visible, but similar local minima are present also for the weakly interacting ($\gamma=-0.01$) 5+5 particle ground state, for which we observe no soliton signatures in $\Psi_{2\text{D}}(x,x)$.

In Fig.~\ref{Sols_Syr18_Fermi_5} one can observe that the two main peaks of $|\Psi_{2\text{D}}|^2$ separated by the dark soliton notch become slightly more distanced when the interparticle attraction increases. Such an effect can be explained by a simple reasoning concerning the {\it yrast} excitation scenarios. Indeed, while for $\gamma\rightarrow 0_-$ the {\it yrast} eigenstate in question corresponds to an excitation of a single fermion, for $\gamma\rightarrow -\infty$ the interparticle 
 attraction is so strong that breaking of $\downarrow$--$\uparrow$ pair is energetically unfavourable and the {\it yrast} eigenstate is a result of an excitation of  the whole $\downarrow$--$\uparrow$ dimer. The total momentum $P$ is unchanged only when the momentum transferred to a single fermion in the former excitation process is double the momentum carried by each fermion constituting  $\downarrow$--$\uparrow$ pair, excited in the second excitation scheme. Thus, we can expect that the stronger attraction, the longer wavelengths in the system are, implying narrower structures visible in the density when we decrease the attraction strength. This may explain why the spatial separation between the two main peaks associated with the soliton notch increases with the intensification of the interparticle attraction \cite{SyrwidFermi2018}.

As for now we have investigated the idealized situation, in which the positions of $N-1$ initially measured $\downarrow$--$\uparrow$ dimers of zero size were assumed to be equidistant. In such a case the resulting wave function $\Psi_{2\text{D}}(x^\downarrow,x^\uparrow)$ reveals very clear dark soliton signatures in a wide range of the attraction strength. This result is in some sense remarkable, but one needs to remember about realistic circumstances, where the positions of $\downarrow$--$\uparrow$ molecules or even single fermions should be chosen randomly from the many-body probability density corresponding to the considered {\it yrast} eigenstate. This kind of a numerical experiment we perform with the help of the Metropolis algorithm. Starting with the same unpolarized 5+5 particle {\it yrast} eigenstate characterized by $P=\hbar\frac{6\pi}{L}$, we carry out plenty of numerical simulations of particle detection process and collect many sets of $N-1=4$ positions of $\downarrow$-$\uparrow$ pairs. Following both the above-mentioned ideas of initial measurements, we perform separate detections of  $\downarrow$--$\uparrow$ pairs of either zero or any size. Due to periodic boundary conditions the system is translationally invariant and the position of the anticipated phase flip signature of a dark soliton, in the diagonal part of $\Psi_{2\text{D}}(x^\downarrow,x^\uparrow)$,  should be completely random. The same is observed in our numerical simulations, where the phase flip position jumps all over the system in consecutive steps of the Markovian walk \cite{SyrwidFermi2018}. 
 Among plenty of realizations we can distinguish two kinds of phase flips corresponding to different winding numbers $J=0$ and $J=1$, where $\phi_{2\text{D}}(x,x)$ has to satisfy the cyclicity condition $\phi_{2\text{D}}(L,L)-\phi_{2\text{D}}(0,0)=2\pi J$ with $J\in \mathbb{Z}$ (see also Sec.~\ref{BoseDynamics}).

 \begin{figure}[h!] 
\begin{center}
\includegraphics[scale=0.315]{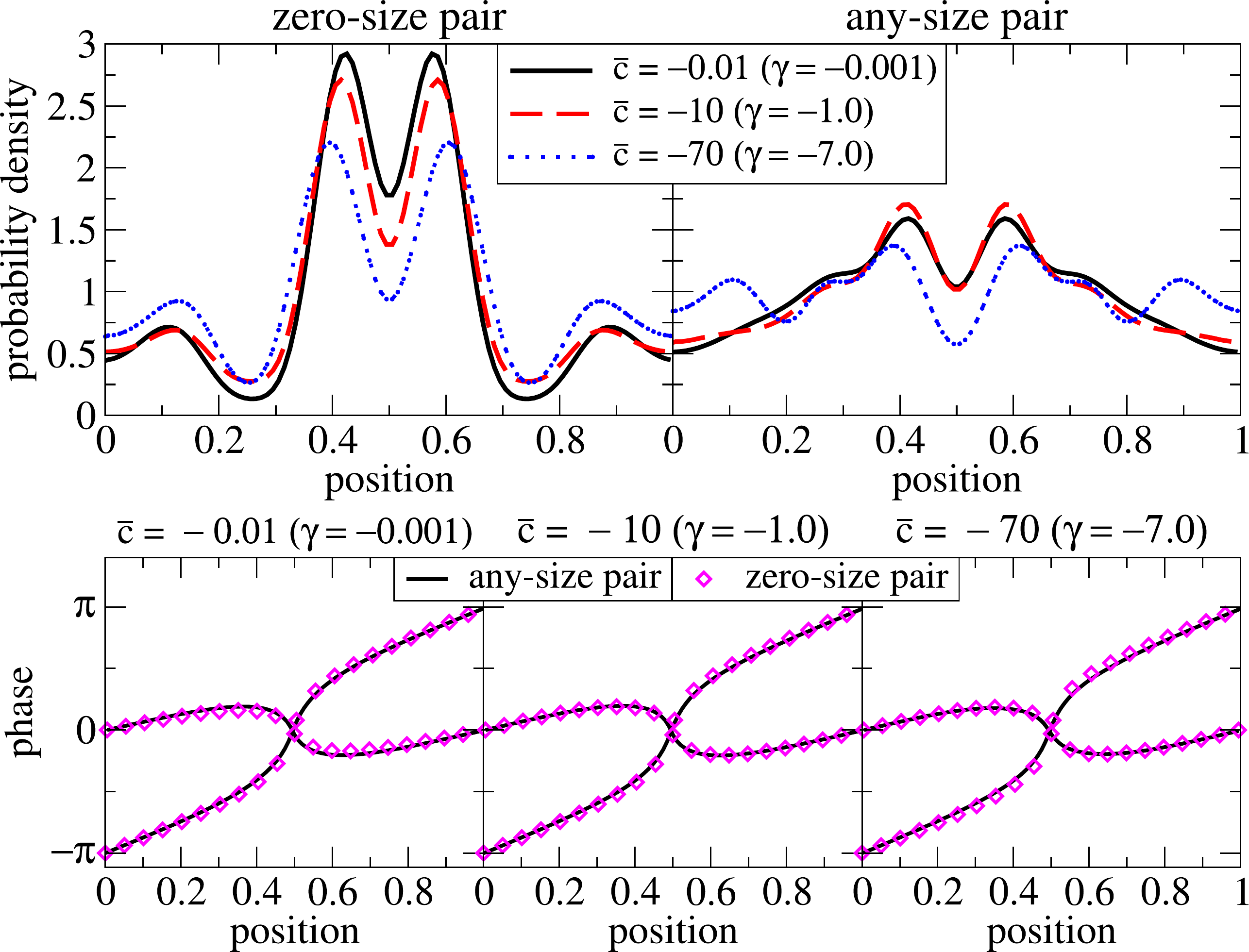}
 \end{center}
\vspace{-0.5cm}
\caption{  Average diagonal probability  densities $|\Psi_{2\text{D}}(x,x)|^2$ (upper panels) and corresponding average phase distributions $\phi_{2\text{D}}(x,x)$ (lower panels) obtained in many realizations of the particle  measurement process (see the main text).
 The system was initially prepared in the {\it yrast} eigenstate with  $P=6\hbar\pi$ ($L=1$). The numerical simulations were performed for weak ($\gamma=-0.001$), moderate ($\gamma=-1$) and strong ($\gamma=-7$)  attraction, as well as assuming two schemes of the $\downarrow$--$\uparrow$ pair detection, in which we measure pairs of zero or any size.
 Reprinted and adapted from \cite{SyrwidFermi2018}. 
}\vspace{-0cm}
\label{Sols_Syr18_Fermi_6}
\end{figure}

All the collected diagonal probability densities and phase distributions have been shifted so that each resulting $\Psi_{2\text{D}}(x,x)$ reveals a phase flip exactly at $x=\frac{L}{2}=0.5$ and then averaged over hundreds of thousands realizations.
The results, presented in Fig.~\ref{Sols_Syr18_Fermi_6}, indicates that the average density profiles $|\Psi_{2\text{D}}(x,x)|^2$ (upper panels) depend on the measurement scheme and are much more oblate for the single particle detection scenario (pairs can have any size) than those obtained in measurements of zero-sized dimers. 
On the other hand, the corresponding average phase distributions (lower panels) with $J=0$ and $J=1$ are almost insensitive to changes of $\gamma<0$ and the detection schemes.
 The resulting dark soliton signatures are not as sharp as in the previously analyzed case of zero-sized  $\downarrow$--$\uparrow$ molecules measured at equidistant positions. That is, in spite of the fact that the average density dips are distinctly visible, they are shallow and far from the zero value of the probability density. Also the corresponding average phase flips are not abrupt, but slightly stretched in space.   Comparing the diagonal average densities obtained for different $\gamma$, we can again observe that the two maxima located around soliton-like notch become more distant when interparticle attraction increases.  We suppose that this is due to the specific character of the {\it yrast} excitation, in which the increase of interparticle attraction leads to a reduction of the momentum per excited fermion \cite{SyrwidFermi2018}.

So far we have analyzed the system containing $N=5$ fermions in each spin component. 
Now, we are going to examine how the increase of $N$ affects dark soliton-like structures visible in diagonal average wave function of the last remaining $\downarrow$--$\uparrow$ pair of fermions in the system.  Such investigations we carry out by employing the idea of numerical diagonalization, where the {\it yrast} eigenstate characterized by the total momentum $P$ corresponds to the lowest energy state in the subspace of chosen $P$ \cite{SyrwidFermi2018}. To monitor the influence of $N$ on the dark soliton signatures, one needs to examine {\it yrast} eigenstates that are generated in a similar way, i.e the states that are expected to share the same general properties.  Namely, we consider the odd number of particles $N$ and the {\it yrast} excitations characterized by  $P=\hbar\frac{\pi(N+1)}{L}$ \cite{SyrwidFermi2018}. Such a choice guarantees that the excitation involves the conjugate pair of quasimomenta with $\Re[k_\pm]=0$. 
Due to a dramatic increase of the computation time with  $N$ and $|\gamma|$, we restrict our considerations to $\gamma=-0.01$, which allowed us to investigate $N\leq 13$. For such a weak attraction the average size of $\downarrow$-$\uparrow$ pairs is larger than $\frac{L}{N}$. Thus, we apply the any size detection scheme only. The successive measurement of particles' positions can be performed by applying field operators $\hat{\Psi}_\sigma(x)$ corresponding to an annihilation of a single fermion of spin $\sigma$ at a position $x$ (for details see Ref.~\cite{SyrwidFermi2018}).

 \begin{figure}[h!] 
\begin{center}
\vspace{-0.cm}
\hspace{-0.0cm}
\includegraphics[scale=0.32]{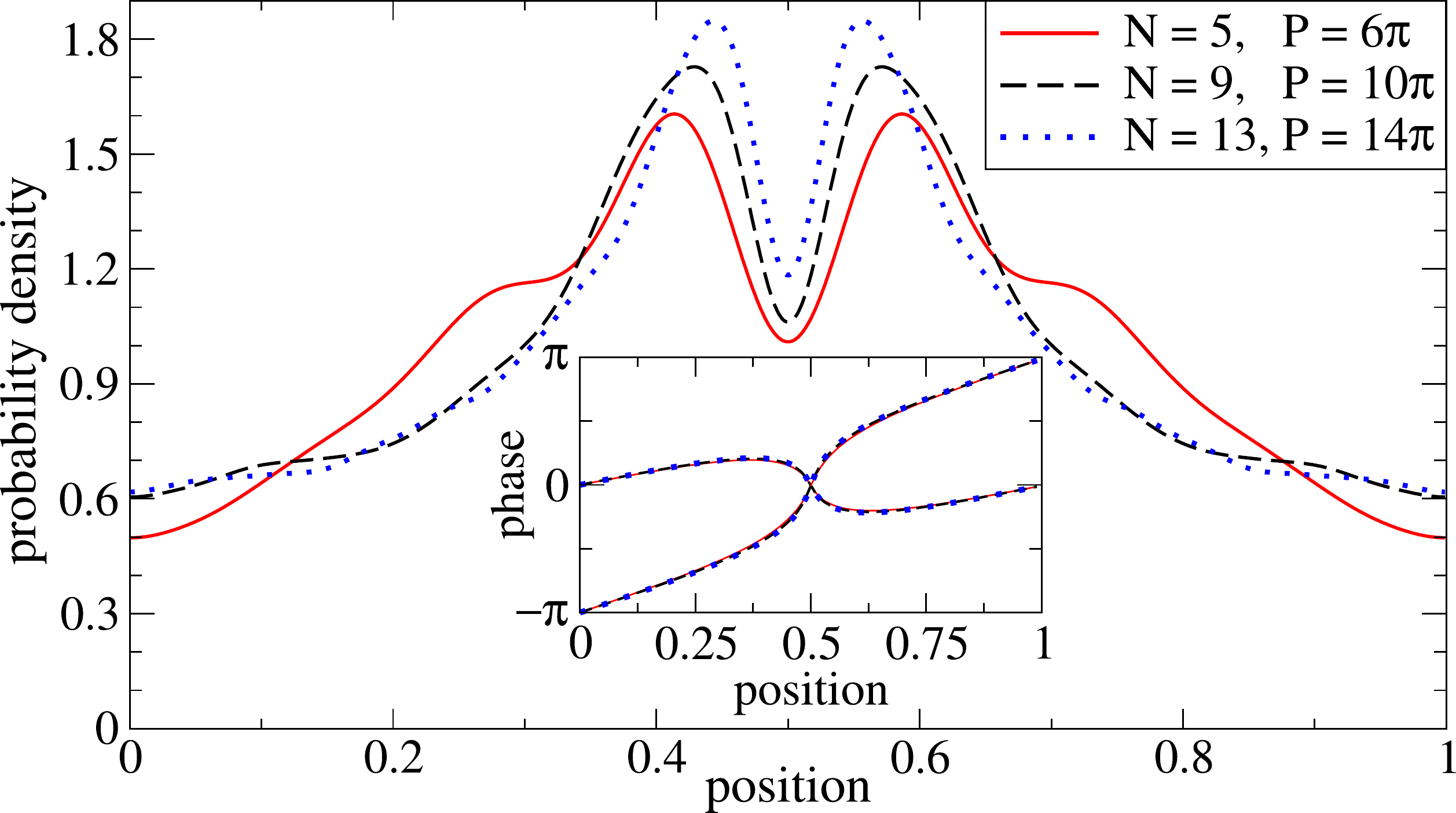}
 \end{center}
\vspace{-0.5cm}
\caption{  Average diagonal densities and corresponding phase distributions of the last  $\downarrow$-$\uparrow$ pair of fermions  for different $N=5,9,13$. The numerical simulations were performed for the weakly interacting ($\gamma=-0.01$) system of size $L=1$, prepared initially in the {\it yrast} eigenstate characterized by $P=\hbar\pi (N+1)$. Note that the average density notch is clearly visible and becomes narrower for larger $N$. The average phase distributions, shown in the inset, reveal two types of the phase flip related to the winding numbers $J=0$ and $J=1$.  
 Reprinted and adapted from \cite{SyrwidFermi2018}. 
}
\label{Sols_Syr18_Fermi_7}
\end{figure}

The resulting average diagonal wave functions are shown in Fig.~\ref{Sols_Syr18_Fermi_7}.
Note that while the average diagonal probability density profile  changes slightly with $N$, the corresponding average phase flips, related to the phase windings $J=0$ and $J=1$, remain almost completely unchanged. 
 Here, we keep the $\gamma$ parameter fixed, and the considered {\it yrast} eigenstates correspond to  $P\propto N$, which seems to be responsible for narrower average density notches visible in Fig.~\ref{Sols_Syr18_Fermi_7} when $N$ is larger \cite{SyrwidFermi2018}.

\subsection{Conclusions}
\label{YG_solitons_concl}

{\it Yrast} states correspond to the lowest energy for a given nonzero total momentum. Thus, among various structures that can appear in physical systems, these associated with the {\it yrast} states seem to be the most stable ones. The same property can also be assigned to solitons. Nevertheless, such a relationship between solitons and many-body {\it yrast} states is only intuitive and requires detailed investigations concerning the nature of the {\it yrast} excitations that were conducted for nearly the last 40 years. 

 Recent studies showed that the {\it yrast} spectrum of the ultracold two-component Fermi gas living in a one-dimensional ring, in the weakly attracting regime closely follows the corresponding mean-field dark soliton dispersion relation. A similar observation in the Lieb-Liniger model triggered the discussion concerning the relationship between {\it yrast} eigenstates and dark solitons. Having such a strong evidence and experience acquired during the analysis of the Lieb-Liniger system, we examined {\it yrast} excitations in the attractively interacting Yang-Gaudin model. Employing methods of the particles' positions detection, similar to those applied to the Bose system, we proved a strict relation between dark solitons and many-body {\it yrast} eigenstates of the investigated Fermi system. Indeed, we showed that the signatures of a dark soliton are clearly visible in the wave function describing the last remaining $\downarrow$--$\uparrow$ pair of fermions, when the system is prepared initially in the many-body {\it yrast} eigenstate.   The dark soliton-like structures are present independently of the interparticle attraction strength and the scheme of the initial measurement of $\downarrow$--$\uparrow$ pairs. 

We have also examined the pairing phenomenon. Our statistical analysis  revealed that in the vicinity of  $\gamma=-1$ one can expect a crossover between the regimes of substantially different physical properties.  For $\gamma\lesssim-1$ we enter the regime in which the average size of $\downarrow$--$\uparrow$ molecules is smaller than their mean spatial separation. It turns out that in such a case the system is dominated by the two-body physics. On the opposite side of the crossover the situation is completely different and the key role is played by the many-body physics. The existence of these two physically different regimes is responsible for a dramatic modification of the {\it yrast} excitation scenario, which is visible as a bifurcation of  quasimomenta for $\gamma\approx-0.9$. The fact that, in the weakly attracting limit the {\it yrast} eigenstate corresponds to a single quasimomentum excitation after a conjugate pair breaking, seems to confirm the idea of phase imprinting, in which the laser radiation is visible only for a single spin component \cite{DelandeSachaProper2014}. 

At the end we also analyzed the influence of the particle number on the dark soliton-like structures. The results show that while the average diagonal density displays a density notch of width decreasing for larger $N$, the corresponding average diagonal phase distribution reveals distinctly visible phase flip that remains almost unchanged for different $N$.  Nevertheless, with the methods we used in our many-body investigations, we are not allowed to explore the BCS regime, which requires much larger number of fermions in each spin component.

\ack 

Foremost, I would like to express my deep gratitude to Krzysztof Sacha, my PhD supervisor, for his very patient guidance, invaluable support  and inspirations.
My sincere appreciation is extended to Dominique Delande, Mariusz Gajda and Mirosław Brewczyk for very fruitful discussions and insightful suggestions during the studies concerning dark solitons in many-body yrast eigenstates.
Special thanks to Jakub Zakrzewski for his everyday commitment and valuable advice. 

Support  of  the  National  Science  Centre,   Poland   via   Project  No.2018/28/T/ST2/00372 and  the  support  of  the  Foundation  for Polish Science (FNP) is acknowledged.

\appendix


\section{Lieb-Liniger model: periodic boundary conditions}
\label{appendixLL_periodic_theorems}

By imposing periodic boundary conditions one assumes that the wave function $\Phi_N$ describing a system of size $L$ satisfies Eqs.~(\ref{LLPeriodicBounds}).
It also means that if we operate in the domain $\mathcal{T}$: $x_{j}<x_{j+1},$ then the wave function satisfies the following cyclicity condition
\begin{eqnarray} 
\Phi_N(x_1,x_2,\ldots,& x_N,\{k\}_N)
\label{LLPeriodicBounds2}   
\\ \nonumber
&=\Phi_N(x_1+L,x_2\ldots, x_N,\{k\}_N)
\\ \nonumber
&=\Phi_N(x_2,x_3,\ldots,x_N,x_1+L,\{k\}_N).
\end{eqnarray} 
Considering the Bethe ansatz solution of the Lieb-Liniger model in Eq.~(\ref{LLBetheWaveFunction}), one gets
\begin{eqnarray}
 \sum_{\pi\in \mathcal{S}_N} \left[ \mathrm{exp}\left( i\sum_{n=1}^N k_{\pi(n)}x_n \right)\prod_{ j>s}\left(1-\frac{i\bar{c}}{k_{\pi(j)}\!-\!k_{\pi(s)}} \right)\right] 
\nonumber 
 \\ \nonumber
 = \sum_{\pi\in \mathcal{S}_N} \left[\mathrm{exp}\left( i\sum_{n=1}^N k_{\pi(n)}x_{n} \right)  \mathrm{e}^{i k_{\pi(1)}L} \right. 
 \\ \nonumber
 \left. \times \prod_{j>s\geq 2}^N\left(1-\frac{i\bar{c}}{k_{\pi(j)}\!-\!k_{\pi(s)}} \right)\prod_{j=2}^N\left(1+\frac{i\bar{c}}{k_{\pi(j)}\!-\!k_{\pi(1)}} \right)\right].
\end{eqnarray}
Hence, it is clear that for each $\pi\in \mathcal{S}_N$ 
\begin{align}
\displaystyle{
  \mathrm{e}^{ik_{\pi(1)}L} = \prod_{s=2}^N\frac{k_{\pi(s)}-k_{\pi(1)}-i\bar{c}}{k_{\pi(s)}-k_{\pi(1)}+i\bar{c}}.
}
\label{LLPeriodicBounds4}
\end{align}
In such a way, starting from $N!$ conditions, one obtains the system of $N$ Bethe Eqs.~(\ref{LLBetheEqsPeriodic}). This is due to the fact that there are $(N-1)!$ permutations $\pi \in \mathcal{S}_N$ satisfying the relation $\pi(1)=r$ for each $r=1,\ldots,N$ and resulting in the same Eq.~(\ref{LLPeriodicBounds4}).

\section{Lieb-Liniger model: open boundary conditions}
\label{appendixLL_open_1}

To find eigenstates of the Lieb-Liniger model with open boundary conditions, we construct a superposition in Eq.~(\ref{LLHWsuperposition}) and impose $\Psi_N(x_1=0,x_2,\ldots,x_N)=0$ getting
\begin{eqnarray}
 \sum_{\epsilon_1,\ldots, \epsilon_N}A(\epsilon_1,\ldots,\epsilon_N)
 &\sum_{\pi\in\mathcal{S}_N}\left[ \mathrm{exp}\left( i \sum_{n=2}^N k_{\pi(n)}x_n \right)
 \right.
 \label{LLHWsuperposition_app}
 \\ \nonumber
& \times\left. \prod_{j>s}\left(1-\frac{i\bar{c}}{k_{\pi(j)}-k_{\pi(s)}}\right) \right] =0,
\end{eqnarray}
where we assumed $x_1<x_2<\ldots<x_N$. Let us now perform the summation in Eq.~(\ref{LLHWsuperposition_app}) only over $\epsilon_{\pi(1)}=\pm 1$.	By doing so, for any $\pi\in \mathcal{S}_N$ and $\{k\}$, one obtains $2^{N-1}$ relations
\begin{eqnarray}
A(\epsilon_1,\ldots,+\epsilon_{\pi(1)},\ldots,\epsilon_N) \!\!\prod_{j \neq \pi(1)}\!\!\left(1-\frac{i\bar{c}}{k_{j}\!-\!k_{\pi(1)}}\right)
\label{LLHWsuperpositionCondition0_app}
\\ \nonumber	 
	 +
	 A(\epsilon_1,\ldots,-\epsilon_{\pi(1)},\ldots,\epsilon_N) \!\!\prod_{j\neq \pi(1)}\!\!\left(1-\frac{i\bar{c}}{k_{j}\!+\!k_{\pi(1)}}\right)
	  =0.
\end{eqnarray}
Noting that for $\{\epsilon \} = \{\epsilon_1, \ldots, \pm\epsilon_{\pi(1)},\ldots, \epsilon_N\}$
\begin{eqnarray}
	\prod_{j>s} \left(1-\frac{i\bar{c}}{k_{j}\!+\!k_{s}}\right)
	=&\left[\prod_{j\neq \pi(1)} \left(1-\frac{i\bar{c}}{k_{j}  \!\pm\!   k_{\pi(1)}}\right) \right] 
	\label{LLHWsuperpositionCondition1_app}
\\	 \nonumber
	 &\times\left[\prod_{\substack{j> s \\ \, j,s\neq \pi(1)}} \left(1-\frac{i\bar{c}}{k_{j}\!+\!k_{s}}\right) \right],
\end{eqnarray}
one easily finds 
\begin{equation}
\displaystyle{
	A(\epsilon_1,\epsilon_2,\ldots,\epsilon_N) = \epsilon_1 \epsilon_2 \dotsb \epsilon_N\prod_{j>s} \left(1-\frac{i\bar{c}}{k_{j}\!+\!k_{s}}\right).
}
\label{LLHWsuperpositionAfunction_app}
\end{equation}
Thus, the solution $\Psi_N$  vanishes at $x_1=0$ and is given by Eq.~(\ref{LLHWPsi}).
Such a solution contains $k_j=\epsilon_j |k_j|$ and is valid when $0\leq x_1 <x_2<\ldots<x_N$.

The requirement that the  wave function vanishes at the other boundary, i.e. $\Psi_N(x_N=L)=0$, leads to the following relation
\begin{eqnarray}
\sum_{\epsilon_{\pi(N)}=\pm 1} \epsilon_{\pi(N)} \mathrm{e}^{i k_{\pi(N)}L}
\label{LLHWPsiVanishingAtL_app}
\\ \nonumber 
	 \qquad\times\prod_{j>s} \! \left( \! 1-\frac{i\bar{c}}{k_{\pi(j)}\!+\!k_{\pi(s)}} \! \right)\!\!\left(\!1-\frac{i\bar{c}}{k_{\pi(j)}\!-\!k_{\pi(s)}} \! \right)=0,
\end{eqnarray}
that has to be valid for all permutations $\pi \in \mathcal{S}_N$ and for an arbitrary set $\{\epsilon\}$. The system of Gaudin's Eqs.~(\ref{GaudinEqs}) can be obtained by noting that 
\begin{eqnarray}
\prod_{j>s}	 \left( \! 1-\frac{i\bar{c}}{k_{\pi(j)}\!+\!k_{\pi(s)}} \right) 
\label{LLHWPsiVanishingAtLHint1_app}
\\ \nonumber
= \prod_{s<N}\!	 \left( \! 1-\frac{i\bar{c}}{\pm k_{\pi(N)}\!+\!k_{\pi(s)}} \right) \!\prod_{N>j>s}\! \left( \! 1-\frac{i\bar{c}}{k_{\pi(j)}\!+\!k_{\pi(s)}} \right),
\\
\prod_{j>s}	 \left( \! 1-\frac{i\bar{c}}{k_{\pi(j)}\!-\!k_{\pi(s)}} \right)
\label{LLHWPsiVanishingAtLHint2_app}
\\ \nonumber
 = \prod_{s<N}	\! \left( \! 1-\frac{i\bar{c}}{\pm k_{\pi(N)}\!-\!k_{\pi(s)}} \right) \!\prod_{N>j>s} \! \left( \! 1-\frac{i\bar{c}}{k_{\pi(j)}\!-\!k_{\pi(s)}} \right).
\end{eqnarray}

At the end of this appendix we prove  the relation (\ref{LLHWWeaklyInteractingLimit3}). For simplicity, we assume that $N,L<\infty$. Knowing that the differences between solutions of the  Gaudin's equations vanishes at most as quickly as $\bar{c}^{1/2}$, we can forget about the factors $\propto \bar{c}/(k_j\pm k_s)$ that are present in the general solution in Eq.~(\ref{LLHWPsi}). Hence,
 \begin{eqnarray}
 \lim_{\bar{c}\rightarrow 0}\Psi_N&(\{k\}_N,\{x\}_N)
 \label{LLHWWeaklyInteractingLimit1_app}  
 \\ \nonumber
& \propto\sum_{\pi\in \mathcal{S}_N, \, \{\epsilon\}}\epsilon_1\epsilon_2\dotsb \epsilon_N \, \mathrm{exp}\left( i\sum_{n=1}^N k_{\pi(n)}x_n \right). 
\end{eqnarray} 
It is clear that the result would not be changed if the indices of $\epsilon$'s are permuted. Indeed, for a given arbitrary permutation $\sigma\in \mathcal{S}_N$ and fixed set $\{\epsilon\}$ the equality $\epsilon_1\epsilon_2\dotsb\epsilon_N=\epsilon_{\sigma(1)}\epsilon_{\sigma(2)}\dotsb\epsilon_{\sigma(N)}$ 
is obviously satisfied. If so, we are allowed to write
\begin{eqnarray}
 \sum_{ \{\epsilon\}}\epsilon_1\epsilon_2\dotsb \epsilon_N \, \mathrm{exp}\left( i\sum_{n=1}^N k_{\pi(n)}x_n \right) 
 \label{LLHWWeaklyInteractingLimit2_app}  
\\ \nonumber
 =\prod_{j=1}^N\sum_{ \epsilon_{\pi(j)}=\pm1}\epsilon_{\pi(j)}\,\mathrm{e}^{i \epsilon_{\pi(j)}|k_{\pi(j)}|x_j}
 \propto \prod_{j=1}^N\sin\left(|k_{\pi(j)}|x_j\right),
\end{eqnarray}
which ends the proof.

\section{Yang-Gaudin model: string hypothesis equations}
\label{appendixYang_Gaudin_String}

The key idea of the derivation of the string hypothesis Eqs.~(\ref{YG_string_4}) is to assume that in the limit $\gamma\rightarrow -\infty$ the paired solutions of the Bethe ansatz Eqs.~(\ref{YG_PBC_conds_1})--(\ref{YG_PBC_conds_2}) approach the string solutions given by Eqs.~(\ref{YG_P_O_string}) (see also Ref.~\cite{Takahashi99}). 
 Let us consider a state corresponding to $M\leq N_\uparrow$ pairs of conjugate quasimomenta and $N-2M$ unpaired ones. Moreover, we take the following numeration of the paired solutions $k_{j=1,\ldots,M}=k_{j,+}, \, k_{j=M+1,\ldots,2M}=k_{j,-}$, then the unpaired real quasimomenta (if they exist) refer to $j=2M+1,\ldots,N$.  In such a case, the Bethe  Eqs.~(\ref{YG_PBC_conds_1}), for the paired quasimomenta ($j=1,2,\ldots,M$) read
\begin{align}
\mathrm{e}^{i k_j L}=\frac{k_j-\Lambda_j+i\frac{\bar{c}}{2}}{k_j-\Lambda_j-i\frac{\bar{c}}{2}}\prod_{\substack{n=1 \\ n\neq j}}^{N_\uparrow}\frac{\Lambda_j-\Lambda_n+i\bar{c}}{\Lambda_j-\Lambda_n} ,
\label{YG_string_1_app00}
 \\ 
\mathrm{e}^{i k_{j+M} L}=\frac{k_{j+M}-\Lambda_j+i\frac{\bar{c}}{2}}{k_{j+M}-\Lambda_j-i\frac{\bar{c}}{2}}\prod_{\substack{n=1 \\ n\neq j}}^{N_\uparrow}\frac{\Lambda_j-\Lambda_n}{\Lambda_j-\Lambda_n-i\bar{c}}.
\label{YG_string_1_app01}
\end{align}
Taking the product of the above equations one obtains
\begin{eqnarray}
\mathrm{e}^{i 2\Lambda_j L}
\label{YG_string_2_app}
\\ \nonumber
\qquad=\frac{k_j\!-\!\Lambda_j\!+\!i\frac{\bar{c}}{2}}{k_j\!-\!\Lambda_j\!-\!i\frac{\bar{c}}{2}}\frac{k_{j+M}\!-\!\Lambda_j\!+\!i\frac{\bar{c}}{2}}{k_{j+M}\!-\!\Lambda_j\!-\!i\frac{\bar{c}}{2}}
\prod_{\substack{n=1 \\ n\neq j}}^{N_\uparrow}\frac{\Lambda_j-\Lambda_n+i\bar{c}}{\Lambda_j-\Lambda_n-i\bar{c}},
\end{eqnarray}
where the prefactor on the right-hand side can be determined from Eqs.~(\ref{YG_PBC_conds_2}). Indeed, one can find that
\begin{eqnarray}
\prod_{s=1}^N\frac{\Lambda_j\!-\!k_s\!+\!i\frac{\bar{c}}{2}}{\Lambda_j\!-\!k_s\!-\!i\frac{\bar{c}}{2}}
&=\frac{k_j\!-\!\Lambda_j\!-\!i\frac{\bar{c}}{2}}{k_j\!-\!\Lambda_j\!+\!i\frac{\bar{c}}{2}}
\frac{k_{j+M}\!-\!\Lambda_j\!-\!i\frac{\bar{c}}{2}}{k_{j+M}\!-\!\Lambda_j\!+\!i\frac{\bar{c}}{2}}
\label{YG_string_01_app}.
\\ \nonumber
&\times\prod_{\substack{s=1 \\ s\neq j}}^M\frac{\Lambda_s\!-\!\Lambda_j\!-\!i\bar{c}}{\Lambda_s\!-\!\Lambda_j\!+\!i\bar{c}}
\prod_{r=2M+1}^N\frac{k_r\!-\!\Lambda_j\!-\!i\frac{\bar{c}}{2}}{k_r\!-\!\Lambda_j\!+\!i\frac{\bar{c}}{2}}.
\end{eqnarray}
In addition,  the right-hand side of Eqs.~(\ref{YG_PBC_conds_2}) reads
\begin{eqnarray}
\prod_{\substack{s=1 \\ s\neq j}}^{N_\uparrow}\!\frac{\Lambda_j\!-\!\Lambda_s\!+\!i\bar{c}}{\Lambda_j\!-\!\Lambda_s\!-\!i\bar{c}}\!=\!\prod_{\substack{s=1 \\ s\neq j}}^{M}\!\frac{\Lambda_s\!-\!\Lambda_j\!-\!i\bar{c}}{\Lambda_s\!-\!\Lambda_j\!+\!i\bar{c}}\!
\prod_{\substack{l=M+1}}^{N_\uparrow}\!\frac{\Lambda_l\!-\!\Lambda_j\!-\!i\bar{c}}{\Lambda_l\!-\!\Lambda_j\!+\!i\bar{c}}.
\label{YG_string_02_app}
\end{eqnarray}
If so, then
\begin{eqnarray}
\frac{k_j-\Lambda_j+i\frac{\bar{c}}{2}}{k_j-\Lambda_j-i\frac{\bar{c}}{2}} \frac{k_{j+M}-\Lambda_j+i\frac{\bar{c}}{2}}{k_{j+M}-\Lambda_j-i\frac{\bar{c}}{2}}
\label{YG_string_3_app}
\\ \nonumber
\qquad\quad=
\prod_{\substack{l=M+1}}^{N_\uparrow}\frac{\Lambda_l-\Lambda_j+i\bar{c}}{\Lambda_l-\Lambda_j-i\bar{c}} \prod_{r=2M+1}^N\frac{k_r-\Lambda_j-i\frac{\bar{c}}{2}}{k_r-\Lambda_j+i\frac{\bar{c}}{2}},
\end{eqnarray}
and thus Eqs.~(\ref{YG_string_2_app}) can be expressed as in Eqs.~(\ref{YG_string_4}).

\section*{References}

\bibliographystyle{iopart-num}

\end{document}